%% file: dissertation_arxiv.tex
\pdfoutput=1
\IfFileExists{XCharter.map}{
    \pdfmapfile{=XCharter.map}
}{}
\IfFileExists{newtx.map}{
    \pdfmapfile{=newtx.map}
}{}
\input{common/documentpreload}     % Настройки файла перед загрузкой documentclass
\input{common/pdfaxsetup}          % Настройки файла для создания PDF/A или PDF/X

\documentclass[a4paper, 14pt, oneside, openany]{memoir} %, draft, verbose

\input{common/setup}               % общие настройки шаблона
\input{common/packages}            % Пакеты общие для диссертации и автореферата

\input{Dissertation/dispackages}   % Пакеты для диссертации
\input{Dissertation/userpackages}  % Пакеты для специфических пользовательских задач
\usepackage[type={CC},
modifier={by-sa},
version={4.0},
lang={english},
]{doclicense}
\input{Dissertation/setup}         % Упрощённые настройки шаблона

\input{common/newnames}            % Новые переменные, которые могут использоваться во всём проекте

\input{common/data}                % Основные сведения
\input{common/styles}              % Стили общие для диссертации и автореферата
\input{common/pdfaxcode}           % Код, отвечающий за атрибуты PDF/A и PDF/X
\input{common/commonuserstyles}    % Пользовательские стили, общие для диссертации и автореферата
\input{Dissertation/disstyles}     % Стили для диссертации
\input{Dissertation/userstyles}    % Стили для специфических пользовательских задач
\input{biblio/biblatex}            % Реализация пакетом biblatex через движок biber
\makeatletter
\DeclareFieldFormat{eprint:arxiv}{%
    arXiv\addcolon\space
    \ifhyperref
    {\href{http://arxiv.org/\abx@arxivpath/#1}{%
            \nolinkurl{#1}%
            \iffieldundef{eprintclass}
            {}
            {\addspace\mkbibbrackets{\thefield{eprintclass}}}}}
    {\nolinkurl{#1}
        \iffieldundef{eprintclass}
        {}
        {\addspace\mkbibbrackets{\thefield{eprintclass}}}}}
\makeatother
\begin{document}
\frontmatter
\input{common/renames}             % Переопределение именований
\renewcommand{\listfigurename}{List of figures}
\renewcommand{\listtablename}{List of tables}

\include{Dissertation/title_en}    % Титульный лист англоязычный
\include{Dissertation/abstract_en}
\clearpage
\ifdefmacro{\microtypesetup}{\microtypesetup{protrusion=false}}{} % не рекомендуется применять пакет микротипографики к автоматически генерируемым спискам
\listoffigures*  % Список изображений

%%% Список таблиц %%%
% (ГОСТ Р 7.0.11-2011, 5.3.10)
\clearpage
\listoftables*   % Список таблиц
\ifdefmacro{\microtypesetup}{\microtypesetup{protrusion=true}}{}
\mainmatter                        % В том числе начинает нумерацию страниц арабскими цифрами с единицы
\addtocounter{TotPages}{-7}
\counterwithout{table}{chapter}    % Убираем связанность номера таблицы с номером главы/раздела
% Структура диссертации (ГОСТ Р 7.0.11-2011, 4)
\include{Dissertation/title}       % Титульный лист
\include{Dissertation/contents}    % Оглавление
\include{Dissertation/introduction}% Введение
\include{Dissertation/part1}       % Глава 1
\include{Dissertation/part2}       % Глава 2
\include{Dissertation/part3}       % Глава 3
\include{Dissertation/part4}       % Глава 4
\include{Dissertation/conclusion}  % Заключение
\include{Dissertation/acronyms}    % Список сокращений и условных обозначений
\include{Dissertation/references}  % Список литературы
\end{document}

%% file: common/packages.tex
\ifnumgreater{\value{pdftype}}{0}{%
    \usepackage[russian]{datetime2} % for date and time fields control % Important for XeTeX to be before hyperxmp and hyperref
    \ifbool{xetex}{%
        \usepackage{atbegshi}
    }{}
}{}

\usepackage{geometry}                               % Для последующего задания полей

\usepackage{amsmath}       % Математические дополнения от AMS
\usepackage{amsfonts,amssymb}       % Математические дополнения от AMS
\usepackage{mathtools}                  % Добавляет окружение multlined

\newlength{\curtextsize}
\newlength{\bigtextsize}

\ifxetexorluatex
\else
    \input glyphtounicode.tex
    \input glyphtounicode-cmr.tex %from pdfx package
\fi

\ifxetexorluatex
    \usepackage{polyglossia}[2014/05/21]            % Поддержка многоязычности (fontspec подгружается автоматически)
\else
    \ifnumequal{\value{usealtfont}}{2}{}{
        \usepackage[noTeX]{mmap}
    }
    \usepackage{cmap}                               % Улучшенный поиск русских слов в полученном pdf-файле
    \ifnumequal{\value{usealtfont}}{2}{}{
        \defaulthyphenchar=127                      % Если стоит до fontenc, то переносы не впишутся в выделяемый текст при копировании его в буфер обмена
    }
    \usepackage{textcomp}
    \usepackage[T1,T2A]{fontenc}                    % Поддержка русских букв
    \usepackage[utf8]{inputenc}[2014/04/30]         % Кодировка utf8
    \usepackage[english, russian]{babel}[2014/03/24]% Языки: русский, английский
    \ifnumequal{\value{usealtfont}}{2}{
        \usepackage[scaled=0.960]{XCharter}[2017/12/19] % Подключение русифицированных шрифтов XCharter
        \usepackage[charter, vvarbb, scaled=1.048]{newtxmath}[2017/12/14]
        \setDisplayskipStretch{-0.078}
    }{}
\fi

\usepackage{indentfirst}                            % Красная строка

\ifboolexpr{test {\ifnumequal{\value{colourmode}}{1}} or test {\ifnumequal{\value{pdftype}}{1}}}{
    \usepackage[dvipsnames, table, hyperref, cmyk]{xcolor}  % cmyk colours --- needed for PDF/X, questionable for PDF/A
}{%
    \ifnumequal{\value{colourmode}}{2}{%
        \usepackage[dvipsnames, table, hyperref, rgb]{xcolor}  % rgb colours
    }{%
        \usepackage[dvipsnames, table, hyperref]{xcolor}
    }
}

\usepackage{multirow,makecell}                      % Улучшенное форматирование таблиц

\usepackage{soulutf8}                               % Поддержка переносоустойчивых подчёркиваний и зачёркиваний
\usepackage{icomma}                                 % Запятая в десятичных дробях

\usepackage{hyperxmp}[2017/02/23] % extended pdf options
\ifnumless{\value{pdftype}}{2}{%
    \usepackage{hyperref}[2012/11/06]
}{%
    \usepackage[pdfa]{hyperref}[2012/11/06] %The default value of the new option 'pdfa' is 'false'. It influences  the loading of the package and cannot be changed after hyperref is  loaded. Hyperxmp also uses it and set pdf/a xmp options
}

\usepackage{graphicx}[2014/04/25]                   % Подключаем пакет работы с графикой

\usepackage{enumitem}

\usepackage[figure,table]{totalcount}               % Счётчик рисунков и таблиц
\usepackage{totcount}                               % Пакет создания счётчиков на основе последнего номера подсчитываемого элемента (может требовать дважды компилировать документ)
\usepackage{totpages}                               % Счётчик страниц, совместимый с hyperref (ссылается на номер последней страницы). Желательно ставить последним пакетом в преамбуле

\ifnumequal{\value{draft}}{1}{% Черновик
    \usepackage[firstpage]{draftwatermark}
    \SetWatermarkText{DRAFT}
    \SetWatermarkFontSize{14pt}
    \SetWatermarkScale{15}
    \SetWatermarkAngle{45}
}{}

\usepackage{siunitx}

\usepackage{tikz}

\usepackage{upgreek} % прямые греческие ради русской традиции

%% file: Dissertation/dispackages.tex
\usepackage{afterpage}
\presentationfalse

%% file: Dissertation/userpackages.tex
\usepackage{tabu, tabulary}  %таблицы с автоматически подбирающейся шириной столбцов

\usepackage{fancyvrb}
\usepackage{listings}
\lccode`\~=0\relax %Без этого хака из-за особенностей пакета listings перестают работать конструкции с \MakeLowercase и т. п. в (xe|lua)latex

\ifnumequal{\value{draft}}{0}{% Только если у нас режим чистовика
    \usepackage[final, babel, shrink=45]{microtype}[2016/05/14] % улучшает представление букв и слов в строках, может помочь при наличии отдельно висящих слов
}{}

\ifnumequal{\value{imgprecompile}}{1}{% Только если у нас включена предкомпиляция
    \usetikzlibrary{external}
    \tikzexternalize[prefix=Dissertation/images_tikz_compiled/] % activate!
    \ifxetex
        \tikzset{external/up to date check={diff}}
    \fi
}{}

%% file: Dissertation/setup.tex
\newcounter{intvl}
\newcounter{otstup}
\newcounter{contnumeq}
\newcounter{contnumfig}
\newcounter{contnumtab}
\newcounter{pgnum}
\newcounter{chapstyle}
\newcounter{headingdelim}
\newcounter{headingalign}
\newcounter{headingsize}
\newcounter{tabcap}
\newcounter{tablaba}
\newcounter{tabtita}
\settocdepth{subsection}            % до какого уровня подразделов выносить в оглавление
\setsecnumdepth{subsection}         % до какого уровня нумеровать подразделы

\newcommand{\tablabelsep}{.} %без пробела потому, что название с новой строки

\newcommand{\figlabelsep}{. }

\definecolor{linkcolor}{rgb}{0.9,0,0}
\definecolor{citecolor}{rgb}{0,0.6,0}
\definecolor{urlcolor}{rgb}{0,0,1}

%% file: common/newnames.tex
\newcommand{\actualityTXT}{Актуальность темы исследования}
\newcommand{\aimTXT}{Целью}
\newcommand{\tasksTXT}{задачи}
\newcommand{\noveltyTXT}{Научная новизна:}
\newcommand{\influenceTXT}{Практическая ценность работы:}
\newcommand{\methodsTXT}{Методы исследований.}
\newcommand{\defpositionsTXT}{Основные положения и~результаты, выносимые на~защиту:}
\newcommand{\reliabilityTXT}{Достоверность полученных результатов}
\newcommand{\probationTXT}{Апробация работы.}

\newcommand{\contributionTXT}{Личный вклад автора.}
\newcommand{\publicationsTXT}{Публикации.}
\newcommand{\realisationTXT}{Реализация результатов.}

\newcommand{\authorbibtitle}{Публикации автора по теме диссертации}
\newcommand{\fullbibtitle}{Список литературы} % (ГОСТ Р 7.0.11-2011, 4)

%% file: common/data.tex
\newcommand{\AuthorURL}{http://orcid.org/0000-0003-2097-9036}
\newcommand{\thesisAuthor}             % Диссертация, ФИО автора
{%
    \texorpdfstring{%
        \hypersetup{urlcolor=black}% гарантия того, что ссылка не раскрашена будет
        \href{\AuthorURL}{Синев Леонид Станиславович}%
        \hypersetup{urlcolor={urlcolor}}%
    }{%
        Леонид Станиславович Синев%
    }%
}

\providecommand{\thesisTitleTXT}{Расчёт и~выбор режимов электростатического соединения кремния со~стеклом по~критерию минимума остаточных напряжений}
\newcommand{\thesisTitle}              % Диссертация, название
{%
    \thesisTitleTXT%
}
\newcommand{\thesisSpecialtyNumber}    % Диссертация, специальность, номер
{05.27.06}
\newcommand{\thesisSpecialtyTitle}     % Диссертация, специальность, название
{Технология и оборудование для~производства полупроводников, материалов и~приборов электронной техники}
\newcommand{\thesisDegree}             % Диссертация, ученая степень
{кандидата технических наук}
\newcommand{\thesisDegreeShort}        % Диссертация, ученая степень, краткая запись
{канд. техн. наук}
\newcommand{\thesisCity}               % Диссертация, город защиты
{Москва}
\newcommand{\thesisYear}               % Диссертация, год защиты
{2016}
\newcommand{\thesisOrganization}       % Диссертация, организация
{Московский государственный технический университет им.~Н.Э.~Баумана}

\newcommand{\thesisInOrganization}     % Диссертация, организация в предложном падеже: Работа выполнена в ...
{Московском государственном техническом
университете им.~Н.Э.~Баумана}

\newcommand{\supervisorFio}            % Научный руководитель, ФИО
{%
    \texorpdfstring{%
        \hypersetup{urlcolor=black}% гарантия того, что ссылка не раскрашена будет
        \href{http://elibrary.ru/author_items.asp?authorid=570330}{Рябов Владимир Тимофеевич}%
        \hypersetup{urlcolor={urlcolor}}%
    }{%
        Рябов Владимир Тимофеевич%
    }%
}
\newcommand{\supervisorRegalia}        % Научный руководитель, регалии
{кандидат технических наук, доцент}
\newcommand{\supervisorFioShort}       % Научный руководитель, ФИО
{В.\,Т.\,Рябов}

\newcommand{\opponentOneFio}           % Оппонент 1, ФИО
{%
    \hypersetup{urlcolor=black}% гарантия того, что ссылка не раскрашена будет
    \href{http://elibrary.ru/author_items.asp?authorid=22067}{Амиров Ильдар Искандерович}%
    \hypersetup{urlcolor={urlcolor}}%
}
\newcommand{\opponentOneRegalia}       % Оппонент 1, регалии
{доктор физико-математических наук}
\newcommand{\opponentOneJobPlace}      % Оппонент 1, место работы
{Ярославского Филиала Федерального государственного бюджетного учреждения науки Физико-технологического института Российской академии наук (ЯФ ФТИАН РАН)}
\newcommand{\opponentOneJobPost}       % Оппонент 1, должность
{заместитель директора по научной работе}

\newcommand{\opponentTwoFio}           % Оппонент 2, ФИО
{%
    \hypersetup{urlcolor=black}% гарантия того, что ссылка не раскрашена будет
    \href{http://elibrary.ru/author_items.asp?authorid=852587}{Жукова Светлана Александровна}%
    \hypersetup{urlcolor={urlcolor}}%
}
\newcommand{\opponentTwoRegalia}       % Оппонент 2, регалии
{кандидат технических наук}
\newcommand{\opponentTwoJobPlace}      % Оппонент 2, место работы
{ФГУП <<Центральный научно-исследовательский институт химии и механики>> (ФГУП <<ЦНИИХМ>>)}
\providecommand{\opponentTwoJobPost}       % Оппонент 2, должность
{заместитель начальника центра~--- начальник научно-технологического комплекса нано- и~микротехнологий НИЦ нанотехнологий}

\newcommand{\leadingOrganizationTitle} % Ведущая организация, дополнительные строки
{Федеральное государственное автономное образовательное учреждение высшего образования «Национальный исследовательский университет «Московский институт электронной техники» (МИЭТ)}

\newcommand{\defenseDate}              % Защита, дата
{<<\blank[\widthof{881}]>>\blank[\widthof{сенсентября}] 201\blank[\widthof{81}]~г.~в~\blank[\widthof{881}]~ч. \blank[\widthof{881}]~мин.}
\newcommand{\defenseCouncilNumber}     % Защита, номер диссертационного совета
{Д\,212.141.18}
\newcommand{\defenseCouncilTitle}      % Защита, учреждение диссертационного совета
{Московском государственном техническом университете им.~Н.Э.~Баумана}
\newcommand{\defenseCouncilAddress}    % Защита, адрес учреждение диссертационного совета
{105005, г.~Москва, ул.~2-я Бауманская д.~5, стр.~1}
\newcommand{\defenseCouncilPhone}      % Телефон для справок
{+7~(499)~267-09-63}

\newcommand{\defenseSecretaryFio}      % Секретарь диссертационного совета, ФИО
{Цветков Ю.\,Б.}
\newcommand{\defenseSecretaryRegalia}  % Секретарь диссертационного совета, регалии
{доктор технических наук, профессор}   % Для сокращений есть ГОСТы, например: ГОСТ Р 7.0.12-2011 + http://base.garant.ru/179724/#block_30000

\newcommand{\synopsisLibrary}          % Автореферат, название библиотеки
{МГТУ им.~Н.Э.~Баумана и~на~сайте
\hypersetup{urlcolor=black}% гарантия того, что ссылка не раскрашена будет
\href{http://www.bmstu.ru/mstu/works/science/degree-candidates/dissertants/?q=dissertation&id=220}{www.bmstu.ru}%
\hypersetup{urlcolor={urlcolor}}%
}
\newcommand{\synopsisDate}             % Автореферат, дата рассылки
{<<\blank[\widthof{888}]>>\blank[\widthof{сенсентября}] 201\blank[\widthof{81}]~года}

\providecommand{\keywords}%            % Ключевые слова для метаданных PDF диссертации и автореферата
{анодная посадка, электростатическое сращивание, электростатическое соединение,
электроадгезионное соединение, термоэлектростимулированное соединение, кремний,
боросиликатное стекло, алюмосиликатное стекло, температурный коэффициент
линейного расширения, ТКЛР, остаточные напряжения, коэффициентные напряжения,
anodic bonding, residual stress, thermal stress, wafer bonding, CTE,
borosilicate glass}

\providecommand{\aimTextContent}%       % цель в именительном падеже с большой буквы
{Научно обоснованный выбор эффективных режимов анодной посадки кремния на~стекло, обеспечивающих минимальный уровень остаточных напряжений}
\providecommand{\aimTextContentRod}%    % цель в родительном (?) падеже с маленькой буквы
{научно обоснованного выбора эффективных режимов анодной посадки кремния на~стекло, обеспечивающих минимальный уровень остаточных напряжений}

%% file: common/styles.tex
\AtBeginDocument{%
    \setlength{\parindent}{2.5em}                   % Абзацный отступ. Должен быть одинаковым по всему тексту и равен пяти знакам (ГОСТ Р 7.0.11-2011, 5.3.7).
}

\ifxetexorluatex
    \setmainlanguage[babelshorthands=true]{russian}  % Язык по-умолчанию русский с поддержкой приятных команд пакета babel
    \setotherlanguage{english}                       % Дополнительный язык = английский (в американской вариации по-умолчанию)
    \newfontfamily\cyrillicfonttt{Courier New}       % моноширинный шрифт для кириллицы
    \defaultfontfeatures{Ligatures=TeX}              % стандартные лигатуры TeX, замены нескольких дефисов на тире и т. п. Настройки моноширинного шрифта должны идти до этой строки, чтобы при врезках кода программ в коде не применялись лигатуры и замены дефисов
    \newfontfamily\cyrillicfont{Times New Roman}     % Шрифт с засечками для кириллицы
    \newfontfamily\cyrillicfontsf{Arial}             % Шрифт без засечек для кириллицы
\else
\fi

\captionwidth{\linewidth}
\normalcaptionwidth

\ifnumequal{\value{tabcap}}{0}{%
    \newcommand{\tabcapalign}{\raggedright}  % по левому краю страницы или аналога parbox
    \renewcommand{\tablabelsep}{~\cyrdash\ } % тире как разделитель идентификатора с номером от наименования
    \newcommand{\tabtitalign}{}
}{%
    \ifnumequal{\value{tablaba}}{0}{%
        \newcommand{\tabcapalign}{\raggedright}  % по левому краю страницы или аналога parbox
    }{}

    \ifnumequal{\value{tablaba}}{1}{%
        \newcommand{\tabcapalign}{\centering}    % по центру страницы или аналога parbox
    }{}

    \ifnumequal{\value{tablaba}}{2}{%
        \newcommand{\tabcapalign}{\raggedleft}   % по правому краю страницы или аналога parbox
    }{}

    \ifnumequal{\value{tabtita}}{0}{%
        \newcommand{\tabtitalign}{\par\raggedright}  % по левому краю страницы или аналога parbox
    }{}

    \ifnumequal{\value{tabtita}}{1}{%
        \newcommand{\tabtitalign}{\par\centering}    % по центру страницы или аналога parbox
    }{}

    \ifnumequal{\value{tabtita}}{2}{%
        \newcommand{\tabtitalign}{\par\raggedleft}   % по правому краю страницы или аналога parbox
    }{}
}

\precaption{\tabcapalign} % всегда идет перед подписью или \legend
\captionnamefont{\normalfont\normalsize} % Шрифт надписи «Таблица #»; также определяет шрифт у \legend
\captiondelim{\tablabelsep} % разделитель идентификатора с номером от наименования
\captionstyle[\tabtitalign]{\tabtitalign}
\captiontitlefont{\normalfont\normalsize} % Шрифт с текстом подписи

\setfloatadjustment{figure}{%
    \setlength{\abovecaptionskip}{-2pt}   % Отбивка над подписью
    \setlength{\belowcaptionskip}{0pt}   % Отбивка под подписью
    \precaption{} % всегда идет перед подписью или \legend
    \captionnamefont{\normalfont\normalsize} % Шрифт надписи «Рисунок #»; также определяет шрифт у \legend
    \captiondelim{\figlabelsep} % разделитель идентификатора с номером от наименования
    \captionstyle[\centering]{\centering} % Центрирование подписей, заданных командой \caption и \legend
    \captiontitlefont{\normalfont\normalsize} % Шрифт с текстом подписи
    \postcaption{} % всегда идет после подписи или \legend, и с новой строки
}

\newsubfloat{figure} % Включает возможность использовать подрисунки у окружений figure
\subcaptionsize{\normalsize} % Шрифт подписи названий подрисунков (не отличается от основного)
\subcaptionlabelfont{\normalfont\itshape}
\subcaptionfont{\normalfont} % Вот так тут добавили скобку после буквы.
\subcaptionstyle{\centering}

\hypersetup{
    unicode,
    linktoc=all,                % both the section and page part are links
    plainpages=false,           % Forces page anchors to be named by the Arabic form  of the page number, rather than the formatted form
    colorlinks,                 % ссылки отображаются раскрашенным текстом, а не раскрашенным прямоугольником, вокруг текста
    linkcolor={linkcolor},      % цвет ссылок типа ref, eqref и подобных
    citecolor={citecolor},      % цвет ссылок-цитат
    urlcolor={urlcolor},        % цвет гиперссылок
    hidelinks,                  % Hide links (removing color and border)
    pdftitle={\thesisTitle},    % Заголовок
    pdfauthor={\thesisAuthor},  % Автор
    pdfsubject={\thesisSpecialtyNumber\ \thesisSpecialtyTitle},      % Тема
    pdfkeywords={\keywords},    % Ключевые слова
    pdflang={ru},
    pdfcopyright={This work is licensed under a Creative Commons        Attribution-ShareAlike 4.0 International license}, %require hyperxmp package
    pdflicenseurl={https://creativecommons.org/licenses/by-sa/4.0/}, %require hyperxmp package
    pdfcontacturl = {\AuthorURL}
}
\ifnumequal{\value{draft}}{1}{% Черновик
    \hypersetup{
        draft,
    }
}{}

\makeatletter
\AddEnumerateCounter{\asbuk}{\russian@alph}{щ}      % Управляем списками/перечислениями через пакет enumitem, а он 'не знает' про asbuk, потому 'учим' его
\makeatother

\setlist{nosep,%                                    % Единый стиль для всех списков (пакет enumitem), без дополнительных интервалов.
    labelindent=\parindent,leftmargin=*%            % Каждый пункт, подпункт и перечисление записывают с абзацного отступа (ГОСТ 2.105-95, 4.1.8)
}

%% file: common/pdfaxcode.tex
\ifnumgreater{\value{pdftype}}{0}{%
    \newcommand{\docModDate}{\DTMnow} %current time, usual situation (today time)
    \newcommand{\docCreationDate}{\docModDate}
    \newcommand{\docCreationDateX}{\docCreationDate}

    \DTMsetstyle{pdf} % sets further dates to be in PDF format

    \hypersetup{
        pdfcreationdate={\docCreationDate}, % hyperref defined, hyperref expects it to be in PDF format
        pdfmoddate={\docModDate}, % hyperref defined, hyperref expects it to be in PDF format
        pdfdate={\docModDate}, % hyperxmp defined, can be in either PDF format or XMP format
        pdfmetadate={\docModDate}, % hyperxmp defined, can be in either PDF format or XMP format
    }

    \pdfstringdef\docTitle{\thesisTitle}
    \pdfstringdef\inputTitle{\docTitle}

    \newcommand\inputxTitle{<< /Title(\inputTitle) >>} % for xelatex

    \newcommand\inputxModDate{<< /ModDate(\docModDate) >>} % for xelatex

    \newcommand\inputxCreationDate{<< /CreationDate(\docCreationDateX) >>} % for xelatex
}{}

\makeatletter
\ifboolexpr{test {\ifnumequal{\value{pdftype}}{2}} and test {\ifnumequal{\value{colourmode}}{1}}}{%
    \patchcmd{\hyper@linkfile}{/C[}{/K[}{}{}
    \patchcmd{\hyper@linkurl}{/C[}{/K[}{}{}
    \patchcmd{\find@pdflink}{/C[}{/K[}{}{}
    \patchcmd{\hyper@linkstart}{/C[}{/K[}{}{}
}{}
\makeatother

\makeatletter
\edef\pwbp{\strip@pt\dimexpr0.996264009963\paperwidth\relax} %paper width in bp (PS points)
\edef\phbp{\strip@pt\dimexpr0.996264009963\paperheight\relax} %paper height in bp (PS points)
\makeatother

\edef\calcpdfpageattr{%
    /TrimBox [0.00000 0.00000 \pwbp\space\phbp]%
}

\ifnumequal{\value{pdftype}}{1}{%
    \hypersetup{pdfstartpage={},% disable openaction of hyperref for PDF/X compliance
    }

    \ifxetex
        \makeatletter
        \AtBeginShipout{% %A hook that is executed for every page
            
        }

        {\DTMsetstyle{pdf}
            
        }

        \ifnumequal{\value{iccinsert}}{1}{
        }{%
        }%
        \makeatother
    \else
        \expandafter\pdfpageattr\expandafter{\calcpdfpageattr} %works

        \pdfinfo{ %for PDF/X %fill manually
          /Title(\inputTitle)
          /GTS_PDFXVersion (PDF/X-1:2001)
          /GTS_PDFXConformance (PDF/X-1a:2001)
        }%

        \ifnumequal{\value{iccinsert}}{1}{
            \immediate\pdfobj stream attr{/N 4^^J/Alternate/DeviceCMYK} file{coated_FOGRA39L_argl.icc}
            \pdfcatalog{%
              /PageMode /UseNone
              /OutputIntents [
                <<
                  /Info (FOGRA39L)
                  /Type /OutputIntent
                  /S /GTS_PDFX
                  /DestOutputProfile \the\pdflastobj\space 0 R
                  /OutputConditionIdentifier (Coated FOGRA39)
                  /RegistryName (http://www.color.org/)
                >>
              ]
            }
        }{%
            \pdfcatalog{ %для PDF/X
              /PageMode /UseNone
              /OutputIntents [
                <<
                  /Info (none)
                  /Type /OutputIntent
                  /S /GTS_PDFX
                  /OutputConditionIdentifier (Custom)
                  /RegistryName (http://www.color.org/)
                >>
              ]
            }%
        }%
    \fi

    \NoHyper%Have to kill all links/annotations for pdf-x compliance
}{}

\ifnumequal{\value{pdftype}}{2}{%

    \ifxetex
        \AtBeginShipout{% %A hook that is executed for every page
            
        }

        {\DTMsetstyle{pdf}
            
        }

        \ifnumequal{\value{colourmode}}{1}{%
        }{%
            \makeatletter
            \makeatother
        }

    \else
        \expandafter\pdfpageattr\expandafter{\calcpdfpageattr} %not really needed for PDF/A

        \ifnumequal{\value{colourmode}}{1}{%
            \immediate\pdfobj stream attr{/N 4} file{coated_FOGRA39L_argl.icc} %loads from pdfx package distribution
            \pdfcatalog{ %for PDF/A
              /PageMode /UseNone
              /OutputIntents [
                <<
                  /Type /OutputIntent
                  /S /GTS_PDFA1
                  /DestOutputProfile \the\pdflastobj\space 0 R
                  /OutputConditionIdentifier (Coated FOGRA39)
                  /Info(FOGRA39 (ISO Coated v2 300\% (ECI)))
                  /RegistryName (http://www.argyllcms.com/)
                >>
              ]
            }%
        }{%
            \immediate\pdfobj stream attr{/N 3^^J/Alternate/DeviceRGB} file{sRGB_IEC61966-2-1_black_scaled.icc}
            \pdfcatalog{ %for PDF/A
              /PageMode /UseNone
              /OutputIntents [
                <<
                  /Type /OutputIntent
                  /S /GTS_PDFA1
                  /DestOutputProfile \the\pdflastobj\space 0 R
                  /OutputConditionIdentifier (sRGB_IEC61966-2-1_black_scale)
                  /Info(sRGB IEC61966 v2.1 with black scaling)
                  /RegistryName (http://www.color.org/)
                >>
              ]
            }%
        }

    \fi

    \hypersetup{%
        pdfapart = {\pdfapart},
        pdfaconformance = {\pdfaconformance},
    }
}{}

\ifbool{xetexorluatex}{}{%
    \ifnumequal{\value{usealtfont}}{2}{}{
        \UndeclareTextCommand{\textnumero}{T2A}
    }
    \UndeclareTextCommand{\S}{T2A}
    \UndeclareTextCommand{\textpertenthousand}{T2A}
    \usepackage{textcomp}%[safe]
}

%% file: common/commonuserstyles.tex
\def\nb-{\nobreak\hskip0pt\hbox{-}\nobreak\hskip0pt}

\def\ellipsiskern{.1em}
\newcommand{\ldotst}{.\kern\ellipsiskern.\kern\ellipsiskern.} %... % \ldotst{}
\newcommand{\ldotse}{!\kern\ellipsiskern.\kern\ellipsiskern.} %!.. % \ldotse{}
\newcommand{\ldotsq}{?\kern\ellipsiskern\kern-.11em.\kern\ellipsiskern.} %?.. % \ldotsq{}

\renewcommand{\epsilon}{\ensuremath{\upvarepsilon}}   %  русская традиция записи
\renewcommand{\phi}{\ensuremath{\upvarphi}}
\renewcommand{\kappa}{\ensuremath{\varkappa}}
\renewcommand{\alpha}{\upalpha}
\renewcommand{\beta}{\upbeta}
\renewcommand{\gamma}{\upgamma}
\renewcommand{\delta}{\updelta}
\renewcommand{\varepsilon}{\upvarepsilon}
\renewcommand{\zeta}{\upzeta}
\renewcommand{\eta}{\upeta}
\renewcommand{\theta}{\uptheta}
\renewcommand{\vartheta}{\upvartheta}
\renewcommand{\iota}{\upiota}
\renewcommand{\kappa}{\upkappa}
\renewcommand{\lambda}{\uplambda}
\renewcommand{\mu}{\upmu}
\renewcommand{\nu}{\upnu}
\renewcommand{\xi}{\upxi}
\renewcommand{\pi}{\uppi}
\renewcommand{\varpi}{\upvarpi}
\renewcommand{\rho}{\uprho}
\renewcommand{\sigma}{\upsigma}
\renewcommand{\tau}{\uptau}
\renewcommand{\upsilon}{\upupsilon}
\renewcommand{\varphi}{\upvarphi}
\renewcommand{\chi}{\upchi}
\renewcommand{\psi}{\uppsi}
\renewcommand{\omega}{\upomega}

\usepackage{letltxmacro} %http://tex.stackexchange.com/a/47372
\LetLtxMacro{\oldint}{\int}

\makeatletter
  \@ifpackagewith{wasysym}{integrals}
  {

  }{%
  \usepackage{scalerel} %http://tex.stackexchange.com/a/222280
    \@ifpackageloaded{newtxmath}{
      \renewcommand{\int}{\mathop{\scalerel*{\rotatebox{15}{$\!\scriptstyle\oldint\!$}}{\oldint}}}
    }{
      \renewcommand{\int}{\mathop{\scalerel*{\rotatebox{12}{$\!\scriptstyle\oldint\!$}}{\oldint}}}
    }
  }
\makeatother

\colorlet{siliconcolour}{black!30} % define color for silicon mass
\colorlet{glasscolour}{white} % define color for glass mass

\newcommand\blank[1][\textwidth]{\noindent\rule[-.2ex]{#1}{.4pt}}

\mathtoolsset{showonlyrefs=true} % Показывать номера только у тех формул, на которые есть \eqref{} в тексте.
\newcommand\mmark{} %Специальное выделение для блоков текста

%% file: Dissertation/disstyles.tex
\graphicspath{{images/}{Dissertation/images/}}         % Пути к изображениям

\OnehalfSpacing*    % Полуторный интервал
\setSpacing{1.42}   % Полуторный интервал, подобный Ворду (возможно, стоит включать вместе с предыдущей строкой)

\vfuzz \hfuzz
\newlength{\otstuplen}
\ifnumequal{\value{headingalign}}{0}{% выравнивание заголовков в тексте
    \newcommand{\hdngalign}{\centering}                % по центру
    \newcommand{\hdngaligni}{}% по центру
    \setlength{\otstuplen}{0pt}
}{%
    \newcommand{\hdngalign}{}                 % по левому краю
    \newcommand{\hdngaligni}{\hspace{\otstuplen}}      % по левому краю
} % В обоих случаях вроде бы без переноса, как и надо (ГОСТ Р 7.0.11-2011, 5.3.5)

\setrmarg{2.55em plus1fil}                             %To have the (sectional) titles in the ToC, etc., typeset ragged right with no hyphenation
\ifnumgreater{\value{headingdelim}}{0}{%
}{}
\ifnumgreater{\value{headingdelim}}{1}{%
    \AtBeginDocument{% без этого polyglossia сама всё переопределяет
        \setsecnumformat{\csname the#1\endcsname.\space}
    }
}{%
    \AtBeginDocument{% без этого polyglossia сама всё переопределяет
        \setsecnumformat{\csname the#1\endcsname\quad}
    }
}

\makeevenhead{plain}{}{\thepage}{}
\makeoddhead{plain}{}{\thepage}{}
\makeevenfoot{plain}{}{}{}
\makeoddfoot{plain}{}{}{}
\newif\ifendTOC

\newcommand*{\tocheader}{
\ifnumequal{\value{pgnum}}{1}{%
    \ifendTOC\else\hbox to \linewidth%
      {\noindent{}~\hfill{Стр.}}\par%
      \ifnumless{\value{page}}{3}{}{%
        \vspace{0.5\onelineskip}
      }
      \afterpage{\tocheader}
    \fi%
}{}%
}%

\newcommand{\basegostsectionfont}{\fontsize{14pt}{16pt}\selectfont\bfseries}

\makechapterstyle{thesisgost}{%
    \chapterstyle{default}
    \setlength{\beforechapskip}{0pt}
    \setlength{\midchapskip}{0pt}
    \setlength{\afterchapskip}{2\onelineskip}
    \renewcommand*{\chapnamefont}{\basegostsectionfont}

    \ifnumgreater{\value{headingdelim}}{0}{%
    }{%
    }
    
    \renewcommand*{\printchaptername}{}
    
}

\makeatletter
\makechapterstyle{thesisgostchapname}{%
    \chapterstyle{thesisgost}
    
    \renewcommand*{\printchaptername}{\hdngaligni\hdngalign\chapnamefont \@chapapp} %
}
\makeatother

\chapterstyle{thesisgost}

\setsecheadstyle{\basegostsectionfont\hdngalign}
\setsecindent{\otstuplen}

\setsubsecheadstyle{\basegostsectionfont\hdngalign}
\setsubsecindent{\otstuplen}

\setsubsubsecheadstyle{\basegostsectionfont\hdngalign}
\setsubsubsecindent{\otstuplen}

\sethangfrom{\noindent #1} %все заголовки подразделов центрируются с учетом номера, как block

\ifnumequal{\value{chapstyle}}{1}{%
    \chapterstyle{thesisgostchapname}
}{}%

\setbeforesecskip{0.8ex}
\setaftersecskip{0.7ex}
\setbeforesubsecskip{0.8ex}
\setaftersubsecskip{0.7ex}
\setbeforesubsubsecskip{0.8ex}
\setaftersubsubsecskip{0.7ex}

\ifnumequal{\value{headingsize}}{1}{% Пропорциональные заголовки и базовый шрифт 14 пт
    \renewcommand{\basegostsectionfont}{\large\bfseries}
    \renewcommand*{\chapnamefont}{\Large\bfseries}

}{}

\ifnumequal{\value{contnumeq}}{1}{%
    \counterwithout{equation}{chapter} % Убираем связанность номера формулы с номером главы/раздела
}{}
\ifnumequal{\value{contnumfig}}{1}{%
    \counterwithout{figure}{chapter}   % Убираем связанность номера рисунка с номером главы/раздела
}{}
\ifnumequal{\value{contnumtab}}{1}{%
    \counterwithout{table}{chapter}    % Убираем связанность номера таблицы с номером главы/раздела
}{}

\makeatletter
\def\formbytotal#1#2#3#4#5{%
    \newcount\@c
    \@c\totvalue{#1}\relax
    \newcount\@last
    \newcount\@pnul
    \@last\@c\relax
    \divide\@last 10
    \@pnul\@last\relax
    \divide\@pnul 10
    \multiply\@pnul-10
    \advance\@pnul\@last
    \multiply\@last-10
    \advance\@last\@c
    \total{#1}~#2%
    \ifnum\@pnul=1#5\else%
    \ifcase\@last#5\or#3\or#4\or#4\or#4\else#5\fi
    \fi
}
\makeatother

\AtBeginDocument{
    \regtotcounter{totalcount@figure}
    \regtotcounter{totalcount@table}       % Если иным способом поставить в преамбуле то ошибка в числе таблиц
    \regtotcounter{TotPages}               % Если иным способом поставить в преамбуле то ошибка в числе страниц
}

\ifxetexorluatex
    \makeatletter
    \def\russian@Alph#1{\ifcase#1\or
       А\or Б\or В\or Г\or Д\or Е\or Ж\or
       И\or К\or Л\or М\or Н\or
       П\or Р\or С\or Т\or У\or Ф\or Х\or
       Ц\or Ш\or Щ\or Э\or Ю\or Я\else\xpg@ill@value{#1}{russian@Alph}\fi}
    \def\russian@alph#1{\ifcase#1\or
       а\or б\or в\or г\or д\or е\or ж\or
       и\or к\or л\or м\or н\or
       п\or р\or с\or т\or у\or ф\or х\or
       ц\or ш\or щ\or э\or ю\or я\else\xpg@ill@value{#1}{russian@alph}\fi}
    \makeatother
\else
    \makeatletter
    \if@uni@ode
      \def\russian@Alph#1{\ifcase#1\or
        А\or Б\or В\or Г\or Д\or Е\or Ж\or
        И\or К\or Л\or М\or Н\or
        П\or Р\or С\or Т\or У\or Ф\or Х\or
        Ц\or Ш\or Щ\or Э\or Ю\or Я\else\@ctrerr\fi}
    \else
      \def\russian@Alph#1{\ifcase#1\or
        \CYRA\or\CYRB\or\CYRV\or\CYRG\or\CYRD\or\CYRE\or\CYRZH\or
        \CYRI\or\CYRK\or\CYRL\or\CYRM\or\CYRN\or
        \CYRP\or\CYRR\or\CYRS\or\CYRT\or\CYRU\or\CYRF\or\CYRH\or
        \CYRC\or\CYRSH\or\CYRSHCH\or\CYREREV\or\CYRYU\or
        \CYRYA\else\@ctrerr\fi}
    \fi
    \if@uni@ode
      \def\russian@alph#1{\ifcase#1\or
        а\or б\or в\or г\or д\or е\or ж\or
        и\or к\or л\or м\or н\or
        п\or р\or с\or т\or у\or ф\or х\or
        ц\or ш\or щ\or э\or ю\or я\else\@ctrerr\fi}
    \else
      \def\russian@alph#1{\ifcase#1\or
        \cyra\or\cyrb\or\cyrv\or\cyrg\or\cyrd\or\cyre\or\cyrzh\or
        \cyri\or\cyrk\or\cyrl\or\cyrm\or\cyrn\or
        \cyrp\or\cyrr\or\cyrs\or\cyrt\or\cyru\or\cyrf\or\cyrh\or
        \cyrc\or\cyrsh\or\cyrshch\or\cyrerev\or\cyryu\or
        \cyrya\else\@ctrerr\fi}
    \fi
    \makeatother
\fi

%% file: Dissertation/userstyles.tex
\def\zz{\ifx\[$\else\aftergroup\zzz\fi}
\def\zzz{\setbox0\lastbox
\dimen0\dimexpr\extrarowheight + \ht0-\dp0\relax
\setbox0\hbox{\raise-.5\dimen0\box0}%
\ht0=\dimexpr\ht0+\extrarowheight\relax
\dp0=\dimexpr\dp0+\extrarowheight\relax
\box0
}

\lstdefinelanguage{Renhanced}%
{keywords={abbreviate,abline,abs,acos,acosh,action,add1,add,%
        aggregate,alias,Alias,alist,all,anova,any,aov,aperm,append,apply,%
        approx,approxfun,apropos,Arg,args,array,arrows,as,asin,asinh,%
        atan,atan2,atanh,attach,attr,attributes,autoload,autoloader,ave,%
        axis,backsolve,barplot,basename,besselI,besselJ,besselK,besselY,%
        beta,binomial,body,box,boxplot,break,browser,bug,builtins,bxp,by,%
        c,C,call,Call,case,cat,category,cbind,ceiling,character,char,%
        charmatch,check,chol,chol2inv,choose,chull,class,close,cm,codes,%
        coef,coefficients,co,col,colnames,colors,colours,commandArgs,%
        comment,complete,complex,conflicts,Conj,contents,contour,%
        contrasts,contr,control,helmert,contrib,convolve,cooks,coords,%
        distance,coplot,cor,cos,cosh,count,fields,cov,covratio,wt,CRAN,%
        create,crossprod,cummax,cummin,cumprod,cumsum,curve,cut,cycle,D,%
        data,dataentry,date,dbeta,dbinom,dcauchy,dchisq,de,debug,%
        debugger,Defunct,default,delay,delete,deltat,demo,de,density,%
        deparse,dependencies,Deprecated,deriv,description,detach,%
        dev2bitmap,dev,cur,deviance,off,prev,,dexp,df,dfbetas,dffits,%
        dgamma,dgeom,dget,dhyper,diag,diff,digamma,dim,dimnames,dir,%
        dirname,dlnorm,dlogis,dnbinom,dnchisq,dnorm,do,dotplot,double,%
        download,dpois,dput,drop,drop1,dsignrank,dt,dummy,dump,dunif,%
        duplicated,dweibull,dwilcox,dyn,edit,eff,effects,eigen,else,%
        emacs,end,environment,env,erase,eval,equal,evalq,example,exists,%
        exit,exp,expand,expression,External,extract,extractAIC,factor,%
        fail,family,fft,file,filled,find,fitted,fivenum,fix,floor,for,%
        For,formals,format,formatC,formula,Fortran,forwardsolve,frame,%
        frequency,ftable,ftable2table,function,gamma,Gamma,gammaCody,%
        gaussian,gc,gcinfo,gctorture,get,getenv,geterrmessage,getOption,%
        getwd,gl,glm,globalenv,gnome,GNOME,graphics,gray,grep,grey,grid,%
        gsub,hasTsp,hat,heat,help,hist,home,hsv,httpclient,I,identify,if,%
        ifelse,Im,image,\%in\%,index,influence,measures,inherits,install,%
        installed,integer,interaction,interactive,Internal,intersect,%
        inverse,invisible,IQR,is,jitter,kappa,kronecker,labels,lapply,%
        layout,lbeta,lchoose,lcm,legend,length,levels,lgamma,library,%
        licence,license,lines,list,lm,load,local,locator,log,log10,log1p,%
        log2,logical,loglin,lower,lowess,ls,lsfit,lsf,ls,machine,Machine,%
        mad,mahalanobis,make,link,margin,match,Math,matlines,mat,matplot,%
        matpoints,matrix,max,mean,median,memory,menu,merge,methods,min,%
        missing,Mod,mode,model,response,mosaicplot,mtext,mvfft,na,nan,%
        names,omit,nargs,nchar,ncol,NCOL,new,next,NextMethod,nextn,%
        nlevels,nlm,noquote,NotYetImplemented,NotYetUsed,nrow,NROW,null,%
        numeric,\%o\%,objects,offset,old,on,Ops,optim,optimise,optimize,%
        options,or,order,ordered,outer,package,packages,page,pairlist,%
        pairs,palette,panel,par,parent,parse,paste,path,pbeta,pbinom,%
        pcauchy,pchisq,pentagamma,persp,pexp,pf,pgamma,pgeom,phyper,pico,%
        pictex,piechart,Platform,plnorm,plogis,plot,pmatch,pmax,pmin,%
        pnbinom,pnchisq,pnorm,points,poisson,poly,polygon,polyroot,pos,%
        postscript,power,ppoints,ppois,predict,preplot,pretty,Primitive,%
        print,prmatrix,proc,prod,profile,proj,prompt,prop,provide,%
        psignrank,ps,pt,ptukey,punif,pweibull,pwilcox,q,qbeta,qbinom,%
        qcauchy,qchisq,qexp,qf,qgamma,qgeom,qhyper,qlnorm,qlogis,qnbinom,%
        qnchisq,qnorm,qpois,qqline,qqnorm,qqplot,qr,Q,qty,qy,qsignrank,%
        qt,qtukey,quantile,quasi,quit,qunif,quote,qweibull,qwilcox,%
        rainbow,range,rank,rbeta,rbind,rbinom,rcauchy,rchisq,Re,read,csv,%
        csv2,fwf,readline,socket,real,Recall,rect,reformulate,regexpr,%
        relevel,remove,rep,repeat,replace,replications,report,require,%
        resid,residuals,restart,return,rev,rexp,rf,rgamma,rgb,rgeom,R,%
        rhyper,rle,rlnorm,rlogis,rm,rnbinom,RNGkind,rnorm,round,row,%
        rownames,rowsum,rpois,rsignrank,rstandard,rstudent,rt,rug,runif,%
        rweibull,rwilcox,sample,sapply,save,scale,scan,scan,screen,sd,se,%
        search,searchpaths,segments,seq,sequence,setdiff,setequal,set,%
        setwd,show,sign,signif,sin,single,sinh,sink,solve,sort,source,%
        spline,splinefun,split,sqrt,stars,start,stat,stem,step,stop,%
        storage,strstrheight,stripplot,strsplit,structure,strwidth,sub,%
        subset,substitute,substr,substring,sum,summary,sunflowerplot,svd,%
        sweep,switch,symbol,symbols,symnum,sys,status,system,t,table,%
        tabulate,tan,tanh,tapply,tempfile,terms,terrain,tetragamma,text,%
        time,title,topo,trace,traceback,transform,tri,trigamma,trunc,try,%
        ts,tsp,typeof,unclass,undebug,undoc,union,unique,uniroot,unix,%
        unlink,unlist,unname,untrace,update,upper,url,UseMethod,var,%
        variable,vector,Version,vi,warning,warnings,weighted,weights,%
        which,while,window,write,\%x\%,x11,X11,xedit,xemacs,xinch,xor,%
        xpdrows,xy,xyinch,yinch,zapsmall,zip},%
    otherkeywords={!,!=,~,$,*,\%,\&,\%/\%,\%*\%,\%\%,<-,<<-},%$
    alsoother={._$},%$
    sensitive,%
    morecomment=[l]\#,%
    morestring=[d]",%
    morestring=[d]'% 2001 Robert Denham
}%

\newlength{\twless}
\newlength{\lmarg}


\DefineVerbatimEnvironment% с шрифтом 12 пт
{Verb}{Verbatim}
{fontsize=\fontsize{12pt}{14pt}\selectfont}

\newfloat[chapter]{ListingEnv}{lol}{Листинг}

\makeatletter
\AtBeginDocument{%
    \let\c@ListingEnv\c@lstlisting
    
    \let\ftype@lstlisting\ftype@ListingEnv % give the floats the same precedence
}
\makeatother

%% file: biblio/biblatex.tex
\usepackage{csquotes} % biblatex рекомендует его подключать. Пакет для оформления сложных блоков цитирования.

\makeatletter
\ifnumequal{\value{draft}}{0}{% Чистовик
\usepackage[%
backend=biber,% движок
bibencoding=utf8,% кодировка bib файла
sorting=none,% настройка сортировки списка литературы
style=gost-numeric,% стиль цитирования и библиографии (по ГОСТ)
language=autobib,% получение языка из babel/polyglossia, default: autobib % если ставить autocite или auto, то цитаты в тексте с указанием страницы, получат указание страницы на языке оригинала
autolang=other,% многоязычная библиография
clearlang=true,% внутренний сброс поля language, если он совпадает с языком из babel/polyglossia
defernumbers=true,% нумерация проставляется после двух компиляций, зато позволяет выцеплять библиографию по ключевым словам и нумеровать не из большего списка
sortcites=true,% сортировать номера затекстовых ссылок при цитировании (если в квадратных скобках несколько ссылок, то отображаться будут отсортированно, а не абы как)
doi=true,% Показывать или нет ссылки на DOI
isbn=false,% Показывать или нет ISBN, ISSN, ISRN
]{biblatex}[2016/09/17]
\ltx@iffilelater{biblatex-gost.def}{2017/05/03}%
{\toggletrue{bbx:gostbibliography}%
}{}
}{%Черновик
\usepackage[%
backend=biber,% движок
bibencoding=utf8,% кодировка bib файла
sorting=none,% настройка сортировки списка литературы
style=gost-numeric,% стиль цитирования и библиографии (по ГОСТ)
]{biblatex}[2016/09/17]%
}
\makeatother

\DeclareSourcemap{ %модификация bib файла перед тем, как им займётся biblatex 
    \maps{
        \map{% перекидываем значения полей language в поля langid, которыми пользуется biblatex
            \step[fieldsource=language, fieldset=langid, origfieldval, final]
            \step[fieldset=language, null]
        }
        \map[overwrite]{% перекидываем значения полей shortjournal, если они есть, в поля journal, которыми пользуется biblatex
            \step[fieldsource=shortjournal, final]
            \step[fieldset=journal, origfieldval]
        }
        \map[overwrite]{% перекидываем значения полей shortbooktitle, если они есть, в поля booktitle, которыми пользуется biblatex
            \step[fieldsource=shortbooktitle, final]
            \step[fieldset=booktitle, origfieldval]
        }
        \map[overwrite, refsection=0]{% стираем значения всех полей addendum
            \perdatasource{biblio/authorpapersVAK.bib}
            \perdatasource{biblio/authorpapers.bib}
            \perdatasource{biblio/authorconferences.bib}
            \step[fieldsource=addendum, final]
            \step[fieldset=addendum, null] %чтобы избавиться от информации об объёме авторских статей, в отличие от автореферата
        }
        \map[overwrite]{
            \perdatasource{biblio/authorpapersVAK.bib}
            \perdatasource{biblio/authorpapers.bib}
            \step[fieldsource=urldate, final]
            \step[fieldset=urldate, null]
        }
        \map[overwrite]{ % добавляем ключевые слова, чтобы различать источники
            \perdatasource{biblio/othercites.bib}
            \step[fieldset=keywords, fieldvalue={biblioother,bibliofull}]
        }
        \map[overwrite]{ % добавляем ключевые слова, чтобы различать источники
            \perdatasource{biblio/authorpapersVAK.bib}
            \step[fieldset=keywords, fieldvalue={biblioauthorvak,biblioauthor,bibliofull}]
        }
        \map[overwrite]{ % добавляем ключевые слова, чтобы различать источники
            \perdatasource{biblio/authorpapers.bib}
            \step[fieldset=keywords, fieldvalue={biblioauthornotvak,biblioauthor,bibliofull}]
        }
        \map[overwrite]{ % добавляем ключевые слова, чтобы различать источники
            \perdatasource{biblio/authorconferences.bib}
            \step[fieldset=keywords, fieldvalue={biblioauthorconf,biblioauthor,bibliofull}]
        }
        \map[overwrite]{% стираем значения всех полей media=eresource
            \step[fieldsource=media,
            match={eresource},
            final]
            \step[fieldset=media, null]
        }
    }
}

\newbibmacro{string+doi}[1]{% новая макрокоманда на простановку ссылки на doi
    \iffieldundef{doi}{#1}{\href{http://dx.doi.org/\thefield{doi}}{#1}}}

\ifnumequal{\value{draft}}{0}{% Чистовик
}{}
\DeclareFieldFormat{title}{\usebibmacro{string+doi}{#1}} % ссылка на doi с названия работы
\DeclareFieldFormat{journaltitle}{\usebibmacro{string+doi}{#1}} % ссылка на doi с названия журнала
\makeatletter
\@ifpackagelater{biblatex}{2016/05/11}% biblatex 3.5+
  {%
    \DefineBibliographyExtras{russian}{%
    }%
  }
  {%
    \DefineBibliographyExtras{russian}{%
    }%
  }
\makeatother

\makeatletter
\@ifpackagelater{biblatex}{2016/05/11}% biblatex 3.5+
  {%
    \DefineBibliographyExtras{english}{%
      \protected\def\accdate{\mainlang дата\addabbrvspace обр\adddot}%
      \restorecommand\mkdaterangecomp% several lines to revert internal US reverse date format modification
      \restorecommand\mkdaterangecompextra
      \restorecommand\mkdaterangeterse
      \restorecommand\mkdaterangeterseextra
    }%
  }
  {%
    \DefineBibliographyExtras{english}{%
      \protected\def\accdate{\mainlang дата\addabbrvspace обр\adddot}%
      \restorecommand\mkbibrangecomp% several lines to revert internal US reverse date format modification
      \restorecommand\mkbibrangecompextra
      \restorecommand\mkbibrangeterse
      \restorecommand\mkbibrangeterseextra
    }%
  }
\makeatother

\DeclareNumChars*{ ()}

\NewBibliographyString{yearsign}
\DefineBibliographyStrings{english}{%
yearsign = {},
urlseen = {\accdate},
}
\DefineBibliographyStrings{russian}{%
yearsign = {\addnbspace г\adddot},
}

\renewbibmacro*{booktitle}{%
  \ifboolexpr{
    test {\iffieldundef{booktitle}}
    and
    test {\iffieldundef{booksubtitle}}
  }
    {}
    {\printtext[booktitle]{%
       \printfield[titlecase]{booktitle}%
       \setunit{\subtitlepunct}%
       \printfield[titlecase]{booksubtitle}}%
     \newunit}%
  \ifboolexpr{
    test {\iffieldundef{eventtitle}}
    and
    not test {\iffieldundef{eventyear}}
  }
    {\setunit{\addspace}%
     \printtext[parens]{%
       \printfield{venue}%
       \setunit*{\addcomma\space}%
       \printeventdate%
       \bibstring{yearsign}%
     }%
    }
    {}
  \setunit{\addcolondelim}%
  \printfield{booktitleaddon}}

\renewbibmacro*{event+venue+date}{%
    \ifboolexpr{
      test {\iffieldundef{eventtitle}}
      or
      (
        test {\iffieldundef{venue}}
        and
        test {\iffieldundef{eventyear}}
      )
    }
      {}
      {\setunit{\addspace}%
       \printtext[parens]{%
         \printfield{eventtitle}%
         \setunit{\addcolondelim}%
         \printfield{eventtitleaddon}%
         \setunit{\addcomma\space}%
         \printfield{venue}%
         \setunit*{\addcomma\space}%
         \printeventdate%
         \iffieldundef{eventyear}{}{\bibstring{yearsign}}%
       }}%
  \newunit}

\makeatletter
\ltx@iffilelater{biblatex-gost.def}{2017/02/05}% biblatex-gost 1.12+
  {%
    
  }{%
  }
\makeatother

\newtotcounter{citenum}
\makeatletter
\defbibenvironment{counter} %Env of bibliography
  {\setcounter{citenum}{0}%
  \renewcommand{\blx@driver}[1]{}%
  } %what is doing at the beginining of bibliography. In your case it's : a. Reset counter b. Say to print nothing when a entry is tested.
  {} %Здесь то, что будет выводиться командой \printbibliography. \thecitenum сюда писать не надо
  {\stepcounter{citenum}} %What is printing / executed at each entry.
\makeatother
\defbibheading{counter}{}

\newtotcounter{citeauthorvak}
\makeatletter
\defbibenvironment{countauthorvak} %Env of bibliography
{\setcounter{citeauthorvak}{0}%
    \renewcommand{\blx@driver}[1]{}%
} %what is doing at the beginining of bibliography. In your case it's : a. Reset counter b. Say to print nothing when a entry is tested.
{} %Здесь то, что будет выводиться командой \printbibliography. Обойдёмся без \theciteauthorvak в нашей реализации
{\stepcounter{citeauthorvak}} %What is printing / executed at each entry.
\makeatother
\defbibheading{countauthorvak}{}

\newtotcounter{citeauthornotvak}
\makeatletter
\defbibenvironment{countauthornotvak} %Env of bibliography
{\setcounter{citeauthornotvak}{0}%
    \renewcommand{\blx@driver}[1]{}%
} %what is doing at the beginining of bibliography. In your case it's : a. Reset counter b. Say to print nothing when a entry is tested.
{} %Здесь то, что будет выводиться командой \printbibliography. Обойдёмся без \theciteauthornotvak в нашей реализации
{\stepcounter{citeauthornotvak}} %What is printing / executed at each entry.
\makeatother
\defbibheading{countauthornotvak}{}

\newtotcounter{citeauthorconf}
\makeatletter
\defbibenvironment{countauthorconf} %Env of bibliography
{\setcounter{citeauthorconf}{0}%
    \renewcommand{\blx@driver}[1]{}%
} %what is doing at the beginining of bibliography. In your case it's : a. Reset counter b. Say to print nothing when a entry is tested.
{} %Здесь то, что будет выводиться командой \printbibliography. Обойдёмся без \theciteauthorconf в нашей реализации
{\stepcounter{citeauthorconf}} %What is printing / executed at each entry.
\makeatother
\defbibheading{countauthorconf}{}

\newtotcounter{citeauthor}
\makeatletter
\defbibenvironment{countauthor} %Env of bibliography
{\setcounter{citeauthor}{0}%
    \renewcommand{\blx@driver}[1]{}%
} %what is doing at the beginining of bibliography. In your case it's : a. Reset counter b. Say to print nothing when a entry is tested.
{} %Здесь то, что будет выводиться командой \printbibliography. Обойдёмся без \theciteauthor в нашей реализации
{\stepcounter{citeauthor}} %What is printing / executed at each entry.
\makeatother
\defbibheading{countauthor}{}

\defbibheading{authorpublications}[\authorbibtitle]{\section*{#1}}
\defbibheading{pubsubgroup}{\noindent\textbf{#1}}
\defbibheading{otherpublications}{\section*{#1}}

\newcommand*{\insertbibliofull}{%
\printbibliography[keyword=bibliofull,section=0]
\printbibliography[heading=counter,env=counter,keyword=bibliofull,section=0]%
}

%% file: common/renames.tex
%%% Переопределение именований %%%
\renewcommand{\alsoname}{см. также}
\renewcommand{\seename}{см.}
\renewcommand{\headtoname}{вх.}
\renewcommand{\ccname}{исх.}
\renewcommand{\enclname}{вкл.}
\renewcommand{\pagename}{Стр.}
\renewcommand{\partname}{Часть}
\renewcommand{\abstractname}{Аннотация}
\renewcommand{\contentsname}{Оглавление} % (ГОСТ Р 7.0.11-2011, 4)
\renewcommand{\figurename}{Рисунок} % (ГОСТ Р 7.0.11-2011, 5.3.9)
\renewcommand{\tablename}{Таблица} % (ГОСТ Р 7.0.11-2011, 5.3.10)
\renewcommand{\indexname}{Предметный указатель}
\renewcommand{\listfigurename}{Список рисунков}
\renewcommand{\listtablename}{Список таблиц}
\renewcommand{\refname}{\fullbibtitle}
\renewcommand{\bibname}{\fullbibtitle}

%% file: Dissertation/title_en.tex
\thispagestyle{empty}%
\begin{center}%
    \hypersetup{urlcolor=black}% гарантия того, что ссылка не раскрашена будет
    \href{http://www.bmstu.ru/en/general-information/about-bmstu/}{Bauman Moscow State Technical University}%
    \hypersetup{urlcolor={urlcolor}}%
\end{center}%
\vspace{0pt plus4fill}
\IfFileExists{images_arxiv/bmstu_logo.pdf}{
  \begin{minipage}[b]{0.495\textwidth}
    \raggedright
      \includegraphics[height=3.5cm]{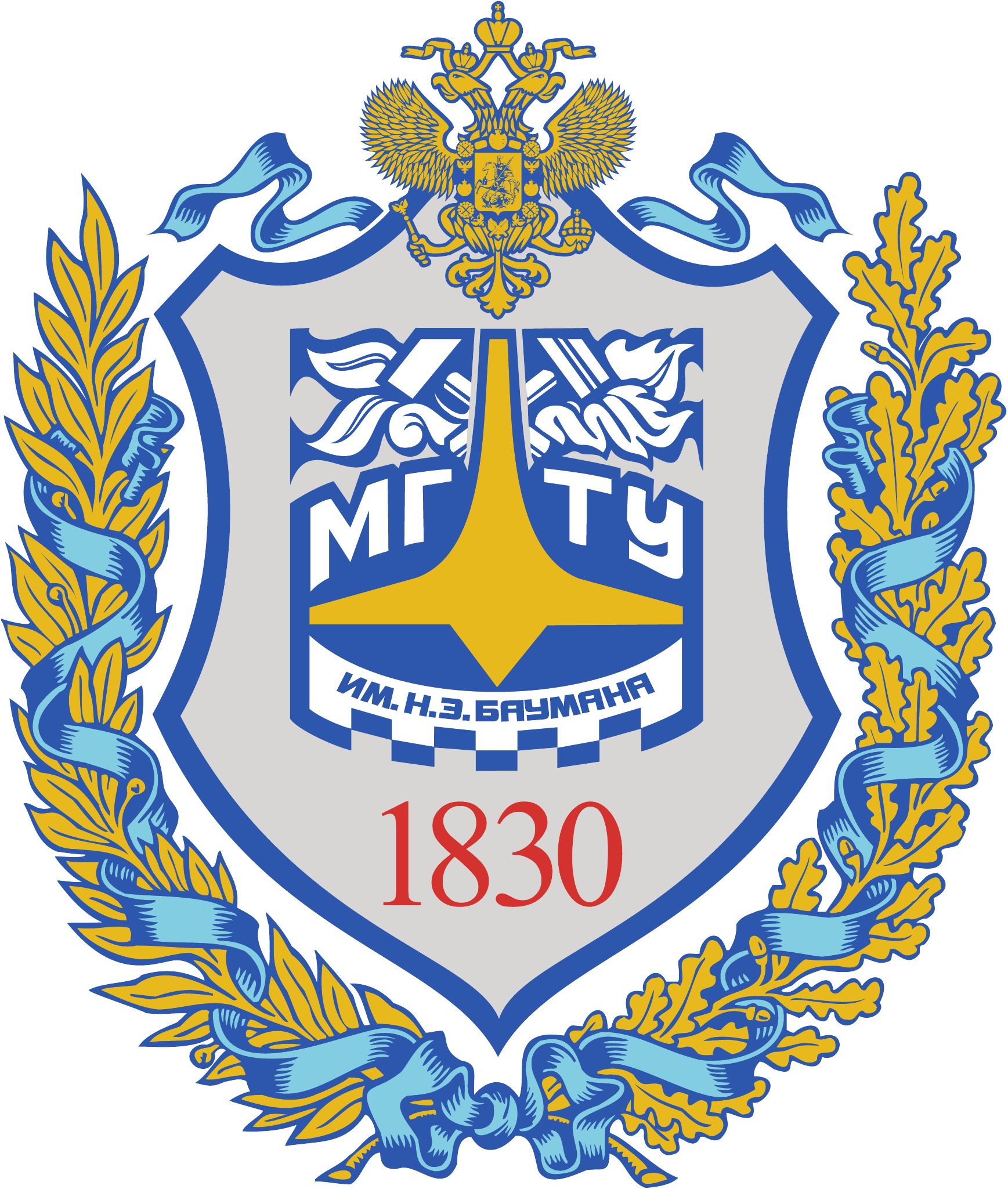}\par
  \end{minipage}
  \begin{minipage}[b]{0.495\textwidth}
    \raggedleft
      Manuscript\par
  \end{minipage}
}{
\raggedleft
Manuscript\par
}
\vspace{0pt plus6fill}
\begin{center}%
{\large
    \hypersetup{urlcolor=black}%
    \href{\AuthorURL}{Leonid Stanislavovich Sinev}%
    \hypersetup{urlcolor={urlcolor}}%
}
\end{center}%
\vspace{0pt plus1fill}
\begin{center}%
\textbf {\large
Calculation and selection of silicon to glass anodic bonding modes based on the criterion of minimum residual stress}

\vspace{0pt plus2fill}
{Speciality \thesisSpecialtyNumber\ "---

<<Technology and Equipment for Production of Semiconductors, Materials
and~Electronic Devices>>%
}

\vspace{0pt plus2fill}
Dissertation in support

of candidate of engineering sciences degree

(in russian)
\end{center}%
\vspace{0pt plus4fill}
\begin{flushright}%
Research supervisor:

candidate of engineering sciences, associate professor

Vladimir Timofeyevich Ryabov
\end{flushright}%
\vspace{0pt plus4fill}
{\centering Moscow, Russia "--- \thesisYear\ (corrected in 2017)\par}
\newpage

%% file: Dissertation/abstract_en.tex
\chapter*{Abstract}
% abstracts longer than 1920 characters will not be accepted at arXiv description

The goal of this work is a scientifically justified choice of efficient
technological parameters of anodic bonding of silicon to glass, ensuring
a minimum level of~residual stresses.
The relevance of this dissertation is determined by the continuous efforts
to~reduce the negative effects of the joint use of dissimilar materials
in~electronic devices. Theoretical and practical results: the temperature
dependence of~the~true values of~the~coefficient of thermal expansion
of popular glass brands in~the~temperature range from minus 100~{\textdegree}C
to 500~{\textdegree}C were obtained; the ratio of~thicknesses of silicon
and glass that minimizes residual stress on the silicon surface was
calculated; the~technique of minimizing the residual stresses in silicon
anodically bonded to~glass was developed.
Results of this research are applicable to~MEMS devices design
and~simulations.

The first chapter provides the analysis of the current state of research
in~the~field of anodic bonding technology application.

The second chapter is devoted to the experimental study of temperature
dependence of the linear thermal expansion coefficient of the glass.
Thermomechanical properties and composition of the following glass brands:
LK5, Schott Borofloat 33, Corning 7740, Hoya SD-2 were measured.

The third chapter describes the existing models of estimating the residual
stresses in the silicon-glass bonds. Two models based on a theory of strength
of materials and the classical lamination theory with an additional temperature
dependence of coefficients of thermal expansion are proposed for further use.
Residual stresses in silicon-glass bonds are estimated with these models
and data from the preceding chapter.

The fourth chapter describes the experiments carried out to confirm claims
from the previous chapters. The proposed technique of minimizing residual
stresses is described as steps for a device designer.

This particular dissertation was minor corrected against the version that
was~presented at the defense procedure.

Author's abstract (in russian) available at
\href{https://doi.org/10.17605/OSF.IO/BG3Q9}{doi.org/10.17605/OSF.IO/BG3Q9}.

\urlstyle{rm}
\ifdefmacro{\microtypesetup}{\microtypesetup{protrusion=false}}{}
\section*{Publications}
Based on the findings presented in this thesis, the following peer-reviewed
articles have been published and were co-authored by Leonid S. Sinev.

\begin{refsection}[wos,rsci]
    \begin{refsegment}
        \nocite{Sinev_Petrov2016_cte_glass_eng, sinev2017reducingstress}%wos
        \printbibliography[heading=pubsubgroup, segment=1,title={Journals listed in Web of Science Core Collection}]
    \end{refsegment}
    \begin{refsegment}
        \nocite{Sinev_Ryabov_NMST_2011, Sinev_Ryabov_rasch_coef_napr_nmst2014}%rsci
        \printbibliography[heading=pubsubgroup, segment=2,
        title={Journals listed in \href{http://wokinfo.com/products_tools/multidisciplinary/rsci/}{Russian Science Citation Index}}]
    \end{refsegment}
\end{refsection}
\ifdefmacro{\microtypesetup}{\microtypesetup{protrusion=true}}{}
\urlstyle{tt}

\doclicenseThis

%% file: Dissertation/title.tex
% Титульный лист (ГОСТ Р 7.0.11-2001, 5.1)
\thispagestyle{empty}%
\begin{center}%
\thesisOrganization
\end{center}%
\vspace{0pt plus4fill} %число перед fill = кратность относительно некоторого расстояния fill, кусками которого заполнены пустые места
\begin{flushright}%
На правах рукописи

\end{flushright}%
\vspace{0pt plus6fill} %число перед fill = кратность относительно некоторого расстояния fill, кусками которого заполнены пустые места
\begin{center}%
{\large \thesisAuthor}
\end{center}%
\vspace{0pt plus1fill} %число перед fill = кратность относительно некоторого расстояния fill, кусками которого заполнены пустые места
\begin{center}%
\textbf {\large \thesisTitle}

\vspace{0pt plus2fill} %число перед fill = кратность относительно некоторого расстояния fill, кусками которого заполнены пустые места
{%
Специальность \thesisSpecialtyNumber\ "---

<<\thesisSpecialtyTitle>>
}

\vspace{0pt plus2fill} %число перед fill = кратность относительно некоторого расстояния fill, кусками которого заполнены пустые места
Диссертация на соискание учёной степени

\thesisDegree
\end{center}%
\vspace{0pt plus4fill} %число перед fill = кратность относительно некоторого расстояния fill, кусками которого заполнены пустые места
\begin{flushright}%
Научный руководитель:

\supervisorRegalia

\supervisorFio
\end{flushright}%
\vspace{0pt plus4fill} %число перед fill = кратность относительно некоторого расстояния fill, кусками которого заполнены пустые места
{\centering\thesisCity\ "--- \thesisYear\ (испр. 2017)\par}

%% file: Dissertation/contents.tex
% Оглавление (ГОСТ Р 7.0.11-2011, 5.2)
\ifdefmacro{\microtypesetup}{\microtypesetup{protrusion=false}}{} % не рекомендуется применять пакет микротипографики к автоматически генерируемому оглавлению
\tableofcontents*
\addtocontents{toc}{\protect\tocheader}
\endTOCtrue
\ifdefmacro{\microtypesetup}{\microtypesetup{protrusion=true}}{}

%% file: Dissertation/introduction.tex
\chapter*{Введение}
\addcontentsline{toc}{chapter}{Введение}

\newcommand{\actuality}{\textbf{\actualityTXT}} %Актуальность проблемы.
\newcommand{\progress}{}
\newcommand{\aim}{\textbf{\aimTXT}}
\newcommand{\tasks}{\tasksTXT}
\newcommand{\novelty}{\textbf{\noveltyTXT}}
\newcommand{\influence}{\textbf{\influenceTXT}}
\newcommand{\methods}{\textbf{\methodsTXT}}
\newcommand{\defpositions}{\textbf{\defpositionsTXT}}
\newcommand{\reliability}{\textbf{\reliabilityTXT}}
\newcommand{\probation}{\textbf{\probationTXT}}
\newcommand{\contribution}{\textbf{\contributionTXT}}
\newcommand{\publications}{\textbf{\publicationsTXT}}
\newcommand{\realisation}{\textbf{\realisationTXT}}

\input{common/characteristic} % Характеристика работы по структуре во введении и в автореферате не отличается (ГОСТ Р 7.0.11, пункты 5.3.1 и 9.2.1), потому её загружаем из одного и того же внешнего файла, предварительно задав форму выделения некоторым параметрам

\textbf{Объём и структура работы.} Диссертация состоит из~введения,
четырёх глав, общих выводов и заключения.
Полный объём диссертации составляет \formbytotal{TotPages}{страниц}{у}{ы}{}
с~\formbytotal{totalcount@figure}{рисунк}{ом}{ами}{ами}
и~\formbytotal{totalcount@table}{таблиц}{ей}{ами}{ами}. Список литературы содержит
\formbytotal{citenum}{наименован}{ие}{ия}{ий}.

В первой главе представлены результаты анализа современного
состояния исследований в области применения технологий
герметичного соединения при изготовлении электронной техники.
Представлен обзор современных технологий, определены их~достоинства
и~недостатки.
Углублённо рассмотрены вопросы исследования и применения технологии
электростатического соединения кремния со~стеклом.

Вторая глава посвящена экспериментальному
исследованию зависимостей
температурных коэффициентов линейного расширения стёкол
от~температуры. Исследованы
термомеханические свойства и~состав основных марок отечественного
и зарубежного стекла, применяемого для электростатического соединения
с кремнием "--- ЛК5, Schott Borofloat~33, Corning~7740, Hoya~SD\nb-2.

В третьей главе описаны существующие модели расчётной оценки
остаточных напряжений в соединениях кремния со стеклом. Предложены две
новые модели на основании общей теории сопротивления материалов
и~классической теории слоистых композитов с дополнительным учётом
температурной зависимости коэффициентов теплового расширения.
Приведены примеры применения этих моделей на исходных данных,
полученных в~предшествующей главе. В~завершении главы~3 даны
рекомендации по~последовательности проведения расчётной оценки
остаточных напряжений.

В четвёртой главе описаны эксперименты,
проведённые для подтверждения утверждений из~предыдущих глав. Описаны
исследования и практический опыт по снижению остаточных напряжений
посредством термообработки. Предложена методика минимизации остаточных
напряжений, представленная в виде шагов для разработчика прибора.
Также описаны сложности, на которые стоит обратить внимание при
разработке современной электронной техники с~применением технологии
электростатического соединения.

В заключении приведены краткие выводы
по~результатам проведённых в~работе исследований.

\begingroup
Автор выражает благодарность И.\,Д.~Петрову и А.\,Ю.~Переяславцеву за~помощь в~организации и~проведении экспериментов по измерению свойств стёкол;
О.\,Л.~Лазаревой за помощь в~организации и~проведении спектроскопии методом комбинационного рассеяния света; М.\,Г.~Лукоперовой за~помощь в~организации и~проведении поляриметрических измерений стекла;
Т.\,К.~Ерофеевой
за плодотворные дискуссии во время работы над диссертацией;
сотрудникам кафедры электронных технологий в~машиностроении \mbox{МГТУ} им.~Н.Э.~Баумана
за~ценные замечания по тексту диссертации и~автореферата.\russianpar
\endgroup

%% file: common/characteristic.tex
{\actuality}

Технология электростатического соединения пластин кремния и~стекла
(анодная посадка) стала одной из ключевых в производстве приборов
электронной техники. Одной из целей применения данной технологии является
изоляция чувствительного элемента от окружающей среды. Это касается
механических и~электрических воздействий, сохранения герметичности
изолируемого объёма в~таких приборах, как чувствительные элементы
датчиков давления, микроакселерометры, высокочувствительные
микромеханические гироскопы, высокочастотные резонаторы.

К преимуществам анодной посадки (по сравнению с~диффузионной сваркой,
и~соединением за~счёт расплавления стекла) относят: температуру
процесса ниже температуры деградации металлизации, возможность
сохранения герметичности соединения при использовании с~технологиями
формирования электрических межсоединений сквозь стекло или кремний,
отсутствие требования механического прижатия.

Вследствие разнородности кремния и стекла после проведения
их~соединения возникают остаточные напряжения, вызванные разницей
в~тепловом расширении этих материалов. Эти напряжения могут влиять
на такие ключевые характеристики, как:
уход нулевого сигнала
у чувствительных элементов датчиков давления, собственная частота
у микроакселерометров и микрогироскопов, геометрическое положение
мембран и~балок у~разнообразных микромеханических приборов.
Таким образом, необходимо, сохранив преимущества технологии сращивания
кремния и~стекла, учесть и~снизить влияние такого недостатка как
описанные остаточные напряжения.

Большой вклад в развитие исследований технологии электростатического
соединения кремния и стекла внесён советскими и российскими учёными
Н.\,Н.~Хоменко, Ю.\,М.~Евдокимовым, С.\,П.~Тимошенковым\ и~др.

\input{common/nedostatki}\beforenedostati{}
\nedostati{}.
\nedostatii{}.
\nedostatiii{}.

Поэтому \MakeLowercase{\thesisTitleTXT}
является актуальной темой исследования.

\aim\ данной работы является
\MakeLowercase{\aimTextContent}.

Для~достижения поставленной цели необходимо было решить следующие {\tasks}:
\begin{enumerate}
\input{common/taskslist}
\end{enumerate}

{\methods}
Теоретические исследования проводились на~основе теории напряжённого
состояния в~композиционных материалах, сопротивления материалов,
математического анализа.
В ходе исследований применялись расчёты напряжённо-деформированного
состояния сборок пластин кремния и~стекла с использованием
компьютерных программных пакетов.
Экспериментальные исследования осуществлялись
на~современных аналитических приборах, обработка результатов велась
с~помощью теории вероятностей и~математической статистики.

\novelty
\begin{enumerate}
\input{common/noveltylist}
\end{enumerate}

\influence\
\begin{enumerate}
    \item Разработаны
    методики снижения
    остаточных напряжений, возникающих в технологии
    электростатического соединения кремния и~стекла, повышающие
    функциональные и~эксплуатационные характеристики приборов
    электронной техники.
    \item Установлена возможность использования инженерной
    методики расчёта для оценки возможностей снижения остаточных
    напряжений и~повышения эффективности применения различных
    марок стекла при сращивании с кремнием.
\end{enumerate}

\defpositions
\begin{enumerate}
\input{common/defpositionslist}
\end{enumerate}

\reliability\ основывается на~проведённом комплексном анализе результатов теоретических данных и экспериментальных исследований.
Результаты экспериментов обработаны и подтверждены статистическими методами.
Сформулированные в диссертации научные положения, выводы и рекомендации
обоснованы теоретическими решениями и~экспериментальными данными
и~не~противоречат известным положениям.

\probation\
Результаты работы докладывались и обсуждались на заседаниях кафедры
электронных технологий в машиностроении \mbox{МГТУ} им.~Н.Э.~Баумана,
на научно-технических конференциях «Молодёжь в~науке» ФГУП \mbox{<<РФЯЦ-ВНИИЭФ>>}
(г. Саров, октябрь 2008~г., октябрь 2014~г.),
на научно-технических конференциях молодых учёных ФГУП <<\mbox{ВНИИА} им.~Н.\,Л.~Духова>> (г. Москва,
март 2010~г.,
март 2013~г., март 2016~г.),
на~научно-технических конференциях молодых специалистов Росатома
<<Высокие технологии атомной отрасли. Молодёжь в~инновационном процессе>> (ФГУП
<<\mbox{ФНПЦ} \mbox{НИИИС} им.~Ю.\,Е.~Седакова>>, г.~Нижний Новгород,
сентябрь 2015~г.
(доклад отмечен дипломом за~активную научную деятельность и~перспективные
исследования),
сентябрь 2016~г.),
на 6-ом международном МЭМС\nb-Форуме 2016
(Курский государственный университет, г.~Курск, июнь 2016~г.).

\contribution\ Диссертация является завершённой работой, в~которой обобщены
результаты исследований, полученные лично автором и~в~соавторстве.
Участие в работе каждого соавтора отражено в~совместных публикациях.
Совместно с научным руководителем был определён план работы,
разработаны основные теоретические положения главы~3.
Личный вклад автора включает:
проведение анализа современного состояния исследований
в~области использования технологии электростатического соединения кремния
со~стеклом;
разработку математических моделей, позволяющих прогнозировать
остаточные напряжения в зависимости от режима проведения
процесса сращивания кремния со~стеклом
и~их~термомеханических свойств;
проведение экспериментов по~измерению температурной
зависимости коэффициентов теплового расширения стёкол;
проведение экспериментов по~спектроскопии комбинационного рассеяния
света на~образцах соединённых пластин стекла и~кремния;
обработку экспериментальных данных
и формулировку рекомендаций по~использованию результатов работы.

\needspace{5\onelineskip}
\begin{refsection}[vak,papers,conf]%
    \printbibliography[heading=countauthornotvak, env=countauthornotvak, keyword=biblioauthornotvak]%
    \printbibliography[heading=countauthorvak, env=countauthorvak, keyword=biblioauthorvak]%
    \printbibliography[heading=countauthorconf, env=countauthorconf, keyword=biblioauthorconf]%
    \printbibliography[heading=countauthor, env=countauthor, keyword=biblioauthor]%
    \publications\ Основные результаты по теме диссертации изложены
    в~\arabic{citeauthor}~научных
    публикациях\nocite{Sinev_Ryabov_Tinyakov_nano2011, Sinev_osoben_primen_inzh_vest201408}
    общим объёмом 3,3~п.~л.,
    \arabic{citeauthorvak} из которых находятся в~изданиях из перечня
    рекомендованных ВАК РФ\nocite{Sinev_Ryabov_NMST_2011, Sinev_Ryabov_rasch_coef_napr_nmst2014, Sinev_technomag2014, Sinev_Petrov2016_cte_glass} (1,3~п.~л.),
    \arabic{citeauthorconf} "--- в тезисах докладов\nocite{sinev2008molodezh, sinev2010otsenka_vniia, sinev2013primenenie_vniia, sinev2014molodezh, sinev2015niiis, sinev2016issledovanie_vniia} (1,4~п.~л.).%
\end{refsection}
\begin{refsection}[vak,papers,conf]%
    \printbibliography[heading=countauthorvak, env=countauthorvak, keyword=biblioauthorvak]%
    \printbibliography[heading=countauthornotvak, env=countauthornotvak, keyword=biblioauthornotvak]%
    \printbibliography[heading=countauthorconf, env=countauthorconf, keyword=biblioauthorconf]%
    \printbibliography[heading=countauthor, env=countauthor, keyword=biblioauthor]%
    \nocite{Sinev_Ryabov_NMST_2011, Sinev_Ryabov_rasch_coef_napr_nmst2014, Sinev_technomag2014, Sinev_Petrov2016_cte_glass}%vak
    \nocite{Sinev_Ryabov_Tinyakov_nano2011, Sinev_osoben_primen_inzh_vest201408}%notvak
    \nocite{sinev2008molodezh, sinev2015niiis}%conf
\end{refsection}

%% file: common/nedostatki.tex
\providecommand{\beforenedostati}%
{Несмотря на~большое количество исследований применения технологии
электростатического соединения кремния со~стеклом,
ряд вопросов остаётся невыясненным.
С повышением требований к~электронным приборам возросли требования
к минимизации остаточных напряжений.
}

\providecommand{\beforenedostatii}%
{Среди вопросов, остающихся нерешёнными,
стоит отметить следующие:}

\providecommand{\nedostati}% %с большой буквы и без точки в конце
{Недостаточно данных по~термомеханическим свойствам стёкол, подходящих
для электростатического соединения с кремнием, в форме, удобной
к применению в аналитических расчётах и в~системах компьютерного
моделирования}

\providecommand{\nedostatii}% %с большой буквы и без точки в конце
{Недостаточно хорошо исследован выбор температуры соединения
и~предпосылки наличия температуры проведения соединения
с минимальными остаточными напряжениями}

\providecommand{\nedostatiii}% %с большой буквы и без точки в конце
{Недостаточно подробно описано влияние соотношения толщин
соединяемых пластин на~возникающие после соединения напряжения}

%% file: common/taskslist.tex
\item {\ifpresentation\label{task1}\else\fi}\mmark{Исследовать зависимость} температурных коэффициентов линейного расширения
(\mmark{ТКЛР}) \mmark{от~температуры для стёкол}, совместимых с~анодной посадкой.
\item {\ifpresentation\label{task2}\else\fi}\mmark{Разработать модели оценки} остаточных напряжений, определить
их~возможности по учёту температурной зависимости ТКЛР, распределению
напряжений по~толщине материалов и~подбору температуры соединения
с~минимальными остаточными напряжениями.
\item {\ifpresentation\label{task3}\else\fi}\mmark{Разработать методику расчёта} остаточных напряжений
при использовании известных марок стёкол.
\item {\ifpresentation\label{task4}\else\fi}\mmark{Предложить способ коррекции} температуры
и~выбора толщины соединяемых деталей, обеспечивающих
минимальную величину остаточных напряжений.
\item {\ifpresentation\label{task5}\else\fi}Разработать методику и~\mmark{провести экспериментальные
исследования} процесса соединения кремния со~стеклом с~измерением
остаточных напряжений в~соединении.

%% file: common/noveltylist.tex
\item \label{novelty_one}%
Впервые \mmark{взаимосвязь температуры}
проведения соединения кремния со~стеклом и~\mmark{остаточных напряжений}
рассмотрена
\mmark{с~использованием температурной зависимости истинных значений}
температурных коэффициентов линейного расширения
материалов.
%Новая постановка известной задачи - приняты новые условия

\item \label{novelty_two}%Рябову нравится этот пункт:
Впервые доказано \mmark{существование соотношения толщин кремния
и~стекла}, минимизирующего остаточные напряжения на поверхности
кремния, за счёт взаимной компенсации деформаций вызванных
тепловым расширением кремния и~присоединяемого стекла.
\input{common/vyvods}\vyvodivmain

\item \label{novelty_three}Впервые предложено
\mmark{рассчитывать температуру} проведения соединения так, чтобы
\mmark{накопленная} за время остывания относительная \mmark{деформация была
минимальной},
что позволит минимизировать остаточные напряжения.
%Новые усовершенствованные критерии, показатели (и их обоснование)

%% file: common/vyvods.tex
\providecommand{\beforevyvods}%
{В диссертации поставлена и решена
научно-техническая задача
исследования методов снижения остаточных напряжений
при~электростатическом соединении кремния со~стеклом.}

%В заключении диссертации излагаются итоги выполненного исследования, рекомендации, перспективы дальнейшей разработки темы.

\newcommand{\vyvodi}%
{%
\item \label{vyvod_one}Расчётным путём выявлено, что для электростатического соединения
с~минимальными остаточными напряжениями \mmark{подбор стекла}, согласованного
с~кремнием по ТКЛР, следует осуществлять \mmark{по~критерию минимальности
накапливаемой разности между ТКЛР}, в~процессе охлаждения с~температуры
соединения до диапазона рабочих температур прибора.
}

\newcommand{\vyvodii}%
{\item \label{vyvod_two}Для моделирования сборок
кремния со~стёклами марок
ЛК5, Borofloat~33, 7740, SD-2
в~диапазоне температур от~минус~100 до~500~{\textdegree}C
следует \mmark{использовать
полученные} автором диссертационной работы
\mmark{температурные зависимости ТКЛР} для этого диапазона температур.}

\newcommand{\vyvodiii}%
{\item \label{vyvod_three}Показано, что
\mmark{предварительный расчёт по моделям}
двух тонких слоёв и~многослойного композиционного материала
\mmark{с учётом динамики накопления}
остаточных напряжений
\mmark{повышает
эффективность} применения
моделирования методом \mmark{конечных элементов}.
}

\newcommand{\vyvodivmain}%
{\mmark{Для минимизации} остаточных \mmark{напряжений на поверхности} сплошного кремния необходимо \mmark{выбирать толщину стекла}:%
    \begin{itemize}
        \item алюмосиликатного (SD-2, SW-YY) в~2,5--2,8~раза больше
        толщины кремния.
        \item боросиликатного (ЛК5, 7740, Borofloat 33) в~3~раза больше
        толщины кремния.
    \end{itemize}
}
\newcommand{\vyvodiv}%
{\item \label{vyvod_four}\vyvodivmain
    Это объясняется разницей в согласованности жёсткости стекла и~кремния.
    Остаточные напряжения на поверхности кремния минимизируются за~счёт компенсации деформаций, вызванных тепловым расширением кремния, деформациями, вызванными воздействием присоединённого стекла.%
}

\newcommand{\vyvodv}%
{\item \label{vyvod_five}\mmark{При охлаждении} сборки
со~скоростью
\mmark{не более
2~{\textdegree}C/мин}
после окончания подачи высокого напряжения
\mmark{происходит снижение остаточных напряжений} в стекле
(например в~ЛК5, соединённом
с~кремнием при температуре 440~{\textdegree}C,
до 63~\% от уровня
напряжений при неконтролируемом охлаждении).
Это объясняется
релаксацией аморфной структуры стекла.%
}

\newcommand{\vyvodvi}%
{\item \label{vyvod_six}\mmark{Рекомендуется проводить} электростатическое соединение
\mmark{в~режиме ограничения тока},
что снижает риски
локального перегрева границы кремний"--~стекло
и~последующего прожига стекла.%
}

\newcommand{\vyvodsall}{\vyvodi\vyvodii\vyvodiii\vyvodiv\vyvodv\vyvodvi}

\providecommand{\aftervyvods}%
{Результаты, полученные в диссертации, могут быть использованы при разработке
и~производстве приборов электронной техники с~использованием технологии
электростатического соединения, таких как микрорезонаторы, микрореле,
микроакселерометры, микрогироскопы, чувствительные элементы датчиков
давления.
Проведённые расчёты и сделанные выводы могут способствовать импортозамещению
зарубежных марок стекла в отечественных приборах электронной техники.

\textbf{Дальнейшая разработка темы}
может состоять в разработке аналитических моделей оценки распределения
остаточных напряжений в микрообработанном кремнии,
с~использованием температурной зависимости истинных значений ТКЛР
материалов. Кроме того, востребованными будут модели, учитывающие
изменение свойств стекла, связанное с переносом ионов в~результате
проведения процесса электростатического соединения. Для этого также
потребуется дополнительное исследование свойств стёкол как по составу,
так и~по~анализу связи подвижности ионов с~температурой и~подводимой
разностью потенциалов.}

%% file: common/defpositionslist.tex
\item \label{defposition_one}\mmark{Результаты исследования} зависимости \mmark{ТКЛР}
различных марок \mmark{стекла от температуры} в интервале от~минус~100
до~500~{\textdegree}C.
\item \label{defposition_two}\mmark{Методика минимизации}
остаточных \mmark{напряжений} за счёт \mmark{выбора марки стекла} для
электростатического соединения с~кремнием и~\mmark{режима} проведения
процесса.
\item \label{defposition_three}\mmark{Конструктивно-технологические
методы производства} чувствительных элементов приборов электронной техники
\mmark{при~наличии требований по~минимизации} механических напряжений,
возникающих в~результате применения технологии электростатического
соединения кремния со~стеклом.

%% file: Dissertation/part1.tex
\chapter{Актуальность исследований технологического процесса электростатического соединения кремния со~стеклом}

Проанализированы данные литературных источников за~предшествующие 45~лет по~технологиям герметичного соединения, применяемым при изготовлении электронных приборов.
Представлен обзор современных технологий, определены их~достоинства и~недостатки. Углублённо рассмотрена технология электростатического соединения кремния со~стеклом.

В~связи с~этим поставлена цель "--- изучить возможности снижения
остаточных напряжений в~сборках, вызванных разницей
в~термомеханических свойствах кремния и~стекла, соединённых анодной
посадкой.  Для достижения этой цели определены задачи по~поиску более
точных способов оценки таких напряжений и~поиску конструктивных
и~технологических решений по~их~снижению, включая подбор марки стекла.

\section{Сравнение способов соединения с~деталями из~кремния}

Микроэлектромеханические системы (МЭМС) "--- это микроминиатюрные интегральные устройства или системы, в~которых комбинируются механические и~электрические компоненты. Они изготовляются на~основе групповой технологии обработки интегральных схем и~могут иметь размеры от~нескольких микрометров до~нескольких миллиметров. Общим методом создания многослойных объёмных микромеханических устройств является соединение двух пластин~\cites[122]{Raspopov_micromechanics2007}.

Выбор подходящего метода соединения определяется рядом факторов,
учитывающих, в том числе, тип (например, оптические или механические
приборы) и назначение (например, коммутационные или измерительные приборы)
изготавливаемого устройства; требования, предъявляемые принципом работы
устройства (например, наличие герметичного шва), а~также различия
в~свойствах материалов деталей и узлов, входящих в~состав прибора
(например, различающимися коэффициентами теплового
расширения)~\cite{New_low_temperature_bonding_tech,Vacuum_Packaging_Technology_and_Appl}.

Соединение, сохраняющее герметичность в~течение всего срока жизни прибора, важно для приборов, функционирование которых зависит от~неизменности параметров среды внутри них, например, работающих в~агрессивной среде~\cite{lit_harpster2002long}. Кроме того, давление внутри прибора определяет характеристики демпфирования и~амплитудно-частотные характеристики рабочих движущихся элементов таких приборов, как микроакселерометры, высокочувствительные МЭМС-гироскопы, высокочастотные резонаторы.

Соединение деталей прибора может использоваться как метод поддержания чистоты внутри собираемого устройства. Например, подложки могут быть соединены перед резкой для предупреждения повреждения хрупких или чувствительных элементов~\cite{lit_Esashi_Wafer2008}.

При выборе способа соединения следует также учитывать температурные ограничения. Это касается как ограничений по~допустимой температуре нагрева металлизации соединяемых деталей (440~{\textdegree}C для алюминиевой металлизации), так и~ограничений, связанных с~разностью коэффициентов температурного расширения.

Также следует учитывать накладываемые способом соединения ограничения
на~возможность создания электрических соединений между деталями
в~сборке. Присутствие свинца (соединение стеклоспаями), натрия
(анодная посадка) или золота (эвтектическое соединение) должно быть
учтено, если одна из~соединяемых деталей содержит полупроводниковые
структуры~\cite{Dragoi_cmos_wafer_bonding}.

В~Таблице~\ref{tab:compare_bonding_methods} приведено сравнение способов соединения с~точки зрения различных требований с~учётом описанных выше ограничений~\cite{New_low_temperature_bonding_tech,Dragoi_cmos_wafer_bonding,Babaevskij_korpusirovanie_NMST2014_3}.

\begin{table} [!bh]%Порядок, в котором заданы опции, позволяющие регулировать положение рисунка на странице, не важен — они всегда будут применяться в порядке h (здесь) — t (вверху) — b (внизу) — p (на отдельной странице). Важно только то, какие именно опции заданы. По умолчанию — [tbp]. Задавать опции по одной (например, просто [t] или просто [b]) не рекомендуется — в некоторых случаях это может приводить к проблемным ситуациям.
	\caption[Wafer bonding methods comparison]{Сравнение способов соединения}%
	\label{tab:compare_bonding_methods}% label всегда желательно идти после caption
    \renewcommand{\arraystretch}{1.3}%% Увеличение расстояния между рядами, для улучшения восприятия.
	\def\tabularxcolumn#1{m{#1}}
	\newlength{\tabcellen}
	\setlength{\tabcellen}{\widthof{соединение}} %
	\begin{SingleSpace}
	\begin{tabularx}{\textwidth}{@{}
	>{\raggedright}X
	>{\centering}m{1.9cm}
	>{\centering}m{1.9cm}
	>{\centering}m{2.0cm}
	>{\centering\arraybackslash}m{1.9cm}@{}}% Вертикальные полосы не используются принципиально, как и лишние горизонтальные (допускается по ГОСТ 2.105 пункт 4.4.5) % @{} позволяет прижиматься к краям
        \toprule     %%% верхняя линейка
    	Критерии сравнения (требования к~процессу) & Анодная посадка &
    	Стекло\-сплав &
    	Эвтекти\-ческое соединение &
    	Прямое сращивание	\\
        \midrule %%% тонкий разделитель. Отделяет названия столбцов. Обязателен по ГОСТ 2.105 пункт 4.4.5
        Совместимость с~КМОП &
        ${\pm}$ &
        ${\pm}$ &
        ${\pm}$ &
        $ + $ \\
        Температура процесса {\textless}440~{\textdegree}C &
        $ + $ &
        $ + $ &
        $ + $ &
        $ - $ \\
        Отсутствие требования механического прижатия &
        $ + $ &
        $ - $ &
        $ - $ &
        ${\pm}$ \\
        Отсутствие требований к~равномерности прижатия &
        $ + $ &
        $ - $ &
        ${\pm}$ &
        $ - $ \\
        Герметичность соединения &
        $ + $ &
        $ + $ &
        $ + $ &
        $ + $ \\
        Возможность формировать вертикальные электрические межсоединения  &
        $ + $ &
        $ - $ &
        $ + $ &
        $ + $ \\
        Допустимая высота выступов на~поверхности, не~более, мкм &
        0,05 &
        2 &
        1 &
        0,004  \\
        \midrule%%% тонкий разделитель
        \multicolumn{5}{@{}p{\textwidth}@{}}{
            \vspace*{-3ex}\hspace*{2.5em}Примечание "---
            <<$+$>> "--- требование полностью выполняется;
            <<$-$>> "--- требование невыполнимо;
            <<${\pm}$>> "--- требование выполняется в~некоторой степени
            с~ограничениями или особенными условиями соединения.
        }
        \\
        \bottomrule %%% нижняя линейка
	\end{tabularx}%
	\end{SingleSpace}
\end{table}

\subsection{Прямое сращивание}
Прямое сращивание пластин это технология соединения двух кремниевых
пластин без приложения электрического напряжения, обычно при комнатной
температуре. Прямое сращивание также называют соединением методом
сплавления кремния~\cites[123]{Raspopov_micromechanics2007}.
Данный процесс основывается на~химической реакции между группами ОН\textsuperscript{\(-\)} или H\textsuperscript{\(+\)}, находящимися на~поверхности исходного кремния или на~образованном на~подложке слое оксида кремния~\cites[487]{lit_madou2002fundamentals}. Прямое сращивание обычно состоит из~трёх этапов: подготовки поверхностей, соединения поверхностей и~термической обработки. Подложки с~высоким качеством полировки после химической обработки предварительно соединяются, выравниваются и~довольно сильно сжимаются в~центральной точке поверхности. Шероховатость соединяемых поверхностей подложек не~должна превышать 4~нм~\cites[487]{lit_madou2002fundamentals}. Неплоскостность пластины не~должна превышать 5~мкм (для пластин диаметром 100 мм)~\cite{schmidt1998wafer}. Соприкосновение двух гидрооксидных плёнок на~подложках приводит к~установлению плотного контакта по~всей поверхности кремниевых пластин. Последующая термообработка проводится при температуре 1100~{\textdegree}C и~создаёт долговременные ковалентные связи~\cite{stoger1999awbtech}. При более тщательной подготовке соединяемых поверхностей температуру термообработки можно понизить до~400~{\textdegree}C~\cite{lit_resnik2000study}.

\subsection{Эвтектическое соединение}
Эвтектика "--- жидкая система (раствор или расплав), находящаяся при данном давлении в~равновесии с~твёрдыми фазами, число которых равно числу компонентов системы~\cite{bse_eutectic}.
Добавляя или отводя тепло, можно изменить пропорцию между кристаллическими фазами и~расплавом в~эвтектической точке без изменения температуры~\cite{wiki_eutectic}.

<<Эвтектика является пересечением поверхностей равновесия расплава с~соответствующими (эвтектическими) фазами. Если отводится соответствующее количество тепла, то~расплав эвтектического состава при кристаллизации в~условиях близких к~равновесным даст все кристаллические фазы, участвующие в~равновесии. Если же~подводится тепло в~достаточном количестве, то~смесь фаз, отвечающая эвтектическому составу, в~условиях близких к~равновесным будет плавиться с~одновременным уменьшением доли каждой из~кристаллических фаз вплоть до~их полного исчезновения>>~\cite{wiki_eutectic}.

Эвтектическое соединение широко распространено в~микроэлектронной
промышленности.
Его применяют как для посадки кристаллов интегральных схем и~дискретных приборов
в~металлокерамические корпуса, так и~для соединения кристаллов
микроэлектромеханических приборов с~подложкой.
Соединяемые материалы нагревают до~температуры эвтектической точки, при которой
начинается процесс взаимодиффузии металлов, входящих в~эвтектический
сплав~\cite{Dragoi_cmos_wafer_bonding}.
После охлаждения сборки получают прочное и~надёжное соединение деталей.
Температура эвтектической точки для системы кремний"--~золото составляет
370~{\textdegree}C~\cites[396]{Ljakishev1996_diag_dvojnyh_met_sis_t1}, для
системы алюминий"--~германий составляет
424~{\textdegree}C~\cites[152]{Ljakishev1996_diag_dvojnyh_met_sis_t1}.

Проблемой эвтектического соединения является получение равномерного прочного соединения больших поверхностей. Даже естественные окислы мешают образованию соединения. Процесс проведения эвтектического соединения требует приложения значительных усилий. Следствием этого является сложность соединения деталей большой площади, так как кроме обеспечения прижима необходимо обеспечить высокую степень чистоты и~равномерности соединяемых поверхностей.

\subsection{Соединение стеклосплавом}
Соединение стеклосплавом проводится следующим образом.
Паста, полученная в~результате смешивания порошкообразных минералов
с~растворителями, наносится на~одну из~соединяемых деталей и, затем,
тщательно высушивается.
После этого, пластины соединяют при температуре большей, чем точка
размягчения высушенной пасты.
Типовая температура процесса находится в~диапазоне от~400
до~600~{\textdegree}C~\cite{stoger1999awbtech}.
Для получения высококачественного соединения необходимо равномерное
приложение механического усилия.

\subsection{Электростатическое соединение}
<<В 1960 г. Н. А. Иофис предложил повышать прочность пайки керамики
с~керамикой и стекла с металлом воздействием внешних
электрополей~\cite{iofis_patent1960}.
Особенно бурное развитие исследований, посвящённых разработке методов
сцепления твёрдых тел под воздействием внешних электрических полей,
начинается с 1967~г., когда в СССР и~США~\cite{AB_patent_Pomerantz_1968}
вышли первые публикации по~указанной проблеме.
Появилась возможность создавать как обратимый адгезионный контакт,
сохраняющийся только при действии внешнего электрического поля, так
и~необратимый>>~\cite{evdokimov_krestov1988}.

Электростатическое соединение "--- это технология герметичного соединения стекла с~высоким содержанием окислов щелочных металлов с~металлом или полупроводником при повышенной температуре и~приложенном высоком напряжении. Будучи изначально разработанной для соединения стекла с~металлами~\cite{AB_patent_Pomerantz_1968}, данная технология получила широкое распространение в~микроэлектронике для соединения кремния со~стеклом~\cite{New_low_temperature_bonding_tech,lit_Esashi_Wafer2008,ushkov_kozlov2007}.

<<В~русскоязычной литературе этот процесс имеет также следующие названия:
анодная посадка, анодное сращивание>>~\cite{Sinev_technomag2014},
анодное соединение~\cites[122]{Raspopov_micromechanics2007},
термоэлектростимулированное соединение, электростимулированное термическое соединение~\cite{timoshenkov2006issledovanija},
<<электростатическая сварка~\cite{andreev2014kremnievye}, электроадгезионное соединение, электродиффузионная сварка>>~\cite{Sinev_technomag2014},
неуправляемый электроадгезионный контакт~\cite{tonkiy1974eak_avtoref}, электрохимическая сварка в~твёрдой фазе~\cite{khomenko2001proizvodstvo}, сварка в~электростатическом поле~\cite{berezin2001}.

\begingroup % ради правильной работы с russianpar
В~англоязычной литературе устоялись следующие синонимичные названия этого
процесса: anodic bonding, field assisted bonding, field-assisted thermal
bonding, electrostatic bonding, Mallory process, electrostatic
welding~\cites[484]{lit_madou2002fundamentals}.\russianpar
\endgroup

Процесс электростатического соединения проводится при температуре от~180 до~500~{\textdegree}C как на~воздухе, так и~в вакууме~\cites[484]{lit_madou2002fundamentals}. Электростатическое притяжение между стеклом и~кремнием исключает необходимость приложения значительного усилия к~соединяемым подложкам в~процессе соединения.

Кремниевую пластину помещают на~стеклянную пластину. Соединяемые
поверхности нагревают. Отрицательный электрод источника высокого напряжения
 прикладывают к~стеклянной пластине, а~положительный электрод "---
к~кремниевой пластине. Значение прикладываемого напряжения составляет от~200
до~1500~В~\cites[484]{lit_madou2002fundamentals}{Khomenko1982useglassproperties, Khomenko1990physprocess}. Процесс в~среднем проходит за~30~минут.

В~отличие от~управляемого электроадгезионного контакта, основанного на~эффекте электростатического притяжения Джонсона"--~Раббека~\cite{johnsen1923physical,atkinson1969simple}, при анодной посадке отведение потенциала не~разрушает уже сформировавшееся соединение~\cite{evdokimov1968issledovanie_avtoref}.

К преимуществам технологии анодной посадки относят:
\begin{enumerate}
    \renewcommand{\labelenumi}{\asbuk{enumi})}
    \item относительно низкую температуру процесса, поэтому нет опасности разрушения
    металлических слоёв (например, алюминиевых)~\cite{Low_temp_wafer_AB}, входящих
    в~состав элементов микросхем;
    \item герметичность соединения и возможность её сохранения при использовании с
    технологиями формирования электрических межсоединений сквозь стекло или
    кремний~\cite{lee_rogers2011ssi_design_rules_wlp};
    \item использование оптически прозрачного стекла, что облегчает процесс
    совмещения соединяемых пластин, а также может быть важным для оптических
    приборов~\cite{shoji1998low_b_quartz};
    \item применение стекла позволяет снижать паразитные ёмкости у чувствительных
    элементов датчиков, работающих на емкостном принципе, а~в~электростатических
    системах полезны диэлектрические свойства стекла~\cite{shoji1998low_b_quartz}.
\end{enumerate}

Недостатком технологии считают возможность выделения кислорода из~стекла
в закрытые полости в процессе проведения соединения кремния
со~стеклом~\cites{lit_Esashi_Wafer2008, lee_rogers2011ssi_design_rules_wlp}[{3--20}]{gad2006mems_applications}{barinov2015_datchik_zhestk, Rogers1992considerations, timoshenkov2010metody}.

\section{Подробное рассмотрение технологии электростатического соединения}

\subsection{Обзор применений процесса}

Электростатическое соединение применяют при изготовлении чувствительных
элементов датчиков давления~\cite{kozin2010_microel_datch_fiz_vel}, а~также
разнообразных приборов электронной техники, таких как микрорезонаторы,
микрореле, микроакселерометры~\cite{lit_Esashi_Wafer2008},
микрогироскопы~\cite{Rogers_current_limited_AB_2005}.
Эту технологию применяют при герметизации оптических приборов, например
инфракрасных датчиков~\cite{lit_Esashi_Wafer2008}, а~также для интеграции
микромеханических систем с~оптическими~\cite{saran2003_ab_opt_fibers}.
В~ряде случаев соединяют полированные поверхности без топологии, например, при
изготовлении чувствительных элементов датчиков
давления~\cite{Andreev2013_Analiz_met_elektrostat_svar}.
Возрастающая сложность приборов обуславливает необходимость соединения деталей,
поверхности которых уже обработаны и покрыты плёнками различных
материалов~\cite{Dragoi_cmos_wafer_bonding}.

Соединяют как отдельные кристаллы стекла и кремния, так и целые пластины. Соединение пластин предпочтительней, потому что оно обеспечивает более точное совмещение. Для получения качественного соединения необходимо обеспечить ряд требований~\cite{Low_temp_wafer_AB, Cozma_Puers_1995, Sinev_osoben_primen_inzh_vest201408}:
\begin{enumerate}[label=\asbuk*)]
    \item качество соединяемых поверхностей должно быть высоким (следует обеспечить
    низкую шероховатость и высокую чистоту поверхностей);
    \item температура процесса должна быть достаточной для обеспечения
    подвижности ионов, и, одновременно, достаточно низкой, чтобы
    не~ухудшить характеристики создаваемого прибора;
    \item распределение температуры и подводимого заряда по
    поверхности должно быть равномерным;
    \item коэффициенты теплового расширения соединяемых материалов
    должны быть согласованы;
    \item число подвижных ионов в стекле при рабочей температуре
    должно быть достаточным для проведения процесса.
\end{enumerate}

\subsection{Основные соединяемые материалы и~критерий их~выбора}
\begingroup
Процесс электростатического соединения может быть применён для
соединения стекла с металлами, сплавами и полупроводниковыми
материалами~\cite{wallis_pomerantz1969fagms}. Практически этот процесс
распространён на~боросиликатные, алюмосиликатные, щелочные и
оптические стекла, а так же~на~керамические
соединения~\cite{wallis_pomerantz1969fagms}. Стекло с достаточно
большим коэффициентом содержания щелочных металлов может быть
соединено с полупроводниками (Si, Ge,
GaAs)~\cite{wallis_pomerantz1969fagms,
Khomenko1996pyrex}, с~металлами и~сплавами (например, Al, Cr, Wo, Ta,
Ti, ковар)~\cite{Khandan2014titanium_ab}
и~с~полупроводниковыми структурами (такими как SiC,
Si\textsubscript{3}N\textsubscript{4}, SiO\textsubscript{2})
\cite{stoger1999awbtech}. Толщина слоя изолирующего материала, такого
как SiO\textsubscript{2}, должна быть минимальна (на уровне
естественного окисла), чтобы пропустить достаточный
ток~\cite{stoger1999awbtech}.
В~работе~\cite{djachkov2000_avtoref} проведено исследование влияния состава ряда
отечественных стёкол (включая ЛК105 и~К8) на~плотность тока при сращивании
с~кремнием.\russianpar
\endgroup

Важным критерием при выборе пары соединяемых материалов является их согласование по значению ТКЛР.

Идея электростатического соединения стекла и кремния может быть заимствована и для соединения кремния с кремнием. Одна кремниевая поверхность с тонкой стеклянной плёнкой будет катодом. Минимальная толщина стекла для удовлетворительного соединения "--- 4 мкм~\cites[487]{lit_madou2002fundamentals}{Wallis_FAGS_1975}. Напряжение необходимое для сращивания составляет 50~В~\cite{Wallis_FAGS_1975}.

Стеклянные пластины могут быть соединены электростатическим сращиванием через слой кремния. На одну из стеклянных пластин наносят тонкую плёнку кремния, через которую будет проходить соединение с другой стеклянной пластиной~\cite{terazaki2012stack_structure_patent}.

\subsection{Требования к~качеству соединяемых поверхностей и~давлению в~камере}
Прочное соединение при электростатическом соединении достигается при хорошо обработанных поверхностях обоих соединяемых объектов~\cite{Wallis_FAGS_1975}. Шероховатость поверхности с целью получения наиболее прочного соединения должна быть не более Ra\,1,0~мкм~\cites[484]{lit_madou2002fundamentals}.
При увеличении шероховатости необходимо повышать температуру нагрева, электрическое напряжение и~время выдержки.

Степень очистки поверхности перед посадкой также влияет на качество
соединения. Например,  очистка раствором
H\textsubscript{2}SO\textsubscript{4}–H\textsubscript{2}O\textsubscript{2}
и HF обеспечивает более качественную посадку~\cite{Lee_Detailed_characterization},
чем очистка ацетоном.
Ацетон используется, чтобы смыть нерастворимые и свободно лежащие органические остатки,
но~не~в~состоянии удалять окись и другие следы загрязнений на поверхности,
которые затрудняют процесс посадки.

Также поверхности перед посадкой при необходимости активируют
плазмой~\cite{choi2002analysis} различного газового состава, которая может
удалять и~естественный окисел.

Чем выше температура проведения процесса, тем меньшее влияние оказывает предварительная очистка поверхностей~\cite{Cozma_Puers_1995}.

Требования к составу окружающей среды определяются, как правило,
требованиями к конкретному прибору или к атмосфере, герметизируемой
в~приборе.
Кроме того, в работе~\cite{Cozma_Puers_1995} показано, что кислород
из~окружающей среды повышает прочность соединения стекло"--~стекло.

\subsection{Описание процесса электростатического соединения}
\begin{figure}[ht]
    \centering
    \begingroup%
      \makeatletter%
      \providecommand\color[2][]{%
        \errmessage{(Inkscape) Color is used for the text in Inkscape, but the package 'color.sty' is not loaded}%
        \renewcommand\color[2][]{}%
      }%
      \providecommand\transparent[1]{%
        \errmessage{(Inkscape) Transparency is used (non-zero) for the text in Inkscape, but the package 'transparent.sty' is not loaded}%
        \renewcommand\transparent[1]{}%
      }%
      \providecommand\rotatebox[2]{#2}%
      \ifx\svgwidth\undefined%
        \setlength{\unitlength}{425.19679633bp}%
        \ifx\svgscale\undefined%
          \relax%
        \else%
          \setlength{\unitlength}{\unitlength * \real{\svgscale}}%
        \fi%
      \else%
        \setlength{\unitlength}{\svgwidth}%
      \fi%
      \global\let\svgwidth\undefined%
      \global\let\svgscale\undefined%
      \makeatother%
      \begin{picture}(1,0.40182386)%
        \put(0,0){\includegraphics[width=\unitlength]{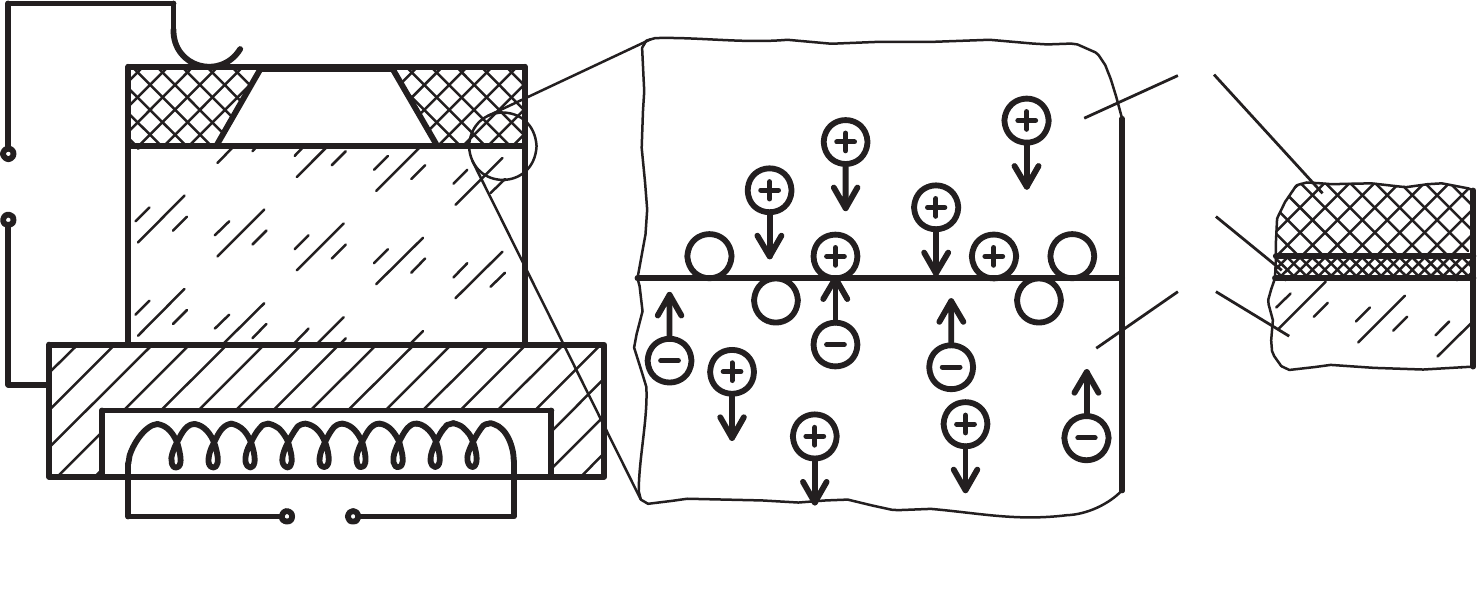}}%
        \put(0.01971018,0.28819333){\color[named]{black}\makebox(0,0)[lb]{\smash{$+$}}}%
        \put(0.02175197,0.24680058){\color[named]{black}\makebox(0,0)[lb]{\smash{$-$}}}%
        \put(0.21694339,0.041859){\color[named]{black}\makebox(0,0)[b]{\smash{$\sim$}}}%
        \put(0.21694339,0.00046624){\color[named]{black}\makebox(0,0)[b]{\smash{\textit{а}}}}%
        \put(0.59324114,0.00046624){\color[named]{black}\makebox(0,0)[b]{\smash{\textit{б}}}}%
        \put(0.93567212,0.00046624){\color[named]{black}\makebox(0,0)[b]{\smash{\textit{в}}}}%
        \put(0.81149385,0.19237811){\color[named]{black}\makebox(0,0)[b]{\smash{\textsl{3}}}}%
        \put(0.81149385,0.2450598){\color[named]{black}\makebox(0,0)[b]{\smash{\textsl{2}}}}%
        \put(0.81149385,0.33913425){\color[named]{black}\makebox(0,0)[b]{\smash{\textsl{1}}}}%
        \put(0.4803518,0.248){\color[named]{black}\makebox(0,0)[b]{\smash{SiO\textsubscript{2}}}}%
        \put(0.58571518,0.33160829){\color[named]{black}\makebox(0,0)[b]{\smash{Si${}^{4+}$}}}%
        \put(0.54055945,0.12464451){\color[named]{black}\makebox(0,0)[lb]{\smash{Na${}^+$}}}%
        \put(0.66,0.155){\color[named]{black}\makebox(0,0)[lb]{\smash{O${}^{2-}$}}}%
      \end{picture}%
    \endgroup%

    \caption[Anodic bonding process scheme]{Иллюстрация процесса электростатического соединения~\cite{Sinev_osoben_primen_inzh_vest201408}:%
    }
    \label{fig:AB_proc_scheme}
    \legend{%
        \textit{а} "--- схема подведения разности потенциалов;
        \textit{б} "--- схема ионного взаимодействия во~время проведения процесса;
        \textit{в} "--- область соединения после завершения процесса.
        \textsl{1} "--- кремний (Si),
        \textsl{2} "--- оксид кремния (SiO\textsubscript{2}),
        \textsl{3}~---~стекло%
    }%
\end{figure}

При анодной посадке кремния на стекло отрицательный заряд подводится к стеклу, а
положительный "--- к кремнию (см. Рисунок \ref{fig:AB_proc_scheme}). Процесс
соединения происходит следующим
образом~\cite{Khomenko1990physprocess, Lee_Detailed_characterization, enikov2006introduction}:
при повышенной температуре и наличии сильного электрического поля положительные
ионы натрия (Na\textsuperscript{$+$}) в~стекле дрейфуют к отрицательному
электроду на стеклянной пластине и~нейтрализуются. Из-за их перемещения на
границе с кремнием образуется избыточный отрицательный
заряд~\cite{rogers1995selection}, сформированный ионами кислорода. Происходит
падение потенциала в области стекла, прилегающего к~кремнию. Электростатическая
сила между отрицательно заряженным слоем в~стекле и~положительным зарядом,
наведённым на аноде, плотно стягивает соединяемые поверхности. Кислород из
стекла соединяется с кремнием на~границе раздела кремний"--~стекло,
формируя тонкий
слой SiO\textsubscript{2}~\cite{Cozma_Puers_1995, Lee_Detailed_characterization,
Wallis_FAGS_1975}.

Если в течение процесса температура и приложенное напряжение поддерживаются
постоянными, то следствием дрейфа Na\textsuperscript{$+$} является скачок
тока, протекающего через соединяемые
поверхности~\cite{Lee_Detailed_characterization}. Чем выше температура, тем
больше амплитуда скачка~\cite{Khomenko1990physprocess, Cozma_Puers_1995,
Low_temp_wafer_AB}. При наблюдении за соединяемыми поверхностями сквозь
стекло, видно, что соединённая область становится
тёмносерой~\cites[484]{lit_madou2002fundamentals}{Khomenko1990physprocess}; когда
эта область расширяется по всей подложке, соединение заканчивается.
Также возможно контролировать прохождение процесса
<<посредством измерения зависимости тока от~времени
в~процессе получения соединения, используя в~качестве
информативных критериев величину прошедшего через единицу
площади электрического заряда и~наличие экстремумов
в~анализируемой зависимости>>~\cites[23]{pshchelko2011_avtoref}.

\Needspace*{6\onelineskip}
\subsection{Основные параметры процесса}
Температура процесса выбирается в~пределах от 180 до 500~{\textdegree}C~\cites[484]{lit_madou2002fundamentals}.
Нижний предел температуры определяется началом ионной проводимости
и~возникновения поляризационных процессов в~стекле.
Верхний предел "--- точкой размягчения стекла
и~температурой плавления материалов топологии на~кремниевом элементе или пластине.
Снижения температуры электростатического соединения можно достичь уменьшением шероховатости поверхности соединяемых материалов, увеличением времени процесса и~прикладываемого электрического напряжения,
а~также применением легкоплавких стёкол с~повышенным содержанием оксида натрия (от 5 до 7~\%).

Прикладываемое электрическое напряжение для электростатического соединения выбирается в~интервале от 200 до 1500~В. Электрическое напряжение можно уменьшить, используя менее шероховатые поверхности, увеличивая температуру и~время выдержки. Верхний предел напряжения ограничивается появлением поверхностных разрядов. Обычно используют источники постоянного или импульсного напряжения. Применения источников переменного напряжения рекомендовано избегать~\cite{Anthony_AB_of_imperfect_surfaces}, поскольку это повышает требования к~шероховатости поверхностей и~к усилию прижатия.

Время процесса электростатического соединения изменяется в~интервале от 1 до 30 мин. Электростатическое соединение осуществляется за~очень короткий промежуток времени, однако, выдержка до~30 минут
повышает прочность соединения.
В~\cite{Rongyan2006Investigation_bond_time} предложена модель оценки
времени требуемого для проведения процесса анодной посадки.

Повышение уровня прилагаемого напряжения снижает длительность процесса~\cite{Lee_Detailed_characterization}. Также время проведения процесса зависит от~характеристик соединяемых пластин и~их покрытий. Необходимое время посадки в~порядке возрастания Si~(p\nb-тип) {\textless} поликремний {\textless} Si\textsubscript{3}N\textsubscript{4} {\textless} SiO\textsubscript{2} {\textless} Si~(n\nb-тип). Si\textsubscript{3}N\textsubscript{4} и~SiO\textsubscript{2} "--- диэлектрические материалы, которые препятствуют дрейфу положительных зарядов к~поверхности соединения~\cite{Lee_Detailed_characterization}.
Вместо этого заряды накапливаются на~поверхности раздела <<нитрид кремния "--- кремний>>.
Это приводит к~возникновению электростатической силы на~поверхности раздела и~замедляет процесс посадки. Кремний p\nb-типа (легированный бором) обеднён электронами по~сравнению с~кремнием n\nb-типа.
Под воздействием электрического поля положительные заряды скапливаются
на~поверхности раздела кремний"--~стекло и~формируют высокую
электростатическую силу способствующую быстрому присоединению.

\subsection{Вывод электрических контактов через область соединения кремния со~стеклом}
В некоторых видах МЭМС возникает необходимость вывода электрического
контакта через зону соединения кремния и стекла (например, подведение
электропитания, шин данных в герметичную
область)~\cite{Vacuum_Packaging_Technology_and_Appl}. Это можно сделать
несколькими способами, у каждого из которых есть свои преимущества
и~недостатки, описанные в таких работах
как~\cite{Jakobsen_AB_for_MEMS_2001_ppt,tanaka2007laterally,Babaevskij_el_vyvody_NMST2014_4,barinov2015_datchik_zhestk}.
Рассмотрим некоторые способы.

\begin{figure}[ht]
    \centering
    \begingroup%
      \makeatletter%
      \providecommand\color[2][]{%
        \errmessage{(Inkscape) Color is used for the text in Inkscape, but the package 'color.sty' is not loaded}%
        \renewcommand\color[2][]{}%
      }%
      \providecommand\transparent[1]{%
        \errmessage{(Inkscape) Transparency is used (non-zero) for the text in Inkscape, but the package 'transparent.sty' is not loaded}%
        \renewcommand\transparent[1]{}%
      }%
      \providecommand\rotatebox[2]{#2}%
      \ifx\svgwidth\undefined%
        \setlength{\unitlength}{424.91000018bp}%
        \ifx\svgscale\undefined%
          \relax%
        \else%
          \setlength{\unitlength}{\unitlength * \real{\svgscale}}%
        \fi%
      \else%
        \setlength{\unitlength}{\svgwidth}%
      \fi%
      \global\let\svgwidth\undefined%
      \global\let\svgscale\undefined%
      \makeatother%
      \begin{picture}(1,0.74055627)(0,-0.048206627878)
        \put(0,0){\includegraphics[width=\unitlength]{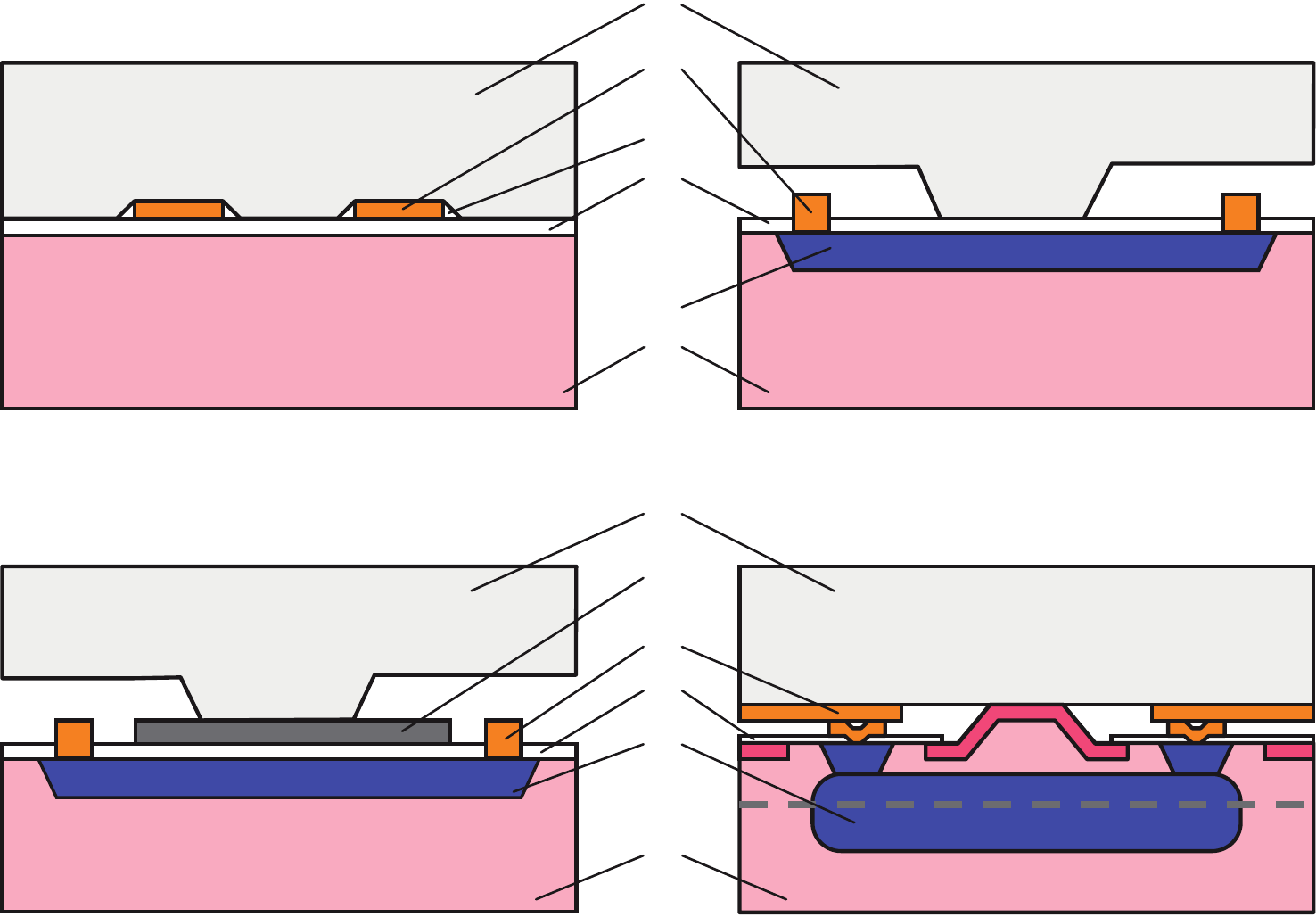}}%
        \put(0.49600326,0.67879616){\color[named]{black}\makebox(0,0)[lb]{\smash{\textsl{1}}}}%
        \put(0.49535263,0.63262167){\color[named]{black}\makebox(0,0)[lb]{\smash{\textsl{4}}}}%
        \put(0.49590687,0.41855278){\color[named]{black}\makebox(0,0)[lb]{\smash{\textsl{2}}}}%
        \put(0.49422532,0.45145387){\color[named]{black}\makebox(0,0)[lb]{\smash{\textsl{6}}}}%
        \put(0.4958667,0.54705054){\color[named]{black}\makebox(0,0)[lb]{\smash{\textsl{3}}}}%
        \put(0.49540884,0.57670397){\color[named]{black}\makebox(0,0)[lb]{\smash{\textsl{5}}}}%
        \put(0.49600326,0.29330271){\color[named]{black}\makebox(0,0)[lb]{\smash{\textsl{1}}}}%
        \put(0.49535263,0.19445824){\color[named]{black}\makebox(0,0)[lb]{\smash{\textsl{4}}}}%
        \put(0.49590685,0.03305937){\color[named]{black}\makebox(0,0)[lb]{\smash{\textsl{2}}}}%
        \put(0.49422532,0.11863041){\color[named]{black}\makebox(0,0)[lb]{\smash{\textsl{6}}}}%
        \put(0.49585521,0.16155713){\color[named]{black}\makebox(0,0)[lb]{\smash{\textsl{3}}}}%
        \put(0.49340073,0.24388048){\color[named]{black}\makebox(0,0)[lb]{\smash{\textsl{7}}}}%
        \put(0.2105012,0.34332804){\color[named]{black}\makebox(0,0)[lb]{\smash{\textit{а}}}}%
        \put(0.21022771,-0.03905968){\color[named]{black}\makebox(0,0)[lb]{\smash{\textit{в}}}}%
        \put(0.76944963,0.34332804){\color[named]{black}\makebox(0,0)[lb]{\smash{\textit{б}}}}%
        \put(0.77107451,-0.03905968){\color[named]{black}\makebox(0,0)[lb]{\smash{\textit{г}}}}%
      \end{picture}%
    \endgroup%
    \caption[Electrical feed-through methods]{Схематическое изображение способов вывода электрических контактов через область соединения~\cite{Jakobsen_AB_for_MEMS_2001_ppt}:}
    \label{fig:contact_through}
    \legend{%
        \textit{а} "--- вытравливанием каналов под проводники; \textit{б} "--- диффузионный <<приповерхностный>> проводник; \textit{в} "--- соединение с~плёнкой поликристаллического кремния; \textit{г} "--- заглублённый диффузионный проводник. \textsl{1} "--- стекло, \textsl{2} "--- кремний (Si), \textsl{3} "--- оксид кремния (SiO\textsubscript{2}),
        \textsl{4}~---~металлизация;
        \textsl{5}~---~вытравленный канал; \textsl{6} "--- легированный кремний (проводимость p+);
        \textsl{7}~---~слой поликристаллического кремния
    }
\end{figure}

В первом из рассматриваемых способов (см. Рисунок
\ref{fig:contact_through}а)
на поверхности стекла или кремния вытравливают углубления, в которых затем
формируют металлические дорожки.
Защитив тем или иным способом дорожки от~электрического контакта с
противоположной деталью, осуществляют соединение.
К~преимуществам таких токоподводов относят их низкое сопротивление, однако,
без дополнительных операций сложно добиться герметичности соединения,
а~также есть опасность повреждения дорожек во~время соединения.
Большое значение для герметичности соединения имеет величина отклонения
толщины нанесённого металла от номинала.

Следующий способ (см. Рисунок \ref{fig:contact_through}б) заключается в
легировании приповерхностной области кремния с нанесением металлизации в
областях будущей разварки и защите оксидом в остальных местах.
Этот способ сохраняет герметичность последующего соединения, однако,
сопротивление токопровода заметно выше, чем в предыдущем варианте.
Кроме того, на~ионных взаимодействиях в~процессе соединения отрицательно
сказывается наличие легированной области.
Снизить негативные последствия возможно осаждением в~области соединения
поверх оксида слоя поликристаллического кремния
(см.~Рисунок~\ref{fig:contact_through}в),
на~который затем производят электростатическое присоединение стекла.

В следующем способе (см. Рисунок \ref{fig:contact_through}г) после легирования приповерхностной
области осуществляют эпитаксиальное наращивание слоя кремния. Таким образом, проводящая область
оказывается заглублена в месте будущего соединения. Токоподвод к ней осуществляют через области ещё
одного легирования, которые обеспечивают её связь с поверхностью.

Металлические дорожки толщиной более 0,2 мкм препятствуют плотному контакту соединяемых пластин, необходимому для качественного соединения поверхностей~\cite{Cozma_Puers_1995}. Толщина дорожек SiO\textsubscript{2} свыше 0,3 мкм также препятствует плотному контакту~\cite{Cozma_Puers_1995}.

\subsection{Моделирование изменения тока в процессе соединения}

Для предсказания экспериментального поведения тока в~зависимости от~времени
начиная с 1970-х годов предпринимаются попытки сформировать достаточно
точную теоретическую модель.

В работе~\cite{Carlson1972polarization}
были предприняты попытки моделирования долговременного поведения тока. В этой модели для ионов была введена мобильность, зависящая от поля, и было сделано предположение, что всё падение напряжения происходит в слое отрицательного заряда в стекле.

\begin{figure}[ht]
    \centering
    \begingroup%
      \makeatletter%
      \providecommand\color[2][]{%
        \errmessage{(Inkscape) Color is used for the text in Inkscape, but the package 'color.sty' is not loaded}%
        \renewcommand\color[2][]{}%
      }%
      \providecommand\transparent[1]{%
        \errmessage{(Inkscape) Transparency is used (non-zero) for the text in Inkscape, but the package 'transparent.sty' is not loaded}%
        \renewcommand\transparent[1]{}%
      }%
      \providecommand\rotatebox[2]{#2}%
      \ifx\svgwidth\undefined%
        \setlength{\unitlength}{120.24470172bp}%
        \ifx\svgscale\undefined%
          \relax%
        \else%
          \setlength{\unitlength}{\unitlength * \real{\svgscale}}%
        \fi%
      \else%
        \setlength{\unitlength}{\svgwidth}%
      \fi%
      \global\let\svgwidth\undefined%
      \global\let\svgscale\undefined%
      \makeatother%
      \begin{picture}(1,1.67141667)%
        \put(0,0){\includegraphics[width=\unitlength]{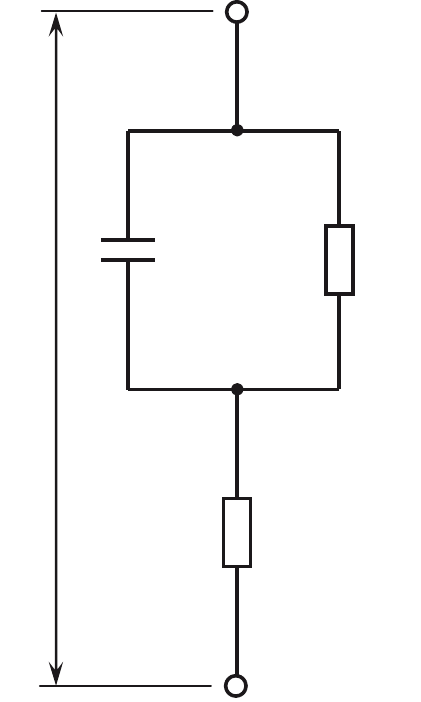}}%
        \put(0.6251985,0.3569699){\color[named]{black}\makebox(0,0)[lb]{\smash{R\textsubscript{1}}}}%
        \put(0.86870192,1.00914014){\color[named]{black}\makebox(0,0)[lb]{\smash{R\textsubscript{2}}}}%
        \put(0.39616544,1.03413581){\color[named]{black}\makebox(0,0)[lb]{\smash{C}}}%
        \put(-0.00102444,0.79774986){\color[named]{black}\makebox(0,0)[lb]{\smash{V}}}%
      \end{picture}%
    \endgroup%
    \caption[Equivalent electrical circuit of the anodic bonding
    process]{Модель эквивалентной электрической цепи процесса
    электростатического
    соединения~\cite{Anthony_AB_of_imperfect_surfaces}}
    \label{fig:process_model}
\end{figure}

В модели эквивалентной электрической цепи
по~\cite{Anthony_AB_of_imperfect_surfaces}
(Рисунок~\ref{fig:process_model}), слой пространственного заряда заменён
постоянной емкостью, вызывающей экспоненциальное затухание тока. Эту модель
используют для оценки зависимости электростатических сил, стягивающих
соединяемые образцы, от~приложенного напряжения, и зависимости тока
от~времени.
Здесь:
R\textsubscript{1} "--- последовательное сопротивление стекла;
R\textsubscript{2} "--- сопротивление утечки;
C "--- ёмкость соединяемой пары;
V "--- приложенная разность потенциалов.
Определив заряд конденсатора в~модели, можно определить получаемую силу
притягивания соединяемых деталей.

Автор работы~\cite{Albaugh1991electrode_phenomena}
указал, что эти модели не годятся для начального промежутка времени, когда
падение напряжения в стекле происходит из-за сопротивления стекла,
а не ёмкости области пространственного заряда. Он~создал модель, которая
включает граничные условия для начального промежутка времени, и~получил
качественное соответствие с~экспериментальной зависимостью тока от~времени.
Когда его уравнения выводятся из экспериментальных характеристик
материалов, его модель имеет значительные расхождения с~соответствующими
экспериментальными результатами~\cite{Rios2000_modelling_current}.

В работе~\cite{Rios2000_modelling_current} была представлена модель, сформулированная на основе изменений во времени слоёв в стекле, обеднённых катионами и анионами.
Аналитическое решение учитывает как кратковременный рост тока, так и~долговременное экспоненциальное падение тока. Переход между двумя режимами происходит, когда электрическое поле на стыке кремния со стеклом достигает величины достаточной для дрейфа анионов~\cite{Rios2000_modelling_current}.

В работе~\cite{Anthony_AB_of_imperfect_surfaces} все элементы эквивалентной
электрической схемы (Рисунок~\ref{fig:process_model}) считаются постоянными
во времени.
В~работе~\cite{He2015_electric_curr_charact} при той же схеме все её
элементы считаются переменными во времени.
Также в работе~\cite{He2015_electric_curr_charact} учитывается площадь
плотного контакта, а полные заряды перемещённых ионов щелочных металлов
и~электрического поля считаются различающимися величинами.
В~этой модели:
R\textsubscript{1} "--- сопротивление стекла и катодного контакта (заметную
роль играет размер катода, точечный или же конечной площади);
R\textsubscript{2} "--- сопротивление слоя обеднённого щелочными ионами, в
связи с переменной толщиной слоя, также является переменным;
C "--- ёмкость слоя обеднённого щелочными ионами, рассчитываемая с учётом
эквивалентной площади плотного контакта и~толщины слоя обеднённого
щелочными ионами.

\subsection{Борьба с~локальным перегревом стыка материалов}\label{chap:local_heating}

Наиболее часто процесс электростатического соединения проводят при постоянном напряжении, однако, в этом случае начальный скачок тока может вызывать локальный перегрев поверхностей и привести к большим остаточным напряжениям~\cite{Sinev_osoben_primen_inzh_vest201408}.
Во избежание такого эффекта применяют ограничение по~току~\cite{Rogers_current_limited_AB_2005}.
В течение процесса поддерживают относительно низкое значение тока за счёт плавного роста напряжения.
Это снижает риск локального перегрева поверхностей, но в то же время может увеличить длительность процесса.

Пиковые значения тока при соединении пластин диаметром 100~мм могут достигать нескольких десятков миллиампер, что, в случае подачи разницы потенциалов в 1~кВ, говорит о рассеянии нескольких десятков ватт тепла~\cite{Rogers_current_limited_AB_2005}.

Однако из-за неидеальной плоскостности поверхностей, кремний и~стекло будут
соприкасаться лишь в некоторых точках, через которые станет протекать ток.
Протекание тока, в соответствии с законом Джоуля"--~Ленца, может вызвать
локальный нагрев частей стыка до~температур выше предполагаемых
по~показаниям термодатчиков на~электродах.
Различие в~локальных температурах на~стыке кремния со~стеклом в момент их
соединения может вызвать локальные вариации остаточных напряжений при
охлаждении до~рабочих температур~\cite{Rogers_current_limited_AB_2005}.

При проведении процесса с~ограничением тока, подводимое напряжение
изначально очень низкое и~затем постепенно возрастает с~увеличением
соединённой площади пластин. При таком режиме происходит лучшее
управление температурой стыка и её воспроизводимостью от процесса
к~процессу~\cite{Rogers_current_limited_AB_2005}.

\subsection{Анализ прочности получаемого соединения}
Дать характеристику проходящему или уже завершившемуся процессу можно
несколькими способами: наблюдением за параметрами процесса (за~током),
визуальным осмотром (область соединения имеет более тёмный оттенок
серого цвета в отличие от несоединённых областей), проводя разрушающие
испытания на отрыв или сдвиг, измеряя изгиб соединённых деталей.
Механическая прочность соединения составляет от 10 до
150~МПа~\cite{guan2006icept06,
Hu2008tensile_tests,
Villanueva2006_Transfer_small}. Величина прочности зависит от
материалов и метода измерения. Обычно соединённые детали разрушаются с
вырывом кремния, либо стекла~\cite{Khomenko1982useglassproperties}.
К~дефектам соединения относят: пустоты из-за посторонних частиц,
несоединённые области из-за неплоскостности соединяемых поверхностей,
погрешности совмещения.

В~\cite{Cozma_Puers_1995} было отмечено повышение твёрдости стекла вследствие миграции ионов и предложено снижать этот эффект применением точечного электрода.

Механизмы различных методов соединения были подробно объяснены к~настоящему
времени, но методы для оценки соединения должны разрабатываться постоянно с
развитием технологии микрообработки в более сложные структуры с высокими
требованиями по надёжности~\cite{Richard2002weibull_fracture_probability}.
Метод, описанный в~работе~\cite{Maszara1988bonding_silicon}, заключается
во~введении тонкого лезвия в область соединения и~последующем вычислении
энергии поверхности исходя из обследования разлома на кончике лезвия.
Этот метод легко применим, но ограничен слабыми соединениями.
Когда анодная посадка формирует прочные соединения и,~более того,
использует как одну из составных частей хрупкий материал "--- стекло, метод
лезвия не подходит.
Метод растягивающего или трёхточечного изгиба неудобен в~применении
и неэффективен при отражении связи прочности соединения с~условиями
соединения~\cite{Richard2002weibull_fracture_probability}.
При изучении влияния параметров процесса соединения на качество соединения,
были разработаны оптические и~неразрушающие
методы~\cite{Go1999Experimental_eval_Taguchi,
TaticLucic1997bond_characterization, Plaza1997nondestructive}.
Они легки в применении и позволяют оценить электростатическое давление во
время анодной посадки~\cite{Richard2002weibull_fracture_probability}, хотя
ещё остаётся в~силе вопрос, можно ли эти методы соотнести с реальной
прочностью соединения~\cite{TaticLucic1997bond_characterization}.
Известен микрошевронный тест для оценки прочности соединённых пластин
кремния~\cite{Petzold1999quality_reliability_chevron}.
Этот метод можно использовать и для анодной
посадки~\cite{Richard2002weibull_fracture_probability}.

Всё ещё существует необходимость в методах оценки для определения
надёжности соединённых микроструктур.
Ранее описанные испытания растягивающим изгибом показали, что трещины
обычно инициируются на~стыке кремний"--~стекло, но распространяются
по~стеклу.
Из этого был сделан вывод, что именно прочность стыка определяет надёжность
всего прибора~\cite{Richard2002weibull_fracture_probability}.
В~работе~\cite{Richard2002weibull_fracture_probability} описаны испытания
прочности стыка посредством подачи газа под давлением и~дальнейший
статистический анализ с применением распределения Вейбулла.

\subsection{Формирование гребенчатых структур в системе «кремний на~стекле»}
В случае проведения электростатического соединения кремния со~стеклом в
вакууме возможен изгиб мембраны над полостью в~кремнии. Если на~этих
мембранах в дальнейшем будут производиться литографические операции,
то~возможно искажение формы получаемых элементов. Это относится, например,
к~гребенчатым структурам.

Существует проблема повреждений высокоаспектных гребенчатых структур во
время их формирования глубинным плазмохимическим травлением над таким
диэлектриком как стекло. Детальное описание проблемы и~возможное решение в
виде формирования дополнительной металлической плёнки на~стекле описано
в~\cite{LeeMC2005_gyro_siog}. В этих работах показано, что введение стадии
напыления плёнки металла перед операцией освобождения гребёнки
чувствительного элемента позволяет резко увеличить коэффициента выхода
годных приборов.

\subsection{Предотвращение нежелательного соединения}\label{chap:unintentional_bond}
При применении анодной посадки для соединения деталей с подвижными
элементами есть опасность того, что из-за сильного электростатического
поля гибкая структура притянется к стеклу и соединится
с~ним~\cite{Cozma_Puers_1995}. Нужно использовать как конструктивные,
так и технологические решения, которые бы~предотвратили контакт
кремния со стеклом в нежелательных
областях~\cite{Yu2008_Yield_improvement_AB}.
Для решения этой задачи есть несколько вариантов:
\begin{itemize}
    \item понижение подводимой разности потенциалов, чтобы снизить стягивающие электростатические силы;
    \item экранирование мест соединения~\cite{lit_Esashi_Wafer2008};
    \item локальная модификация поверхности с~целью препятствования соединению~\cites[6]{ushkov2008_avtoref}.
\end{itemize}

\section{Следствия тепловой несогласованности кремния и стекла}\label{chap_temp_inconsistence}
В результате соединения образуются остаточные напряжения.
Напряжения, возникающие вследствие разности ТКЛР стекла и кремния называют
коэффициентными напряжениями~\cite{ost_Steklo_terminy}. Также, механические
напряжения, возникающие в результате различия ТКЛР соединяемых материалов после
их охлаждения, называют термическими~\cites[{19, 26}]{mehan_napr_plenki1981obzor}.
До начала нагрева стеклянная и кремниевая детали имеют одинаковые размеры, при
нагреве детали расширяются неравномерно и~при~температуре соединения $T_b$ имеют
отличающиеся размеры. После соединения детали, охлаждаясь до рабочей температуры
$T_w$, взаимно деформируются~\cite{Sinev_Ryabov_rasch_coef_napr_nmst2014}.
В работе~\cites[13]{matuzov2008_avtoref} говорится, что <<для плоских
мембран чувствительность уменьшается с~увеличением значения внутренних
напряжений>>.
Согласно~\cite{Cozma_Puers_1995} формирующиеся после
соединения механические напряжения в~стекле являются важным критерием
оптимизации режима процесса анодной посадки.

Наглядно деформации выражаются в прогибе соединённых пластин.
В~работе~\cite{Rogers1992considerations} говорится о прогибе порядка 30~мкм
даже когда кремний соединён со~стеклом толщиной 3~мм.
В работе~\cite{rogers1995selection} говорится о прогибе 100~мкм в~случае
соединения пластин кремния с пластинами стекла Corning~7070 толщиной 1~мм
при 460~{\textdegree}C.
В~\cite{Rogers_current_limited_AB_2005} выдвинуто требование прогиба не
более 70~мкм для производства микрогироскопов.
В работе~\cite{LeeMC2005_gyro_siog} говорится о прогибе до~110~мкм
в~случае соединения пластин кремния толщиной 500~мкм с~пластинами стекла
Corning~7740 толщиной 350~мкм при 460~{\textdegree}C.

Согласно~\cite{Rogers1992considerations, rogers1995selection} прогиб
пластины стекла также может быть вызван изменением распределения её состава
по толщине вследствие дрейфа ионов.
В~\cite{Sadaba2006CompositionalGradients} проведено
исследование влияния смоделированной неравномерности распределения состава
также и по площади пластины вследствие применения как точечного, так
и плоского электродов.
В~\cite{Rogers1992considerations} предлагается снижать возникающий прогиб
пластин подведением по завершении соединения тока обратной полярности для
изменения стехиометрии стекла или же проводить процесс соединения так,
чтобы в момент соединения пластины кремния и~стекла имели разные,
специально подобранные температуры.
Второй из~этих подходов был подробно проверен
в работе~\cite{Inzinga2012_infrared_characterization}.

В~\cite{maj2014influence_proceedings} показано средствами конечно-элементного моделирования, что сборка чувствительного элемента датчика давления в оптимальных условиях из~\cite{ettouhami1996thermal} даёт разброс механических напряжений не~менее 10~\% в~мембране, работающей в~интервале температур от~0~до~50~{\textdegree}C.

В работе~\cite{Kim2015warpage} описаны различные способы оценки остаточных напряжений,
возникающих в стекле, соединяемом с кремнием, с учётом исследований вязкости
и моделей вязкоупругости материалов.
В ней разработаны модели оценки остаточных напряжений и возникающего прогиба пластин, которые могут учитывать термообработку в виде выдержки соединённых деталей в~течение нескольких часов при температурах от 450 до 560~{\textdegree}C.
Также рассмотрено влияние скорости охлаждения на температуру стеклования.
Рассчитана возможность управления прогибом пластин за счёт подобной
термообработки.
Практические эксперименты в этой области были описаны ранее в
работе~\cite{Harz1996Curvature_changing}, и~некоторые реализации были
запатентованы в~\cite{engelke1998process_bend_patent} (в~настоящее время
патент не~поддерживается).

В~\cite{Cozma_Puers_1995} были представлены результаты измерений деформаций
при разных температурах соединения.
Было показано, что с ростом температуры, растут и~деформации.
Также там же было показано, что чем толще стеклянная пластина, тем больше
вызываемые механические напряжения в кремнии.
В~зависимости от~толщины стекла температура ненапряженного соединения
находится в~интервале от 265 до 315~{\textdegree}C~\cite{Cozma_Puers_1995}.
В~\cite{ettouhami1996thermal} было заявлено, что оптимальная температура
посадки с пирексом 270~{\textdegree}C.
В связи с этим, в целях минимизации последствий различия ТКЛР
электростатическое соединение кремния со~стеклом стараются проводить при
температурах около 300~{\textdegree}C~\cite{Low_temp_wafer_AB}.

Существует температура, при которой формируется соединение, не вызывающее механических напряжений~\cite{Cozma_Puers_1995}. В зависимости от температуры соединения можно получить сжимающие напряжения или растягивающие напряжения~\cite{Cozma_Puers_1995}. При сжимающих напряжениях, их носитель стремится расшириться, при растягивающих напряжениях "--- сжаться~\cites[19]{mehan_napr_plenki1981obzor}. Растягивающие напряжения считают положительными, сжимающие "--- отрицательными~\cites[5]{mehan_napr_plenki1981obzor}.

Снижение температуры электростатического соединения требует повышения прилагаемой разности потенциалов, поскольку в связи со снижением мобильности ионов, требуется большее усилие для их смещения~\cite{Low_temp_wafer_AB, Cozma_Puers_1995}.

\section{Постановка цели и~задач исследования}

\input{common/nedostatki}\beforenedostati{}
\beforenedostatii{}
\begin{enumerate}
    \item \nedostati{}.
    \item \nedostatii{}.
    \item \nedostatiii{}.
\end{enumerate}

На основании вышесказанного с целью
\MakeLowercase{\protect\aimTextContentRod{}}
задачи исследования можно сформулировать следующим образом:
\begin{enumerate}
\input{common/taskslist}
\end{enumerate}

%% file: Dissertation/part2.tex
\chapter{Исследование свойств применяемых стёкол}

В данной главе представлены результаты исследований состава стёкол марок
Borofloat~33, Corning~7740, ЛК5 и~Hoya~SD\nb-2.
Затем описана процедура определения их температурных коэффициентов линейного
расширения. Полученные значения аппроксимированы полиномиальными функциями
и~сравнены с~данными производителей.

\section{Литературные данные по маркам стекла, совместимым с~анодной посадкой} \label{GlassInfoOffic}

\subsection{Температурная зависимость коэффициентов теплового расширения}

Для описания температурной зависимости температурных коэффициентов линейного расширения используется приближение в виде полиномов:
\begin{equation*}
    P(T) = a + b \cdot T + c \cdot T^2 + d \cdot T^3 + e \cdot T^4 + f \cdot T^5,
\end{equation*}
где $a$, $b$, $c$, $d$, $e$, $f$ "--- коэффициенты полинома.
Подобным представлением удобно пользоваться как в аналитических моделях,
так и~при применении конечноэлементных вычислительных комплексов.

Коэффициенты полиномов для рассматриваемых стёкол с диапазоном применимости сведены в Таблицу~\ref{tab:glass_polynoms_official}. Производители не приводят данных по~температурной зависимости температурных коэффициентов линейного расширения своих стёкол в аналитической форме. Поэтому, при наличии среди данных производителей графиков зависимостей, полиномиальные зависимости получались аппроксимацией точек графиков.

\begin{table} [!ht]
    \centering%
    \caption[Polynomial coefficients of glass's CTE approximations, based
    on~published manufacturers' information]{Коэффициенты полиномов $\alpha (T)$, 1/K, на
    основании данных производителей стёкол}%
    \label{tab:glass_polynoms_official}% label всегда желательно идти после caption
    \tabulinesep=2.1mm
    \begin{SingleSpace}
    \begin{tabu} to 0.99\textwidth {@{}
    X[l,m]@{}
    S[table-format=-1.3, table-column-width = 1.9cm]@{}
    S[table-format=-1.3, table-column-width = 1.9cm]@{}
    S[table-format=-2.3, table-column-width = 2.1cm]@{}
    S[table-format=-2.3, table-column-width = 2.1cm]@{}
    S[table-format=-1.3, table-column-width = 2.0cm]@{}
    S[table-format=-1.3, table-column-width = 1.7879cm]
    >{\raggedleft}m{0.83cm}@{---}>{\raggedright\arraybackslash}m{0.85cm}%
    @{}}
        \toprule     %%% верхняя линейка
        {Марка стекла} &
        {\(a\), 10\textsuperscript{\textminus 6}} &
        {\(b\), 10\textsuperscript{\textminus 8}} &
        {\(c\), 10\textsuperscript{\textminus 11}}&
        {\(d\), 10\textsuperscript{\textminus 14}}&
        {\(e\), 10\textsuperscript{\textminus 16}} &
        {\(f\), 10\textsuperscript{\textminus 19}} &
        \multicolumn{2}{>{\centering\arraybackslash}m{2.3cm}@{}}{Диапа\-зон, K}\\
        \midrule
        Corning 7740 &
        6,975 &
        -5,405 &
        28,730 &
        -69,350 &
        7,694 &
        -3,167 &
        323&673\\
        Schott Borofloat~33 &
        -1.195 &
        3.574 &
        -11.760 &
        20.540 &
        -1.919 &
        0.748 &
        153&773\\
        ЛК5 & 2.737 & 0.222 & 0.000 & 0.000 & 0.000 & 0.000 & 213&393\\
        Hoya SD\nobreakdash-2 &
        -3,382 &
        2,750 &
        -2,723 &
        -2,496 &
        0,628 &
        -0,296 &
        373&773\\
        Asahi SW\nobreakdash-YY &
        -3,879 &
        6,526 &
        -27,290 &
        59,860 &
        -6,408 &
        2,638 &
        303&723\\
        \bottomrule %%% нижняя линейка
    \end{tabu}%
    \end{SingleSpace}
\end{table}

Зависимости ТКЛР от температуры для стёкол марок SD\nb-2 и SW\nb-YY получены из графиков в~\cite{SD_2_properties} и~\cite{swyy_properties}, соответственно.
Зависимости ТКЛР от~температуры для стёкол марок 7740 и Borofloat~33 получены расчётным путём из~графиков температурных зависимостей относительного удлинения от~температуры, приведённых в~\cite{corning7740_wafersheet} и~\cite{bf33_properties}.
Расчёт полиномиальной зависимости ТКЛР для стекла марки ЛК5 был проведён на
основании данных по средним \mbox{ТКЛР} в~двух температурных диапазонах,
указанных производителем~\cite{LK5_properties}, исходя из~предположения,
что температурный ход значения коэффициента в~области «нормального»
состояния стёкол, от минус 120~{\textdegree}C до температуры нижней границы
зоны отжига, практически выражается линейным
уравнением~\cites[180]{Mazurin1969_Tepl_rassh_stekla}.

На основании тех же литературных источников в Таблице~\ref{tab:glass_cteavg_official} представлены средние ТКЛР для рассмотренных стёкол.

\begin{figure}[!htb]
    \centering
    \begingroup%
      \makeatletter%
      \providecommand\color[2][]{%
        \errmessage{(Inkscape) Color is used for the text in Inkscape, but the package 'color.sty' is not loaded}%
        \renewcommand\color[2][]{}%
      }%
      \providecommand\transparent[1]{%
        \errmessage{(Inkscape) Transparency is used (non-zero) for the text in Inkscape, but the package 'transparent.sty' is not loaded}%
        \renewcommand\transparent[1]{}%
      }%
      \providecommand\rotatebox[2]{#2}%
      \ifx\svgwidth\undefined%
        \setlength{\unitlength}{0.6\textwidth}%
        \ifx\svgscale\undefined%
          \relax%
        \else%
          \setlength{\unitlength}{\unitlength * \real{\svgscale}}%
        \fi%
      \else%
        \setlength{\unitlength}{\svgwidth}%
      \fi%
      \global\let\svgwidth\undefined%
      \global\let\svgscale\undefined%
      \makeatother%
      \begin{picture}(1,0.91245741)%
        \put(0,0){\includegraphics[width=\unitlength]{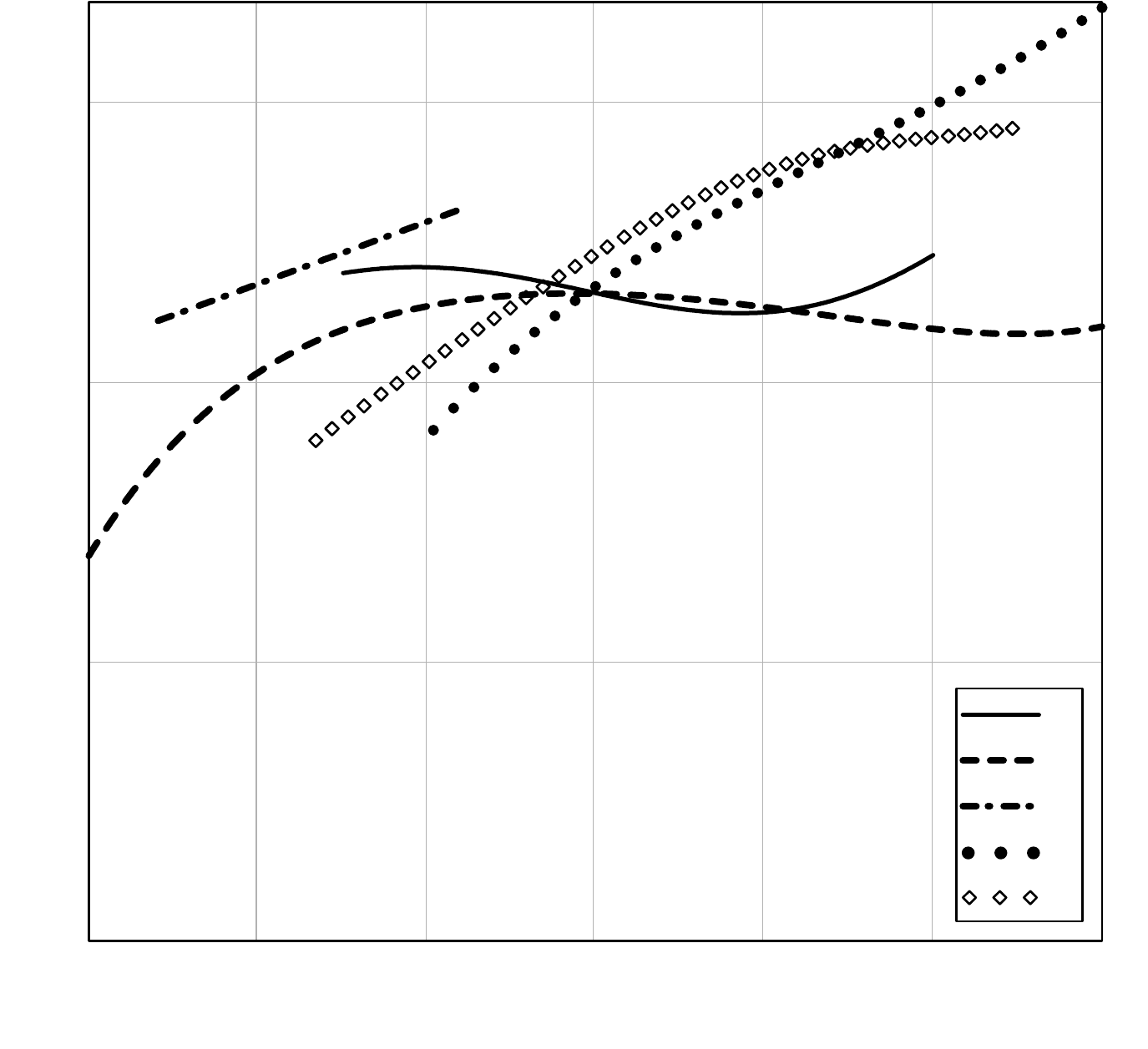}}%
        \put(0.911,0.279){\color[named]{black}\makebox(0,0)[lb]{\smash{\textsl{1}}}}%
        \put(0.911,0.239){\color[named]{black}\makebox(0,0)[lb]{\smash{\textsl{2}}}}%
        \put(0.911,0.199){\color[named]{black}\makebox(0,0)[lb]{\smash{\textsl{3}}}}%
        \put(0.911,0.159){\color[named]{black}\makebox(0,0)[lb]{\smash{\textsl{4}}}}%
        \put(0.911,0.120){\color[named]{black}\makebox(0,0)[lb]{\smash{\textsl{5}}}}%
        \put(0.04942711,0.05401028){\color[named]{black}\makebox(0,0)[lb]{\smash{$-$100}}}%
        \put(0.22365622,0.05401028){\color[named]{black}\makebox(0,0)[lb]{\smash{0}}}%
        \put(0.3564527,0.05401028){\color[named]{black}\makebox(0,0)[lb]{\smash{100}}}%
        \put(0.50245287,0.05401028){\color[named]{black}\makebox(0,0)[lb]{\smash{200}}}%
        \put(0.65042609,0.05401028){\color[named]{black}\makebox(0,0)[lb]{\smash{300}}}%
        \put(0.79839931,0.05401028){\color[named]{black}\makebox(0,0)[lb]{\smash{400}}}%
        \put(0.066,0.32142437){\color[named]{black}\makebox(0,0)[rb]{\smash{2}}}%
        \put(0.066,0.56622527){\color[named]{black}\makebox(0,0)[rb]{\smash{3}}}%
        \put(0.066,0.81087425){\color[named]{black}\makebox(0,0)[rb]{\smash{4}}}%
        \put(0.02458798,0.47333211){\color[named]{black}\rotatebox{90}{\makebox(0,0)[b]{\smash{$\alpha$, 10\textsuperscript{$-$6}~{\textdegree}C\textsuperscript{$-$1}}}}}%
        \put(0.94814372,0.05401028){\color[named]{black}\makebox(0,0)[lb]{\smash{500}}}%
        \put(0.066,0.08264282){\color[named]{black}\makebox(0,0)[rb]{\smash{1}}}%
        \put(0.51958915,0.0075735){\color[named]{black}\makebox(0,0)[b]{\smash{$T$,~{\textdegree}C}}}%
      \end{picture}%
    \endgroup%
    \caption[Coefficients of thermal expansion of several glass brands
    and silicon]{Зависимости температурных коэффициентов линейного
    расширения от температуры для нескольких марок стекла и кремния:}
    \label{fig:cte_offic}
    \legend{%
        \textsl{1} "--- Corning 7740,  \textsl{2} "--- Schott Borofloat 33,  \textsl{3} "--- ЛК5,  \textsl{4} "--- Hoya~SD-2,  \textsl{5}~---~Asahi~SW\nb-YY%
    }
\end{figure}

На Рисунке~\ref{fig:cte_offic}
приведены графики зависимостей ТКЛР
от температуры из~Таблицы~\ref{tab:glass_polynoms_official}.

\begin{table} [!htb]
    \centering%
    \parbox{0.8\textwidth}{
	\caption[Average CTE of glass brands, based on published data]{Средние ТКЛР для стёкол по литературным данным}%
	\label{tab:glass_cteavg_official}% label всегда желательно идти после caption
	}
    \tabulinesep=1.8mm
	\begin{SingleSpace}
	\begin{tabu} to 0.8\textwidth {@{}
	X[l]
	>{\raggedleft}m{0.9cm}@{---}>{\raggedright}m{0.9cm}%
	S[table-format=1.2]
	@{}}
        \toprule     %%% верхняя линейка
        {Марка стекла} &
        \multicolumn{2}{>{\centering}m{3.0cm}}{Диапа\-зон,~{\textdegree}C} &
        {Средний ТКЛР, 10\textsuperscript{$-$6}~{\textdegree}C\textsuperscript{$-$1}}\\
        \midrule
        Corning 7740 &
        0&300 &
        3,25\\
        Schott Borofloat~33 &
        20&300 &
        3,25\\
        \multirow{2}{*}{ЛК5} &
        $-$60&20 &
        3,30\\
         &
        20&120 &
        3,50\\
        Hoya SD\nobreakdash-2 &
        20&300 &
        3,20\\
        Asahi SW\nobreakdash-YY &
        30&300 &
        3,30\\
        \bottomrule %%% нижняя линейка
	\end{tabu}%
	\end{SingleSpace}
\end{table}

\subsection{Упругие свойства}

В Таблице~\ref{table_glass_elasticity} приведены упругие свойства (модуль Юнга, \(E\), ГПа, и коэффициент Пуассона,  \(\mu\))  стёкол Hoya SD\nobreakdash-2~\cite{SD_2_properties}, Asahi SW\nobreakdash-YY~\cite{swyy_properties}, Corning~7740~\cite{corning7740_wafersheet}, Schott Borofloat~33~\cite{bf33_properties} и ЛК5~\cite{LK5_properties}.

\begin{table} [!htb]%
    \centering
    \parbox{0.5\textwidth}{
    	\caption[Elasticity characteristics of glass brands discussed]{Упругие свойства рассматриваемых марок стёкол}%
    	\label{table_glass_elasticity}% label всегда желательно идти после caption
	}
    \renewcommand{\arraystretch}{1.15}%% Увеличение расстояния между рядами, для улучшения восприятия.
	\def\tabularxcolumn#1{m{#1}}
    \sisetup{round-mode = places}
	\begin{SingleSpace}
	\begin{tabularx}{0.5\textwidth}{@{}
	>{\raggedright}X
	S[round-precision = 2]
	S[round-precision = 3]
	@{}}
        \toprule     %%% верхняя линейка
        {Марка стекла} &
        {\(E\), ГПа} &
        {$\mu$}\\
        \midrule %%% тонкий разделитель. Отделяет названия столбцов. Обязателен по ГОСТ 2.105 пункт 4.4.5

        Corning 7740 &
        62,75 &
        0,2\\
        Schott Borofloat 33 &
        64,0 &
        0,2\\
        ЛК5 &
        68,45 &
        0,184\\
        Hoya~SD-2 &
        86,89 &
        0,244 \\
        Asahi SW-YY &
        82,0 &
        0,2\\
        \bottomrule %%% нижняя линейка
	\end{tabularx}%
	\end{SingleSpace}
\end{table}

\Needspace*{6\onelineskip}
\section{Исследование состава стёкол}

<<Рентгеновская фотоэлектронная спектроскопия (РФЭС) "--- количественный
спектроскопический метод исследования элементного состава, эмпирической
формулы, химического и электронного состояния атомов, присутствующих в
материале.
Он основан на явлении внешнего фотоэффекта.
Спектры \mbox{РФЭС} получают облучением материала пучком рентгеновских
лучей с регистрацией зависимости количества испускаемых электронов
от~их~кинетической энергии.
Исследуемые электроны испускаются верхним слоем исследуемого материала
толщиной от 1 до 10 нм.
РФЭС проводится в~сверхвысоком вакууме>>~\cite{wiki_arxps}.

<<РФЭС "--- метод химического анализа поверхности, который может быть
использован для анализа химического состояния материала как в~его
первоначальном состоянии, так и после некоторой обработки, например скола,
разреза или очистки в воздухе или сверхвысоком вакууме для исследования
внутреннего химического состава образца, облучения высокоэнергетическим
пучком ионов для очистки поверхности от загрязнений, нагрева образца, чтобы
изучить изменения вследствие нагревания, помещения в~атмосферу реактивного
газа или раствора, облучения ионами с целью их~внедрения, облучения
ультрафиолетовым светом>>~\cite{wiki_arxps}.

На установке Theta Probe {Angle-Resolved} {X-ray} Photoelectron Spectrometer (ARXPS) System методом рентгеновской фотоэлектронной спектроскопии были проведены измерения образцов стёкол марок Borofloat~33, Corning~7740, ЛК5 и~Hoya~SD\nb-2.
Измерения проводились
научным сотрудником ФГУП <<ВНИИА им.~Н.\,Л.~Духова>>
А.\:Ю.~Переяславцевым.
Полученный химический поэлементный состав приведён
в~Таблице~\ref{tab:results_arxps_glass}.

\begin{table} [!ht]
    \centering%
    \parbox{0.86\textwidth}{
        \caption[Experimentally obtained chemical composition of glass brands]{Результаты измерения химического состава стёкол}%
        \label{tab:results_arxps_glass}% label всегда желательно идти после caption
    }
    \renewcommand{\arraystretch}{1.3}%% Увеличение расстояния между рядами, для улучшения восприятия.
    \begin{SingleSpace}
    \begin{tabularx}{0.86\textwidth}{@{}
    >{\raggedright}X
    llllllll@{}}
    \toprule
        \multirow{2}{*}{Марка стекла}
        &
        \multicolumn{8}{c}{Компоненты стекла, ат.~\%} \\
    \cmidrule(l){2-9}
        & O & Si & B & Al & Na & K & Mg & Zn \\
    \midrule
        Borofloat 33 & 64,44 &  30,60 &  3,37 &   1,12 &   0,33 &  0,14 &   0,00 &   0,00 \\
        Corning 7740 & 63,68 &  29,20 &  5,24 &   1,26 &   0,47 &  0,15 &   0,00 &   0,00 \\
        ЛК5          & 63,19 &  31,28 &  2,55 &   1,34 &   0,93 &  0,07 &   0,64 &   0,00 \\
        Hoya SD\nb-2 & 63,59 &  23,72 &  1,11 &   9,29 &   0,56 &  0,12 &   0,95 &   0,67 \\
    \bottomrule %%% нижняя линейка
    \end{tabularx}%
    \end{SingleSpace}
\end{table}

Для удобства сравнения с данными литературных источников был проведён
пересчёт результатов из Таблицы~\ref{tab:results_arxps_glass} в формат
оксидного состава с~допущением о простом строении оксидов.
Описание процедуры пересчёта приведено в~подразделе~\ref{GlassCompositionProgram}.
Результат пересчёта приведён в Таблице~\ref{tab:glass_oxides_calc}.

\begin{table} [!ht]
    \centering%
    \parbox{0.8\textwidth}{
    	\caption[Glass chemical composition recalculated for oxide form]{Химический состав стёкол, пересчитанный в оксидной форме}%
    	\label{tab:glass_oxides_calc}% label всегда желательно идти после caption
	}
    \renewcommand{\arraystretch}{1.3}%% Увеличение расстояния между рядами, для улучшения восприятия.
    \sisetup{
        table-number-alignment = center,
        table-text-alignment = center,
        table-format=2.1,
        round-mode = places,
        round-precision = 1
    }%
    \begin{SingleSpace}
    \begin{tabularx}{0.8\textwidth}{@{}
    >{\raggedright}X
    SSSSSSS@{}}
    \toprule
        \multirow{2}{*}{Марка стекла}
        &
        \multicolumn{7}{c}{Оксидный состав, вес.~\%} \\
    \cmidrule(l){2-8}
        & {SiO\textsubscript{x}} & {B\textsubscript{y}O\textsubscript{z}} & {Al\textsubscript{2}O\textsubscript{3}} & {Na\textsubscript{2}O} & {K\textsubscript{2}O} & {MgO} & {ZnO} \\
    \midrule
        Borofloat 33 & 83.5 & 12.8 &  2.9 & 0.5 & 0.3 & 0.0 & 0.0 \\
        Corning 7740 & 75.4 & 20.1 &  3.3 & 0.7 & 0.4 & 0.0 & 0.0 \\
        ЛК5          & 84.1 &  9.6 &  3.4 & 1.4 & 0.2 & 1.3 & 0.0 \\
        Hoya SD\nb-2 & 66.9 &  4.1 & 23.3 & 0.9 & 0.3 & 1.9 & 2.7 \\
    \bottomrule %%% нижняя линейка
	\end{tabularx}%
    \end{SingleSpace}
\end{table}

\subsection{Программа пересчёта состава стекла}\label{GlassCompositionProgram}

Пересчёт состава стекла проводился при помощи программы на языке \verb|R|. Ниже приводится листинг программы с комментариями и выдачей, сформированный с помощью пакета \verb|knitr|, и запускавшийся в среде  \verb|RStudio|.

\subsubsection{Загрузка исходных данных}

\begin{lstlisting}[language=Renhanced]
data <- read.csv2("composition20151006.csv",stringsAsFactors=FALSE)
\end{lstlisting}

Заведение атомных весов компонентов как констант.

\begin{lstlisting}[language=Renhanced]
aem_o  <- 15.999
aem_si <- 28.086
aem_b  <- 10.811
aem_al <- 26.982
aem_na <- 22.99
aem_k  <- 39.098
aem_mg  <- 24.304
aem_zn  <- 65.382
\end{lstlisting}

Вывод исходных измерений:

\begin{lstlisting}[language=Renhanced]
data
\end{lstlisting}

\begin{Verb}
##          Glass O.atomic Si.atomic B.atomic Al.atomic
## 1 Borofloat 33    64.44     30.60     3.37      1.12
## 2 Corning 7740    63.68     29.20     5.24      1.26
## 3    Hoya SD-2    63.59     23.72     1.11      9.29
## 4          LK5    63.19     31.28     2.55      1.34
##   Na.atomic K.atomic Mg.atomic Zn.atomic
## 1      0.33     0.14      0.00      0.00
## 2      0.47     0.15      0.00      0.00
## 3      0.56     0.12      0.95      0.67
## 4      0.93     0.07      0.64      0.00
\end{Verb}

\subsubsection{Обработка данных}

Оставим только данные по атомарным процентам элементов. Весовые можно из
них высчитать.

\begin{lstlisting}[language=Renhanced]
# оставляем только те столбцы, которые содержат слова atomic и Glass
data_atom <- data[,grep("atomic|Glass",names(data))]
data_atom
\end{lstlisting}

\begin{Verb}
##          Glass O.atomic Si.atomic B.atomic Al.atomic
## 1 Borofloat 33    64.44     30.60     3.37      1.12
## 2 Corning 7740    63.68     29.20     5.24      1.26
## 3    Hoya SD-2    63.59     23.72     1.11      9.29
## 4          LK5    63.19     31.28     2.55      1.34
##   Na.atomic K.atomic Mg.atomic Zn.atomic
## 1      0.33     0.14      0.00      0.00
## 2      0.47     0.15      0.00      0.00
## 3      0.56     0.12      0.95      0.67
## 4      0.93     0.07      0.64      0.00
\end{Verb}

Посчитаем общую массу в а. е. м. (через доли)

\begin{lstlisting}[language=Renhanced]
# Для удобства собираем вектор атомных масс в том же порядке, что и столбцы (посмотрели это в выдаче выше). Вообще можно анализировать названия столбцов и подставлять.
aem <- c(aem_o,aem_si,aem_b,aem_al,aem_na,aem_k,aem_mg,aem_zn)

#Загрузка библиотеки для некоторых дальнейших функций
library(dplyr, warn.conflicts = FALSE)

#Разделяем по стёклам состав
splitted <- split(data_atom[,grep("atomic",names(data))],data_atom$Glass)

# Поэлементно множим на атомную массу
mass <- lapply(splitted, function(x) x*aem/100)
mass <- unsplit(mass,data_atom$Glass)
names(mass) <- gsub(".atomic","", names(mass))

#Собираем таблицу (dataframe) из сумм масс элементов постекольно (rowSums)
#Сначала прицепляем к имеющейся таблице
total_atomic_mass <- cbind(data_atom, total_mass = rowSums(mass))

#оставляем только то, что нужно (можно обойтись и без функции select из dplyr)
total_atomic_mass <- select(total_atomic_mass,Glass,total_mass)

total_atomic_mass
\end{lstlisting}

\begin{lstlisting}
##          Glass total_mass
## 1 Borofloat 33   19.70120
## 2 Corning 7740   19.46244
## 3    Hoya SD-2   20.30700
## 4          LK5   19.92903
\end{lstlisting}

Весовые проценты компонентов.

\begin{lstlisting}[language=Renhanced]
data.frame(mass/rowSums(mass)*100, row.names = data$Glass, stringsAsFactors = FALSE)
\end{lstlisting}

\begin{Verb}
##                     O       Si         B        Al
## Borofloat 33 52.33058 43.62330 1.8492813  1.533908
## Corning 7740 52.34781 42.13814 2.9107155  1.746817
## Hoya SD-2    50.09978 32.80641 0.5909395 12.343662
## LK5          50.72885 44.08293 1.3833112  1.814232
##                     Na         K        Mg       Zn
## Borofloat 33 0.3850881 0.2778368 0.0000000 0.000000
## Corning 7740 0.5551872 0.3013342 0.0000000 0.000000
## Hoya SD-2    0.6339882 0.2310415 1.1369871 2.157184
## LK5          1.0728420 0.1373303 0.7804976 0.000000
\end{Verb}

\subsubsection{Таблица оксидов в весовых процентах}

Сформируем таблицу оксидов в весовых процентах.

\begin{lstlisting}[language=Renhanced]
# названия оксидов
oxide_names <- c("SiOx","BxOx","Al2O3","Na2O","K2O","MgO","ZnO")

# Коэффициент процента атомов кислорода уходящих на соединение с K
ok <- 1/2

# Коэффициент процента атомов кислорода уходящих на соединение с Na
ona <- 1/2

# Коэффициент процента атомов кислорода уходящих на соединение с Al
oal <- 3/2

# Коэффициент процента атомов кислорода уходящих на соединение с B
ob <- 4

# Коэффициент процента атомов кислорода уходящих на соединение с Mg
omg <- 1

# Коэффициент процента атомов кислорода уходящих на соединение с Zn
ozn <- 1

# Вектор коэффициентов для всего, кроме кремния
oxide_coef <- c(ob, oal, ona, ok, omg, ozn)

# на что тратим кислород (без кремния)
oxygen_split <- data_atom[,grep("Glass|Si|O",names(data_atom),invert = TRUE)] %>% split(data_atom$Glass) %>%
    lapply(function(x) x*oxide_coef/100) %>% unsplit(data_atom$Glass)
# правильные имена столбцам
names(oxygen_split) <- gsub(".atomic","", names(oxygen_split))

# Остальное уйдёт на кремний
# Проценты кислорода на кремний
oxygen_split_si <- data_atom[,grep("O",names(data_atom))]/100 - rowSums(oxygen_split)

# на что тратим кислород (полный)
oxygen_split <- cbind(Si=oxygen_split_si,oxygen_split)

#(Массу элементов + с массой распределенного кислорода)/полная масса * 100
oxides <- (select(mass,-O)+oxygen_split*aem_o)/total_atomic_mass[,2]*100
# правильные имена столбцам
names(oxides) <- oxide_names
# оформляем в таблицу (data frame)
oxides <- data.frame(oxides, row.names = data_atom$Glass, stringsAsFactors = FALSE)
#oxides
print("Таблица оксидов в весовых процентах (wt. %)")
print(round(oxides, digits = 1))
\end{lstlisting}

\begin{Verb}
## [1] "Таблица оксидов в весовых процентах (wt. %)"
##              SiOx BxOx Al2O3 Na2O K2O MgO ZnO
## Borofloat 33 83.5 12.8   2.9  0.5 0.3 0.0 0.0
## Corning 7740 75.4 20.1   3.3  0.7 0.4 0.0 0.0
## Hoya SD-2    66.9  4.1  23.3  0.9 0.3 1.9 2.7
## LK5          84.1  9.6   3.4  1.4 0.2 1.3 0.0
\end{Verb}

\subsubsection{Таблица оксидов в молярных процентах}

Сформируем таблицу оксидов в молярных процентах.

\begin{lstlisting}[language=Renhanced]
#молярная масса компонентов (без изменяющегося для каждого стекла SiOx)
molar_mass_wo_siox <- c(aem_b*1+aem_o*4, aem_al*2+aem_o*3, aem_na*2+aem_o, aem_k*2+aem_o, aem_mg+aem_o, aem_zn+aem_o)

#молярная масса компонентов
molar_mass<-as.data.frame(
        t(#приходится транспонировать, потому что unsplit не хочет срабатывать, а as.data.frame выдает плохой результат
            as.data.frame(
                lapply(#в цикле дополняем постоянной частью
                        split(aem_si+aem_o*oxygen_split$Si*100/data_atom$Si, data_atom$Glass)
                    ,function(x) c(x,molar_mass_wo_siox))
            ,optional = TRUE)
        )
    )

# правильные имена столбцам
names(molar_mass) <- oxide_names

# Условно Число моль в 100 г всего вещества
# Весовой процент делим на молярную массу оксидов, затем относим к сумме получившихся молей и превращаем в проценты
molar_weight <- (oxides/molar_mass)/rowSums(oxides/molar_mass)*100

#molar_weight
print("Таблица оксидов в молярных процентах (mol. %)")
print(round(molar_weight, digits = 1))
\end{lstlisting}

\begin{Verb}
## [1] "Таблица оксидов в молярных процентах (mol. %)"
##              SiOx BxOx Al2O3 Na2O K2O MgO ZnO
## Borofloat 33 88.0  9.7   1.6  0.5 0.2 0.0 0.0
## Corning 7740 82.5 14.8   1.8  0.7 0.2 0.0 0.0
## Hoya SD-2    75.5  3.5  14.8  0.9 0.2 3.0 2.1
## LK5          87.8  7.2   1.9  1.3 0.1 1.8 0.0
\end{Verb}

\subsubsection{Израсходовано кислорода (в процентах к общему числу
атомов)}

\begin{lstlisting}[language=Renhanced]
#print("А вот проценты трат кислорода (at. %)")
#print(oxygen_split, digits=1)
oxygen_split*100
\end{lstlisting}

\begin{Verb}
##       Si     B     Al    Na     K   Mg   Zn
## 1 49.045 13.48  1.680 0.165 0.070 0.00 0.00
## 2 40.520 20.96  1.890 0.235 0.075 0.00 0.00
## 3 43.255  4.44 13.935 0.280 0.060 0.95 0.67
## 4 49.840 10.20  2.010 0.465 0.035 0.64 0.00
\end{Verb}

\subsubsection{Дополнительная информация по использованному программному комплекту}
\begin{lstlisting}[language=Renhanced]
sessionInfo()
\end{lstlisting}

\begin{Verb}
## R version 3.2.2 (2015-08-14)
## Platform: x86_64-w64-mingw32/x64 (64-bit)
## Running under: Windows 7 x64 (build 7601) Service Pack 1
##
## locale:
## [1] LC_COLLATE=Russian_Russia.1251
## [2] LC_CTYPE=Russian_Russia.1251
## [3] LC_MONETARY=Russian_Russia.1251
## [4] LC_NUMERIC=C
## [5] LC_TIME=Russian_Russia.1251
##
## attached base packages:
## [1] stats     graphics  grDevices utils     datasets
## [6] methods   base
##
## other attached packages:
## [1] dplyr_0.4.3
##
## loaded via a namespace (and not attached):
##  [1] Rcpp_0.12.1     digest_0.6.8    assertthat_0.1
##  [4] R6_2.1.1        DBI_0.3.1       formatR_1.2.1
##  [7] magrittr_1.5    evaluate_0.8    stringi_0.5-5
## [10] lazyeval_0.1.10 rmarkdown_0.8   tools_3.2.2
## [13] stringr_1.0.0   yaml_2.1.13     parallel_3.2.2
## [16] htmltools_0.2.6 knitr_1.11
\end{Verb}

\section{Определение температурных коэффициентов линейного расширения стёкол}\label{chap_cte_measure}
\subsection{Общее описание процедуры исследования}
\begingroup
Для получения экспериментальных данных в интервале температур от~130 до~800~K (от~минус~143~до~526~{\textdegree}C) для стёкол Borofloat 33 и ЛК5 использовался термомеханический анализатор TMA 7100 SII Nanotechnology.
Прибор измеряет линейные размеры образца в условиях тепловых и
механических нагрузок в~кварцевой дилатометрической ячейке типа
Хеннинга~\cites[58]{novikova1974}.\russianpar
\endgroup

\begin{figure}[htb]
    \centering
    \begingroup%
      \makeatletter%
      \providecommand\color[2][]{%
        \errmessage{(Inkscape) Color is used for the text in Inkscape, but the package 'color.sty' is not loaded}%
        \renewcommand\color[2][]{}%
      }%
      \providecommand\transparent[1]{%
        \errmessage{(Inkscape) Transparency is used (non-zero) for the text in Inkscape, but the package 'transparent.sty' is not loaded}%
        \renewcommand\transparent[1]{}%
      }%
      \providecommand\rotatebox[2]{#2}%
      \ifx\svgwidth\undefined%
        \setlength{\unitlength}{102.0841175bp}%
        \ifx\svgscale\undefined%
          \relax%
        \else%
          \setlength{\unitlength}{\unitlength * \real{\svgscale}}%
        \fi%
      \else%
        \setlength{\unitlength}{\svgwidth}%
      \fi%
      \global\let\svgwidth\undefined%
      \global\let\svgscale\undefined%
      \makeatother%
      \begin{picture}(1,2.78781632)%
        \put(0,0){\includegraphics[width=\unitlength]{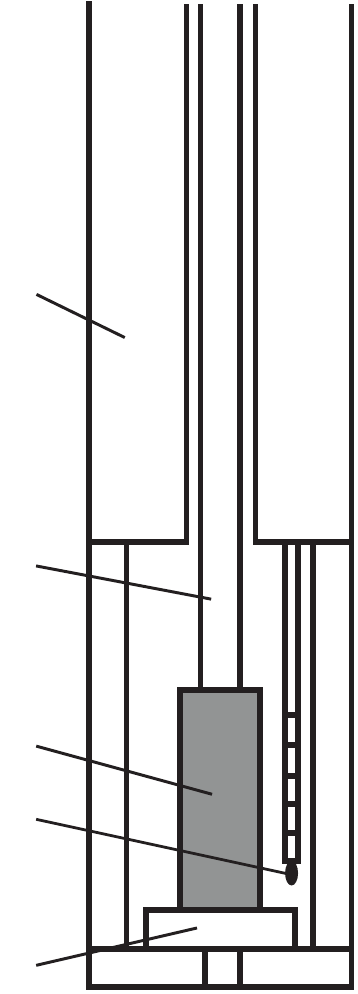}}%
        \put(0.08226702,1.93639265){\color[named]{black}\makebox(0,0)[rb]{\smash{\textsl{1}}}}%
        \put(0.07624028,1.1771024){\color[named]{black}\makebox(0,0)[rb]{\smash{\textsl{2}}}}%
        \put(0.0780196,0.65991058){\color[named]{black}\makebox(0,0)[rb]{\smash{\textsl{3}}}}%
        \put(0.07411657,0.44885082){\color[named]{black}\makebox(0,0)[rb]{\smash{\textsl{4}}}}%
        \put(0.07503493,0.033652){\color[named]{black}\makebox(0,0)[rb]{\smash{\textsl{5}}}}%
      \end{picture}%
    \endgroup%
    \caption[Arrangement of the measured sample in a thermomechanical
    analyzer]{Схема размещения измеряемого образца в термомеханическом
    анализаторе~\cite{Sinev_Petrov2016_cte_glass}:}
    \label{fig:scheme_sample_measure}
    \legend{%
        \textsl{1} "--- колба, \textsl{2} "--- зонд, \textsl{3} "--- измеряемый образец, \textsl{4} "--- термопара, \textsl{5} "--- пьедестал%
    }
\end{figure}

Измеряли по три образца длиной (20$\pm$0,1)~мм для каждой марки стекла, результаты измерений затем усредняли~\cite{Sinev_Petrov2016_cte_glass}.
Образцы устанавливали согласно Рисунку~\ref{fig:scheme_sample_measure} в соответствии с инструкцией на термомеханический анализатор.
Линейное приращение длины и длину образца при комнатной температуре измеряли с помощью зонда из кварцевого стекла, опирающегося на~образец с~силой 0,2~Н.
Перед началом измерений термокамеру анализатора охлаждали.
Последующий нагрев проводили со скоростью 10 K/мин.
Значения температуры и~линейного приращения длины регистрировались автоматически программным обеспечением термомеханического анализатора через равные промежутки времени.
Для каждого образца было получено по~20~экспериментальных точек в~интервале
от~129 до~800~K с~шагом не~менее 35~K.
По~окончании измерений для каждого образца проводили программную коррекцию
искажений, вносимых материалом колбы и зонда, с~использованием эталона
боросиликатного стекла SRM731.

Аналогичным образом были исследованы образцы стёкол марок Corning~7740 и SD\nb-2  в диапазоне от 170 до 780~K (от минус 100 до 500~{\textdegree}C).

\subsection{Расчёт погрешностей}
Значения
температурных коэффициентов линейного расширения
$\alpha (T)$, 1/K, были рассчитаны следующим образом в соответствии с определением \mbox{ТКЛР}~\cites[150]{Mazurin1969_Tepl_rassh_stekla}[44]{novikova1974}:
\begin{equation}
  \alpha (T)=\frac{\Delta L}{L_0 \Delta T},
\end{equation}
\[
    T=\frac{T_i+T_{i+1}}{2},
\]
\[
    \Delta T=T_{i+1}-T_i,
\]
\[
    \Delta L=L(T_{i+1})-L(T_i),
\]
где $T$ "--- средняя температура между $i$ и $i+1$ измеренными значениями, K;  $\Delta L$ "--- разница между $i+1$ и $i$ измеренными значениями длины, м; $L_0$ "--- исходная длина образца при комнатной температуре $T_0 =$~293,15~K, м;  $\Delta T$ "--- разница между $i+1$ и $i$ измеренными значениями температуры, K; $T_i$ "--- $i$\nb-ое измеренное значение температуры, K; $L(T_i)$ "--- измеренное значение длины образца при температуре $T_i$, м.

Температурная зависимость относительного теплового удлинения образца,  $\epsilon^T(T_i)$, рассчитывалась по следующей формуле:

\begin{equation}
\epsilon^T(T_i)=\frac{L(T_i)-L_0}{L_0}
\end{equation}

Согласно~\cite{ref_gosreestr_tma_ss} для данной измерительной установки пределы допускаемой относительной погрешности измерений линейных приращений длины составляют ${\pm}$3~\%, пределы допускаемой абсолютной погрешности измерений температуры равны ${\pm}$1~{\textdegree}C.

Была проведена оценка относительных погрешностей косвенных измерений ТКЛР
и относительного удлинения в точках их максимальных значений в~рассматриваемом
интервале температур.
По формуле~\eqref{eq:cte_uncertainty} была рассчитана относительная погрешность
ТКЛР согласно~\cites[58]{KassandrovaLebedev1970Obrabotka}:
\begin{equation}
    \label{eq:cte_uncertainty}
    \frac{\delta \alpha}{\alpha}=
    \sqrt{
        \left(
            \frac{\delta (\Delta L)}{\Delta L}
        \right)^{\!\!2} %притягиваем степень к скобкам
        +
        \left(
            \frac{\delta (\Delta T)}{\Delta T}
        \right)^{\!\!2} %притягиваем степень к скобкам
        +
        \left(
            \frac{\delta L_0}{L_0}
        \right)^{\!\!2} %притягиваем степень к скобкам
    },
\end{equation}
где относительные погрешности измерений:
$\frac{\delta (\Delta L)}{\Delta L}$ "--- линейных приращений длины;
$\frac{\delta (\Delta T)}{\Delta T}$ "--- разницы температур;
$\frac{\delta L_0}{L_0}$ "--- длины образца при комнатной температуре.

Значения относительной погрешности измерений линейных приращений длины были
взяты из данных, приведённых в~\cite{ref_gosreestr_tma_ss}. Относительная
погрешность измерений разницы температур была определена в соответствии
с~\cites[57]{KassandrovaLebedev1970Obrabotka} по~формуле:
\begin{equation}
    \frac{\delta (\Delta T)}{\Delta T}=\frac{\delta T\cdot \sqrt 2}{\Delta T},
\end{equation}
где  $\delta T$ "--- абсолютная погрешность измерений температуры, взятая из~\cite{ref_gosreestr_tma_ss}, а~в~качестве величины  $\Delta T$ было взято минимальное значение разницы температур из~всей последовательности измерений в заявленном температурном интервале.

Относительная погрешность измерений длины образца при комнатной температуре с
учётом данных по относительной погрешности измерений линейных приращений длины
измерительной установкой~\cite{ref_gosreestr_tma_ss} была определена
в~соответствии с~\cites[57]{KassandrovaLebedev1970Obrabotka} следующим образом:
\begin{equation}
    \label{eq:otn_deltaL0}
    \frac{\delta L_0}{L_0}=\frac{\delta (\Delta L)}{\Delta L}\cdot \frac{\Delta L}{\sqrt 2L_0},
\end{equation}
где в качестве величины  $\Delta L$ было взято максимальное значение удлинений образца из всей последовательности измерений в заявленном температурном интервале.

В формуле~\eqref{eq:elongation_uncertainty} приведена оценка относительной
погрешности измерения относительного удлинения образца
согласно~\cites[58]{KassandrovaLebedev1970Obrabotka}:
\begin{equation}
    \label{eq:elongation_uncertainty}
    \frac{\delta \epsilon^T}{\epsilon^T}
    =
    \sqrt{
        \left(
            \frac{\delta (\Delta L)}{\Delta L}
        \right)^{\!\!2} %притягиваем степень к скобкам
        +
        \left(
            \frac{\alpha \delta T}{\epsilon^T}
        \right)^{\!\!2} %притягиваем степень к скобкам
        +
        \left(
            \frac{\delta L_0}{L_0}
        \right)^{\!\!2} %притягиваем степень к скобкам
    },
\end{equation}
где выражение  $\displaystyle\frac{\alpha \delta T}{\epsilon^T}$
введено аналогично описанному в~\cite{LTEC_si_293_1000}
для учёта зависимости  $\epsilon^T$ от температуры.
В нём использованы расчётные значения  $\alpha$~и~$\epsilon^T$,
полученные из аппроксимации экспериментальных данных для температуры 800~K.
Третий подкоренной элемент формулы~\eqref{eq:elongation_uncertainty}
был рассчитан по~формуле~\eqref{eq:otn_deltaL0},
но~в~качестве  $\Delta L$ было принято:
\begin{equation}
    \Delta L=\epsilon^T\cdot L_0.
\end{equation}

Относительная погрешность косвенных измерений ТКЛР стёкол, рассчитанная по
формуле~\eqref{eq:cte_uncertainty}, не превышает ${\pm}$5~\%.
Эта величина использована для отображения погрешности измерений на графиках на
Рисунках~\ref{fig:cte_bf33+official} и~\ref{fig:cte_lk5+official}.
Относительная погрешность косвенных измерений относительного удлинения стёкол,
рассчитанная по формуле~\eqref{eq:elongation_uncertainty}, не превышает
${\pm}$3~\%.

\subsection{Аппроксимация результатов полиномиальными функциями}
Функции, аппроксимирующие экспериментальные данные, получены методом наименьших квадратов. Данные функции имеют вид полиномов:
\begin{equation}\label{eq:polynom_sample}
    P(T) = a + b \cdot T + c \cdot T^2 + d \cdot T^3 + e \cdot T^4,
\end{equation}
где $a$, $b$, $c$, $d$, $e$ "--- коэффициенты полинома.

Результаты аппроксимации экспериментальных данных ТКЛР и относительного
теплового удлинения полиномами не выше 4-го порядка сведены
в~Таблицы~\ref{tab:results_approx_cte} и~\ref{tab:results_approx_rel_exp},
соответсвенно.

Полученные стандартные ошибки регрессий не превышают рассчитанных относительных
погрешностей косвенных измерений кроме аппроксимаций относительного удлинения в
интервале температур от~235 до~340~K. Кривые полученных зависимостей $\alpha(T)$
стёкол показаны на рисунках: Borofloat~33 "---
на~Рисунке~\ref{fig:cte_bf33+official}, ЛК5 "--- на Рисунке~\ref{fig:cte_lk5+official},
Hoya SD-2 "--- на~Рисунке~\ref{fig:cte_SD-2+official}, Corning 7740 "---
на~Рисунке~\ref{fig:cte_7740+official}.

\begin{table} [!ht]
    \centering%
    \caption[Polynomial coefficients of glass's CTE approximations $\alpha (T)$, 1/K, obtained experimentally]{Результаты аппроксимации экспериментальных значений температурной зависимости ТКЛР $\alpha (T)$, 1/K}%
    \label{tab:results_approx_cte}% label всегда желательно идти после caption
    \renewcommand{\arraystretch}{1.5}%% Увеличение расстояния между рядами, для улучшения восприятия.
    \sisetup{
        table-number-alignment = center,
    }%
    \begin{SingleSpace}
    \begin{tabularx}{\textwidth}{@{}
    >{\raggedright}X
    S[table-format=-1.3]
    S[table-format=1.3]
    S[table-format=-1.3]
    S[table-format=1.3]
    >{\raggedleft}m{0.069\textwidth}
    >{\raggedleft}m{0.069\textwidth}
    >{\raggedleft\arraybackslash}m{0.069\textwidth}
    @{}%
    }
        \toprule     %%% верхняя линейка
        Марка стекла &
        {$a$, 10\textsuperscript{$-$6}} &
        {$b$, 10\textsuperscript{$-$8}} &
        {$c$, 10\textsuperscript{$-$11}}&
        {$d$, 10\textsuperscript{$-$14}}&
        {RSS, 10\textsuperscript{$-$14}} &
        {MSE, 10\textsuperscript{$-$15}} &
        {SER, 10\textsuperscript{$-$8}}\\
        \midrule
        Corning~7740 &
        2,893 &
        0,500 &
        -1,231 &
        0,740 &
        23,0 &
        14,0 &
        12,0\\
        Schott Borofloat~33 &
        1,628 &
        1,040 &
        -2,198 &
        1,398 &
        5,9 &
        3,9 &
        6,2\\
        ЛК5 &
        1,123 &
        1,607 &
        -3,496 &
        2,435 &
        25,0 &
        17,0 &
        13,0\\
        Hoya SD\nobreakdash-2 &
        -0,193 &
        1,453 &
        -1,847 &
        0,910 &
        9,8 &
        5,5 &
        7,4\\
        \bottomrule %%% нижняя линейка
    \end{tabularx}%
    \end{SingleSpace}
\end{table}

\begin{table} [ht]
    \centering%
    \caption[Polynomial coefficients of glass's relative expansion approximations $\epsilon^T (T)$ obtained experimentally]{Результаты аппроксимации экспериментальных значений температурной зависимости относительного удлинения $\epsilon^T (T)$}%
    \label{tab:results_approx_rel_exp}% label всегда желательно идти после caption
    \renewcommand{\arraystretch}{1.6}%% Увеличение расстояния между рядами, для улучшения восприятия.
    \sisetup{
        table-number-alignment = center,
    }%
    \tabulinesep = 1ex
    \begin{SingleSpace}
    \begin{tabu} to \textwidth {@{}
    X[l,m]@{}
    S[table-format=-2.3]
    S[table-format=-1.3]
    S[table-format=1.3]
    S[table-format=-2.3]
    S[table-format=1.3]
    >{\raggedleft}m{0.0680\textwidth}
    >{\raggedleft}m{0.0680\textwidth}
    >{\raggedleft\arraybackslash}m{0.0600\textwidth}
    @{}}
        \toprule     %%% верхняя линейка
        Марка стекла&
        {$a$, 10\textsuperscript{$-$4}}&
        {$b$, 10\textsuperscript{$-$6}}&
        {$c$, 10\textsuperscript{$-$9}}&
        {$d$, 10\textsuperscript{$-$12}}&
        {$e$, 10\textsuperscript{$-$15}}&
        {RSS, 10\textsuperscript{$-$11}}&
        {MSE, 10\textsuperscript{$-$12}}&
        {SER, 10\textsuperscript{$-$6}}\\
        \midrule
        Corning 7740 &
        -10,015 &
          3,297 &
          0,782 &
        -1,196 &
         0,139 &
        60,0 &
        36,0 &
        6,0\\
        Schott Borofloat 33 &
        -7,468 &
        1,451 &
        5,880 &
        -8,426 &
        4,136 &
        4,8 &
        3,2 &
        1,8\\
        ЛК5 &
        -7,494 &
        0,957 &
        8,460 &
        -12,080 &
        6,246 &
        36,0 &
        24,0 &
        4,9\\
        Hoya~SD\nobreakdash-2 &
        -4,208 &
        -0,292 &
        7,691 &
        -6,860 &
        2,657 &
        16,0 &
        9,1 &
        3,0\\
        \bottomrule %%% нижняя линейка
    \end{tabu}%
    \end{SingleSpace}
\end{table}

\begin{samepage}
\begin{figure}[!ht]
    \centering
    \begingroup%
      \makeatletter%
      \providecommand\color[2][]{%
        \errmessage{(Inkscape) Color is used for the text in Inkscape, but the package 'color.sty' is not loaded}%
        \renewcommand\color[2][]{}%
      }%
      \providecommand\transparent[1]{%
        \errmessage{(Inkscape) Transparency is used (non-zero) for the text in Inkscape, but the package 'transparent.sty' is not loaded}%
        \renewcommand\transparent[1]{}%
      }%
      \providecommand\rotatebox[2]{#2}%
      \ifx\svgwidth\undefined%
        \setlength{\unitlength}{0.5\textwidth}%
        \ifx\svgscale\undefined%
          \relax%
        \else%
          \setlength{\unitlength}{\unitlength * \real{\svgscale}}%
        \fi%
      \else%
        \setlength{\unitlength}{\svgwidth}%
      \fi%
      \global\let\svgwidth\undefined%
      \global\let\svgscale\undefined%
      \makeatother%
      \begin{picture}(1,0.69801469)%
        \put(0,0){\includegraphics[width=\unitlength]{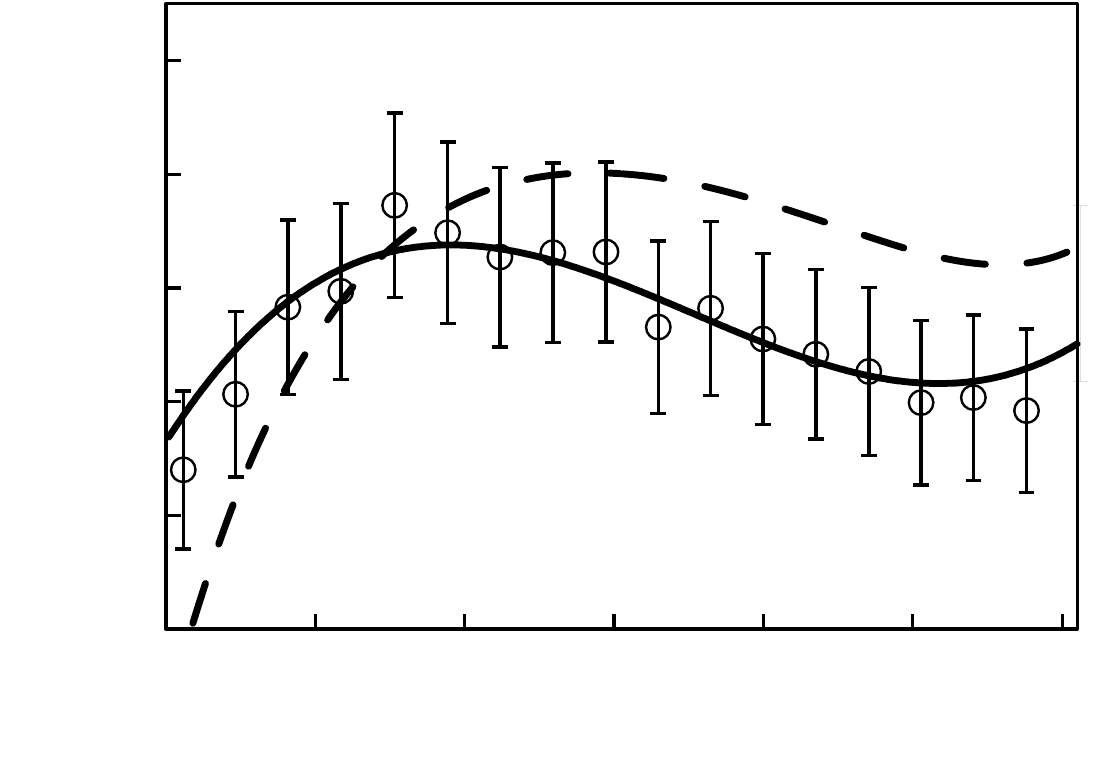}}%
        \put(0.05707158,0.11269191){\color[named]{black}\makebox(0,0)[lb]{\smash{2,5}}}%
        \put(0.05707158,0.21630763){\color[named]{black}\makebox(0,0)[lb]{\smash{2,7}}}%
        \put(0.05707158,0.31988995){\color[named]{black}\makebox(0,0)[lb]{\smash{2,9}}}%
        \put(0.05707158,0.42350566){\color[named]{black}\makebox(0,0)[lb]{\smash{3,1}}}%
        \put(0.05707158,0.52712137){\color[named]{black}\makebox(0,0)[lb]{\smash{3,3}}}%
        \put(0.05707158,0.63070369){\color[named]{black}\makebox(0,0)[lb]{\smash{3,5}}}%
        \put(0.11744589,0.06020709){\color[named]{black}\makebox(0,0)[lb]{\smash{170}}}%
        \put(0.253639,0.06020709){\color[named]{black}\makebox(0,0)[lb]{\smash{270}}}%
        \put(0.38983211,0.06020709){\color[named]{black}\makebox(0,0)[lb]{\smash{370}}}%
        \put(0.52605861,0.06020709){\color[named]{black}\makebox(0,0)[lb]{\smash{470}}}%
        \put(0.66225172,0.06020709){\color[named]{black}\makebox(0,0)[lb]{\smash{570}}}%
        \put(0.79844483,0.06020709){\color[named]{black}\makebox(0,0)[lb]{\smash{670}}}%
        \put(0.93463794,0.06020709){\color[named]{black}\makebox(0,0)[lb]{\smash{770}}}%
        \put(0.03073901,0.25971627){\color[named]{black}\rotatebox{90}{\makebox(0,0)[lb]{\smash{$\alpha$, $10^{-6}$ 1/K}}}}%
        \put(0.50717079,0.00737128){\color[named]{black}\makebox(0,0)[lb]{\smash{$T$, K}}}%
      \end{picture}%
    \endgroup%
    \caption[Fit and experimental values of the linear thermal
    expansion coefficient of Borofloat 33 glass]{Аппроксимация и
    экспериментальные значения коэффициента теплового линейного расширения
    Borofloat 33:}
    \label{fig:cte_bf33+official}
    \legend{%
        Штриховая линия "--- аппроксимация данных производителя
    }
\end{figure}

\begin{figure}[!hb]
    \centering
    \begingroup%
      \makeatletter%
      \providecommand\color[2][]{%
        \errmessage{(Inkscape) Color is used for the text in Inkscape, but the package 'color.sty' is not loaded}%
        \renewcommand\color[2][]{}%
      }%
      \providecommand\transparent[1]{%
        \errmessage{(Inkscape) Transparency is used (non-zero) for the text in Inkscape, but the package 'transparent.sty' is not loaded}%
        \renewcommand\transparent[1]{}%
      }%
      \providecommand\rotatebox[2]{#2}%
      \ifx\svgwidth\undefined%
        \setlength{\unitlength}{0.5\textwidth}%
        \ifx\svgscale\undefined%
          \relax%
        \else%
          \setlength{\unitlength}{\unitlength * \real{\svgscale}}%
        \fi%
      \else%
        \setlength{\unitlength}{\svgwidth}%
      \fi%
      \global\let\svgwidth\undefined%
      \global\let\svgscale\undefined%
      \makeatother%
      \begin{picture}(1,0.76900229)%
        \put(0,0){\includegraphics[width=\unitlength]{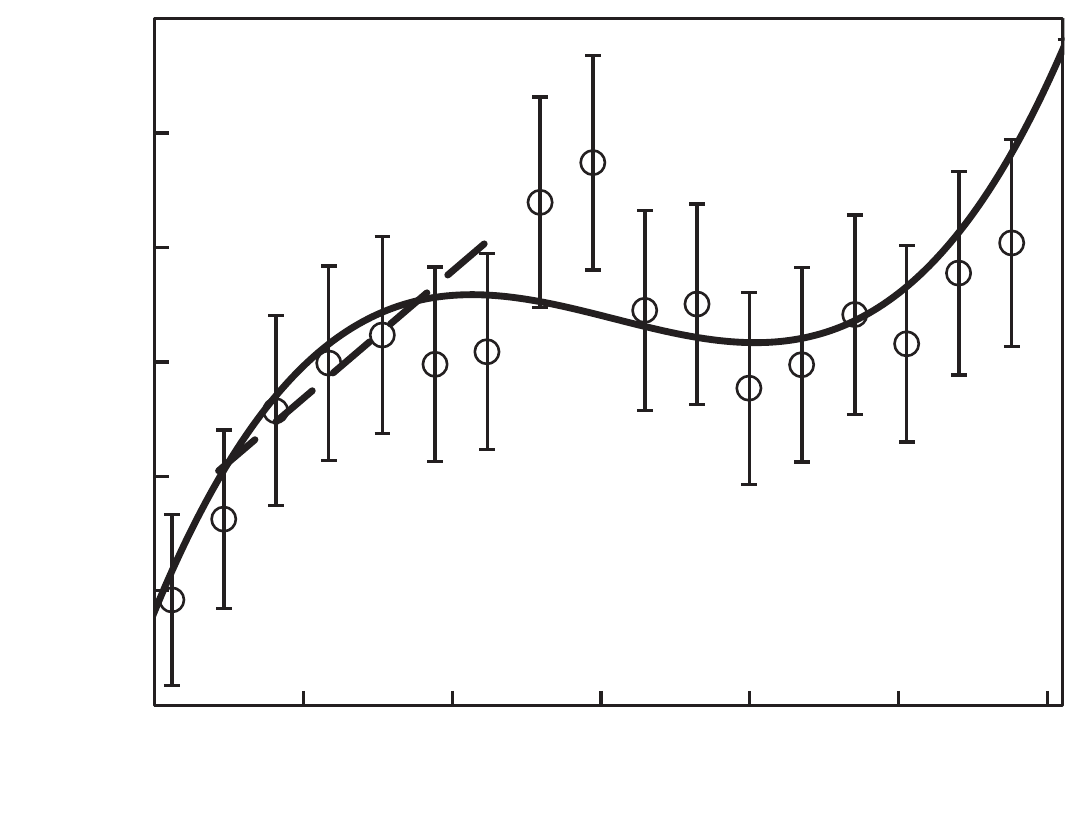}}%
        \put(0.05896161,0.10384021){\color[named]{black}\makebox(0,0)[lb]{\smash{2,8}}}%
        \put(0.05896161,0.20966433){\color[named]{black}\makebox(0,0)[lb]{\smash{3,0}}}%
        \put(0.05896161,0.31545445){\color[named]{black}\makebox(0,0)[lb]{\smash{3,2}}}%
        \put(0.05896161,0.42124457){\color[named]{black}\makebox(0,0)[lb]{\smash{3,4}}}%
        \put(0.05896161,0.52706876){\color[named]{black}\makebox(0,0)[lb]{\smash{3,6}}}%
        \put(0.05896161,0.63285888){\color[named]{black}\makebox(0,0)[lb]{\smash{3,8}}}%
        \put(0.05896161,0.73864897){\color[named]{black}\makebox(0,0)[lb]{\smash{4,0}}}%
        \put(0.1090145,0.05902576){\color[named]{black}\makebox(0,0)[lb]{\smash{170}}}%
        \put(0.24651103,0.05902576){\color[named]{black}\makebox(0,0)[lb]{\smash{270}}}%
        \put(0.38399049,0.05902576){\color[named]{black}\makebox(0,0)[lb]{\smash{370}}}%
        \put(0.52147002,0.05902576){\color[named]{black}\makebox(0,0)[lb]{\smash{470}}}%
        \put(0.65896647,0.05902576){\color[named]{black}\makebox(0,0)[lb]{\smash{570}}}%
        \put(0.79644601,0.05902576){\color[named]{black}\makebox(0,0)[lb]{\smash{670}}}%
        \put(0.93397646,0.05902576){\color[named]{black}\makebox(0,0)[lb]{\smash{770}}}%
        \put(0.0311866,0.26509523){\color[named]{black}\rotatebox{90}{\makebox(0,0)[lb]{\smash{$\alpha$, $10^{-6}$ 1/K}}}}%
        \put(0.53009853,0.00747856){\color[named]{black}\makebox(0,0)[lb]{\smash{$T$, K}}}%
      \end{picture}%
    \endgroup%
    \caption[Fit and experimental values of the linear thermal expansion
    coefficient of LK5 glass]{Аппроксимация и экспериментальные значения
    коэффициента теплового линейного расширения ЛК5:}
    \label{fig:cte_lk5+official}
    \legend{%
        Штриховая линия "--- аппроксимация данных производителя
    }
\end{figure}
\end{samepage}

\begin{samepage}
\begin{figure}[!htb]
    \centering
    \begingroup%
      \makeatletter%
      \providecommand\color[2][]{%
        \errmessage{(Inkscape) Color is used for the text in Inkscape, but the package 'color.sty' is not loaded}%
        \renewcommand\color[2][]{}%
      }%
      \providecommand\transparent[1]{%
        \errmessage{(Inkscape) Transparency is used (non-zero) for the text in Inkscape, but the package 'transparent.sty' is not loaded}%
        \renewcommand\transparent[1]{}%
      }%
      \providecommand\rotatebox[2]{#2}%
      \ifx\svgwidth\undefined%
        \setlength{\unitlength}{0.5\textwidth}%
        \ifx\svgscale\undefined%
          \relax%
        \else%
          \setlength{\unitlength}{\unitlength * \real{\svgscale}}%
        \fi%
      \else%
        \setlength{\unitlength}{\svgwidth}%
      \fi%
      \global\let\svgwidth\undefined%
      \global\let\svgscale\undefined%
      \makeatother%
      \begin{picture}(1,0.73146883)%
        \put(0,0){\includegraphics[width=\unitlength]{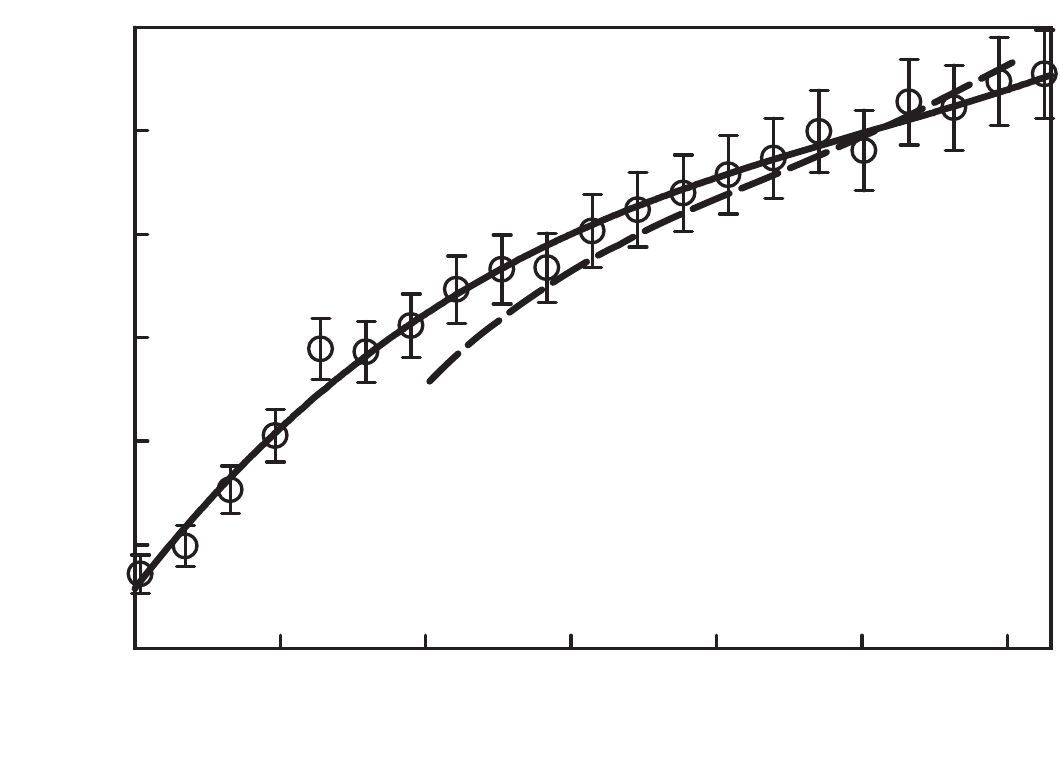}}%
        \put(0.52181269,0.00655847){\color[named]{black}\makebox(0,0)[lb]{\smash{$T$, K}}}%
        \put(0.02734947,0.2858294){\color[named]{black}\rotatebox{90}{\makebox(0,0)[lb]{\smash{$\alpha$, $10^{-6}$ 1/K}}}}%
        \put(0.04882056,0.1174594){\color[named]{black}\makebox(0,0)[lb]{\smash{1,5}}}%
        \put(0.05321852,0.21461843){\color[named]{black}\makebox(0,0)[lb]{\smash{2,0}}}%
        \put(0.05321852,0.31177746){\color[named]{black}\makebox(0,0)[lb]{\smash{2,5}}}%
        \put(0.05234342,0.40893649){\color[named]{black}\makebox(0,0)[lb]{\smash{3,0}}}%
        \put(0.05234342,0.50609572){\color[named]{black}\makebox(0,0)[lb]{\smash{3,5}}}%
        \put(0.05348779,0.60325475){\color[named]{black}\makebox(0,0)[lb]{\smash{4,0}}}%
        \put(0.05348779,0.70041378){\color[named]{black}\makebox(0,0)[lb]{\smash{4,5}}}%
        \put(0.09083587,0.064941){\color[named]{black}\makebox(0,0)[lb]{\smash{170}}}%
        \put(0.23102922,0.064941){\color[named]{black}\makebox(0,0)[lb]{\smash{270}}}%
        \put(0.36858599,0.064941){\color[named]{black}\makebox(0,0)[lb]{\smash{370}}}%
        \put(0.50715254,0.064941){\color[named]{black}\makebox(0,0)[lb]{\smash{470}}}%
        \put(0.64440638,0.064941){\color[named]{black}\makebox(0,0)[lb]{\smash{570}}}%
        \put(0.78251299,0.064941){\color[named]{black}\makebox(0,0)[lb]{\smash{670}}}%
        \put(0.92064194,0.064941){\color[named]{black}\makebox(0,0)[lb]{\smash{770}}}%
      \end{picture}%
    \endgroup%
    \caption[Fit and experimental values of the linear thermal
    expansion coefficient of Hoya SD-2 glass]{Аппроксимация и
    экспериментальные значения коэффициента теплового линейного расширения
    Hoya SD-2:}
    \label{fig:cte_SD-2+official}
    \legend{%
        Штриховая линия "--- аппроксимация данных производителя
    }
\end{figure}

\begin{figure}[!htb]
    \centering
    \begingroup%
      \makeatletter%
      \providecommand\color[2][]{%
        \errmessage{(Inkscape) Color is used for the text in Inkscape, but the package 'color.sty' is not loaded}%
        \renewcommand\color[2][]{}%
      }%
      \providecommand\transparent[1]{%
        \errmessage{(Inkscape) Transparency is used (non-zero) for the text in Inkscape, but the package 'transparent.sty' is not loaded}%
        \renewcommand\transparent[1]{}%
      }%
      \providecommand\rotatebox[2]{#2}%
      \ifx\svgwidth\undefined%
        \setlength{\unitlength}{0.5\textwidth}%
        \ifx\svgscale\undefined%
          \relax%
        \else%
          \setlength{\unitlength}{\unitlength * \real{\svgscale}}%
        \fi%
      \else%
        \setlength{\unitlength}{\svgwidth}%
      \fi%
      \global\let\svgwidth\undefined%
      \global\let\svgscale\undefined%
      \makeatother%
      \begin{picture}(1,0.71082807)%
        \put(0,0){\includegraphics[width=\unitlength]{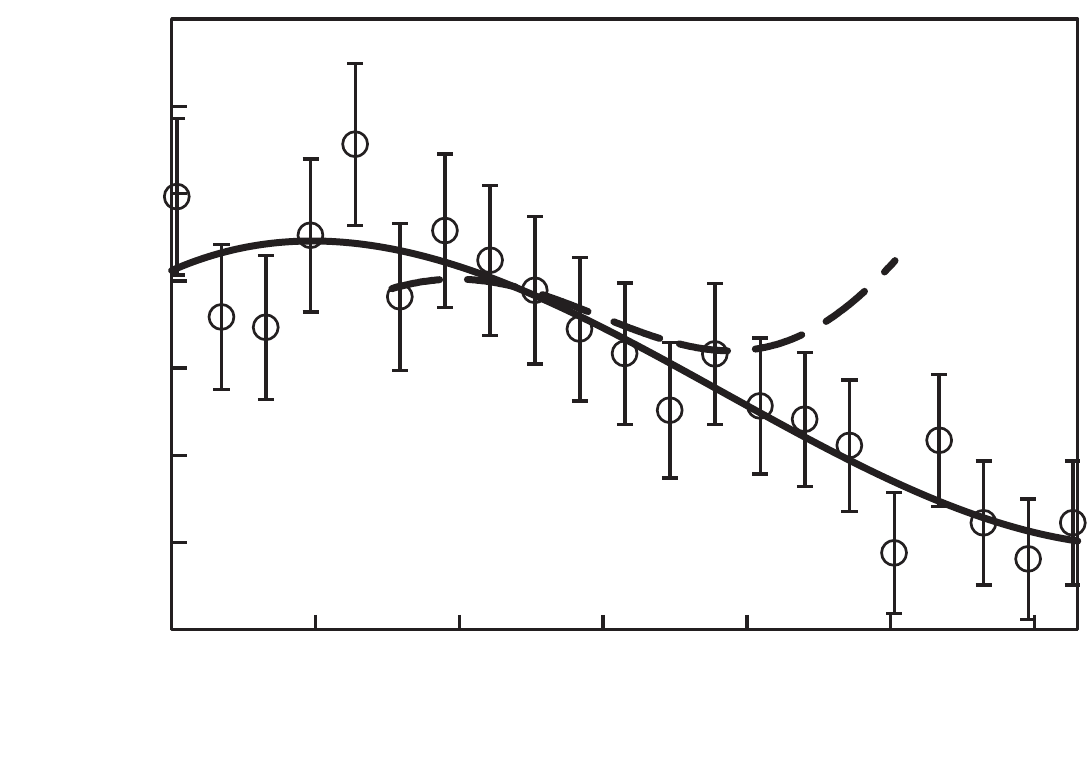}}%
        \put(0.06109288,0.11853542){\color[named]{black}\makebox(0,0)[lb]{\smash{2,6}}}%
        \put(0.06109288,0.1988594){\color[named]{black}\makebox(0,0)[lb]{\smash{2,8}}}%
        \put(0.06109288,0.27913159){\color[named]{black}\makebox(0,0)[lb]{\smash{3,0}}}%
        \put(0.06109288,0.35945557){\color[named]{black}\makebox(0,0)[lb]{\smash{3,2}}}%
        \put(0.06109288,0.43972776){\color[named]{black}\makebox(0,0)[lb]{\smash{3,4}}}%
        \put(0.06109288,0.51999994){\color[named]{black}\makebox(0,0)[lb]{\smash{3,6}}}%
        \put(0.06109288,0.60032392){\color[named]{black}\makebox(0,0)[lb]{\smash{3,8}}}%
        \put(0.06109288,0.68059611){\color[named]{black}\makebox(0,0)[lb]{\smash{4,0}}}%
        \put(0.12343288,0.06432454){\color[named]{black}\makebox(0,0)[lb]{\smash{170}}}%
        \put(0.25581035,0.06432454){\color[named]{black}\makebox(0,0)[lb]{\smash{270}}}%
        \put(0.38815334,0.06432454){\color[named]{black}\makebox(0,0)[lb]{\smash{370}}}%
        \put(0.52054813,0.06432454){\color[named]{black}\makebox(0,0)[lb]{\smash{470}}}%
        \put(0.6529256,0.06432454){\color[named]{black}\makebox(0,0)[lb]{\smash{570}}}%
        \put(0.78530307,0.06432454){\color[named]{black}\makebox(0,0)[lb]{\smash{670}}}%
        \put(0.91769787,0.06432454){\color[named]{black}\makebox(0,0)[lb]{\smash{770}}}%
        \put(0.031062,0.2772676){\color[named]{black}\rotatebox{90}{\makebox(0,0)[lb]{\smash{$\alpha$, $10^{-6}$ 1/K}}}}%
        \put(0.53822139,0.00744878){\color[named]{black}\makebox(0,0)[lb]{\smash{$T$, K}}}%
      \end{picture}%
    \endgroup%
    \caption[Fit and experimental values of the linear thermal
    expansion coefficient of Corning 7740 glass]{Аппроксимация и
    экспериментальные значения коэффициента теплового линейного расширения
    Corning 7740:}
    \label{fig:cte_7740+official}
    \legend{%
        Штриховая линия "--- аппроксимация данных производителя
    }
\end{figure}
\end{samepage}

\subsection{Сравнение с данными производителей}
Сравним данные производителей по средним ТКЛР со средними значениями на тех же температурных интервалах рассчитанными по полученным аппроксимациям.

Средний ТКЛР стекла марки ЛК5 по полученным экспериментальным данным в диапазонах от минус 60 до плюс 20~{\textdegree}C и от 20 до 120~{\textdegree}C  совпадает с~данными производителя~\cite{LK5_properties}.

Средний ТКЛР стекла марки Borofloat 33 в диапазоне от 20 до
300~{\textdegree}C по~данным производителя~\cite{bf33_properties}
составляет 3,25$\:\cdot\:$10\textsuperscript{$-$6}~1/{\textdegree}C, по
полученным данным "---
3,12$\:\cdot\:$10\textsuperscript{$-$6}~1/{\textdegree}C.
Такое различие укладывается в погрешность проведённых измерений.

Средний ТКЛР стекла марки Corning 7740 в диапазоне от 0 до
300~{\textdegree}C по~данным производителя~\cite{corning7740_wafersheet}
составляет 3,25$\:\cdot\:$10\textsuperscript{$-$6}~1/{\textdegree}C, по
полученным данным "---
3,34$\:\cdot\:$10\textsuperscript{$-$6}~1/{\textdegree}C.
Такое различие укладывается в погрешность проведённых измерений.

Средний ТКЛР стекла марки Hoya SD\nb-2 в диапазоне от 20 до
300~{\textdegree}C по~данным производителя~\cite{SD_2_properties}
составляет 3,20$\:\cdot\:$10\textsuperscript{$-$6}~1/{\textdegree}C, по
полученным данным "---
3,33$\:\cdot\:$10\textsuperscript{$-$6}~1/{\textdegree}C.
Такое различие укладывается в погрешность проведённых измерений.

\clearpage
\section{Выводы по главе 2}

\begin{enumerate}
\setlist{midpenalty=5000}
    \item Исследован состав образцов четырёх марок стёкол, применяемых в
    отечественной и мировой практике для соединения с кремнием: Borofloat~33,
    Corning~7740, ЛК5 и~Hoya~SD\nb-2.
    Это позволит сопоставить данные, полученные в~этой работе,
    с~исследованиями других авторов.
    \item Анализ погрешностей проведённых измерений ТКЛР в
    диапазоне от~170 до 780~K (от минус 100 до
    500~{\textdegree}C) показал, что относительная погрешность
    измерений не превышает ${\pm}$5~\%.
    Полученные стандартные ошибки регрессий не превышают
    рассчитанные относительные погрешности.
    \item Сравнение с данными производителей по средним ТКЛР
    в диапазоне от~20 до 300~{\textdegree}C показало расхождение
    не превышающее погрешность проведённых измерений.
\end{enumerate}

%% file: Dissertation/part3.tex
\chapter{Разработка методики оценки остаточных напряжений в~деталях собранных электростатическим соединением}

Для планирования процесса соединения стекла с кремнием,
обеспечивающего минимальные остаточные напряжения или же
выдерживающего их~в~определённых пределах, необходимо иметь доступные
способы расчётной оценки таких напряжений. В этой главе описано
несколько моделей оценки. Приведены примеры применения этих моделей
на~исходных данных, полученных в~предшествующей главе. В~завершении даны
рекомендации по~проведению оценки напряжений.

\section{Оценка остаточных напряжений, вызванных неоднородностью теплового расширения}
Прежде, чем приступать к описанию моделей, следует отметить
терминологический нюанс.
Истинным коэффициентом теплового линейного расширения  $\alpha $,
1/{\textdegree}C, называется отношение изменения линейного размера тела 
$\mathrm{d}l$, м, делённого на его начальный размер  $l_0$, м, к малому
изменению температуры  $\mathrm{d}T$,~{\textdegree}C, вызвавшему изменение
размера тела~\cites[6]{Mazurin1969_Tepl_rassh_stekla}.
\begin{equation*}
    \alpha =\frac 1{l_0} \cdot \frac{\mathrm{d}l}{\mathrm{d}T}
\end{equation*}

\subsection{Упрощённая оценка остаточных напряжений}
В литературе по электронной технике~\cite{zemen2010impact}, посвящённой соединению материалов,
напряжения в материалах вызванные разницей их коэффициентов расширения предлагается оценивать по
формуле:
\[
\sigma =E\left(\alpha_1-\alpha_2\right)\Delta T,
\]
где $\sigma$ "--- напряжения в детали, вызванные разницей между
коэффициентами теплового расширения материалов, Па;
$E$ "--- модуль упругости первого рода материала, в котором исследуются
напряжения, Па;
$\alpha_1$, $\alpha_2$ "--- средние коэффициенты теплового линейного
расширения каждого из пары соединяемых материалов, 1/{\textdegree}C;
$\Delta T$ "--- разница между температурой соединения материалов и
температурой, при которой исследуются остаточные
напряжения,~{\textdegree}C.

При этом коэффициенты теплового расширения на практике берутся средние для рассматриваемого интервала температур. В формуле используется модуль Юнга того материала, в котором рассчитывают напряжения.

В этой оценке не приняты во внимание влияние толщины материалов и~нелинейность
температурной зависимости ТКЛР материалов.

\subsection{Двухосное напряжённое состояние в осаждённой плёнке}
Двухосное напряжённое состояние в плёнке, осаждённой на подложку~\cite{Adams_Elasticity_ppt} рассчитывается по следующей формуле в предположении, что плёнка не оказывает влияния на подложку.
\begin{equation}
    \label{eq_sigma_film}
    \sigma _f=\frac{E_f}{1-\mu _f}\left(\alpha _f-\alpha _s\right)(T_d-T_r),
\end{equation}
где $ E_f $ "--- модуль упругости первого рода осаждённой плёнки, Па;
$\mu_f$ "--- коэффициент Пуассона осаждённой плёнки;
$ T_d $ "--- температура осаждения плёнки,~{\textdegree}C;
$ T_r $ "--- температура, для которой определяются
напряжения,~{\textdegree}C;
$\alpha_f$, $\alpha_s$ "--- средние коэффициенты теплового линейного
расширения плёнки и~подложки в~рассматриваемом интервале температур,
соответственно, 1/{\textdegree}C.

Формула~\eqref{eq_sigma_film} является фактически упрощённой версией расчёта,
полученного на основе формулы для определения относительных деформаций в~плёнке,
которая учитывает нелинейный характер ТКЛР
материалов~\cite{gleskova2009mechanical}:
\[
    \epsilon_f
    =
    \frac{1}{1-\mu_f}
    \int\limits_{T_r}^{T_d}
    (\alpha_f-\alpha_s)\:\mathrm{d}T,
\]
где $\epsilon_f$ "--- относительная деформация плёнки, вызванная разницей
коэффициентов теплового расширения плёнки и подложки; $\alpha_f$,
$\alpha_s$ "--- истинные коэффициенты теплового линейного расширения плёнки
и подложки, 1/{\textdegree}C.

\subsection{Двухслойный материал под тепловой нагрузкой}

По следующей формуле в работе~\cite{li2007stress} предлагается оценивать
напряжения, возникающие на свободной поверхности кремния после сборки
с~текстолитовым основанием методом перевёрнутого кристалла (flip\nb-chip).
Данная оценка основана на модели двухслойного материала, где каждый слой
изотропен, находящегося под тепловой нагрузкой:
\[
\sigma = \frac{\epsilon_{m} M_1 h m (2 + 3 h + h^3 m) }{1 + h m (4 + 6 h + 4 h^2 )},
\]
\[
M_1=\frac{E_1}{1-\mu_1},
\]
\[
M_2=\frac{E_2}{1-\mu_2},
\]
\[
m = \frac{M_2}{M_1},
\]
\[
h = \frac{h_2}{h_1},
\]
\[
\epsilon_{m} = (T_{w} - T_{b})(\alpha_{1} - \alpha_{2}),
\]
где $h_1$, $h_2$ "--- толщины верхней и~нижней пластин, соответственно,~м;
$E_1$, $E_2$ "--- модули упругости первого рода верхней пластины и~нижней
пластины, соответственно,~Па;
$\mu_1$, $\mu_2$ "--- коэффициенты Пуассона верхней пластины и~нижней
пластины, соответственно;
$T_{b}$ "--- температура соединения пластин,~{\textdegree}C;
$T_{w}$ "--- температура определения напряжений,~{\textdegree}C;
$\alpha_{1}$, $\alpha_{2}$ "--- средние ТКЛР верхней пластины и~нижней
пластины в~рассматриваемом интервале температур,
соответственно,~1/{\textdegree}C;
$\epsilon_{m}$ "--- относительная деформация вызванная разницей ТКЛР
соединённых пластин.

\subsection{Модель двух тонких слоёв}

В данной модели приняты следующие допущения:
\begin{itemize}
    \item обе соединяемых детали сплошные, однородные,
    изотропны и непрерывны, представляют по форме прямоугольные
    параллелепипеды;
    \item нагрев деталей равномерен и источник тепла расположен
    вне области соединения;
    \item область соединения представляет собой плоскость;
    \item удлинения и напряжения в области соединения равны удлинениям
    и~напряжениям во всей детали;
    \item влияние краевых эффектов и разницы в коэффициентах теплопроводности материалов
    исключены из рассмотрения;
    \item толщина кремниевой детали примерно в 10 раз меньше каждого
    из~двух других её размеров, и в 5"--~10 раз меньше толщины стеклянной детали;
    \item изгиб деталей под действием возникающих деформаций пренебрежимо
    мал.
\end{itemize}

Изменения размеров в этой модели оценивают в плоскости перпендикулярной плоскости
соединения.

Поскольку детали соединены, то в каждый момент времени изменения длины должны быть равны:
\begin{equation*}
    \mathrm{d}l_g=\mathrm{d}l_{si}
\end{equation*}
где  $ \mathrm{d}l_g $ "--- изменение длины стеклянной детали, м;
$ \mathrm{d}l_{si} $ "--- изменение длины кремниевой детали, м.

Изменения длины, определяются как тепловым расширением самой детали, так и влиянием присоединённой детали:
\[
    \mathrm{d}l_{si}
    =
    \alpha_{si}(T) \cdot l_0 \mathrm{d}T
    -
    \frac{\mathrm{d}Z \cdot l_0}{E_{si} (T) A_{si}},
\]
\[
    \mathrm{d}l_g
    =
    \alpha_g(T) \cdot l_0\mathrm{d}T
    +
    \frac{\mathrm{d}Z \cdot l_0}{E_g (T) A_g},
\]
\[
    \mathrm{d}Z
    =
    \frac{E_g(T) A_g E_{si}(T) A_{si}}{E_{si}(T) A_{si} + E_g(T) A_g}
    (
        \alpha_{si}(T) - \alpha_g(T)
    )
    \:\mathrm{d}T,
\]
\[
    A_{si}
    =
    b h_{si},
\]
\[
    A_g
    =
    b h_g,
\]
где
$\alpha_g(T)$,  $\alpha_{si}(T)$ "--- ТКЛР стекла и~кремния,
соответственно, 1/{\textdegree}C;
$l_0$ "--- начальная длина кремниевой и стеклянной деталей, м;
$E_g(T)$,  $E_{si}(T)$ "--- модули упругости первого рода стекла и кремния,
соответственно, Па;
$A_g$,  $A_{si}$ "--- площади сечений, расположенных поперёк оси удлинения
стекла и кремния,~м\textsuperscript{2};
$b$ "--- общая ширина соединяемых стекла и кремния, если смотреть поперёк
оси удлинения стекла и кремния,~м;
$h_g$,  $h_{si}$ "--- толщины соединяемых стекла и~кремния,
соответственно,~м;
$\mathrm{d}Z$ "--- изменение растягивающей силы в~стекле из-за изменения
температуры, Н.

Учитывая, что поперечные сечения деталей имеют общую линию соединения, то от площадей поперечных сечений можно перейти к толщинам соединяемых деталей.
\begin{equation}
    \label{eq:sigma_g}
    \sigma_g(T)
    =
    \int\limits_{T_b}^{T_w}
    \frac{\mathrm{d}Z}{A_g}
    \mathrm{d}T
    =
    \int\limits_{T_b}^{T_w}
    \frac{E_g(T)\, E_{si}(T) h_{si}}{E_{si}(T) h_{si} + E_g(T) h_g}
    (
        \alpha_{si}(T) - \alpha_g(T)
    )
    \:\mathrm{d}T,
\end{equation}
\begin{equation}
    \label{eq:sigma_si}
    \sigma_{si}(T)
    =
    \int\limits_{T_b}^{T_w}
    \frac{\mathrm{d}Z}{A_{si}}
    \mathrm{d}T
    =
    \int\limits_{T_b}^{T_w}
    \frac{E_g(T) h_{g}\, E_{si}(T)}{E_{si}(T) h_{si} + E_g(T) h_g}
    (
        \alpha_{si}(T) - \alpha_g(T)
    )
    \:\mathrm{d}T,
\end{equation}
где  $T_b$ "--- температура соединения,~{\textdegree}C;  $T_w$ "--- температура, при которой оценивают напряжения, рабочая температура прибора,~{\textdegree}C;
$\sigma_g$,  $\sigma_{si}$ "--- напряжения в~деталях из стекла и кремния, соответственно, соединённых при температуре  $T_b$, возникающие при температуре  $T_w$, Па.

Интеграл в формулах~\eqref{eq:sigma_g} и~\eqref{eq:sigma_si} представляет собой разницу между относительными удлинениями кремния и стекла, если бы они не были соединены. Также видно, что соотношение напряжений в кремнии и стекле обратно пропорционально толщинам соединяемых деталей.

Преобразуем формулы~\eqref{eq:sigma_g} и~\eqref{eq:sigma_si}, чтобы упростить запись и положительными напряжениями в стекле и кремнии считались растягивающие напряжения:%
\[
    \sigma_g(T)
    =
    \int\limits_{T_b}^{T_w}
    \frac{E_g(T)}%
        {1 +
            \left(
                \dfrac{E_g(T)}{E_{si}(T)}
            \right)
        \cdot
            \left(
                \dfrac{h_g}{h_{si}}
            \right)
        }
    (
        \alpha_{si}(T) - \alpha_g(T)
    )
    \:\mathrm{d}T,
\]
\begin{equation}
    \label{eq:sigma_siupdated}
    \sigma_{si}(T)
    =
    \int\limits_{T_b}^{T_w}
    \frac{E_{si}(T)}%
        {1 +
            \left(
                \dfrac{E_{si}(T)}{E_g(T)}
            \right)
        \cdot
            \left(
                \dfrac{h_{si}}{h_g}
            \right)
        }
    (
         \alpha_g(T) - \alpha_{si}(T)
    )
    \:\mathrm{d}T.
\end{equation}

\subsection{Модель многослойного композиционного материала}

Для оценки распределения остаточных напряжений по~толщине соединяемых
кремния и стекла, воспользуемся теорией слоистых
композитов~\cite{gigliotti2007assessment}.

Соединённые детали рассматриваем как многослойный композиционный материал
под тепловой нагрузкой.
В качестве координатной плоскости $xy$ принимаем плоскость,
лежащую до нагружения сборки посередине между её~верхней и нижней поверхностями, то есть срединную
плоскость многослойной пластины.
В данной модели приняты следующие допущения:
\begin{itemize}
    \item срединная плоскость не меняет своего положения в процессе нагружения;
    \item пластина состоит из произвольного числа слоёв, соединённых друг с~другом;
    \item закон Гука справедлив для каждого слоя;
    \item слои не оказывают сдавливающего воздействия один на другой;
    \item толщина пластины не изменяется;
    \item нагрев пластины равномерен;
    \item длина и ширина пластины значительно превышают её толщину;
    \item изменение жёсткости рассматриваемых материалов от температуры незначительно;
    \item влияние краевых эффектов и разницы в коэффициентах теплопроводности
    материалов исключены из рассмотрения;
    \item изменение свойств стекла, связанное с переносом ионов в
    результате проведения процесса электростатического соединения
    \cite{Rogers1992considerations, Cozma_Puers_1995, Sadaba2006CompositionalGradients} не~учитываем.
\end{itemize}

\begin{figure}[!htbp]
    \centering
    \includegraphics[width=85.0mm, height=42.0mm]{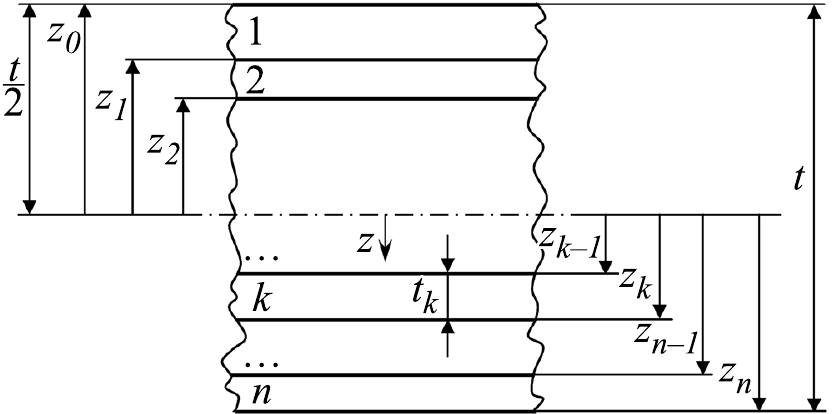}
    \caption[\(n\)-layered laminate model]{Иллюстрация модели
    многослойного композиционного материала (слоистого композита):}
    \label{fig:sloisty_kompozit}
    \legend{%
        $t$ "--- толщина многослойной пластины; 1, 2, \dots, $k$, \dots, $n$ "--- номер слоя
    }
\end{figure}

Рисунок~\ref{fig:sloisty_kompozit} иллюстрирует модель многослойного композиционного материала.
Положительным направлением оси $z$ будем считать направление вниз.

Напряжения в~каждом слое при механическом и~тепловом нагружении можно выразить
следующим уравнением:
\begin{equation}\label{eq:form_sigma_Q}
    \boldsymbol{\sigma}
    =
    \mathbf{Q} (\boldsymbol{\epsilon} - \boldsymbol{\epsilon}^{T}),
\end{equation}
\[
    \boldsymbol{\epsilon}^{T} = \int\limits_{T_{b}}^{T_{w}} \boldsymbol{\alpha}(T)\mathrm{d}T,
\]
\[
    \boldsymbol{\epsilon} = \boldsymbol{\epsilon^0} + z \boldsymbol{\kappa},
\]
где $\boldsymbol{\sigma}$ "--- вектор напряжений,~Па;
$\mathbf{Q}$ "--- преобразованная матрица жёсткости каждого слоя,~Па;
$\boldsymbol{\epsilon}$ "--- вектор индуцированных деформаций (растяжения), вызванных механической нагрузкой;
$\boldsymbol{\epsilon}^{T}$ "--- вектор индуцированных деформаций (растяжения), вызванных тепловой нагрузкой;
$T_{b}$ "--- температура соединения,~{\textdegree}C;
$T_{w}$~---~рабочая температура,~{\textdegree}C;
\(\boldsymbol{\alpha}(T)\) "--- вектор ТКЛР материала, 1/K;
$\boldsymbol{\epsilon^0}$ "--- относительное удлинение срединной поверхности многослойной композитной пластины (по~осям);
$\boldsymbol{\kappa}$ "--- радиус кривизны
срединной поверхности
многослойной композитной пластины (по~осям),~1/м;
$z$~---~расстояние, измеряемое от~срединной поверхности,~м.
Матрицы $\mathbf{Q}$~получены из~матриц жёсткости в~соответствии с~формулами
поворота системы координат \cites[{18,~215,~224}]{Alfutov1984_povorot_matric} так,
чтобы они отражали упругие свойства слоёв в~исследуемых направлениях
координатных осей.

\begingroup
Напряжённое состояние многослойной пластины будет рассмотрено как плоское
напряжённое состояние \cites[14]{Alfutov1984_povorot_matric}, поскольку ранее
было принято допущение, что размеры пластины много больше её толщины.
Векторы и~матрицы в~уравнениях будут представлены в~сокращённой форме за~счёт
отбрасывания незадействованных компонентов
(сокращения до~размерности 3\(\,\times\,\)3).\russianpar
\endgroup

Поэтому уравнение~\eqref{eq:form_sigma_Q} может быть записано в~матричном
представлении следующим образом:
\[
\begin{pmatrix}
\sigma_{x}\\
\sigma_{y}\\
\tau_{xy}
\end{pmatrix}
=
\begin{pmatrix}
Q_{11} & Q_{12} & Q_{16}\\
Q_{12} & Q_{22} & Q_{26}\\
Q_{16} & Q_{26} & Q_{66}
\end{pmatrix}
\left(
    \begin{pmatrix}
    \epsilon_{x}^0\\
    \epsilon_{y}^0\\
    \gamma_{xy}^0
    \end{pmatrix}
+ z
    \begin{pmatrix}
    \kappa_{x}\\
    \kappa_{y}\\
    \kappa_{xy}
    \end{pmatrix}
-
    \begin{pmatrix}
    \epsilon_{x}^T\\
    \epsilon_{y}^T\\
    \gamma_{xy}^T
    \end{pmatrix}
\right)\!,
\]
где $\sigma_{x}$,  $\sigma_{y}$,  $\tau_{xy}$ "--- элементы вектора напряжений,~Па;
$Q_{ij}$ "--- элементы преобразованной матрицы жёсткости каждого слоя,~Па;
$\epsilon_{x}^0$, $\epsilon_{y}^0$, $\gamma_{xy}^0$ "--- элементы вектора относительного удлинения срединной поверхности многослойной композитной пластины;
$\kappa_{x}$, $\kappa_{y}$, $\kappa_{xy}$ "--- элементы векторного представления радиуса кривизны срединной поверхности многослойной композитной пластины,~1/м;
$\epsilon_{x}^T$, $\epsilon_{y}^T$, $\gamma_{xy}^T$~---~элементы вектора индуцированных деформаций, вызванных тепловой нагрузкой.

\begingroup%
Обобщённые силовые факторы, воздействующие на~пластину, определяются посредством интегрирования уравнения~\eqref{eq:form_sigma_Q} по~всей толщине пластины:\russianpar
\endgroup%
\[
\mathbf{N} = \int\limits_t \boldsymbol{\sigma} \mathrm{d}z,
\]
\[
\mathbf{M} = \int\limits_t \boldsymbol{\sigma} z \mathrm{d}z,
\]
где $t$ "--- толщина многослойной пластины, м;
$ \mathbf{N} $ "--- результирующая нагрузка, отнесённая к~единице длины линий, ограничивающих элемент рассматриваемой поверхности \cites[69]{Vasiljev1988_Meh_konstr_kompozit}, \mbox{Н/м};
$ \mathbf{M} $ "--- результирующий момент, отнесённый к~единице длины линий, ограничивающих элемент рассматриваемой поверхности,~Н.

Жёсткость каждого слоя $\mathbf{Q}$~неизменна по~толщине слоя, поэтому можно записать:
\[
A_{ij} = \sum\limits_{k=1}^n (Q_{ij})_k (z_k - z_{k-1}),
\]
\[
B_{ij} = \frac{1}{2} \sum\limits_{k=1}^n (Q_{ij})_k (z_k^2 - z_{k-1}^2),
\]
\ifnumequal{\value{usealtfont}}{2}{\setDisplayskipStretch{0.5}}{}%
\[
D_{ij} = \frac{1}{3} \sum\limits_{k=1}^n (Q_{ij})_k (z_k^3 - z_{k-1}^3),
\]
где $ A_{ij} $ "--- элементы матрицы жёсткости при растяжении (мембранной жёсткости) \cites[157]{kristensen1982_per_Vved_v_kompozit}[{70,~79}]{Vasiljev1988_Meh_konstr_kompozit},~Н/м;
$ B_{ij} $ "--- элементы матрицы жёсткости изгиб\nb-растяжение (смешанной жёсткости), Н;
$ D_{ij} $ "--- элементы матрицы жёсткости при изгибе (изгибной жёсткости), Н$\cdot$м;
$ z_k $ "--- расстояние до~текущего слоя, измеряемое от~срединной поверхности (см.~Рисунок~\ref{fig:sloisty_kompozit}).

Тогда взаимосвязь нагрузок и~деформаций может быть представлена в~уравнениях в~матричной форме:
\begin{equation}\label{eq:n_m_eps_kap_simp_matrix_comb}
\left(
    \begin{array}{@{}c@{}}
        \mathbf{N} \\
        \mathbf{M}
    \end{array}
\right)
=
\left(
    \begin{array}{@{}cc@{}}
        \mathbf{A} & \mathbf{B} \\
        \mathbf{B} & \mathbf{D}
    \end{array}
\right)
\left(
    \begin{array}{@{}c@{}}
        \boldsymbol{\epsilon^0} \\
        \boldsymbol{\kappa}
    \end{array}
\right)
-
\left(
    \begin{array}{@{}c@{}}
        \mathbf{N}^T \\
        \mathbf{M}^T
    \end{array}
\right)\!\!,
\end{equation}
где вызванные тепловым воздействием:
$ \mathbf{N}^T $ "--- усилие, отнесённое к~единице длины линий,
ограничивающих элемент рассматриваемой поверхности, Н/м;
$ \mathbf{M}^T $ "--- момент силы, отнесённый к~единице длины линий,
ограничивающих элемент рассматриваемой поверхности, Н.

Преобразовав запись, получим:
\begin{equation}\label{eq:N_matrix}%
\ifnumequal{\value{usealtfont}}{2}{}{%
\advance\thickmuskip -2mu%Only do this as a last resort http://tex.stackexchange.com/a/85091
\advance\medmuskip -1mu minus -1mu%Only do this as a last resort http://tex.stackexchange.com/a/85091
\hspace{-0.8em}%
}%
    \begin{pmatrix}
    N_{x}\\
    N_{y}\\
    N_{xy}
    \end{pmatrix}
    =
    \begin{pmatrix}
    A_{11} & A_{12} & A_{16}\\
    A_{12} & A_{22} & A_{26}\\
    A_{16} & A_{26} & A_{66}
    \end{pmatrix}
    \begin{pmatrix}
    \epsilon_{x}^0\\
    \epsilon_{y}^0\\
    \epsilon_{xy}^0
    \end{pmatrix}
    +
    \begin{pmatrix}
    B_{11} & B_{12} & B_{16}\\
    B_{12} & B_{22} & B_{26}\\
    B_{16} & B_{26} & B_{66}
    \end{pmatrix}
    \begin{pmatrix}
    \kappa_{x}\\
    \kappa_{y}\\
    \kappa_{xy}
    \end{pmatrix}
    -
    \begin{pmatrix}
    N_{x}^T\\
    N_{y}^T\\
    N_{xy}^T
    \end{pmatrix}\!,%
\end{equation}%
\begin{equation}\label{eq:M_matrix}%
\ifnumequal{\value{usealtfont}}{2}{}{%
\advance\thickmuskip -3mu %Only do this as a last resort http://tex.stackexchange.com/a/85091
\advance\medmuskip -2mu minus -1mu%Only do this as a last resort http://tex.stackexchange.com/a/85091
\hspace{-0.8em}%
}%
    \begin{pmatrix}
    M_{x}\\
    M_{y}\\
    M_{xy}
    \end{pmatrix}
    =
    \begin{pmatrix}
    B_{11} & B_{12} & B_{16}\\
    B_{12} & B_{22} & B_{26}\\
    B_{16} & B_{26} & B_{66}
    \end{pmatrix}
    \begin{pmatrix}
    \epsilon_{x}^0\\
    \epsilon_{y}^0\\
    \epsilon_{xy}^0
    \end{pmatrix}
    +
    \begin{pmatrix}
    D_{11} & D_{12} & D_{16}\\
    D_{12} & D_{22} & D_{26}\\
    D_{16} & D_{26} & D_{66}
    \end{pmatrix}
    \begin{pmatrix}
    \kappa_{x}\\
    \kappa_{y}\\
    \kappa_{xy}
    \end{pmatrix}
    -
    \begin{pmatrix}
    M_{x}^T\\
    M_{y}^T\\
    M_{xy}^T
    \end{pmatrix}\!,
\end{equation}
где $ N_{ij} $ "--- элементы вектора результирующей нагрузки, отнесённой к~единице длины линий, ограничивающих элемент рассматриваемой поверхности, Н/м;
$ M_{ij} $ "--- элементы вектора результирующего момента, отнесённого к~единице длины линий, ограничивающих элемент рассматриваемой поверхности, Н;
$ N_{ij}^T $~---~элементы векторной записи усилия, вызванного тепловым воздействием, отнесённого к~единице длины линий, ограничивающих элемент рассматриваемой поверхности, \mbox{Н/м};
$ M_{ij}^T $ "--- элементы векторной записи момента силы, вызванного тепловым воздействием, отнесённого к~единице длины линий, ограничивающих элемент рассматриваемой поверхности, Н.

Силы и~моменты, вызванные тепловым воздействием, определяются следующими уравнениями:
\begin{equation}\label{eq:NT_integral}
\mathbf{N}^T = \int\limits_t \mathbf{Q} \boldsymbol{\epsilon}^T \mathrm{d}z,
\end{equation}
\begin{equation}\label{eq:MT_integral}\noeqref{eq:MT_integral}
\mathbf{M}^T = \int\limits_t \mathbf{Q} \boldsymbol{\epsilon}^T z \mathrm{d}z.
\end{equation}

Уравнение~\eqref{eq:n_m_eps_kap_simp_matrix_comb} можно переписать следующим
образом, так как внешняя механическая нагрузка не приложена:
\[
\left(
    \begin{array}{c}
        \mathbf{N}^T \\
        \mathbf{M}^T
    \end{array}
\right)
=
\left(
    \begin{array}{cc}
        \mathbf{A} & \mathbf{B} \\
        \mathbf{B} & \mathbf{D}
    \end{array}
\right)
\left(
    \begin{array}{c}
        \boldsymbol{\epsilon^0} \\
        \boldsymbol{\kappa}
    \end{array}
\right)\!\!.
\]
Уравнения (\refeq{eq:N_matrix},~\refeq{eq:M_matrix}) можно переписать:
\[
    \begin{pmatrix}
    N_{x}^T\\
    N_{y}^T\\
    N_{xy}^T
    \end{pmatrix}
    =
    \begin{pmatrix}
    A_{11} & A_{12} & A_{16}\\
    A_{12} & A_{22} & A_{26}\\
    A_{16} & A_{26} & A_{66}
    \end{pmatrix}
    \begin{pmatrix}
    \epsilon_{x}^0\\
    \epsilon_{y}^0\\
    \epsilon_{xy}^0
    \end{pmatrix}
    +
    \begin{pmatrix}
    B_{11} & B_{12} & B_{16}\\
    B_{12} & B_{22} & B_{26}\\
    B_{16} & B_{26} & B_{66}
    \end{pmatrix}
    \begin{pmatrix}
    \kappa_{x}\\
    \kappa_{y}\\
    \kappa_{xy}
    \end{pmatrix}\!,
\]
\[
    \begin{pmatrix}
    M_{x}^T\\
    M_{y}^T\\
    M_{xy}^T
    \end{pmatrix}
    =
    \begin{pmatrix}
    B_{11} & B_{12} & B_{16}\\
    B_{12} & B_{22} & B_{26}\\
    B_{16} & B_{26} & B_{66}
    \end{pmatrix}
    \begin{pmatrix}
    \epsilon_{x}^0\\
    \epsilon_{y}^0\\
    \epsilon_{xy}^0
    \end{pmatrix}
    +
    \begin{pmatrix}
    D_{11} & D_{12} & D_{16}\\
    D_{12} & D_{22} & D_{26}\\
    D_{16} & D_{26} & D_{66}
    \end{pmatrix}
    \begin{pmatrix}
    \kappa_{x}\\
    \kappa_{y}\\
    \kappa_{xy}
    \end{pmatrix}\!.
\]
Преобразуем эту систему уравнений:
\begin{multline}%
\label{eq:eps0_simpmatrix}\noeqref{eq:eps0_simpmatrix}
\boldsymbol{\epsilon^0} =
\left(
\mathbf{A^{\!\!-1}} %притягиваем показатель обратности матрицы
+
(\mathbf{A^{\!\!-1}B}) %притягиваем показатель обратности матрицы
(\mathbf{D} - \mathbf{B A^{\!\!-1}B})\mathbf{{}^{\!-1}} %\! притягиваем показатель обратности матрицы
(\mathbf{B A^{\!\!-1}})
\right)
\mathbf{N}^T
\ifnumequal{\value{usealtfont}}{2}{-}{- \\ -} %перенос формулы
(\mathbf{A^{\!\!-1}B}) %притягиваем показатель обратности матрицы
(\mathbf{D} - \mathbf{B A^{\!\!-1}B})\mathbf{{}^{\!-1}} %притягиваем показатель обратности матрицы
\mathbf{M}^T,
\end{multline}
\begin{equation}\label{eq:kappa_simpmatrix}
\boldsymbol{\kappa} =
-
(\mathbf{D} - \mathbf{B A^{\!\!-1}B})\mathbf{{}^{\!-1}}
(\mathbf{B A^{\!\!-1}})
\mathbf{N}^T
+
(\mathbf{D} - \mathbf{B A^{\!\!-1}B})\mathbf{{}^{\!-1}}
\mathbf{M}^T,
\end{equation}
где $ \mathbf{A} $ "--- матрица жёсткости при растяжении (мембранная жёсткость), Н/м;
$ \mathbf{B} $~---~матрица жёсткости изгиб\nb-растяжение (смешанная жёсткость), Н;
$ \mathbf{D} $~---~матрица жёсткости при изгибе (изгибная жёсткость), Н$\cdot$м.

Подстановкой уравнений (\refeq{eq:NT_integral}\textemdash\refeq{eq:kappa_simpmatrix})
в~уравнение~\eqref{eq:form_sigma_Q} получаем зависимость для определения
остаточных напряжений в~любой плоскости внутри рассматриваемой сборки
параллельной срединной поверхности:
\begin{multline*}
    \boldsymbol{\sigma}
    =
    \mathbf{Q}
    \left(
        \boldsymbol{\epsilon^0}
        +
        z
        \left(
            -
            (\mathbf{D} - \mathbf{B A^{\!\!-1}B})\mathbf{{}^{\!-1}} %притягиваем показатель обратности матрицы
            (\mathbf{B A^{\!\!-1}}) %притягиваем показатель обратности матрицы
            \mathbf{N}^T
            \ifnumequal{\value{usealtfont}}{2}{+}{%
            \right.\right. % фейк ради переноса
            + \\ +
            \left.} % фейк ради переноса
            (\mathbf{D} - \mathbf{B A^{\!\!-1}B})\mathbf{{}^{\!-1}} %притягиваем показатель обратности матрицы
            \mathbf{M}^T
        \right)
        -
        \int\limits_{T_{b}}^{T_{w}}
        \ifnumequal{\value{usealtfont}}{2}{}{\left.} % фейк ради переноса
        \boldsymbol{\alpha}(T)\:\mathrm{d}T
    \right).
\end{multline*}

Напряжения в~любых направлениях можно определять зная расположение главных осей
кремния, преобразуя значения элементов матриц жёсткости в~элементы матрицы
$\mathbf{Q}$~в~соответствии с~формулами поворота системы координат \cites[{18,~215,~224}]{Alfutov1984_povorot_matric}.

Описанные модели, в частности модель двух тонких слоёв и модель
многослойного композиционного материала, позволяют, последовательно
увеличивая вычислительную сложность, оценивать остаточные напряжения
в~сплошных соединённых пластинах кремния и стекла. Уменьшение числа допущений
приводит к усложнению и дальнейшему развитию методов расчёта
напряжённо\nb-деформированного состояния многослойных
пластин~\cites[{165,~170}]{kristensen1982_per_Vved_v_kompozit}{bitkina2013_avtoref}.
Однако, для наиболее точной оценки
остаточных напряжений в приборах с~объёмной микрообработкой кремния
следует применять конечно-элементное
моделирование. Предварительное использование простых моделей, описанных
ранее в этой главе облегчает поиск оптимальных макроскопических параметров
моделей в~комплексах численного конечно-элементного моделирования.

\ifnumequal{\value{usealtfont}}{2}{%
    \Needspace*{10\onelineskip}
}{}
\section{Результаты применения моделей}

\subsection{Исходные данные (характеристики применяемых материалов)}\label{chap_source_data}

Зависимости ТКЛР от температуры для стёкол марок Borofloat 33, ЛК5,
7740 и SD\nb-2 были получены автором экспериментально
термомеханическим анализатором TMA7100 с относительной погрешностью
${\pm}$5~\% (см.~подраздел~\ref{chap_cte_measure}). Данные
производителя по температурной зависимости ТКЛР стекла марки SW\nb-YY
приведены в~подразделе~\ref{GlassInfoOffic}. Для тех диапазонов
температур, которые выходили за границы, рассмотренные производителем,
использовались зависимости ТКЛР подобные имеющимся в известном
диапазоне температур. Упругие свойства вышеназванных стёкол также
приведены в~подразделе~\ref{GlassInfoOffic}.

\begin{figure}[!hb]%Порядок, в котором заданы опции, позволяющие регулировать положение рисунка на странице, не важен — они всегда будут применяться в порядке h (здесь) — t (вверху) — b (внизу) — p (на отдельной странице). Важно только то, какие именно опции заданы. По умолчанию — [tbp]. Задавать опции по одной (например, просто [t] или просто [b]) не рекомендуется — в некоторых случаях это может приводить к проблемным ситуациям.
    \centering
    \begingroup%
    \makeatletter%
    \providecommand\color[2][]{%
        \errmessage{(Inkscape) Color is used for the text in Inkscape, but the package 'color.sty' is not loaded}%
        \renewcommand\color[2][]{}%
    }%
    \providecommand\transparent[1]{%
        \errmessage{(Inkscape) Transparency is used (non-zero) for the text in Inkscape, but the package 'transparent.sty' is not loaded}%
        \renewcommand\transparent[1]{}%
    }%
    \providecommand\rotatebox[2]{#2}%
    \ifx\svgwidth\undefined%
    \setlength{\unitlength}{0.7\textwidth}%
    \ifx\svgscale\undefined%
    \relax%
    \else%
    \setlength{\unitlength}{\unitlength * \real{\svgscale}}%
    \fi%
    \else%
    \setlength{\unitlength}{\svgwidth}%
    \fi%
    \global\let\svgwidth\undefined%
    \global\let\svgscale\undefined%
    \makeatother%
    \begin{picture}(1,0.91078009)%
    \put(0,0){\includegraphics[width=\unitlength]{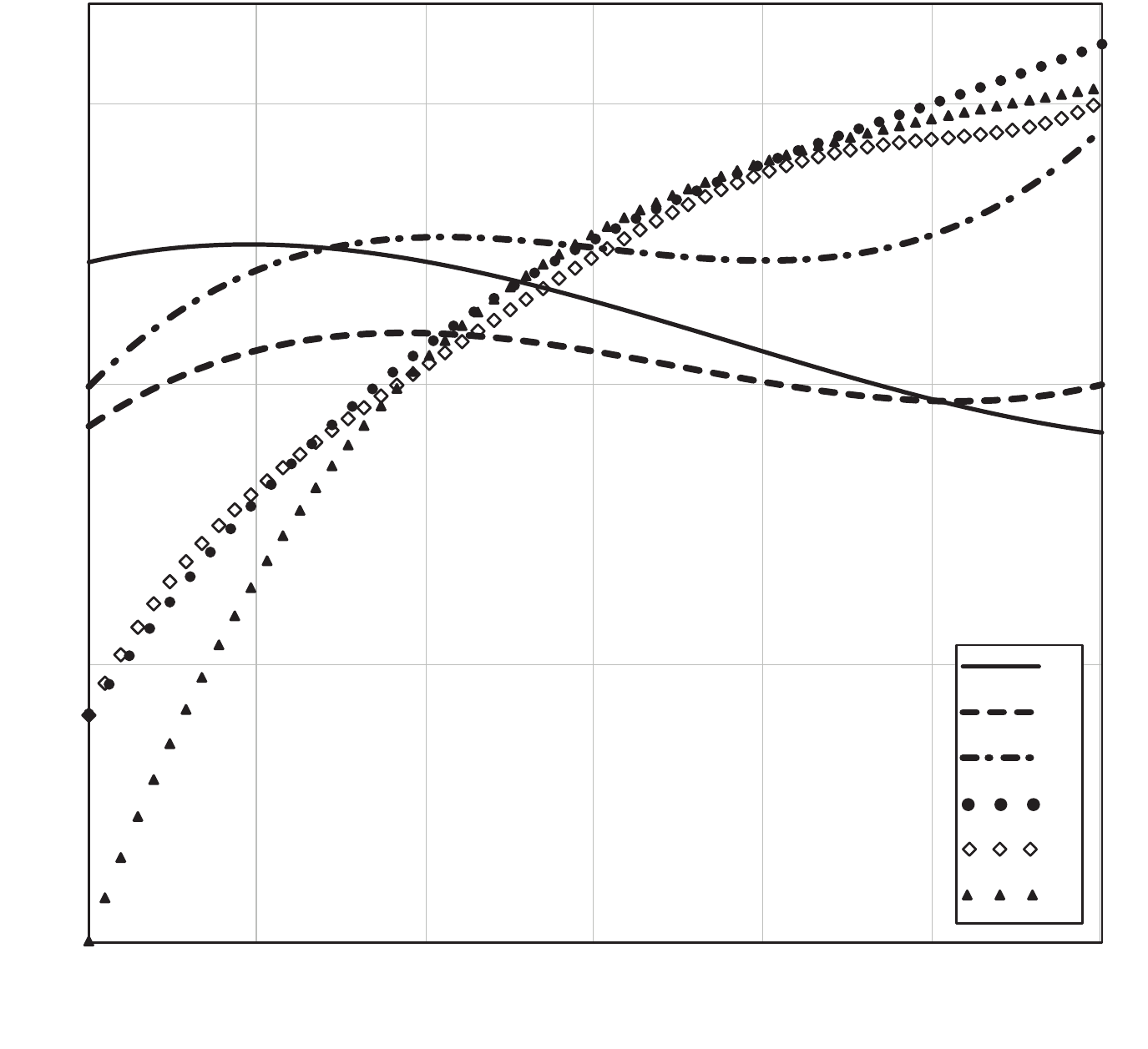}}%
    \put(0.04942721,0.05233306){\color[named]{black}\makebox(0,0)[lb]{\smash{$-$100}}}%
    \put(0.22365626,0.05233304){\color[named]{black}\makebox(0,0)[lb]{\smash{0}}}%
    \put(0.35645273,0.05233304){\color[named]{black}\makebox(0,0)[lb]{\smash{100}}}%
    \put(0.50245295,0.05233304){\color[named]{black}\makebox(0,0)[lb]{\smash{200}}}%
    \put(0.65042616,0.05233304){\color[named]{black}\makebox(0,0)[lb]{\smash{300}}}%
    \put(0.79839936,0.05233304){\color[named]{black}\makebox(0,0)[lb]{\smash{400}}}%
    \put(0.0542787,0.3197472){\color[named]{black}\makebox(0,0)[lb]{\smash{2}}}%
    \put(0.05360435,0.56454799){\color[named]{black}\makebox(0,0)[lb]{\smash{3}}}%
    \put(0.0544862,0.80919695){\color[named]{black}\makebox(0,0)[lb]{\smash{4}}}%
    \put(0.92608635,0.32172017){\color[named]{black}\makebox(0,0)[b]{\smash{\textsl{1}}}}%
    \put(0.92608635,0.28180535){\color[named]{black}\makebox(0,0)[b]{\smash{\textsl{2}}}}%
    \put(0.92608635,0.24204229){\color[named]{black}\makebox(0,0)[b]{\smash{\textsl{3}}}}%
    \put(0.92608635,0.20212747){\color[named]{black}\makebox(0,0)[b]{\smash{\textsl{4}}}}%
    \put(0.92608635,0.16236441){\color[named]{black}\makebox(0,0)[b]{\smash{\textsl{5}}}}%
    \put(0.92608635,0.12244959){\color[named]{black}\makebox(0,0)[b]{\smash{\textsl{6}}}}%
    \put(0.02458804,0.38804245){\color[named]{black}\rotatebox{90}{\makebox(0,0)[lb]{\smash{$\alpha$, 10\textsuperscript{$-$6}~{\textdegree}C\textsuperscript{$-$1}}}}}%
    \put(0.94814375,0.05233306){\color[named]{black}\makebox(0,0)[lb]{\smash{500}}}%
    \put(0.05088963,0.0809656){\color[named]{black}\makebox(0,0)[lb]{\smash{1}}}%
    \put(0.48230067,0.00589629){\color[named]{black}\makebox(0,0)[lb]{\smash{$T$,~{\textdegree}C}}}%
    \end{picture}%
    \endgroup%
    \caption[Coefficients of thermal expansion of several glass brands
    and silicon used for estimations]{Зависимости коэффициентов теплового
    линейного расширения от~температуры для нескольких марок стекла
    и кремния:}
    \label{fig:cte}
    \legend{%
        \textsl{1} "--- Corning 7740,  \textsl{2} "--- Schott Borofloat 33,  \textsl{3} "--- ЛК5,  \textsl{4} "--- Hoya~SD-2,  \textsl{5}~---~Asahi~SW\nb-YY,  \textsl{6} "--- кремний%
    }
\end{figure}

На Рисунке~\ref{fig:cte}
приведены графики зависимостей ТКЛР от температуры для стёкол нескольких марок и кремния, использованные для расчётов в~этой главе.

Упругие свойства пластины кремния ориентации \{100\} взяты из
\cites[42]{Bao_part_Mech_Beam_Diaphragm_Structures}.
Зависимость ТКЛР кремния от температуры приведена в~\cite{OkadaTokumaru_precise_silicon1984}. Для удобства использования в расчётах, она была переведена в полиномиальную форму:
\begin{multline}
    \alpha_{si}(T) = -3,268 \cdot 10^{-6} + 3,469 \cdot 10^{-8} \cdot T - 6,889 \cdot 10^{-11} \cdot T^2
    + \\  %перенос формулы
   + 6,991 \cdot 10^{-14} \cdot T^3 - 3,51 \cdot 10^{-17} \cdot T^4 + 6,916 \cdot 10^{-21} \cdot T^5,
\end{multline}

Во всех материалах с кубической кристаллической решёткой, в том числе и в кремнии, ТКЛР не зависит от направления измерения~\cite{slack1975thermal}.

\subsection{Методика определения температуры соединения из~анализа кривых зависимостей температурных коэффициентов линейного расширения}
\begin{figure}[!htb]%Порядок, в котором заданы опции, позволяющие регулировать положение рисунка на странице, не важен — они всегда будут применяться в порядке h (здесь) — t (вверху) — b (внизу) — p (на отдельной странице). Важно только то, какие именно опции заданы. По умолчанию — [tbp]. Задавать опции по одной (например, просто [t] или просто [b]) не рекомендуется — в некоторых случаях это может приводить к проблемным ситуациям.
    \centering
    \noindent%
    \begingroup%
      \makeatletter%
      \ifx\svgwidth\undefined%
        \setlength{\unitlength}{0.54\textwidth}%
        \ifx\svgscale\undefined%
          \relax%
        \else%
          \setlength{\unitlength}{\unitlength * \real{\svgscale}}%
        \fi%
      \else%
        \setlength{\unitlength}{\svgwidth}%
      \fi%
      \global\let\svgwidth\undefined%
      \global\let\svgscale\undefined%
      \makeatother%
      \begin{picture}(1,0.75870968)%
        \put(0,0){\includegraphics[width=\unitlength]{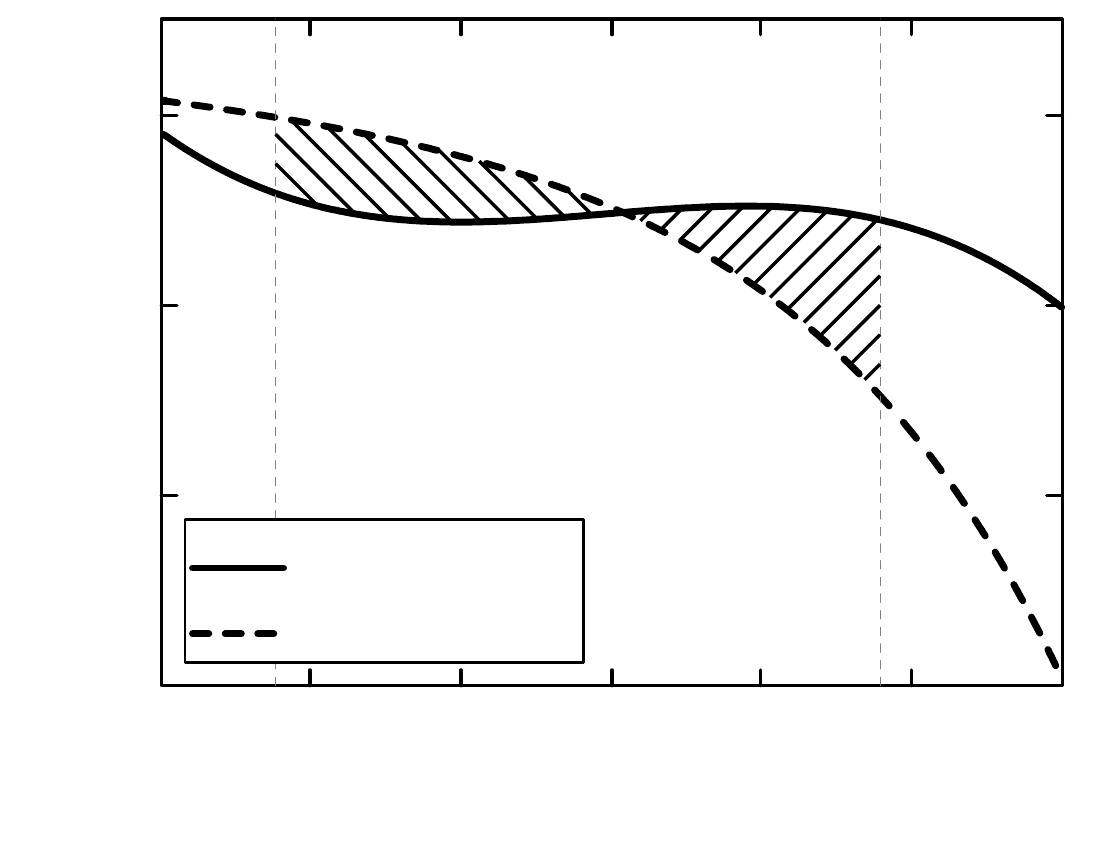}}%
        \put(0.13379818,0.13843943){\color[named]{black}\makebox(0,0)[rb]{\smash{1}}}%
        \put(0.13198206,0.30434553){\color[named]{black}\makebox(0,0)[rb]{\smash{2}}}%
        \put(0.13243075,0.47250155){\color[named]{black}\makebox(0,0)[rb]{\smash{3}}}%
        \put(0.13125561,0.64322094){\color[named]{black}\makebox(0,0)[rb]{\smash{4}}}%
        \put(0.14691889,0.08513952){\color[named]{black}\makebox(0,0)[b]{\smash{500}}}%
        \put(0.28063032,0.08513952){\color[named]{black}\makebox(0,0)[b]{\smash{400}}}%
        \put(0.41659216,0.08513952){\color[named]{black}\makebox(0,0)[b]{\smash{300}}}%
        \put(0.552554,0.08513952){\color[named]{black}\makebox(0,0)[b]{\smash{200}}}%
        \put(0.68607791,0.08513952){\color[named]{black}\makebox(0,0)[b]{\smash{100}}}%
        \put(0.82078945,0.08513952){\color[named]{black}\makebox(0,0)[b]{\smash{0}}}%
        \put(0.94206112,0.08513952){\color[named]{black}\makebox(0,0)[b]{\smash{$-$100}}}%
        \put(0.06309799,0.44107293){\color[named]{black}\rotatebox{90}{\makebox(0,0)[b]{\smash{ТКЛР, 10\textsuperscript{$-$6}~{\textdegree}C\textsuperscript{$-$1}}}}}%
        \put(0.26708896,0.23391665){\color[named]{black}\makebox(0,0)[lb]{\smash{ЛК5}}}%
        \put(0.26447009,0.17756749){\color[named]{black}\makebox(0,0)[lb]{\smash{Кремний}}}%
        \put(0.79069249,0.685){\color[named]{black}\makebox(0,0)[b]{\smash{20}}}%
        \put(0.24981493,0.685){\color[named]{black}\makebox(0,0)[b]{\smash{422}}}%
      \end{picture}%
    \endgroup%

    \noindent%
    \begingroup%
      \makeatletter%
      \ifx\svgwidth\undefined%
        \setlength{\unitlength}{0.54\textwidth}%
        \ifx\svgscale\undefined%
          \relax%
        \else%
          \setlength{\unitlength}{\unitlength * \real{\svgscale}}%
        \fi%
      \else%
        \setlength{\unitlength}{\svgwidth}%
      \fi%
      \global\let\svgwidth\undefined%
      \global\let\svgscale\undefined%
      \makeatother%
      \begin{picture}(1,0.77746311)%
        \put(0,0){\includegraphics[width=\unitlength]{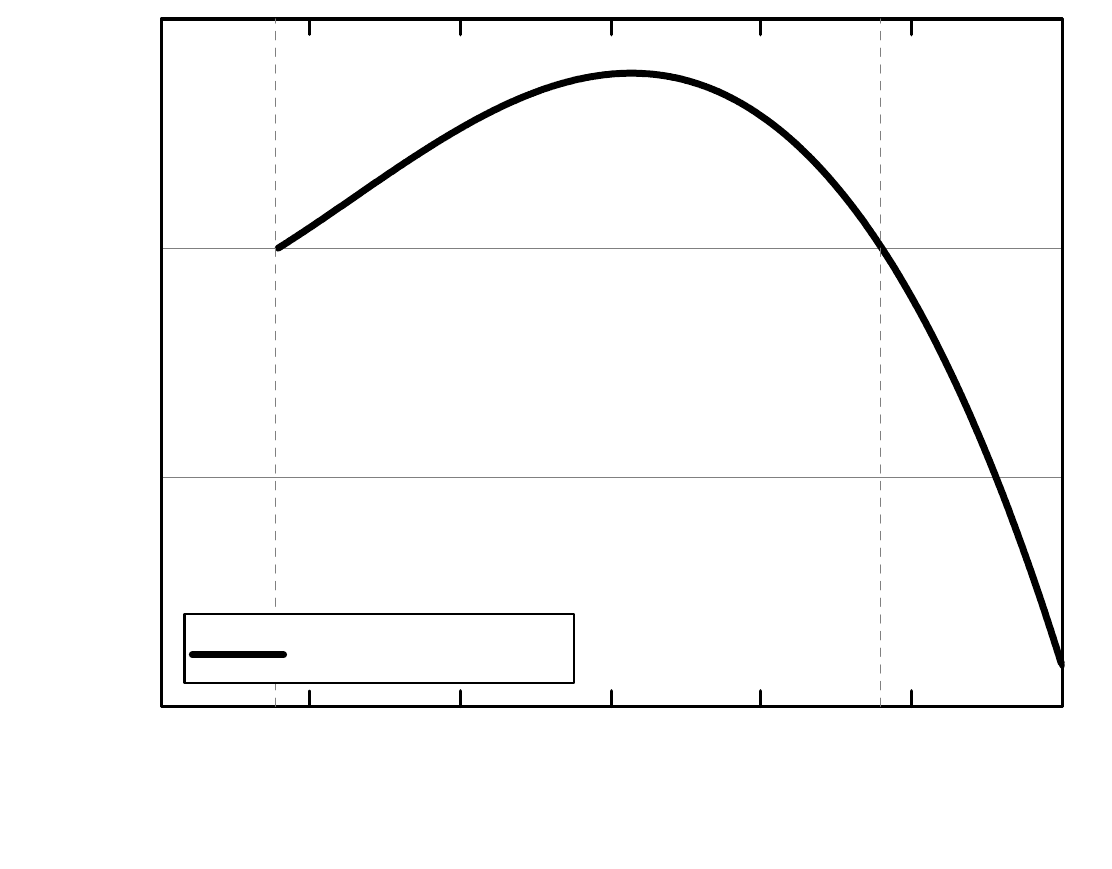}}%
        \put(0.1330892,0.12928924){\color[named]{black}\makebox(0,0)[rb]{\smash{$-$10}}}%
        \put(0.13274734,0.33651364){\color[named]{black}\makebox(0,0)[rb]{\smash{$-$5}}}%
        \put(0.1330892,0.54117533){\color[named]{black}\makebox(0,0)[rb]{\smash{0}}}%
        \put(0.13274734,0.74564893){\color[named]{black}\makebox(0,0)[rb]{\smash{5}}}%
        \put(0.03769707,0.46873699){\color[named]{black}\rotatebox{90}{\makebox(0,0)[b]{\smash{Напряжения, МПа}}}}%
        \put(0.55242838,0.01292082){\color[named]{black}\makebox(0,0)[b]{\smash{Температура, {\textdegree}C}}}%
        \put(0.14705445,0.08375692){\color[named]{black}\makebox(0,0)[b]{\smash{500}}}%
        \put(0.28076591,0.08375692){\color[named]{black}\makebox(0,0)[b]{\smash{400}}}%
        \put(0.41672777,0.08375692){\color[named]{black}\makebox(0,0)[b]{\smash{300}}}%
        \put(0.55268964,0.08375692){\color[named]{black}\makebox(0,0)[b]{\smash{200}}}%
        \put(0.68621509,0.08375692){\color[named]{black}\makebox(0,0)[b]{\smash{100}}}%
        \put(0.82092923,0.08375692){\color[named]{black}\makebox(0,0)[b]{\smash{0}}}%
        \put(0.94720432,0.08375692){\color[named]{black}\makebox(0,0)[b]{\smash{$-$100}}}%
        \put(0.2457595,0.69886941){\color[named]{black}\makebox(0,0)[b]{\smash{422}}}%
        \put(0.79085926,0.69886941){\color[named]{black}\makebox(0,0)[b]{\smash{20}}}%
        \put(0.26933091,0.17707933){\color[named]{black}\makebox(0,0)[lb]{\smash{Кремний}}}%
      \end{picture}%
    \endgroup%
    \caption[Graphical way of bonding temperature determination
    example]{Иллюстрация графического интегрирования на примере соединения
    стекла ЛК5 с кремнием}
    \label{graphic_integ}
\end{figure}

Рассмотрим  простой способ определения оптимальной температуры соединения с
помощью модели двух тонких слоёв. Для примера возьмём случай, когда в одном
диапазоне температур коэффициент теплового расширения стекла больше коэффициента
теплового расширения кремния, а в другом диапазоне меньше. Таким образом, график
зависимости ТКЛР стекла от~температуры пересекает график зависимости ТКЛР
кремния. Иллюстрацией данной ситуации послужат графики для стекла ЛК5 и кремния
(см.~Рисунок~\ref{graphic_integ}).

При температуре соединения, в данном примере 422~{\textdegree}C,
в соединённых пластинах не возникает остаточных напряжений.
Несоединённые кремний и~стекло имеют одинаковую длину.

В процессе дальнейшего охлаждения, пока $\alpha_{si} > \alpha_{g}$
(кремний сжимается быстрее стекла до температуры примерно 190~{\textdegree}C в случае стекла ЛК5, см.~Рисунок~\ref{graphic_integ}), увеличивается прогиб пластин (вогнутостью
в сторону пластины кремния). Также увеличиваются растягивающие напряжения внутри
пластины кремния.

Когда при дальнейшем охлаждении приходим к точке совпадения значений истинных
коэффициентов теплового расширения кремния и стекла $\alpha_{si} = \alpha_{g}$ (примерно около температуры 190~{\textdegree}C для стекла ЛК5), прогиб перестаёт увеличиваться,
достигнув максимального значения. Перестают увеличиваться и~растягивающие
напряжения в кремнии.
Несоединённый кремний был бы всё ещё короче стекла.

При продолжении охлаждения кремний сжимается медленнее стекла
\mbox{($\alpha_{si} < \alpha_{g}$)}. Величина прогиба начинает уменьшаться.
Растягивающие напряжения в кремнии так же пропорционально снижаются.

При достижении комнатной температуры (для данного примера,
см.~Рисунок~\ref{graphic_integ}), совокупная накопленная деформация становится
нулевой. Отсутствует прогиб пластин и остаточные напряжения в кремнии и~стекле
равны нулю.
Если бы кремний и стекло не были соединены, они снова были бы~одинаковой длины.

Если продолжить охлаждать, то за счёт того, что кремний сжимается медленнее
стекла ($\alpha_{si} < \alpha_{g}$), пластины продолжат изгибаться. Теперь
вогнутость будет в сторону пластины стекла. При этом в кремнии будут формироваться
остаточные сжимающие напряжения.

Определить температуру проведения процесса электростатического соединения,
обеспечивающую отсутствие напряжений при заданной рабочей температуре, можно
графически. Области, отсечённые линиями температур и~графиков ТКЛР стекла и
кремния по обе стороны от точки пересечения графиков ТКЛР, должны быть равными
(см. Рисунок~\ref{graphic_integ}). Это утверждение также вытекает
из~формулы~\eqref{eq:sigma_siupdated}. Для ЛК5 и рабочей температуры
20~{\textdegree}C температура соединения, обеспечивающая отсутствие остаточных
напряжений, составляет 422~{\textdegree}C.

Для рассматриваемых марок боросиликатных стёкол (ЛК5, 7740, Borofloat~33) точка
пересечения кривых температурных зависимостей ТКЛР располагается при
температурах ниже типичных температур проведения анодной посадки. Таким образом,
для определения температуры ненапряжённого соединения можно заключить, что
интеграл разности температурных зависимостей ТКЛР стекла и кремния от рабочей
температуры до~температуры в~момент подачи высокого напряжения должен равняться
нулю.

Расчётные значения накапливаемой относительной деформации приведены на
Рисунке~\ref{fig:nakop_deform}.
Графики построены на основе следующей формулы:
\begin{equation}
    \label{eq:nakop_deform}
    \frac {\Delta l}{l_0}
    =
    \int\limits_{T_b}^{T_w}
    (
         \alpha_g(T) - \alpha_{si}(T)
    )
    \:\mathrm{d}T.
\end{equation}

Из Рисунка~\ref{fig:nakop_deform} видно, что каждое стекло влияет по-разному
на напряжения в кремнии в рабочем диапазоне температур. Увеличение температуры
сращивания кремния со стеклом \(T_b\) увеличивает растягивающие напряжения
в~стекле для четырёх из пяти рассматриваемых марок стёкол (исключая Hoya SD-2).
Из-за различий в ТКЛР (главным образом из-за разной температуры пересечения
с~ТКЛР кремния, см. Рисунок~\ref{fig:cte}) остаточные напряжения, вызываемые
стеклом ЛК5, более сжимающие и, в то же время, меньше зависят от выбора
температуры сращивания, чем напряжения, вызываемые стёклами Borofloat~33 и
Corning 7740. Изменение температуры сращивания с алюмосиликатными стёклами (SD-2
и~SW\nb-YY) оказывает заметно меньшее влияние на остаточные напряжения, чем с
боросиликатными стёклами (Borofloat~33, Corning 7740 и~ЛК5). Также
Рисунок~\ref{fig:nakop_deform} иллюстрирует, что идея «снижение температуры
сращивания снижает остаточные напряжения» верна лишь для определённых марок
стекла или даже для определённых партий стекла, поскольку зависимость ТКЛР от
температуры может отличаться от~партии к~партии. Например, снижение температуры
соединения кремния со~стеклом ЛК5 лишь повышает сжимающие напряжения в кремнии
(согласно модели двух тонких слоёв). Видно, что утверждение о~наличии
температуры ненапряжённого сращивания~\cite{Cozma_Puers_1995} верно лишь для
конкретной заданной рабочей температуры \(T_w\). В~диапазоне температур
остаточные напряжения будут изменяться.

\begin{figure}[!htb]%Порядок, в котором заданы опции, позволяющие регулировать положение рисунка на странице, не важен — они всегда будут применяться в порядке h (здесь) — t (вверху) — b (внизу) — p (на отдельной странице). Важно только то, какие именно опции заданы. По умолчанию — [tbp]. Задавать опции по одной (например, просто [t] или просто [b]) не рекомендуется — в некоторых случаях это может приводить к проблемным ситуациям.
    \centering
    \ifdefmacro{\tikzsetnextfilename}{\tikzsetnextfilename{nakop_deform}}{}%
    \input{Dissertation/images_tikz/disser_nakop_deform.tikz}%
    \caption[Relative deformation in silicon bonded to different
    glasses estimated by~the~thin-film model]{Расчётная оценка
    накапливаемой относительной деформации в~области рабочих температур
    сборки кремний"--~стекло для стёкол разных марок}
    \label{fig:nakop_deform}
\end{figure}

\subsection{Сравнение результатов применения среднего и~истинного температурных коэффициентов линейного расширения}
Рассмотрим результаты применения формулы~\eqref{eq:sigma_siupdated} для оценки напряжений в кремнии, соединённом электростатическим соединением (анодной посадкой) с несколькими марками стёкол.

\begin{figure}[!htb]%Порядок, в котором заданы опции, позволяющие регулировать положение рисунка на странице, не важен — они всегда будут применяться в порядке h (здесь) — t (вверху) — b (внизу) — p (на отдельной странице). Важно только то, какие именно опции заданы. По умолчанию — [tbp]. Задавать опции по одной (например, просто [t] или просто [b]) не рекомендуется — в некоторых случаях это может приводить к проблемным ситуациям.
    \centering
    \begingroup%
      \makeatletter%
      \providecommand\color[2][]{%
        \errmessage{(Inkscape) Color is used for the text in Inkscape, but the package 'color.sty' is not loaded}%
        \renewcommand\color[2][]{}%
      }%
      \providecommand\transparent[1]{%
        \errmessage{(Inkscape) Transparency is used (non-zero) for the text in Inkscape, but the package 'transparent.sty' is not loaded}%
        \renewcommand\transparent[1]{}%
      }%
      \providecommand\rotatebox[2]{#2}%
      \ifx\svgwidth\undefined%
        \setlength{\unitlength}{0.60\textwidth}%
        \ifx\svgscale\undefined%
          \relax%
        \else%
          \setlength{\unitlength}{\unitlength * \real{\svgscale}}%
        \fi%
      \else%
        \setlength{\unitlength}{\svgwidth}%
      \fi%
      \global\let\svgwidth\undefined%
      \global\let\svgscale\undefined%
      \makeatother%
      \begin{picture}(1,0.92674309)%
        \put(0,0){\includegraphics[width=\unitlength]{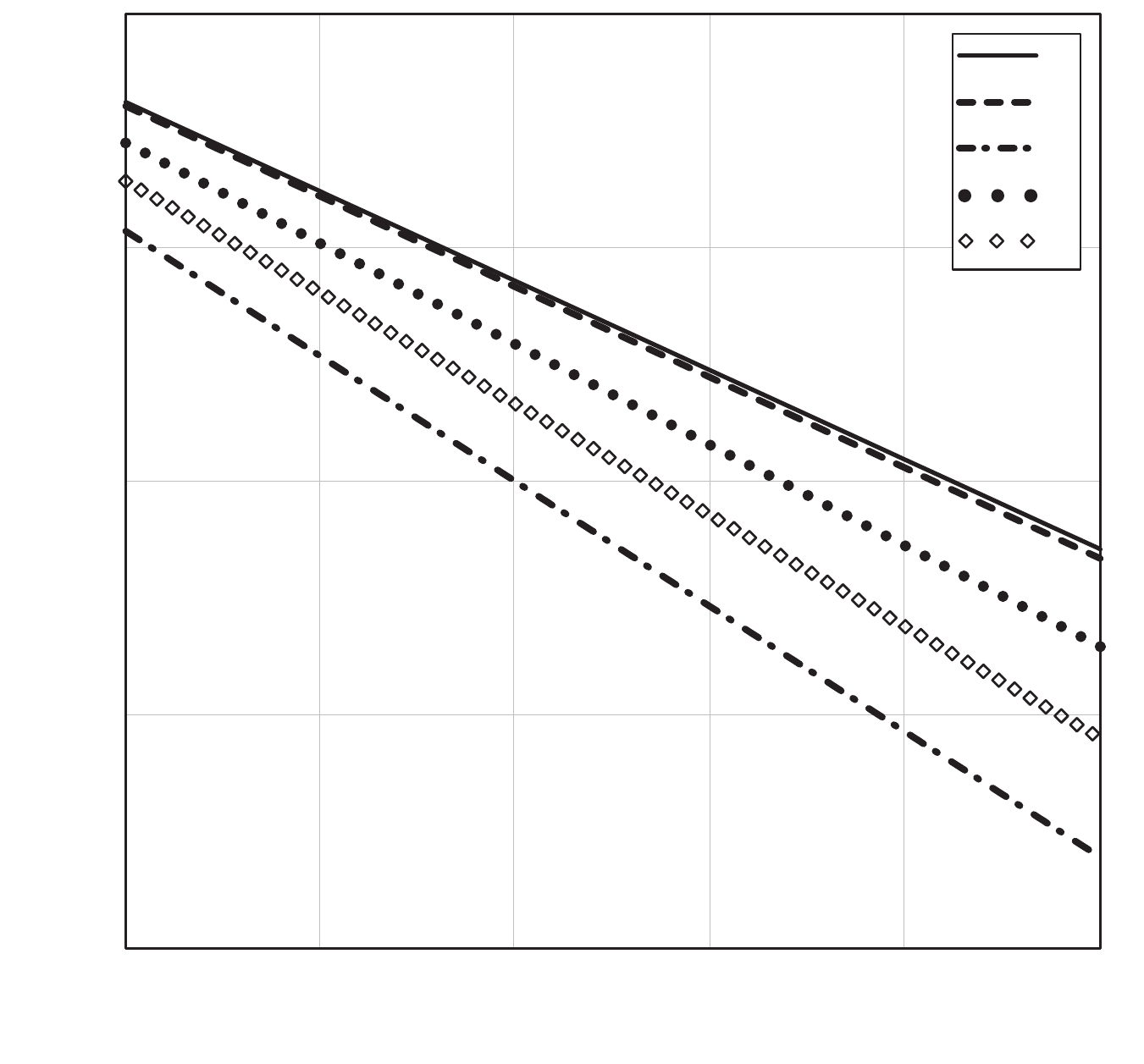}}%
        \put(0.09451188,0.05987052){\color[named]{black}\makebox(0,0)[lb]{\smash{200}}}%
        \put(0.26429042,0.05987052){\color[named]{black}\makebox(0,0)[lb]{\smash{250}}}%
        \put(0.43406896,0.05987052){\color[named]{black}\makebox(0,0)[lb]{\smash{300}}}%
        \put(0.60585397,0.05987052){\color[named]{black}\makebox(0,0)[lb]{\smash{350}}}%
        \put(0.77563251,0.05987052){\color[named]{black}\makebox(0,0)[lb]{\smash{400}}}%
        \put(0.94726318,0.05987052){\color[named]{black}\makebox(0,0)[lb]{\smash{450}}}%
        \put(0.10293962,0.28938027){\color[named]{black}\makebox(0,0)[rb]{\smash{$-$20}}}%
        \put(0.10402987,0.49388617){\color[named]{black}\makebox(0,0)[rb]{\smash{$-$15}}}%
        \put(0.10293962,0.69839215){\color[named]{black}\makebox(0,0)[rb]{\smash{$-$10}}}%
        \put(0.10402987,0.90289813){\color[named]{black}\makebox(0,0)[rb]{\smash{$-$5}}}%
        \put(0.92663254,0.870){\color[named]{black}\makebox(0,0)[b]{\smash{\textsl{1}}}}%
        \put(0.92663254,0.829){\color[named]{black}\makebox(0,0)[b]{\smash{\textsl{2}}}}%
        \put(0.92663254,0.789){\color[named]{black}\makebox(0,0)[b]{\smash{\textsl{3}}}}%
        \put(0.92663254,0.748){\color[named]{black}\makebox(0,0)[b]{\smash{\textsl{4}}}}%
        \put(0.92663254,0.708){\color[named]{black}\makebox(0,0)[b]{\smash{\textsl{5}}}}%
        \put(0.02500558,0.45122627){\color[named]{black}\rotatebox{90}{\makebox(0,0)[b]{\smash{$\sigma_{si}$, МПа}}}}%
        \put(0.10402987,0.08770316){\color[named]{black}\makebox(0,0)[rb]{\smash{$-$25}}}%
        \put(0.53641175,0.00777256){\color[named]{black}\makebox(0,0)[b]{\smash{$T_b$,~{\textdegree}C}}}%
      \end{picture}%
    \endgroup%
    \caption[Residual stress at 20~{\textdegree}C estimation by~the~thin-film
    model of~silicon--glass bonds made at different temperatures (average
    CTEs used)]{Расчётная оценка остаточных напряжений по модели двух тонких
    слоёв в кремнии при температуре 20~{\textdegree}C в~зависимости
    от~температуры соединения с разными марками стекла (для расчёта взяты
    средние ТКЛР по~данным производителей):}
    \label{fig:thermal_mismatch_stress_siavg}
    \legend{%
        \textsl{1} "--- Corning 7740,  \textsl{2} "--- Schott Borofloat 33,  \textsl{3} "--- ЛК5,  \textsl{4} "--- Hoya~SD\nobreakdash-2,  \textsl{5}~---~Asahi~SW\nobreakdash-YY%
    }
\end{figure}

\begin{figure}[!htb]
    \centering
    \begingroup%
      \makeatletter%
      \providecommand\color[2][]{%
        \errmessage{(Inkscape) Color is used for the text in Inkscape, but the package 'color.sty' is not loaded}%
        \renewcommand\color[2][]{}%
      }%
      \providecommand\transparent[1]{%
        \errmessage{(Inkscape) Transparency is used (non-zero) for the text in Inkscape, but the package 'transparent.sty' is not loaded}%
        \renewcommand\transparent[1]{}%
      }%
      \providecommand\rotatebox[2]{#2}%
      \ifx\svgwidth\undefined%
        \setlength{\unitlength}{0.60\textwidth}%
        \ifx\svgscale\undefined%
          \relax%
        \else%
          \setlength{\unitlength}{\unitlength * \real{\svgscale}}%
        \fi%
      \else%
        \setlength{\unitlength}{\svgwidth}%
      \fi%
      \global\let\svgwidth\undefined%
      \global\let\svgscale\undefined%
      \makeatother%
      \begin{picture}(1,0.95586421)%
        \put(0,0){\includegraphics[width=\unitlength]{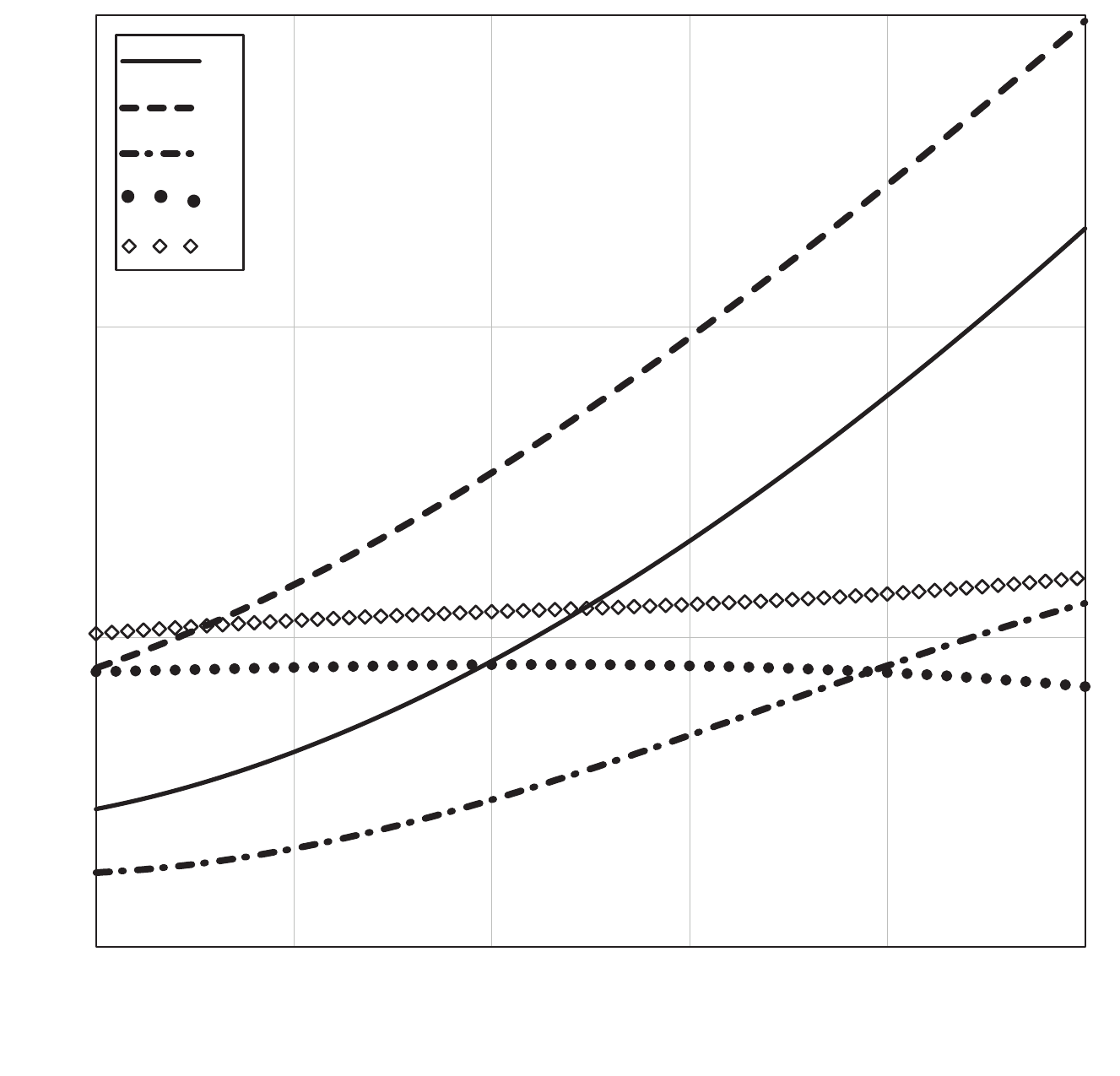}}%
        \put(0.07468862,0.37533739){\color[named]{black}\makebox(0,0)[rb]{\smash{0}}}%
        \put(0.07468862,0.65219154){\color[named]{black}\makebox(0,0)[rb]{\smash{5}}}%
        \put(0.07468862,0.9310884){\color[named]{black}\makebox(0,0)[rb]{\smash{10}}}%
        \put(0.19834598,0.89358804){\color[named]{black}\makebox(0,0)[b]{\smash{\textsl{1}}}}%
        \put(0.19834598,0.85226409){\color[named]{black}\makebox(0,0)[b]{\smash{\textsl{2}}}}%
        \put(0.19834598,0.81109742){\color[named]{black}\makebox(0,0)[b]{\smash{\textsl{3}}}}%
        \put(0.19834598,0.76977347){\color[named]{black}\makebox(0,0)[b]{\smash{\textsl{4}}}}%
        \put(0.19834598,0.7286068){\color[named]{black}\makebox(0,0)[b]{\smash{\textsl{5}}}}%
        \put(0.02545609,0.47127867){\color[named]{black}\rotatebox{90}{\makebox(0,0)[b]{\smash{$\sigma_{si}$, МПа}}}}%
        \put(0.07819749,0.06094908){\color[named]{black}\makebox(0,0)[lb]{\smash{200}}}%
        \put(0.25103498,0.06094908){\color[named]{black}\makebox(0,0)[lb]{\smash{250}}}%
        \put(0.42387246,0.06094908){\color[named]{black}\makebox(0,0)[lb]{\smash{300}}}%
        \put(0.59875257,0.06094908){\color[named]{black}\makebox(0,0)[lb]{\smash{350}}}%
        \put(0.77158998,0.06094908){\color[named]{black}\makebox(0,0)[lb]{\smash{400}}}%
        \put(0.94631305,0.06094908){\color[named]{black}\makebox(0,0)[lb]{\smash{450}}}%
        \put(0.52795474,0.00791253){\color[named]{black}\makebox(0,0)[b]{\smash{$T_b$,~{\textdegree}C}}}%
        \put(0.07468862,0.10330381){\color[named]{black}\makebox(0,0)[rb]{\smash{$-$5}}}%
      \end{picture}%
    \endgroup%
    \caption[Residual stress at 20~{\textdegree}C estimation by~the~thin-film
    model of~silicon--glass bonds made at different temperatures]{Расчётная
    оценка остаточных напряжений по модели двух тонких слоёв в кремнии при
    температуре 20~{\textdegree}C в зависимости от~температуры соединения с
    разными марками стекла:}
    \label{fig:thermal_mismatch_stress_si}
    \legend{%
        \textsl{1} "--- Corning 7740,  \textsl{2} "--- Schott Borofloat 33,  \textsl{3} "--- ЛК5,  \textsl{4} "--- Hoya~SD\nobreakdash-2,  \textsl{5}~---~Asahi~SW\nobreakdash-YY%
    }
\end{figure}

\begingroup
На Рисунках~\ref{fig:thermal_mismatch_stress_siavg}
и~\ref{fig:thermal_mismatch_stress_si} графиками показана расчётная оценка
по формуле~\eqref{eq:sigma_siupdated} остаточных напряжений в кремнии при
температуре 20~{\textdegree}C в зависимости от температуры соединения с
разными марками стекла. Видно, что при применении для оценки среднего ТКЛР
тенденции изменения напряжений в~зависимости от~температуры соединения для
всех стёкол примерно похожи. Чего нельзя сказать о случае применения
истинных нелинейных значений ТКЛР.\russianpar
\endgroup

В этих расчётах использовалась толщина кремния 0,46~мм и толщина стекла 0,5~мм.
Зависимости модулей Юнга стёкол от температуры производителями не приводятся, поэтому в расчётах использовались приведённые
в~тех же~источниках, что и ТКЛР, постоянные значения.
Зависимость модуля Юнга кремния в~направлении [100] от температуры взята
из работы~\cite{swarnakar2014determination}. Учёт зависимости модуля Юнга от
температуры приводит к разнице в оценке остаточных напряжений не
превышающей диапазон~${\pm}$0,1~МПа.


Расчётная температура минимальных остаточных напряжений для Corning~7740 приведённая на Рисунке~\ref{fig:thermal_mismatch_stress_si} подтверждается экспериментальными данными из работы~\cite{LeeMC2005_gyro_siog}.

\begin{figure}[!htbp]
    \centering
    \begingroup%
      \makeatletter%
      \providecommand\color[2][]{%
        \errmessage{(Inkscape) Color is used for the text in Inkscape, but the package 'color.sty' is not loaded}%
        \renewcommand\color[2][]{}%
      }%
      \providecommand\transparent[1]{%
        \errmessage{(Inkscape) Transparency is used (non-zero) for the text in Inkscape, but the package 'transparent.sty' is not loaded}%
        \renewcommand\transparent[1]{}%
      }%
      \providecommand\rotatebox[2]{#2}%
      \ifx\svgwidth\undefined%
        \setlength{\unitlength}{393.24069286bp}%
        \ifx\svgscale\undefined%
          \relax%
        \else%
          \setlength{\unitlength}{\unitlength * \real{\svgscale}}%
        \fi%
      \else%
        \setlength{\unitlength}{\svgwidth}%
      \fi%
      \global\let\svgwidth\undefined%
      \global\let\svgscale\undefined%
      \makeatother%
      \begin{picture}(1,0.57786925)(0,-0.04)%
        \put(0,0){\includegraphics[width=\unitlength]{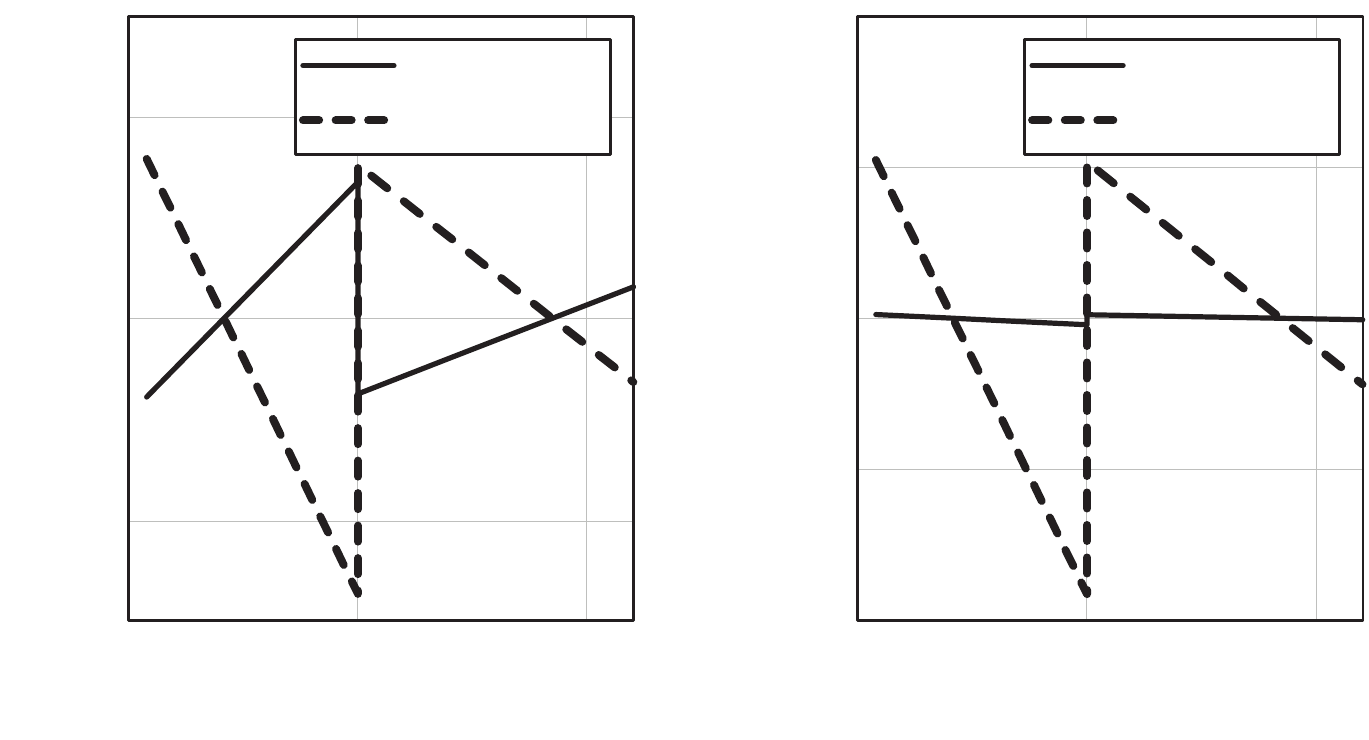}}%
        \put(0.08193627,0.04661595){\color[named]{black}\makebox(0,0)[lb]{\smash{$-$0.5}}}%
        \put(0.26197873,0.04661588){\color[named]{black}\makebox(0,0)[lb]{\smash{0}}}%
        \put(0.41837149,0.04661588){\color[named]{black}\makebox(0,0)[lb]{\smash{0.5}}}%
        \put(0.08422509,0.14380795){\color[named]{black}\makebox(0,0)[rb]{\smash{$-$20}}}%
        \put(0.08422509,0.2925718){\color[named]{black}\makebox(0,0)[rb]{\smash{0}}}%
        \put(0.08422509,0.43950478){\color[named]{black}\makebox(0,0)[rb]{\smash{20}}}%
        \put(0.3001233,0.48146381){\color[named]{black}\makebox(0,0)[lb]{\smash{\small{Истинный}}}}%
        \put(0.3001233,0.44133571){\color[named]{black}\makebox(0,0)[lb]{\smash{{\small Средний}}}}%
        \put(0.27917176,0.00592778){\color[named]{black}\makebox(0,0)[b]{\smash{$z$, мм}}}%
        \put(0.02471946,0.30516171){\color[named]{black}\rotatebox{90}{\makebox(0,0)[b]{\smash{$\sigma_x^T$, МПа}}}}%
        \put(0.61596035,0.04661595){\color[named]{black}\makebox(0,0)[lb]{\smash{$-$0.5}}}%
        \put(0.79600281,0.04661588){\color[named]{black}\makebox(0,0)[lb]{\smash{0}}}%
        \put(0.95239558,0.04661588){\color[named]{black}\makebox(0,0)[lb]{\smash{0.5}}}%
        \put(0.61824917,0.18195252){\color[named]{black}\makebox(0,0)[rb]{\smash{$-$20}}}%
        \put(0.61824917,0.2925718){\color[named]{black}\makebox(0,0)[rb]{\smash{0}}}%
        \put(0.61824917,0.40319107){\color[named]{black}\makebox(0,0)[rb]{\smash{20}}}%
        \put(0.61824917,0.51381035){\color[named]{black}\makebox(0,0)[rb]{\smash{40}}}%
        \put(0.83414739,0.48146381){\color[named]{black}\makebox(0,0)[lb]{\small{Истинный}}}%
        \put(0.83414739,0.44133571){\color[named]{black}\makebox(0,0)[lb]{\smash{{\small Средний}}}}%
        \put(0.81319588,0.00592778){\color[named]{black}\makebox(0,0)[b]{\smash{$z$, мм}}}%
        \put(0.55874372,0.30516171){\color[named]{black}\rotatebox{90}{\makebox(0,0)[b]{\smash{$\sigma_x^T$, МПа}}}}%
        \put(0.61824917,0.07209782){\color[named]{black}\makebox(0,0)[rb]{\smash{$-$40}}}%
        \put(0.108,-0.04){%
        \fcolorbox{black}{siliconcolour}{\makebox[54pt]{Si}}%
        \fcolorbox{black}{glasscolour}{\makebox[74pt]{Borofloat 33}}%
        }%
        \put(0.64,-0.04){%
        \fcolorbox{black}{siliconcolour}{\makebox[55pt]{Si}}%
        \fcolorbox{black}{glasscolour}{\rule[1pt]{0pt}{7pt}\makebox[73pt]{ЛК5}}%
        }%
      \end{picture}%
    \endgroup%
    \caption[Estimation by the two-layer laminated plate model of residual
    stresses in the silicon--glass bond at 20~{\textdegree}C that was made at
    400~{\textdegree}C (using average CTE and instantaneous CTE)]{Расчётная
    оценка остаточных напряжений в~кремнии для двух стёкол (модель
    многослойного композиционного материала для истинных и~средних ТКЛР)
    при $T_w$ = 20~{\textdegree}C, $T_b$ = 400~{\textdegree}C}
    \label{fig:tm_stress_si_compos_2fig_bf33_lk5}
\end{figure}

\begingroup
Иллюстрация, сравнивающая остаточные напряжения, рассчитанные по~средним значениям ТКЛР и рассчитанные по истинным значениям ТКЛР для модели многослойного композиционного материала, приведена
на~Рисунке~\ref{fig:tm_stress_si_compos_2fig_bf33_lk5}.\russianpar
\endgroup

\subsection{Оценка вариации напряжений по модели двух тонких слоёв}
\begin{figure}[htbp]
    \centering
    \ifdefmacro{\tikzsetnextfilename}{\tikzsetnextfilename{sigma_workt_bf33_lk5_simple}}{}%
    \input{Dissertation/images_tikz/disser_sigma_workt_bf33_lk5.tikz}%
    \caption[Residual stresses in silicon bonded to Borofloat 33 and LK5
    glasses estimated by~the~thin-layer model]{Расчётные остаточные напряжения
    в кремнии в рабочем диапазоне температур (от~минус~60
    до~85~{\textdegree}C) прибора при разных фиксированных температурах
    $ T_b $ проведения процесса соединения со~стёклами Borofloat~33 и~ЛК5}
    \label{fig:sigma_workt_bf33_lk5_simple}
\end{figure}
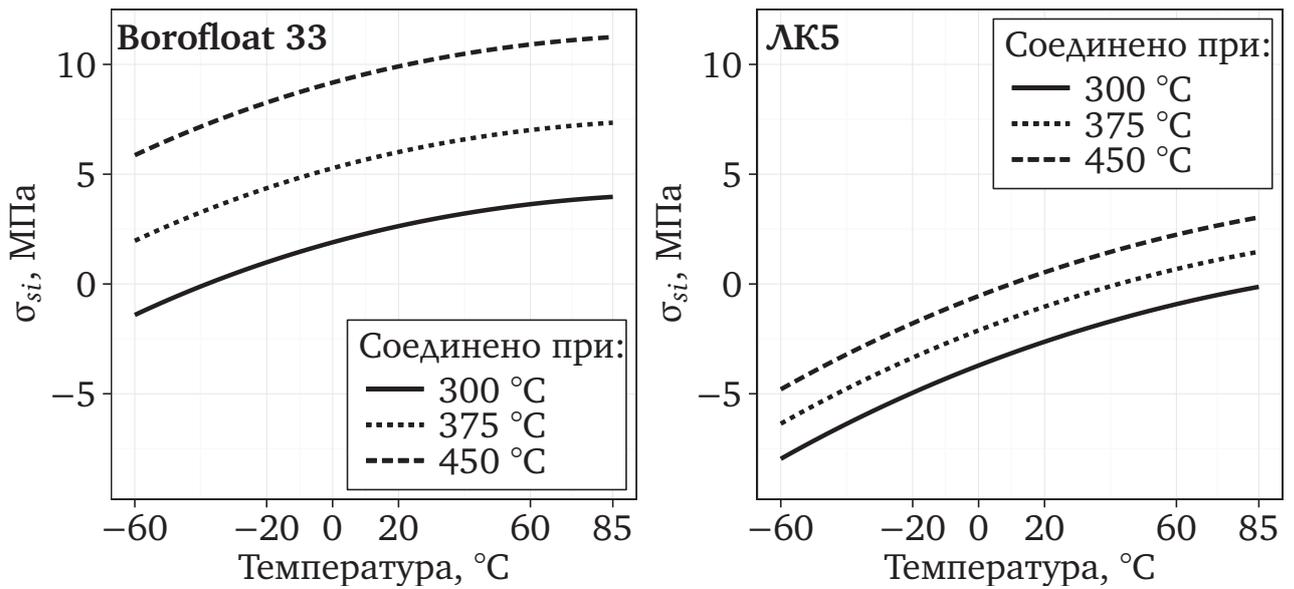

На Рисунке~\ref{fig:sigma_workt_bf33_lk5_simple} графиками показана расчётная
оценка по формуле~\eqref{eq:sigma_siupdated} остаточных напряжений в кремнии в
рабочем диапазоне температур (от~минус~60 до~85~{\textdegree}C) прибора при
разных фиксированных температурах $ T_b $ проведения процесса соединения со
стёклами марок Borofloat~33 и ЛК5. Расчёт проводился исходя из толщины кремния
460~мкм и толщины стекла 600~мкм.

Заметно различие как в разбросе (вариации) остаточных напряжений в~зависимости от рабочей температуры, так и в разбросе (вариации) остаточных напряжений в~зависимости от температуры соединения.

В Таблицу~\ref{tab_sigma_var} сведены рассчитанные разбросы для пяти марок стёкол, применяемых при проведении анодной посадки в разных странах мира.

\begingroup
Низкие значения разброса напряжений, связанного с изменением температуры
соединения, \(\sigma_b\), показывают, что изменением температуры соединения
для данных марок стёкол можно очень слабо влиять на получаемые
напряжения.\russianpar
\endgroup

\begin{table} [!htb]
    \centering%
    \parbox{0.7\textwidth}{% чтобы лучше смотрелось, подбирается самостоятельно
        \caption[Variations of residual stresses in silicon bonded to
        different glasses estimated by the thin-film model]{Расчётные
        вариации остаточных напряжений для различных марок стёкол}%
        \label{tab_sigma_var}% label всегда желательно идти после caption
    }%
    \renewcommand{\arraystretch}{1.3}%% Увеличение расстояния между рядами, для улучшения восприятия.
    \def\tabularxcolumn#1{m{#1}}
    \begin{SingleSpace}
        \begin{tabularx}{0.7\textwidth}{@{}
                >{\raggedright}X
                S
                S
            }
            \toprule     %%% верхняя линейка
            {Марка стекла} &
            {$\sigma_w$, МПа} &
            {$\sigma_b$, МПа}\\
            \midrule%%% тонкий разделитель
            ЛК5 &
            7,5 &
            3,8\\
            Corning 7740 &
            6,5 &
            4,9\\
            Schott Borofloat 33 &
            5,1 &
            8,7\\
            Hoya SD\nobreakdash-2 &
            3,5 &
            0,3\\
            Asahi SW\nobreakdash-YY &
            2,2 &
            0,4\\
            \midrule%%% тонкий разделитель
            \multicolumn{3}{@{}p{0.7\textwidth}@{}}{
                \hspace*{2.5em}Примечание "---
                $\sigma_w=|\sigma_{si}^{-60} - \sigma_{si}^{85}|$ "--- разброс (вариация) остаточных напряжений по модели двух тонких слоёв в рабочем диапазоне температур, МПа;  $\sigma_b=|\sigma_{si}^{250} - \sigma_{si}^{450}|$ "--- разброс (вариация) остаточных напряжений по~модели двух тонких слоёв в диапазоне допустимых температур соединения, МПа.
            }
            \\
            \bottomrule %%% нижняя линейка
        \end{tabularx}%
    \end{SingleSpace}
\end{table}

\subsection{Оценка влияния толщины пластины стекла}\label{chap_optim_glass_thickness}
Рассмотрим влияние выбора толщины стеклянной пластины на остаточные напряжения на свободной поверхности кремния. Для этого воспользуемся моделью многослойного композиционного материала.

\begin{figure}[!hb]
    \label{fig:sigma_z_bf33_comp_3thickness}
    \centering
    \begingroup%
      \makeatletter%
      \providecommand\color[2][]{%
        \errmessage{(Inkscape) Color is used for the text in Inkscape, but the package 'color.sty' is not loaded}%
        \renewcommand\color[2][]{}%
      }%
      \providecommand\transparent[1]{%
        \errmessage{(Inkscape) Transparency is used (non-zero) for the text in Inkscape, but the package 'transparent.sty' is not loaded}%
        \renewcommand\transparent[1]{}%
      }%
      \providecommand\rotatebox[2]{#2}%
      \ifx\svgwidth\undefined%
        \setlength{\unitlength}{0.6\textwidth}%
        \ifx\svgscale\undefined%
          \relax%
        \else%
          \setlength{\unitlength}{\unitlength * \real{\svgscale}}%
        \fi%
      \else%
        \setlength{\unitlength}{\svgwidth}%
      \fi%
      \global\let\svgwidth\undefined%
      \global\let\svgscale\undefined%
      \makeatother%
      \begin{picture}(1,0.79787736)%
        \put(0,0){\includegraphics[width=\unitlength]{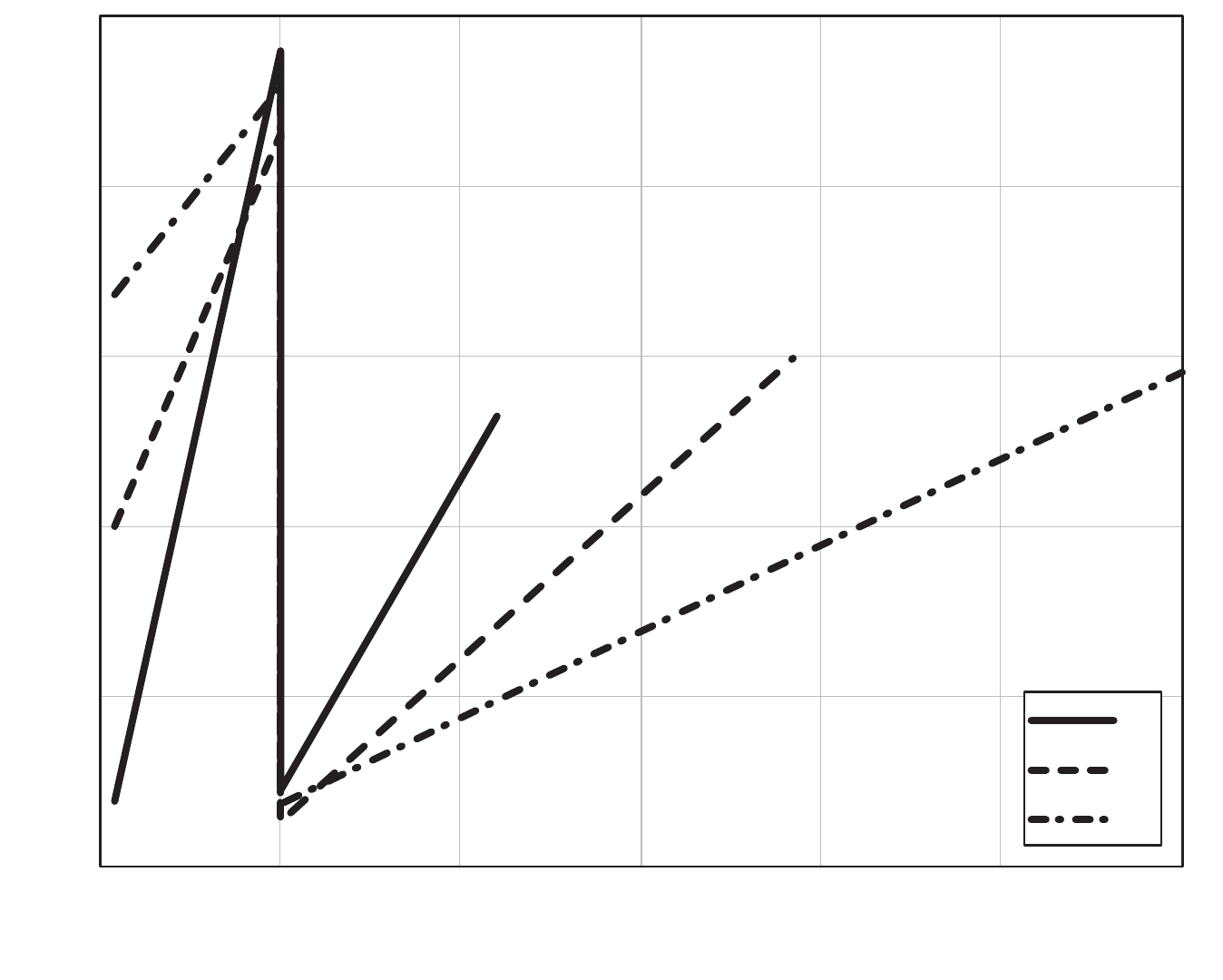}}%
        \put(0.06979721,0.0514458){\color[named]{black}\makebox(0,0)[lb]{\smash{$ - $0,5}}}%
        \put(0.22931495,0.05144579){\color[named]{black}\makebox(0,0)[lb]{\smash{0}}}%
        \put(0.36482755,0.05144579){\color[named]{black}\makebox(0,0)[lb]{\smash{0,5}}}%
        \put(0.52558421,0.05144579){\color[named]{black}\makebox(0,0)[lb]{\smash{1}}}%
        \put(0.6610968,0.05144579){\color[named]{black}\makebox(0,0)[lb]{\smash{1,5}}}%
        \put(0.81984013,0.05144579){\color[named]{black}\makebox(0,0)[lb]{\smash{2}}}%
        \put(0.95721119,0.05144579){\color[named]{black}\makebox(0,0)[lb]{\smash{2,5}}}%
        \put(0.07249948,0.21591931){\color[named]{black}\makebox(0,0)[rb]{\smash{$ - $1}}}%
        \put(0.06937633,0.35530369){\color[named]{black}\makebox(0,0)[rb]{\smash{0}}}%
        \put(0.07249948,0.49468808){\color[named]{black}\makebox(0,0)[rb]{\smash{1}}}%
        \put(0.06958807,0.63407247){\color[named]{black}\makebox(0,0)[rb]{\smash{2}}}%
        \put(0.07108789,0.77345685){\color[named]{black}\makebox(0,0)[rb]{\smash{3}}}%
        \put(0.92437842,0.20017409){\color[named]{black}\makebox(0,0)[lb]{\smash{\textsl{1}}}}%
        \put(0.92437842,0.15959775){\color[named]{black}\makebox(0,0)[lb]{\smash{\textsl{2}}}}%
        \put(0.92437842,0.11886654){\color[named]{black}\makebox(0,0)[lb]{\smash{\textsl{3}}}}%
        \put(0.52553513,0.00601688){\color[named]{black}\makebox(0,0)[b]{\smash{$ z $, мм}}}%
        \put(0.025091,0.43777503){\color[named]{black}\rotatebox{90}{\makebox(0,0)[b]{\smash{$\sigma_x^T$, МПа}}}}%
        \put(0.06958807,0.0837627){\color[named]{black}\makebox(0,0)[rb]{\smash{$ - $2}}}%
      \end{picture}%
    \endgroup%
    \caption[Estimation by the two-layer laminated plate model of residual
    stresses in the silicon--glass bond at 20~{\textdegree}C that was made at
    270~{\textdegree}C for three thicknesses of Borofloat~33 glass
    brand]{Графики расчётного распределения остаточных напряжений
    по~толщине сборки при $T_w =$~20~{\textdegree}C ($T_b=$~270~{\textdegree}C),
    для случая сборки кремния со~стеклом Borofloat~33 нескольких толщин:}
    \legend{%
        \textsl{1} "--- толщина стекла 0,6 мм; \textsl{2} "--- толщина стекла 1,4 мм; \textsl{3} "--- толщина стекла~2,5~мм%
    }
\end{figure}

\begin{figure}[!ht]
    \label{fig:sigma_z_lk5_comp_3thickness}
    \centering
    \begingroup%
      \makeatletter%
      \providecommand\color[2][]{%
        \errmessage{(Inkscape) Color is used for the text in Inkscape, but the package 'color.sty' is not loaded}%
        \renewcommand\color[2][]{}%
      }%
      \providecommand\transparent[1]{%
        \errmessage{(Inkscape) Transparency is used (non-zero) for the text in Inkscape, but the package 'transparent.sty' is not loaded}%
        \renewcommand\transparent[1]{}%
      }%
      \providecommand\rotatebox[2]{#2}%
      \ifx\svgwidth\undefined%
        \setlength{\unitlength}{0.6\textwidth}%
        \ifx\svgscale\undefined%
          \relax%
        \else%
          \setlength{\unitlength}{\unitlength * \real{\svgscale}}%
        \fi%
      \else%
        \setlength{\unitlength}{\svgwidth}%
      \fi%
      \global\let\svgwidth\undefined%
      \global\let\svgscale\undefined%
      \makeatother%
      \begin{picture}(1,0.79779126)%
        \put(0,0){\includegraphics[width=\unitlength]{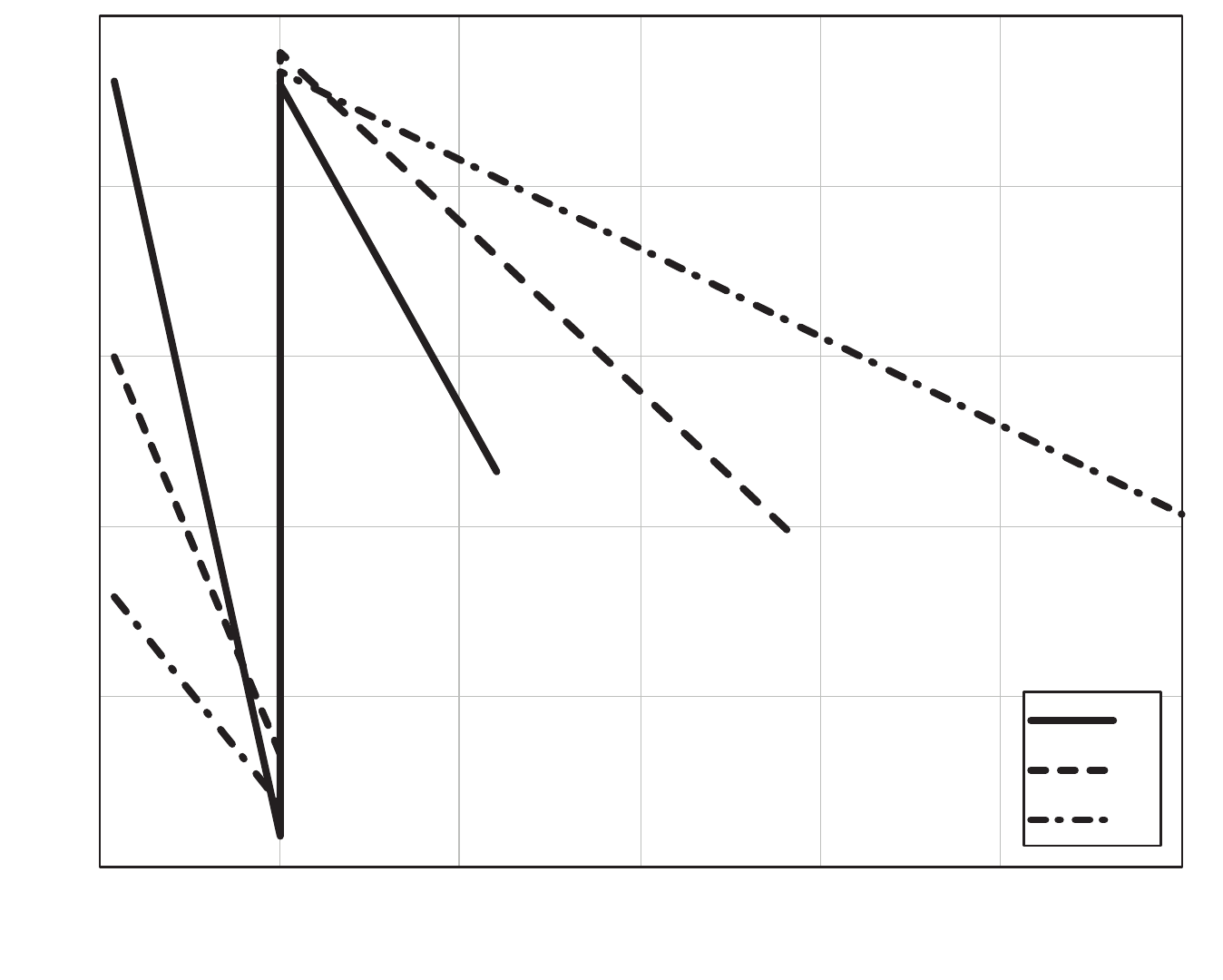}}%
        \put(0.06843194,0.21583314){\color[named]{black}\makebox(0,0)[rb]{\smash{$ - $4}}}%
        \put(0.06867897,0.35521754){\color[named]{black}\makebox(0,0)[rb]{\smash{$ - $2}}}%
        \put(0.06846723,0.49460195){\color[named]{black}\makebox(0,0)[rb]{\smash{0}}}%
        \put(0.06867897,0.63398635){\color[named]{black}\makebox(0,0)[rb]{\smash{2}}}%
        \put(0.06843194,0.77337075){\color[named]{black}\makebox(0,0)[rb]{\smash{4}}}%
        \put(0.06979721,0.0514458){\color[named]{black}\makebox(0,0)[lb]{\smash{$ - $0,5}}}%
        \put(0.22931493,0.0514458){\color[named]{black}\makebox(0,0)[lb]{\smash{0}}}%
        \put(0.36482762,0.0514458){\color[named]{black}\makebox(0,0)[lb]{\smash{0,5}}}%
        \put(0.52558418,0.0514458){\color[named]{black}\makebox(0,0)[lb]{\smash{1}}}%
        \put(0.66109679,0.0514458){\color[named]{black}\makebox(0,0)[lb]{\smash{1,5}}}%
        \put(0.81984014,0.0514458){\color[named]{black}\makebox(0,0)[lb]{\smash{2}}}%
        \put(0.95721122,0.0514458){\color[named]{black}\makebox(0,0)[lb]{\smash{2,5}}}%
        \put(0.92437852,0.2001741){\color[named]{black}\makebox(0,0)[lb]{\smash{\textsl{1}}}}%
        \put(0.92437852,0.15959783){\color[named]{black}\makebox(0,0)[lb]{\smash{\textsl{2}}}}%
        \put(0.92437852,0.11886654){\color[named]{black}\makebox(0,0)[lb]{\smash{\textsl{3}}}}%
        \put(0.52525113,0.00601687){\color[named]{black}\makebox(0,0)[b]{\smash{$z$, мм}}}%
        \put(0,0.40712119){\color[named]{black}\rotatebox{90}{\makebox(0,0)[b]{\smash{$\sigma_x^T$, МПа}}}}%
        \put(0.0685731,0.07622534){\color[named]{black}\makebox(0,0)[rb]{\smash{$-$6}}}%
      \end{picture}%
    \endgroup%
    \caption[Estimation by the two-layer laminated plate model of residual
    stresses in the silicon--glass bond at 20~{\textdegree}C that was made at
    270~{\textdegree}C for three thicknesses of LK5 glass brand]{Графики
    расчётного распределения остаточных напряжений по~толщине сборки при
    $T_w =$~20~{\textdegree}C ($T_b=$~270~{\textdegree}C), для случая
    сборки кремния со~стеклом ЛК5 нескольких толщин:}
    \legend{%
        \textsl{1} "--- толщина стекла 0,6 мм; \textsl{2} "--- толщина стекла 1,4 мм; \textsl{3} "--- толщина стекла~2,5~мм%
    }
\end{figure}

На Рисунках~\ref{fig:sigma_z_bf33_comp_3thickness}
и~\ref{fig:sigma_z_lk5_comp_3thickness} представлены графики расчётного
распределения остаточных напряжений в сборках при рабочей температуре
20~{\textdegree}C (температура соединения 270~{\textdegree}C), рассчитанные
по модели многослойного композиционного материала для случаев разных толщин
стеклянного слоя.
Марки стёкол, использованные в расчёте: Borofloat 33
(Рисунок~\ref{fig:sigma_z_bf33_comp_3thickness}) и~ЛК5
(Рисунок~\ref{fig:sigma_z_lk5_comp_3thickness}).
Толщина пластины кремния в этих расчётах составляет 0,46 мм. За~плоскость
отсчёта координаты по оси $ z $ взята плоскость соединения кремния со
стеклом.
Показательно, что напряжения, как в кремнии, так и~в~стекле, могут менять
свой знак на протяжении толщины материала.
Из~Рисунков~\ref{fig:sigma_z_bf33_comp_3thickness}
и~\ref{fig:sigma_z_lk5_comp_3thickness} видно, что, варьируя толщину
стекла, можно получить минимальные остаточные напряжения на~поверхности
сплошного кремния.

\begin{figure}[!htb]
    \centering
    \begingroup%
      \makeatletter%
      \providecommand\color[2][]{%
        \errmessage{(Inkscape) Color is used for the text in Inkscape, but the package 'color.sty' is not loaded}%
        \renewcommand\color[2][]{}%
      }%
      \providecommand\transparent[1]{%
        \errmessage{(Inkscape) Transparency is used (non-zero) for the text in Inkscape, but the package 'transparent.sty' is not loaded}%
        \renewcommand\transparent[1]{}%
      }%
      \providecommand\rotatebox[2]{#2}%
      \ifx\svgwidth\undefined%
        \setlength{\unitlength}{0.75\textwidth}%
        \ifx\svgscale\undefined%
          \relax%
        \else%
          \setlength{\unitlength}{\unitlength * \real{\svgscale}}%
        \fi%
      \else%
        \setlength{\unitlength}{\svgwidth}%
      \fi%
      \global\let\svgwidth\undefined%
      \global\let\svgscale\undefined%
      \makeatother%
      \begin{picture}(1,0.94457367)%
        \put(0,0){\includegraphics[width=\unitlength]{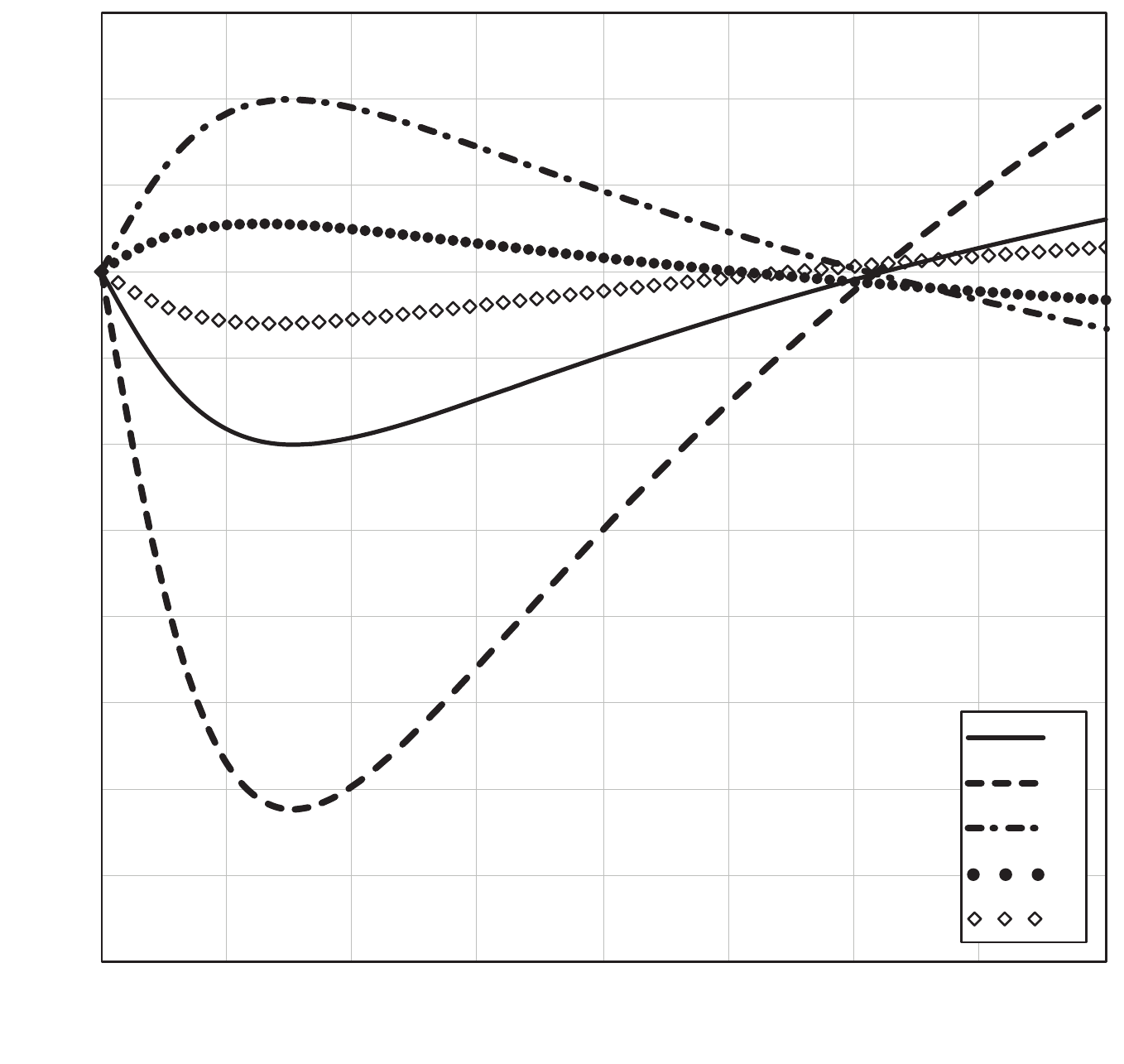}}%
        \put(0.08055673,0.15171806){\color[named]{black}\makebox(0,0)[rb]{\smash{$ - $7}}}%
        \put(0.08055673,0.22857968){\color[named]{black}\makebox(0,0)[rb]{\smash{$ - $6}}}%
        \put(0.08055673,0.3054413){\color[named]{black}\makebox(0,0)[rb]{\smash{$ - $5}}}%
        \put(0.08055673,0.38230285){\color[named]{black}\makebox(0,0)[rb]{\smash{$ - $4}}}%
        \put(0.08055673,0.45916447){\color[named]{black}\makebox(0,0)[rb]{\smash{$ - $3}}}%
        \put(0.08055673,0.53602609){\color[named]{black}\makebox(0,0)[rb]{\smash{$ - $2}}}%
        \put(0.08055673,0.61288771){\color[named]{black}\makebox(0,0)[rb]{\smash{$ - $1}}}%
        \put(0.08055673,0.68974934){\color[named]{black}\makebox(0,0)[rb]{\smash{0}}}%
        \put(0.08055673,0.76661096){\color[named]{black}\makebox(0,0)[rb]{\smash{1}}}%
        \put(0.08055673,0.84347258){\color[named]{black}\makebox(0,0)[rb]{\smash{2}}}%
        \put(0.08055673,0.9203342){\color[named]{black}\makebox(0,0)[rb]{\smash{3}}}%
        \put(0.94177486,0.27828351){\color[named]{black}\makebox(0,0)[lb]{\smash{\textsl{1}}}}%
        \put(0.94177486,0.23800802){\color[named]{black}\makebox(0,0)[lb]{\smash{\textsl{2}}}}%
        \put(0.94177486,0.19757881){\color[named]{black}\makebox(0,0)[lb]{\smash{\textsl{3}}}}%
        \put(0.94177486,0.15730331){\color[named]{black}\makebox(0,0)[lb]{\smash{\textsl{4}}}}%
        \put(0.08992072,0.05199177){\color[named]{black}\makebox(0,0)[lb]{\smash{0}}}%
        \put(0.19056419,0.05199177){\color[named]{black}\makebox(0,0)[lb]{\smash{0,5}}}%
        \put(0.31327655,0.05199177){\color[named]{black}\makebox(0,0)[lb]{\smash{1}}}%
        \put(0.4123701,0.05199177){\color[named]{black}\makebox(0,0)[lb]{\smash{1,5}}}%
        \put(0.5368076,0.05199177){\color[named]{black}\makebox(0,0)[lb]{\smash{2}}}%
        \put(0.63734594,0.05199177){\color[named]{black}\makebox(0,0)[lb]{\smash{2,5}}}%
        \put(0.76046997,0.05199177){\color[named]{black}\makebox(0,0)[lb]{\smash{3}}}%
        \put(0.86026396,0.05199177){\color[named]{black}\makebox(0,0)[lb]{\smash{3,5}}}%
        \put(0.98330909,0.05199177){\color[named]{black}\makebox(0,0)[lb]{\smash{4}}}%
        \put(0.53839145,-0.00000007){\color[named]{black}\makebox(0,0)[b]{\smash{$ H $}}}%
        \put(0.01490498,0.51027967){\color[named]{black}\rotatebox{90}{\makebox(0,0)[b]{\smash{$\sigma_x^T$, МПа}}}}%
        \put(0.08055673,0.07793062){\color[named]{black}\makebox(0,0)[rb]{\smash{\textsl{$-$}8}}}%
        \put(0.94177486,0.11631029){\color[named]{black}\makebox(0,0)[lb]{\smash{\textsl{5}}}}%
      \end{picture}%
    \endgroup%

    \caption[Estimation by the two-layer laminated plate model of residual
    stresses at free side of the silicon layer (bonded to different glass
    brands) at~working temperature \(T_w\)~=~20~{\textdegree}C
    and bonding temperature \(T_b\)~=~350~{\textdegree}C]{Расчётные
    остаточные напряжения на свободной поверхности кремния в зависимости
    от отношения \(H\) толщины стекла к толщине кремния для рабочей температуры
    \(T_w\)~=~20~{\textdegree}C и температуры соединения
    \(T_b\)~=~350~{\textdegree}C:}
    \label{fig:sigma_ot_h_rus}
    \legend{%
    \textsl{1} "--- Corning 7740,  \textsl{2} "--- Schott Borofloat 33,
    \textsl{3} "--- ЛК5,  \textsl{4} "--- Hoya~SD-2,
    \textsl{5}~---~Asahi~SW\nb-YY%
    }
\end{figure}

\begin{figure}[!htb]
    \centering
    \begingroup%
      \makeatletter%
      \providecommand\color[2][]{%
        \errmessage{(Inkscape) Color is used for the text in Inkscape, but the package 'color.sty' is not loaded}%
        \renewcommand\color[2][]{}%
      }%
      \providecommand\transparent[1]{%
        \errmessage{(Inkscape) Transparency is used (non-zero) for the text in Inkscape, but the package 'transparent.sty' is not loaded}%
        \renewcommand\transparent[1]{}%
      }%
      \providecommand\rotatebox[2]{#2}%
      \ifx\svgwidth\undefined%
        \setlength{\unitlength}{0.75\textwidth}%
        \ifx\svgscale\undefined%
          \relax%
        \else%
          \setlength{\unitlength}{\unitlength * \real{\svgscale}}%
        \fi%
      \else%
        \setlength{\unitlength}{\svgwidth}%
      \fi%
      \global\let\svgwidth\undefined%
      \global\let\svgscale\undefined%
      \makeatother%
      \begin{picture}(1,0.94457378)%
        \put(0,0){\includegraphics[width=\unitlength]{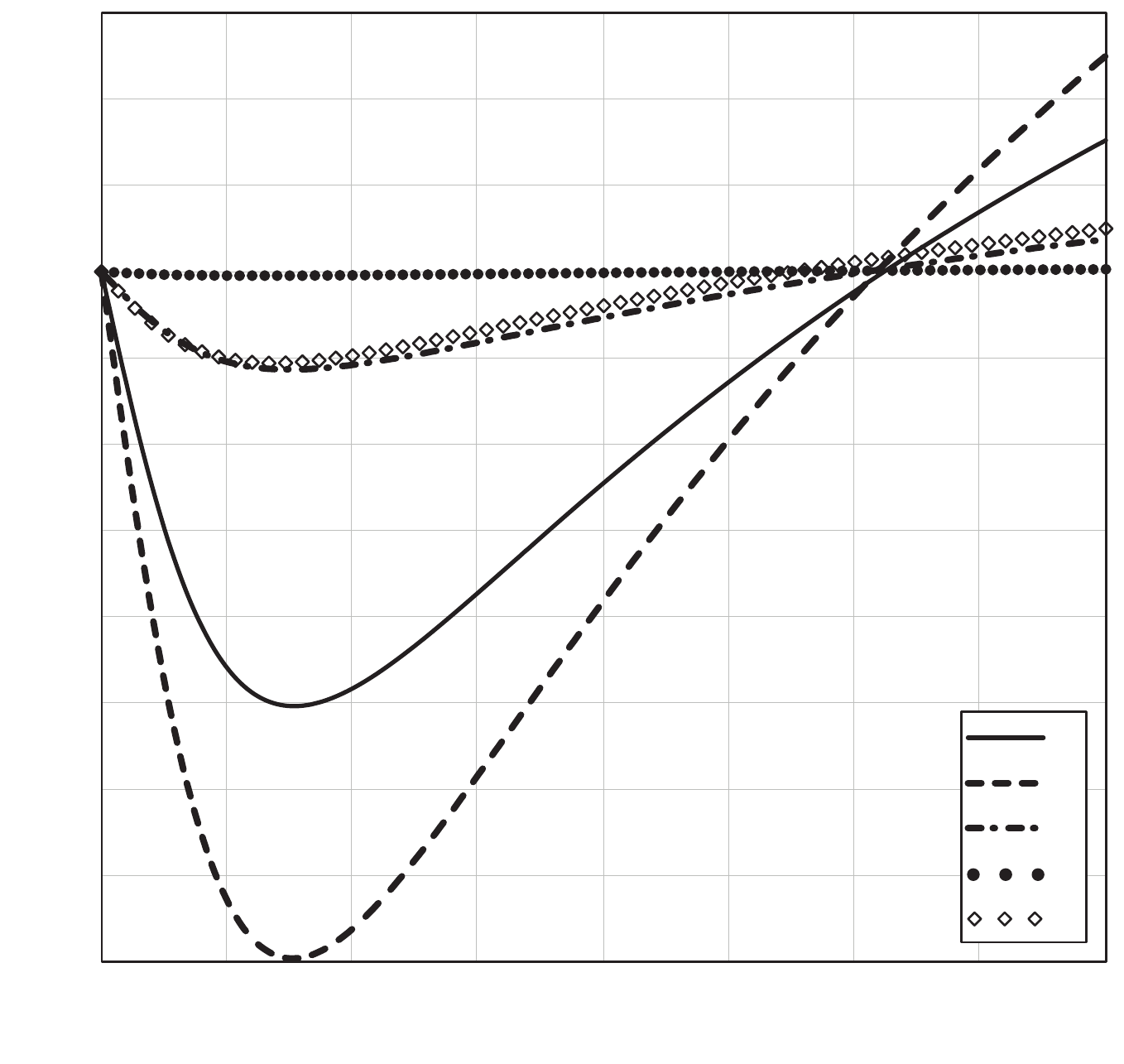}}%
        \put(0.08055675,0.15171806){\color[named]{black}\makebox(0,0)[rb]{\smash{$-$7}}}%
        \put(0.08055675,0.22857969){\color[named]{black}\makebox(0,0)[rb]{\smash{$-$6}}}%
        \put(0.08055675,0.30544124){\color[named]{black}\makebox(0,0)[rb]{\smash{$-$5}}}%
        \put(0.08055675,0.38230295){\color[named]{black}\makebox(0,0)[rb]{\smash{$-$4}}}%
        \put(0.08055675,0.45916457){\color[named]{black}\makebox(0,0)[rb]{\smash{$-$3}}}%
        \put(0.08055675,0.5360262){\color[named]{black}\makebox(0,0)[rb]{\smash{$-$2}}}%
        \put(0.08055675,0.61288783){\color[named]{black}\makebox(0,0)[rb]{\smash{$-$1}}}%
        \put(0.08055675,0.68974946){\color[named]{black}\makebox(0,0)[rb]{\smash{0}}}%
        \put(0.08055675,0.76661108){\color[named]{black}\makebox(0,0)[rb]{\smash{1}}}%
        \put(0.08055675,0.84347271){\color[named]{black}\makebox(0,0)[rb]{\smash{2}}}%
        \put(0.08055675,0.92033434){\color[named]{black}\makebox(0,0)[rb]{\smash{3}}}%
        \put(0.94177492,0.27828353){\color[named]{black}\makebox(0,0)[lb]{\smash{\textsl{1}}}}%
        \put(0.94177492,0.23800803){\color[named]{black}\makebox(0,0)[lb]{\smash{\textsl{2}}}}%
        \put(0.94177492,0.19757882){\color[named]{black}\makebox(0,0)[lb]{\smash{\textsl{3}}}}%
        \put(0.94177492,0.15730332){\color[named]{black}\makebox(0,0)[lb]{\smash{\textsl{4}}}}%
        \put(0.08992074,0.05199178){\color[named]{black}\makebox(0,0)[lb]{\smash{0}}}%
        \put(0.19056421,0.05199178){\color[named]{black}\makebox(0,0)[lb]{\smash{0,5}}}%
        \put(0.31327658,0.05199178){\color[named]{black}\makebox(0,0)[lb]{\smash{1}}}%
        \put(0.41237014,0.05199178){\color[named]{black}\makebox(0,0)[lb]{\smash{1,5}}}%
        \put(0.53680764,0.05199178){\color[named]{black}\makebox(0,0)[lb]{\smash{2}}}%
        \put(0.63734607,0.05199178){\color[named]{black}\makebox(0,0)[lb]{\smash{2,5}}}%
        \put(0.76047002,0.05199178){\color[named]{black}\makebox(0,0)[lb]{\smash{3}}}%
        \put(0.86026402,0.05199178){\color[named]{black}\makebox(0,0)[lb]{\smash{3,5}}}%
        \put(0.98330916,0.05199178){\color[named]{black}\makebox(0,0)[lb]{\smash{4}}}%
        \put(0.53839149,-0.00000007){\color[named]{black}\makebox(0,0)[b]{\smash{$ H $}}}%
        \put(0.01490501,0.51027978){\color[named]{black}\rotatebox{90}{\makebox(0,0)[b]{\smash{$\sigma_x^T$, МПа}}}}%
        \put(0.08055675,0.07793062){\color[named]{black}\makebox(0,0)[rb]{\smash{$-$8}}}%
        \put(0.94177492,0.1163103){\color[named]{black}\makebox(0,0)[lb]{\smash{\textsl{5}}}}%
      \end{picture}%
    \endgroup%
    \caption[Estimation by the two-layer laminated plate model of residual
    stresses at free side of the silicon layer (bonded to different glass
    brands) at~working temperature \(T_w\)~=~85~{\textdegree}C
    and bonding temperature \(T_b\)~=~350~{\textdegree}C]{Расчётные остаточные
    напряжения на свободной поверхности кремния в зависимости от отношения
    \(H\) толщины стекла к толщине кремния для случая рабочей температуры
    \(T_w\)~=~85~{\textdegree}C и температуры соединения
    \(T_b\)~=~350~{\textdegree}C:}
    \label{fig:sigma_ot_h85_rus}
    \legend{%
    \textsl{1} "--- Corning 7740,  \textsl{2} "--- Schott Borofloat 33,
    \textsl{3} "--- ЛК5,  \textsl{4} "--- Hoya~SD-2,
    \textsl{5}~---~Asahi~SW\nb-YY%
    }
\end{figure}

Если проводить расчёт с допущением об отсутствии зависимости жёсткостей
стекла и кремния  от температуры, то из
Рисунков~\ref{fig:glass_stress_multilayer_ot_h_tw},~\ref{fig:glass_stress_multilayer_ot_h_tb},~\ref{fig:sigma_ot_h_rus}
и~\ref{fig:sigma_ot_h85_rus} видно, что для каждой из рассматриваемых марок
стекла есть такое отношение толщины стекла к толщине кремния, при котором
остаточные напряжения на~несоединённой поверхности кремния будут равны нулю
во~всём рабочем диапазоне температур.

Это отношение зависит от жёсткости стекла и его расчётное значение приведено в Таблице~\ref{tab_h0_stekla}.

Подбирать толщину стекла с целью получения минимальных напряжений
на~некоторой глубине от~свободной поверхности кремния возможно среди
тех толщин стекла, при которых напряжения в~кремнии меняют знак
по~толщине пластины. Это приводит к следующим расчётным ограничениям
на~глубину возможного зануления напряжений:
\begin{itemize}
    \item не превышает трети от толщины пластины кремния;
    \item уменьшается с увеличением толщины стекла.
\end{itemize}%
\begingroup%
В случае
несплошного контакта соединяемых
кремния и стекла, моделирование конечными элементами показало, что
искомая толщина будет меньше пропорционально уменьшению площади
границы соединения кремния со~стеклом.\russianpar
\endgroup

\begin{figure}[!ht]%
    \namerefoff %refs inside legend don't work without it (some weird xref errors)
    \centering
    \subbottom[\label{fig:glass_stress_multilayer_ot_h_tw}]%
    {%
        \ifdefmacro{\tikzsetnextfilename}{\tikzsetnextfilename{glass_stress_multilayer_ot_h_tw}}{}%
        \input{Dissertation/images_tikz/disser_glass_stress_multilayer_ot_h_tw.tikz}%
    }%
    \subbottom[\label{fig:glass_stress_multilayer_ot_h_tb}]%
    {%
        \ifdefmacro{\tikzsetnextfilename}{\tikzsetnextfilename{glass_stress_multilayer_ot_h_tb}}{}%
        \input{Dissertation/images_tikz/disser_glass_stress_multilayer_ot_h_tb.tikz}%
    }%

    \caption[Estimation by the two-layer laminated plate model of residual
    stresses at free side of the silicon layer (bonded to either Borofloat~33
    or LK5 glass) at several bonding and working temperatures]{Расчётные
    остаточные напряжения на свободной поверхности кремния для разных
    толщин стёкол Borofloat~33 и~ЛК5 при различных температурах:}
    \legend{%
        \textit{\subcaptionref*{fig:glass_stress_multilayer_ot_h_tw}} "--- при
        нескольких рабочих температурах \(T_w\) (\(T_b\) = 300~{\textdegree}C);
    
        \textit{\subcaptionref*{fig:glass_stress_multilayer_ot_h_tb}} "--- при
        нескольких температурах соединения \(T_b\) (\(T_w\) = 20~{\textdegree}C)
    }
\end{figure}

\begin{table} [!hb]
    \centering%
    \caption[Estimated proportion of glass wafer thickness to silicon (both
    wafers without cavities), which provides a no-stress state at free side
    of the silicon]{Расчётное значение оптимального отношения толщины стекла
    к~толщине кремния (для неструктурированных материалов)}%
    \label{tab_h0_stekla}% label всегда желательно идти после caption
    \renewcommand{\arraystretch}{1.3}%% Увеличение расстояния между рядами, для улучшения восприятия.
    \def\tabularxcolumn#1{m{#1}}
    \begin{SingleSpace}
    \begin{tabular}{@{}lc}
        \toprule     %%% верхняя линейка
        Марка стекла &
        $H_0$\\
        \midrule
        Corning 7740 &
        3,12\\
        Schott Borofloat 33 &
        3,09\\
        ЛК5 &
        3,05\\
        Asahi SW\nobreakdash-YY &
        2,76\\
        Hoya SD\nobreakdash-2 &
        2,54\\
        \bottomrule %%% нижняя линейка
    \end{tabular}%
    \end{SingleSpace}
\end{table}

Если учитывать температурную зависимость модуля Юнга материалов, то~однозначно говорить о наличии наиболее оптимальной толщины стекла нельзя.
Для определения оптимальной толщины нужно рассматривать каждую конкретную комбинацию стекла и кремния с учётом всех допустимых диапазонов температур соединения и их влияния на напряжения в рабочем диапазоне.
При~этом может быть найден узкий диапазон соотношений толщин стекла и~кремния, удовлетворяющий целям оптимизации.
Для повышения качества оценки в этом случае следует подобрать такую расчётную модель, которая бы~позволила без дополнительных усложнений расчётов учитывать температурную зависимость жёсткости.

\ifnumequal{\value{usealtfont}}{2}{%
\clearpage
}{}
\subsection[Оценка напряжений в сборках стекло---кремний---стекло]{Оценка напряжений в сборках стекло"--~кремний"--~стекло}

\begin{figure}[!htb]
    \centering
    \begingroup%
     \makeatletter%
      \providecommand\color[2][]{%
        \errmessage{(Inkscape) Color is used for the text in Inkscape, but the package 'color.sty' is not loaded}%
        \renewcommand\color[2][]{}%
      }%
      \providecommand\transparent[1]{%
        \errmessage{(Inkscape) Transparency is used (non-zero) for the text in Inkscape, but the package 'transparent.sty' is not loaded}%
        \renewcommand\transparent[1]{}%
      }%
      \providecommand\rotatebox[2]{#2}%
      \ifx\svgwidth\undefined%
        \setlength{\unitlength}{0.5\textwidth}%
        \ifx\svgscale\undefined%
          \relax%
        \else%
          \setlength{\unitlength}{\unitlength * \real{\svgscale}}%
        \fi%
      \else%
        \setlength{\unitlength}{\svgwidth}%
      \fi%
      \global\let\svgwidth\undefined%
      \global\let\svgscale\undefined%
      \makeatother%
      \begin{picture}(1,0.79291385)%
        \put(0,0){\includegraphics[width=\unitlength]{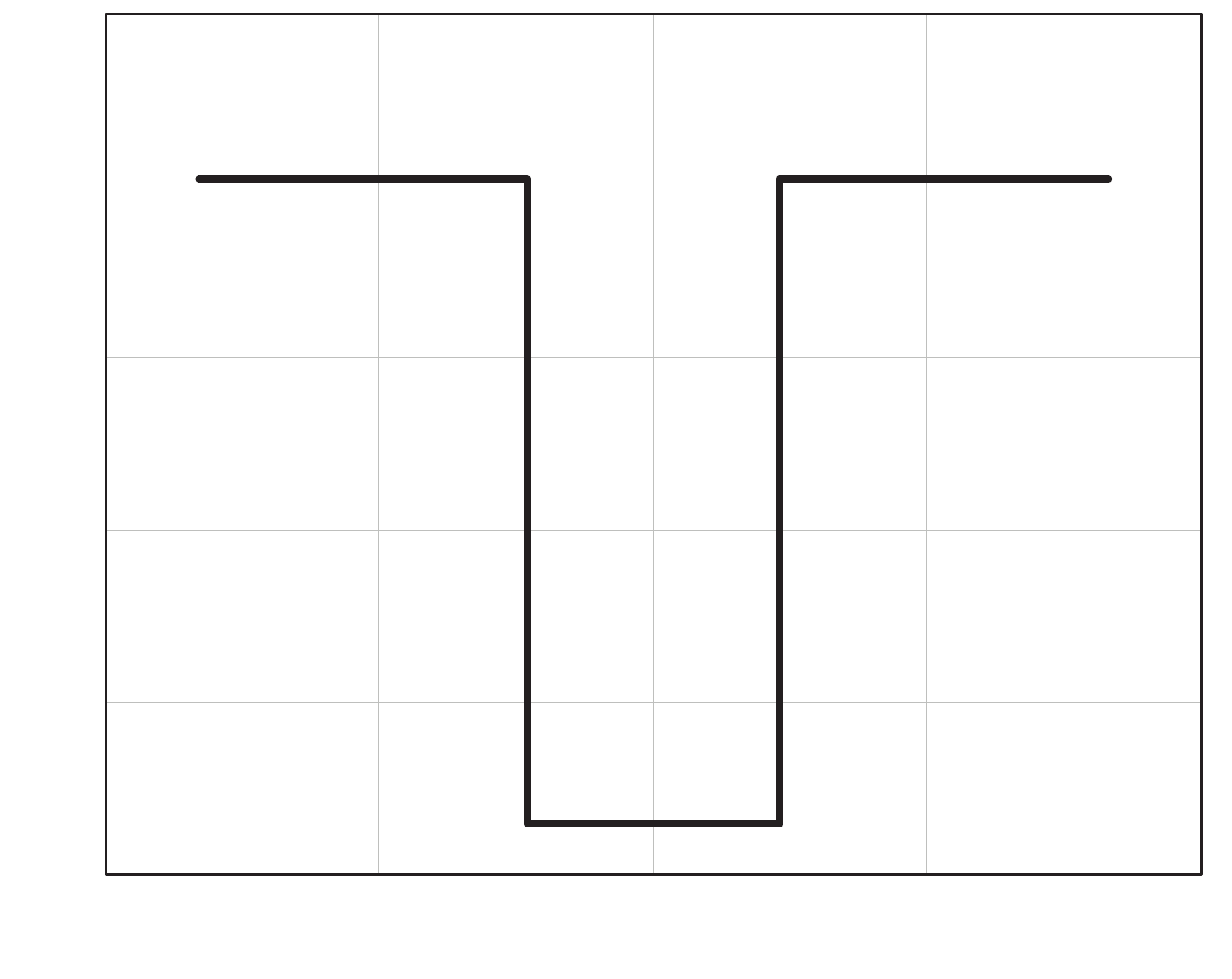}}%
        \put(0.11408864,0.04336622){\color[named]{black}\makebox(0,0)[rb]{\smash{$-$1}}}%
        \put(0.34892102,0.04336622){\color[named]{black}\makebox(0,0)[rb]{\smash{$-$0,5}}}%
        \put(0.54767302,0.04336622){\color[named]{black}\makebox(0,0)[rb]{\smash{0}}}%
        \put(0.79909406,0.04336622){\color[named]{black}\makebox(0,0)[rb]{\smash{0,5}}}%
        \put(1.0044296,0.04336622){\color[named]{black}\makebox(0,0)[rb]{\smash{1}}}%
        \put(0.07054353,0.07167086){\color[named]{black}\makebox(0,0)[rb]{\smash{$-$5}}}%
        \put(0.07054353,0.20852634){\color[named]{black}\makebox(0,0)[rb]{\smash{$-$4}}}%
        \put(0.07054353,0.34849261){\color[named]{black}\makebox(0,0)[rb]{\smash{$-$2}}}%
        \put(0.07054353,0.48845887){\color[named]{black}\makebox(0,0)[rb]{\smash{0}}}%
        \put(0.07054353,0.62842514){\color[named]{black}\makebox(0,0)[rb]{\smash{2}}}%
        \put(0.07054353,0.7683914){\color[named]{black}\makebox(0,0)[rb]{\smash{4}}}%
        \put(-0.05,0.43130823){\color[named]{black}\rotatebox{90}{\makebox(0,0)[b]{\smash{$\sigma_x^T$, МПа}}}}%
        \put(0.53103509,0.00604198){\color[named]{black}\makebox(0,0)[b]{\smash{$z$, мм}}}%
      \end{picture}%
    \endgroup%
    \caption[Estimation by the two-layer laminated plate model of residual
    stresses at working temperature 20~{\textdegree}C and bonding temperature
    310~{\textdegree}C through the thickness of glass--silicon--glass stack (LK5
    glass 0.6~mm thickness and silicon 0.46~mm thickness)]{Расчётный график
    распределения остаточных напряжений при рабочей температуре
    20~{\textdegree}C (температура соединения 310~{\textdegree}C) в~сборке
    стекло"--~кремний"--~стекло, где кремний толщиной 0,46~мм и~стекло ЛК5
    толщиной 0,6 мм}
    \label{fig:sigma_z_gsg_lk5}
\end{figure}

На Рисунке~\ref{fig:sigma_z_gsg_lk5} представлен график распределения
остаточных напряжений в сборке при рабочей температуре 20~{\textdegree}C
(температура соединения 310~{\textdegree}C), рассчитанный по модели
многослойного композиционного материала для случая трёхслойной модели
стекло"--~кремний"--~стекло.
Марка стекла, использованная в расчёте "--- ЛК5.
За плоскость отсчёта координаты по оси z взята срединная плоскость пластины
кремния.
Толщина пластины кремния равна 0,46~мм, толщина пластин стекла равна
0,6~мм.
В симметричных структурах, как, например, стекло"--~кремний"--~стекло,
соединённых при единой температуре, согласно применённой модели напряжения
постоянны по~толщине слоёв.
Причиной тому являются допущения применяемой модели расчёта.

\section{Моделирование методом конечных элементов}
На основе исходных данных по свойствам материалов, описанных
в~подразделе \ref{chap_source_data}, было проведено моделирование
методом конечных элементов. Использовалось программное обеспечение
CoventorWare Turbo 2012, которое в~своей расчётной части опирается
на универсальную систему конечно-элементного анализа Abaqus 6.9.
Эти результаты и особенности моделирования можно распространить
на~прочие программы и~комплексы конечно-элементного моделирования.

\subsection{Задание температурной зависимости температурного коэффициента линейного расширения в~программах конечно-элементного моделирования}

Согласно инструкции для пользователя CoventorWare Turbo 2012 температурную
зависимость истинного температурного
коэффициента линейного расширения необходимо задавать в интегральной форме
вместо дифференциальной, приводимой в большинстве литературных источников,
а~также в~данной работе. Для такой формы представления важно знать пределы
интегрирования, в первую очередь температуру начала интегрирования. Для случая
анодной посадки это будет температура соединения \(T_b\).

\ifnumequal{\value{usealtfont}}{2}{\setDisplayskipStretch{-0.078}}{}%
Приведение записи ТКЛР в дифференциальной форме к записи в интегральной форме
производится согласно следующей формуле:
\begin{equation}
    \alpha^\mathrm{int}(T_w, T_b) =
    \int\limits_{T_b}^{T_w}
    \frac{1}{T_w-T_b}
    \alpha^\mathrm{dif}(T)\:\mathrm{d}T
\end{equation}

После такого преобразования зависимость ТКЛР, представленная
в~дифференциальной форме полиномом вида \eqref{eq:polynom_sample}, будет
выражаться полиномом того же порядка вида
\begin{multline}
    \alpha^\mathrm{int}(T_w, T_b) =
    \left(a + \frac{b}{2} \cdot T_b + \frac{c}{3} \cdot T_b^2 + \frac{d}{4} \cdot T_b^3 + \frac{e}{5} \cdot T_b^4\right)
    + \\ +
    \left(\frac{b}{2} + \frac{c}{3} \cdot T_b + \frac{d}{4} \cdot T_b^2 + \frac{e}{5} \cdot T_b^3\right) \cdot T_w
    +
    \left(\frac{c}{3}+\frac{d}{4} \cdot T_b + \frac{e}{5} \cdot T_b^2\right) \cdot T_w^2
    + \\ +
    \left(\frac{d}{4} + \frac{e}{5} \cdot T_b\right) \cdot T_w^3 +
    \frac{e}{5} \cdot T_w^4
\end{multline}

Автоматизация преобразования полиномов возможна
при помощи следующей программы на языке \verb|R|:
\begin{Verb}
cteIntFormCoef <-
  function(coeforig = c(0, 1),
           temperature0 = 273.15) {
    a <- as.numeric(coeforig)
    coefresult <- a
    la <- length(a)
    for (i in la:1) {
      if (i == la)
        coefresult[i] <- a[i] / i
      if (i != la)
        coefresult[i] <- a[i] / i + coefresult[i + 1] * temperature0
    }
    coefresult
  }
\end{Verb}

\subsection{Результаты моделирования}

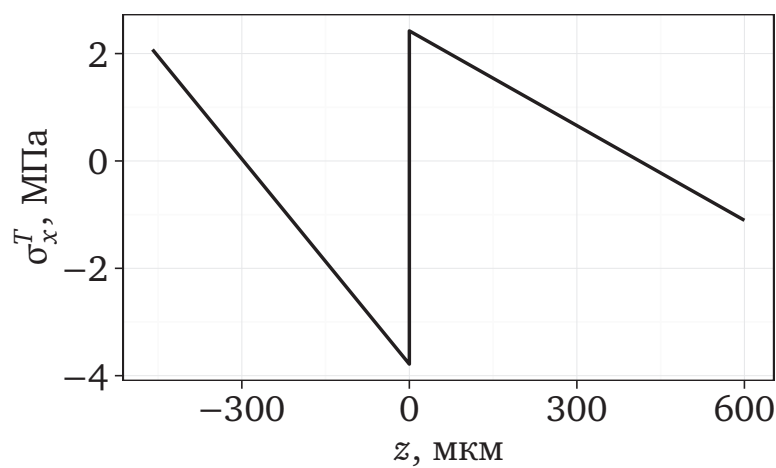
\begin{figure}[!hbt]
    \centering
    \ifdefmacro{\tikzsetnextfilename}{\tikzsetnextfilename{stressX_distrib}}{}%
    \input{Dissertation/images_tikz/disser_stressX_distrib_600.tikz}%
    \caption[Residual stresses through the thickness of the silicon--glass bond
    at~working temperature 20~{\textdegree}C and bonding temperature
    300~{\textdegree}C (LK5 glass 0.6~mm thickness and silicon 0.46~mm
    thickness) simulated with finite-element modeling]{Распределение остаточных
    напряжений при \(T_w\) = 20~{\textdegree}C  в~сборке кремния толщиной
    460~мкм и стекла ЛК5 толщиной 600 мкм, соединёнными при температуре
    \(T_b\) = 300~{\textdegree}C, смоделированное в~программе
    конечно-элементного моделирования}
    \label{fig:stressX_distrib}
\end{figure}

На Рисунке~\ref{fig:stressX_distrib} показан результат моделирования конечными
элементами распределения напряжений в соединённых пластинах кремния и стекла.
Использовалась модель из двух соединённых пластин диаметром 100 мм:
кремний толщиной 460 мкм и стекло толщиной 600 мкм.
ТКЛР материалов модели устанавливались с использованием возможности задать
температуру, при которой в~материале отсутствуют напряжения.
Эта температура задавалась равной температуре соединения $T_b$.
Граничное условие температуры моделируемых объёмов задавалось равным рабочей температуре $T_w$.

Моделирование методом конечных элементов в комплексе для моделирования МЭМС
Coventorware Turbo подтвердило форму эпюр расчётных напряжений.

\begin{figure}[!hb]
    \centering
    \includegraphics[height=0.2\textheight]{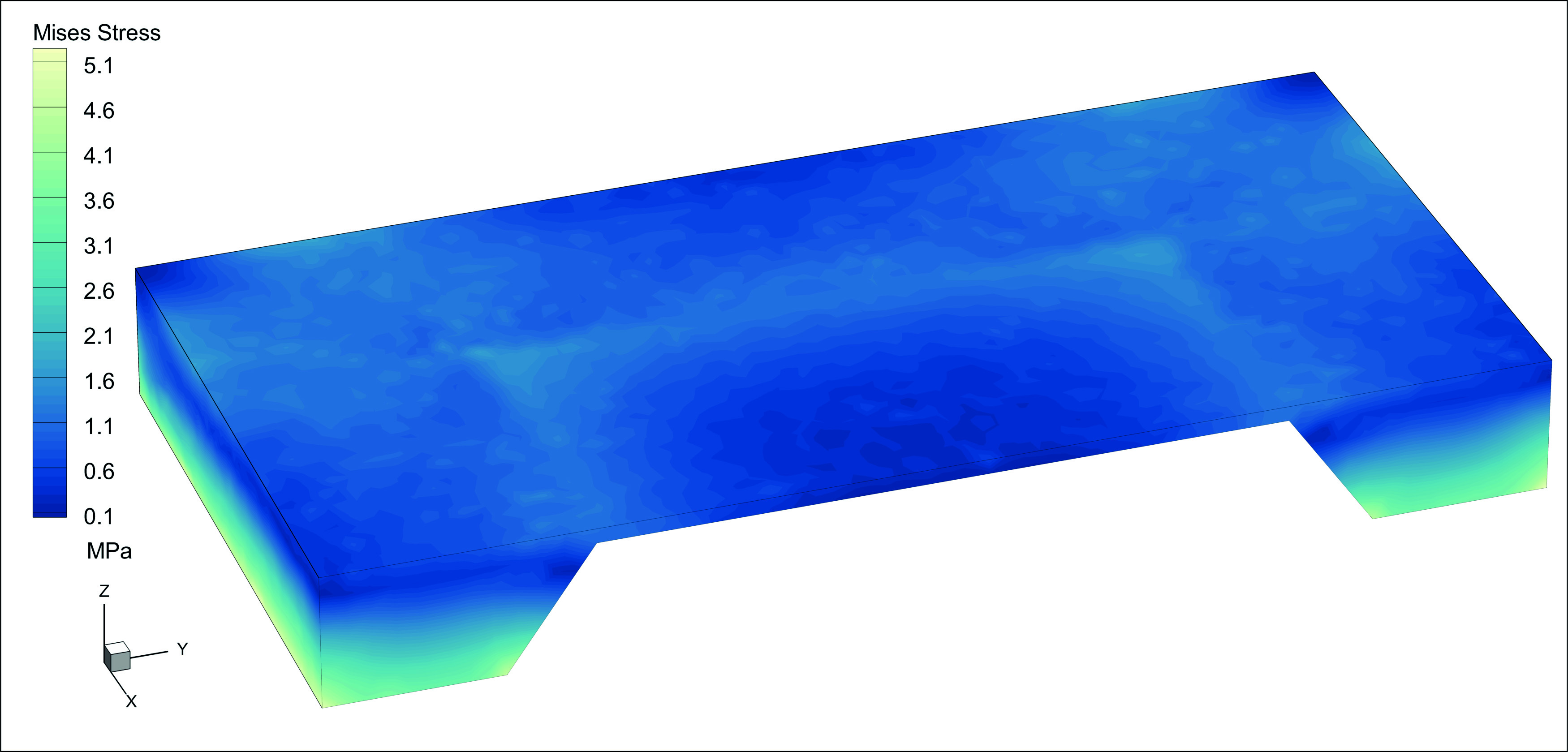}\par%
    \includegraphics[height=0.2\textheight]{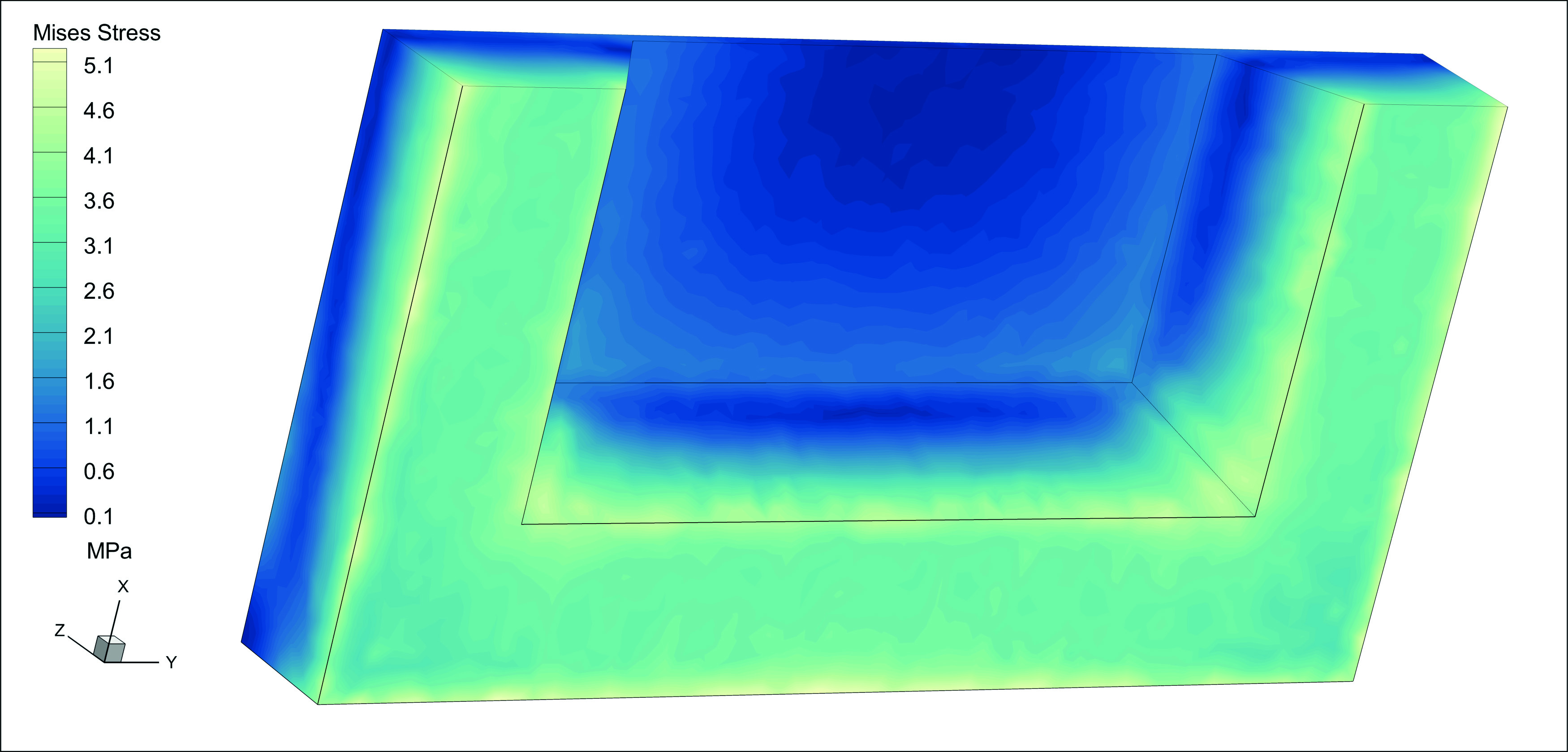}%
    \caption[Residual stresses at 20~{\textdegree}C in the structured silicon
    after bonding with the LK5 glass at 300~{\textdegree}C simulated with
    a finite-element modeling]{Смоделированная картина распределения напряжений
    в~структурированном кремнии после анодной посадки на стекло ЛК5}
    \label{fig:si_membrane_bonded_3d}
\end{figure}

На Рисунках~\ref{fig:si_membrane_bonded_3d}
и~\ref{fig:si_membrane_third_inv_stress} показаны результаты  моделирования
конечными элементами сборки кремниевого элемента размерами
4\(\,\times\,\)4\(\,\times\,\)0,46~мм со~структурированным элементом толщиной 50~мкм
и стеклянного элемента размерами
4\(\,\times\,\)4\(\,\times\,\)0,6~мм. Площадь области соединения со стеклом
составляет половину от площади поверхности стеклянного элемента. Температура
проведения смоделированного соединения 300~{\textdegree}C. Марка стекла "---
ЛК5.

Видно, что в данном случае также существует оптимальная толщина стекла,
описанная в подразделе~\ref{chap_optim_glass_thickness}, минимизирующая
остаточные напряжения на поверхности кремния. Отношение оптимальной толщины
стекла к~толщине кремния в данной модели примерно в два раза меньше, чем
представлено ранее на Рисунке~\ref{fig:glass_stress_multilayer_ot_h_tw} на
странице~\pageref{fig:glass_stress_multilayer_ot_h_tw},
поскольку площадь соединения с~поверхностью стекла
в два раза меньше, чем было бы в случае отсутствия
структуры, сформированной вытравливанием кремния.

\begin{figure}[!hb]
    \hspace{1.5em}%
    \ifdefmacro{\tikzsetnextfilename}{\tikzsetnextfilename{si_membrane_third_inv_stress}}{}%
    \input{Dissertation/images_tikz/disser_tis_g.tikz}%
    \caption[Residual stresses at the free surface of the structured
    silicon after bonding with the LK5 glass at 300~{\textdegree}C
    simulated at different working temperatures with a finite-element
    modeling]{Смоделированная оценка остаточных напряжений на~открытой
    поверхности структурированного кремния для различных толщин стекла
    ЛК5 при нескольких рабочих температурах}
    \label{fig:si_membrane_third_inv_stress}
\end{figure}

\section{Методика оценки остаточных напряжений, вызванных неоднородностью теплового расширения кремния и стекла}
Рассмотренные марки стекла разными источниками рекомендованы
к~соединению с кремнием. Тем не менее представленная оценка напряжений
показывает значительные отличия в результатах такого соединения.
Не стоит ограничиваться выбором конкретной
марки стекла лишь на основе сходства средних значений коэффициентов
расширения с кремнием в рабочем диапазоне температур.
Следует получить
зависимость истинных значений ТКЛР рассматриваемых кандидатур стёкол
от~температуры в диапазоне от нижней границы рабочего диапазона
конечного прибора до верхней температуры, допустимой технологическим
процессом соединения кремния со стеклом. Затем нужно проанализировать
её совместно с~аналогичной зависимостью для кремния, чтобы
определиться с~маркой стекла и режимом проведения процесса соединения.

Исходя из данных о ТКЛР применяемых стекла и кремния на основании вышеописанной последовательности расчётов можно определить значения остаточных напряжений при рабочей температуре $T_w$ в деталях, соединённых при температуре $T_b$.
Проведя предварительный расчёт, можно спланировать процесс соединения,
обеспечивающий минимальные остаточные напряжения или же~выдерживающий их в
определённых пределах.
Также, зная температуру, при которой провели соединение деталей, можно оценить характер изменения остаточных напряжений в рабочем диапазоне температур получаемого изделия.
В~связи со сложностью описанной методики расчёта рекомендуется проводить
его при помощи ЭВМ.

\section{Выводы по главе 3}
\begin{enumerate}
    \item Предлагается применять модель двух тонких слоёв и модель
    многослойного композиционного материала с~использованием температурной зависимости
    истинных значений
    ТКЛР материалов. Их~применение также облегчает
    подготовку параметров конечно-элементных моделей и расчётов.
    \item Расчёты с применением средних ТКЛР и истинных ТКЛР приводят
    к~кардинально различающимся выводам. Результаты, полученные с~использованием
    температурной зависимости истинных значений ТКЛР материалов, подтверждаются
    экспериментальными исследованиями других авторов.
     
    \item На основании проведённых расчётов можно заключить, что для
    снижения остаточных напряжений в результате процесса анодной
    посадки в~кремнии на~глубинах до~100~мкм необходимо согласовывать
    толщины стекла и~кремния.
    \item Для формирования на~кремниевой поверхности сборки
    не~подвергавшегося объёмной микрообработке кремния со~стеклом
    минимальных остаточных напряжений во всём рабочем диапазоне
    температур необходимо выбирать толщину стекла пропорционально
    толщине кремния. Для алюмосиликатных стёкол (SD\nb-2, SW\nb-YY)
    толщина стекла должна быть примерно в 2,5--2,8 раза больше
    толщины кремния. Для боросиликатных стёкол (7740, ЛК5, Borofloat~33)
    толщина стекла должна быть примерно в 3 раза больше толщины
    кремния.
    \item Проиллюстрировано, что идея «снижение температуры
    сращивания снижает остаточные напряжения» верна лишь для определённых марок
    стекла или даже для определённых партий стекла, поскольку зависимость ТКЛР от
    температуры может отличаться от~партии к~партии. Например, снижение температуры
    соединения кремния со~стеклом ЛК5 лишь повышает сжимающие напряжения в кремнии
    (согласно модели двух тонких слоёв).
\end{enumerate}

%% file: Dissertation/images_tikz/disser_sigma_workt_bf33_lk5.tikz
\begin{tikzpicture}[x=1pt,y=1pt]
\definecolor{fillColor}{RGB}{255,255,255}
\path[use as bounding box,fill=fillColor] (0,0) rectangle (483.70,227.34);
\begin{scope}
\path[clip] (  0.00,  0.00) rectangle (483.70,227.34);

\path[] (  0.00,  0.00) rectangle (483.70,227.34);
\end{scope}
\begin{scope}
\path[clip] (  0.00,  0.00) rectangle (241.85,227.34);
\definecolor{drawColor}{RGB}{255,255,255}
\definecolor{fillColor}{RGB}{255,255,255}

\path[draw=drawColor,line width= 0.6pt,line join=round,line cap=round,fill=fillColor] (  0.00,  0.00) rectangle (241.85,227.34);
\end{scope}
\begin{scope}
\path[clip] ( 40.77, 32.43) rectangle (237.63,217.40);
\definecolor{fillColor}{RGB}{255,255,255}

\path[fill=fillColor] ( 40.77, 32.43) rectangle (237.63,217.40);
\definecolor{drawColor}{gray}{0.98}

\path[draw=drawColor,line width= 0.6pt,line join=round] ( 40.77, 51.61) --
	(237.63, 51.61);

\path[draw=drawColor,line width= 0.6pt,line join=round] ( 40.77, 93.03) --
	(237.63, 93.03);

\path[draw=drawColor,line width= 0.6pt,line join=round] ( 40.77,134.44) --
	(237.63,134.44);

\path[draw=drawColor,line width= 0.6pt,line join=round] ( 40.77,175.86) --
	(237.63,175.86);

\path[draw=drawColor,line width= 0.6pt,line join=round] ( 40.77,217.28) --
	(237.63,217.28);

\path[draw=drawColor,line width= 0.6pt,line join=round] ( 74.40, 32.43) --
	( 74.40,217.40);

\path[draw=drawColor,line width= 0.6pt,line join=round] (111.43, 32.43) --
	(111.43,217.40);

\path[draw=drawColor,line width= 0.6pt,line join=round] (136.11, 32.43) --
	(136.11,217.40);

\path[draw=drawColor,line width= 0.6pt,line join=round] (173.14, 32.43) --
	(173.14,217.40);

\path[draw=drawColor,line width= 0.6pt,line join=round] (213.26, 32.43) --
	(213.26,217.40);
\definecolor{drawColor}{gray}{0.90}

\path[draw=drawColor,line width= 0.2pt,line join=round] ( 40.77, 72.32) --
	(237.63, 72.32);

\path[draw=drawColor,line width= 0.2pt,line join=round] ( 40.77,113.73) --
	(237.63,113.73);

\path[draw=drawColor,line width= 0.2pt,line join=round] ( 40.77,155.15) --
	(237.63,155.15);

\path[draw=drawColor,line width= 0.2pt,line join=round] ( 40.77,196.57) --
	(237.63,196.57);

\path[draw=drawColor,line width= 0.2pt,line join=round] ( 49.72, 32.43) --
	( 49.72,217.40);

\path[draw=drawColor,line width= 0.2pt,line join=round] ( 99.09, 32.43) --
	( 99.09,217.40);

\path[draw=drawColor,line width= 0.2pt,line join=round] (123.77, 32.43) --
	(123.77,217.40);

\path[draw=drawColor,line width= 0.2pt,line join=round] (148.46, 32.43) --
	(148.46,217.40);

\path[draw=drawColor,line width= 0.2pt,line join=round] (197.83, 32.43) --
	(197.83,217.40);

\path[draw=drawColor,line width= 0.2pt,line join=round] (228.68, 32.43) --
	(228.68,217.40);
\definecolor{drawColor}{RGB}{0,0,0}

\path[draw=drawColor,line width= 1.7pt,line join=round] ( 49.72,102.05) --
	( 51.50,102.89) --
	( 53.29,103.72) --
	( 55.08,104.54) --
	( 56.87,105.35) --
	( 58.66,106.15) --
	( 60.45,106.95) --
	( 62.24,107.73) --
	( 64.03,108.50) --
	( 65.82,109.27) --
	( 67.61,110.02) --
	( 69.40,110.77) --
	( 71.19,111.51) --
	( 72.98,112.24) --
	( 74.77,112.96) --
	( 76.56,113.67) --
	( 78.35,114.37) --
	( 80.14,115.07) --
	( 81.93,115.75) --
	( 83.72,116.43) --
	( 85.51,117.10) --
	( 87.30,117.76) --
	( 89.09,118.41) --
	( 90.88,119.06) --
	( 92.67,119.69) --
	( 94.46,120.32) --
	( 96.25,120.94) --
	( 98.04,121.55) --
	( 99.83,122.15) --
	(101.62,122.74) --
	(103.41,123.33) --
	(105.20,123.91) --
	(106.99,124.48) --
	(108.77,125.04) --
	(110.56,125.59) --
	(112.35,126.14) --
	(114.14,126.68) --
	(115.93,127.21) --
	(117.72,127.73) --
	(119.51,128.25) --
	(121.30,128.76) --
	(123.09,129.26) --
	(124.88,129.75) --
	(126.67,130.23) --
	(128.46,130.71) --
	(130.25,131.18) --
	(132.04,131.64) --
	(133.83,132.10) --
	(135.62,132.55) --
	(137.41,132.99) --
	(139.20,133.42) --
	(140.99,133.85) --
	(142.78,134.27) --
	(144.57,134.68) --
	(146.36,135.08) --
	(148.15,135.48) --
	(149.94,135.87) --
	(151.73,136.26) --
	(153.52,136.63) --
	(155.31,137.00) --
	(157.10,137.37) --
	(158.89,137.72) --
	(160.68,138.07) --
	(162.47,138.41) --
	(164.26,138.75) --
	(166.05,139.08) --
	(167.83,139.40) --
	(169.62,139.71) --
	(171.41,140.02) --
	(173.20,140.33) --
	(174.99,140.62) --
	(176.78,140.91) --
	(178.57,141.19) --
	(180.36,141.47) --
	(182.15,141.74) --
	(183.94,142.00) --
	(185.73,142.26) --
	(187.52,142.51) --
	(189.31,142.76) --
	(191.10,142.99) --
	(192.89,143.23) --
	(194.68,143.45) --
	(196.47,143.67) --
	(198.26,143.89) --
	(200.05,144.09) --
	(201.84,144.29) --
	(203.63,144.49) --
	(205.42,144.68) --
	(207.21,144.86) --
	(209.00,145.04) --
	(210.79,145.21) --
	(212.58,145.38) --
	(214.37,145.54) --
	(216.16,145.69) --
	(217.95,145.84) --
	(219.74,145.98) --
	(221.53,146.12) --
	(223.32,146.25) --
	(225.10,146.37) --
	(226.89,146.49) --
	(228.68,146.60);

\path[draw=drawColor,line width= 1.7pt,dash pattern=on 2pt off 2pt ,line join=round] ( 49.72,130.02) --
	( 51.50,130.85) --
	( 53.29,131.68) --
	( 55.08,132.50) --
	( 56.87,133.31) --
	( 58.66,134.12) --
	( 60.45,134.91) --
	( 62.24,135.69) --
	( 64.03,136.46) --
	( 65.82,137.23) --
	( 67.61,137.99) --
	( 69.40,138.73) --
	( 71.19,139.47) --
	( 72.98,140.20) --
	( 74.77,140.92) --
	( 76.56,141.63) --
	( 78.35,142.34) --
	( 80.14,143.03) --
	( 81.93,143.72) --
	( 83.72,144.39) --
	( 85.51,145.06) --
	( 87.30,145.72) --
	( 89.09,146.37) --
	( 90.88,147.02) --
	( 92.67,147.65) --
	( 94.46,148.28) --
	( 96.25,148.90) --
	( 98.04,149.51) --
	( 99.83,150.11) --
	(101.62,150.71) --
	(103.41,151.29) --
	(105.20,151.87) --
	(106.99,152.44) --
	(108.77,153.00) --
	(110.56,153.56) --
	(112.35,154.10) --
	(114.14,154.64) --
	(115.93,155.17) --
	(117.72,155.69) --
	(119.51,156.21) --
	(121.30,156.72) --
	(123.09,157.22) --
	(124.88,157.71) --
	(126.67,158.20) --
	(128.46,158.67) --
	(130.25,159.14) --
	(132.04,159.61) --
	(133.83,160.06) --
	(135.62,160.51) --
	(137.41,160.95) --
	(139.20,161.38) --
	(140.99,161.81) --
	(142.78,162.23) --
	(144.57,162.64) --
	(146.36,163.05) --
	(148.15,163.44) --
	(149.94,163.83) --
	(151.73,164.22) --
	(153.52,164.59) --
	(155.31,164.96) --
	(157.10,165.33) --
	(158.89,165.68) --
	(160.68,166.03) --
	(162.47,166.37) --
	(164.26,166.71) --
	(166.05,167.04) --
	(167.83,167.36) --
	(169.62,167.68) --
	(171.41,167.99) --
	(173.20,168.29) --
	(174.99,168.58) --
	(176.78,168.87) --
	(178.57,169.16) --
	(180.36,169.43) --
	(182.15,169.70) --
	(183.94,169.97) --
	(185.73,170.22) --
	(187.52,170.47) --
	(189.31,170.72) --
	(191.10,170.96) --
	(192.89,171.19) --
	(194.68,171.41) --
	(196.47,171.63) --
	(198.26,171.85) --
	(200.05,172.05) --
	(201.84,172.26) --
	(203.63,172.45) --
	(205.42,172.64) --
	(207.21,172.82) --
	(209.00,173.00) --
	(210.79,173.17) --
	(212.58,173.34) --
	(214.37,173.50) --
	(216.16,173.65) --
	(217.95,173.80) --
	(219.74,173.94) --
	(221.53,174.08) --
	(223.32,174.21) --
	(225.10,174.33) --
	(226.89,174.45) --
	(228.68,174.57);

\path[draw=drawColor,line width= 1.7pt,dash pattern=on 4pt off 2pt ,line join=round] ( 49.72,162.31) --
	( 51.50,163.15) --
	( 53.29,163.98) --
	( 55.08,164.80) --
	( 56.87,165.61) --
	( 58.66,166.41) --
	( 60.45,167.20) --
	( 62.24,167.99) --
	( 64.03,168.76) --
	( 65.82,169.53) --
	( 67.61,170.28) --
	( 69.40,171.03) --
	( 71.19,171.77) --
	( 72.98,172.50) --
	( 74.77,173.22) --
	( 76.56,173.93) --
	( 78.35,174.63) --
	( 80.14,175.33) --
	( 81.93,176.01) --
	( 83.72,176.69) --
	( 85.51,177.36) --
	( 87.30,178.02) --
	( 89.09,178.67) --
	( 90.88,179.31) --
	( 92.67,179.95) --
	( 94.46,180.58) --
	( 96.25,181.19) --
	( 98.04,181.81) --
	( 99.83,182.41) --
	(101.62,183.00) --
	(103.41,183.59) --
	(105.20,184.17) --
	(106.99,184.74) --
	(108.77,185.30) --
	(110.56,185.85) --
	(112.35,186.40) --
	(114.14,186.94) --
	(115.93,187.47) --
	(117.72,187.99) --
	(119.51,188.51) --
	(121.30,189.01) --
	(123.09,189.52) --
	(124.88,190.01) --
	(126.67,190.49) --
	(128.46,190.97) --
	(130.25,191.44) --
	(132.04,191.90) --
	(133.83,192.36) --
	(135.62,192.81) --
	(137.41,193.25) --
	(139.20,193.68) --
	(140.99,194.11) --
	(142.78,194.53) --
	(144.57,194.94) --
	(146.36,195.34) --
	(148.15,195.74) --
	(149.94,196.13) --
	(151.73,196.51) --
	(153.52,196.89) --
	(155.31,197.26) --
	(157.10,197.62) --
	(158.89,197.98) --
	(160.68,198.33) --
	(162.47,198.67) --
	(164.26,199.01) --
	(166.05,199.34) --
	(167.83,199.66) --
	(169.62,199.97) --
	(171.41,200.28) --
	(173.20,200.58) --
	(174.99,200.88) --
	(176.78,201.17) --
	(178.57,201.45) --
	(180.36,201.73) --
	(182.15,202.00) --
	(183.94,202.26) --
	(185.73,202.52) --
	(187.52,202.77) --
	(189.31,203.01) --
	(191.10,203.25) --
	(192.89,203.48) --
	(194.68,203.71) --
	(196.47,203.93) --
	(198.26,204.14) --
	(200.05,204.35) --
	(201.84,204.55) --
	(203.63,204.75) --
	(205.42,204.94) --
	(207.21,205.12) --
	(209.00,205.30) --
	(210.79,205.47) --
	(212.58,205.63) --
	(214.37,205.79) --
	(216.16,205.95) --
	(217.95,206.10) --
	(219.74,206.24) --
	(221.53,206.37) --
	(223.32,206.50) --
	(225.10,206.63) --
	(226.89,206.75) --
	(228.68,206.86);

\path[draw=drawColor,line width= 0.9pt,line join=round,line cap=round] ( 40.77, 32.43) rectangle (237.63,217.40);
\end{scope}
\begin{scope}
\path[clip] (  0.00,  0.00) rectangle (483.70,227.34);
\definecolor{drawColor}{RGB}{0,0,0}

\node[text=drawColor,anchor=base east,inner sep=0pt, outer sep=0pt, scale=  1.00] at ( 35.37, 67.36) {\(-5\)};

\node[text=drawColor,anchor=base east,inner sep=0pt, outer sep=0pt, scale=  1.00] at ( 35.37,108.78) {\(0\)};

\node[text=drawColor,anchor=base east,inner sep=0pt, outer sep=0pt, scale=  1.00] at ( 35.37,150.19) {\(5\)};

\node[text=drawColor,anchor=base east,inner sep=0pt, outer sep=0pt, scale=  1.00] at ( 35.37,191.61) {\(10\)};
\end{scope}
\begin{scope}
\path[clip] (  0.00,  0.00) rectangle (483.70,227.34);
\definecolor{drawColor}{RGB}{0,0,0}

\path[draw=drawColor,line width= 0.6pt,line join=round] ( 37.77, 72.32) --
	( 40.77, 72.32);

\path[draw=drawColor,line width= 0.6pt,line join=round] ( 37.77,113.73) --
	( 40.77,113.73);

\path[draw=drawColor,line width= 0.6pt,line join=round] ( 37.77,155.15) --
	( 40.77,155.15);

\path[draw=drawColor,line width= 0.6pt,line join=round] ( 37.77,196.57) --
	( 40.77,196.57);
\end{scope}
\begin{scope}
\path[clip] (  0.00,  0.00) rectangle (483.70,227.34);
\definecolor{drawColor}{RGB}{0,0,0}

\path[draw=drawColor,line width= 0.6pt,line join=round] ( 49.72, 29.43) --
	( 49.72, 32.43);

\path[draw=drawColor,line width= 0.6pt,line join=round] ( 99.09, 29.43) --
	( 99.09, 32.43);

\path[draw=drawColor,line width= 0.6pt,line join=round] (123.77, 29.43) --
	(123.77, 32.43);

\path[draw=drawColor,line width= 0.6pt,line join=round] (148.46, 29.43) --
	(148.46, 32.43);

\path[draw=drawColor,line width= 0.6pt,line join=round] (197.83, 29.43) --
	(197.83, 32.43);

\path[draw=drawColor,line width= 0.6pt,line join=round] (228.68, 29.43) --
	(228.68, 32.43);
\end{scope}
\begin{scope}
\path[clip] (  0.00,  0.00) rectangle (483.70,227.34);
\definecolor{drawColor}{RGB}{0,0,0}

\node[text=drawColor,anchor=base,inner sep=0pt, outer sep=0pt, scale=  1.00] at ( 49.72, 17.12) {\(-60\)};

\node[text=drawColor,anchor=base,inner sep=0pt, outer sep=0pt, scale=  1.00] at ( 99.09, 17.12) {\(-20\)};

\node[text=drawColor,anchor=base,inner sep=0pt, outer sep=0pt, scale=  1.00] at (123.77, 17.12) {\(0\)};

\node[text=drawColor,anchor=base,inner sep=0pt, outer sep=0pt, scale=  1.00] at (148.46, 17.12) {\(20\)};

\node[text=drawColor,anchor=base,inner sep=0pt, outer sep=0pt, scale=  1.00] at (197.83, 17.12) {\(60\)};

\node[text=drawColor,anchor=base,inner sep=0pt, outer sep=0pt, scale=  1.00] at (228.68, 17.12) {\(85\)};
\end{scope}
\begin{scope}
\path[clip] (  0.00,  0.00) rectangle (483.70,227.34);
\definecolor{drawColor}{RGB}{0,0,0}

\node[text=drawColor,anchor=base,inner sep=0pt, outer sep=0pt, scale=  1.00] at (139.20,  2.40) {Температура, {\textdegree}C};
\end{scope}
\begin{scope}
\path[clip] (  0.00,  0.00) rectangle (483.70,227.34);
\definecolor{drawColor}{RGB}{0,0,0}

\node[text=drawColor,rotate= 90.00,anchor=base,inner sep=0pt, outer sep=0pt, scale=  1.00] at ( 12.32,124.92) {\(\sigma_{si}\), МПа};
\end{scope}
\begin{scope}
\path[clip] (  0.00,  0.00) rectangle (483.70,227.34);
\definecolor{drawColor}{RGB}{0,0,0}
\definecolor{fillColor}{RGB}{255,255,255}

\path[draw=drawColor,line width= 0.6pt,line join=round,line cap=round,fill=fillColor] (129.31, 36.13) rectangle (233.70,100.12);
\end{scope}
\begin{scope}
\path[clip] (  0.00,  0.00) rectangle (483.70,227.34);
\definecolor{drawColor}{RGB}{0,0,0}

\node[text=drawColor,anchor=base west,inner sep=0pt, outer sep=0pt, scale=  1.00] at (133.58, 85.93) {Соединено при:\hspace*{1.3em}};
\end{scope}
\begin{scope}
\path[clip] (  0.00,  0.00) rectangle (483.70,227.34);
\definecolor{drawColor}{RGB}{255,255,255}
\definecolor{fillColor}{RGB}{255,255,255}

\path[draw=drawColor,line width= 0.6pt,line join=round,line cap=round,fill=fillColor] (133.58, 67.38) rectangle (160.56, 80.87);
\end{scope}
\begin{scope}
\path[clip] (  0.00,  0.00) rectangle (483.70,227.34);
\definecolor{drawColor}{RGB}{0,0,0}

\path[draw=drawColor,line width= 1.7pt,line join=round] (136.27, 74.13) -- (157.86, 74.13);
\end{scope}
\begin{scope}
\path[clip] (  0.00,  0.00) rectangle (483.70,227.34);
\definecolor{drawColor}{RGB}{0,0,0}

\path[draw=drawColor,line width= 1.7pt,line join=round] (136.27, 74.13) -- (157.86, 74.13);
\end{scope}
\begin{scope}
\path[clip] (  0.00,  0.00) rectangle (483.70,227.34);
\definecolor{drawColor}{RGB}{0,0,0}

\path[draw=drawColor,line width= 1.7pt,line join=round] (136.27, 74.13) -- (157.86, 74.13);
\end{scope}
\begin{scope}
\path[clip] (  0.00,  0.00) rectangle (483.70,227.34);
\definecolor{drawColor}{RGB}{255,255,255}
\definecolor{fillColor}{RGB}{255,255,255}

\path[draw=drawColor,line width= 0.6pt,line join=round,line cap=round,fill=fillColor] (133.58, 53.89) rectangle (160.56, 67.38);
\end{scope}
\begin{scope}
\path[clip] (  0.00,  0.00) rectangle (483.70,227.34);
\definecolor{drawColor}{RGB}{0,0,0}

\path[draw=drawColor,line width= 1.7pt,dash pattern=on 2pt off 2pt ,line join=round] (136.27, 60.64) -- (157.86, 60.64);
\end{scope}
\begin{scope}
\path[clip] (  0.00,  0.00) rectangle (483.70,227.34);
\definecolor{drawColor}{RGB}{0,0,0}

\path[draw=drawColor,line width= 1.7pt,dash pattern=on 2pt off 2pt ,line join=round] (136.27, 60.64) -- (157.86, 60.64);
\end{scope}
\begin{scope}
\path[clip] (  0.00,  0.00) rectangle (483.70,227.34);
\definecolor{drawColor}{RGB}{0,0,0}

\path[draw=drawColor,line width= 1.7pt,dash pattern=on 2pt off 2pt ,line join=round] (136.27, 60.64) -- (157.86, 60.64);
\end{scope}
\begin{scope}
\path[clip] (  0.00,  0.00) rectangle (483.70,227.34);
\definecolor{drawColor}{RGB}{255,255,255}
\definecolor{fillColor}{RGB}{255,255,255}

\path[draw=drawColor,line width= 0.6pt,line join=round,line cap=round,fill=fillColor] (133.58, 40.40) rectangle (160.56, 53.89);
\end{scope}
\begin{scope}
\path[clip] (  0.00,  0.00) rectangle (483.70,227.34);
\definecolor{drawColor}{RGB}{0,0,0}

\path[draw=drawColor,line width= 1.7pt,dash pattern=on 4pt off 2pt ,line join=round] (136.27, 47.15) -- (157.86, 47.15);
\end{scope}
\begin{scope}
\path[clip] (  0.00,  0.00) rectangle (483.70,227.34);
\definecolor{drawColor}{RGB}{0,0,0}

\path[draw=drawColor,line width= 1.7pt,dash pattern=on 4pt off 2pt ,line join=round] (136.27, 47.15) -- (157.86, 47.15);
\end{scope}
\begin{scope}
\path[clip] (  0.00,  0.00) rectangle (483.70,227.34);
\definecolor{drawColor}{RGB}{0,0,0}

\path[draw=drawColor,line width= 1.7pt,dash pattern=on 4pt off 2pt ,line join=round] (136.27, 47.15) -- (157.86, 47.15);
\end{scope}
\begin{scope}
\path[clip] (  0.00,  0.00) rectangle (483.70,227.34);
\definecolor{drawColor}{RGB}{0,0,0}

\node[text=drawColor,anchor=base east,inner sep=0pt, outer sep=0pt, scale=  1.00] at (203.62, 69.17) {300 {\textdegree}C};
\end{scope}
\begin{scope}
\path[clip] (  0.00,  0.00) rectangle (483.70,227.34);
\definecolor{drawColor}{RGB}{0,0,0}

\node[text=drawColor,anchor=base east,inner sep=0pt, outer sep=0pt, scale=  1.00] at (203.62, 55.68) {375 {\textdegree}C};
\end{scope}
\begin{scope}
\path[clip] (  0.00,  0.00) rectangle (483.70,227.34);
\definecolor{drawColor}{RGB}{0,0,0}

\node[text=drawColor,anchor=base east,inner sep=0pt, outer sep=0pt, scale=  1.00] at (203.62, 42.19) {450 {\textdegree}C};
\end{scope}
\begin{scope}
\path[clip] (  0.00,  0.00) rectangle (483.70,227.34);
\definecolor{drawColor}{RGB}{0,0,0}

\node[text=drawColor,anchor=base west,inner sep=0pt, outer sep=0pt, scale=  1.00] at ( 43.00,202.22) {\bfseries Borofloat 33};
\end{scope}
\begin{scope}
\path[clip] (241.85,  0.00) rectangle (483.70,227.34);
\definecolor{drawColor}{RGB}{255,255,255}
\definecolor{fillColor}{RGB}{255,255,255}

\path[draw=drawColor,line width= 0.6pt,line join=round,line cap=round,fill=fillColor] (241.85,  0.00) rectangle (483.70,227.34);
\end{scope}
\begin{scope}
\path[clip] (282.62, 32.43) rectangle (479.48,217.40);
\definecolor{fillColor}{RGB}{255,255,255}

\path[fill=fillColor] (282.62, 32.43) rectangle (479.48,217.40);
\definecolor{drawColor}{gray}{0.98}

\path[draw=drawColor,line width= 0.6pt,line join=round] (282.62, 51.61) --
	(479.48, 51.61);

\path[draw=drawColor,line width= 0.6pt,line join=round] (282.62, 93.03) --
	(479.48, 93.03);

\path[draw=drawColor,line width= 0.6pt,line join=round] (282.62,134.44) --
	(479.48,134.44);

\path[draw=drawColor,line width= 0.6pt,line join=round] (282.62,175.86) --
	(479.48,175.86);

\path[draw=drawColor,line width= 0.6pt,line join=round] (282.62,217.28) --
	(479.48,217.28);

\path[draw=drawColor,line width= 0.6pt,line join=round] (316.25, 32.43) --
	(316.25,217.40);

\path[draw=drawColor,line width= 0.6pt,line join=round] (353.28, 32.43) --
	(353.28,217.40);

\path[draw=drawColor,line width= 0.6pt,line join=round] (377.96, 32.43) --
	(377.96,217.40);

\path[draw=drawColor,line width= 0.6pt,line join=round] (414.99, 32.43) --
	(414.99,217.40);

\path[draw=drawColor,line width= 0.6pt,line join=round] (455.10, 32.43) --
	(455.10,217.40);
\definecolor{drawColor}{gray}{0.90}

\path[draw=drawColor,line width= 0.2pt,line join=round] (282.62, 72.32) --
	(479.48, 72.32);

\path[draw=drawColor,line width= 0.2pt,line join=round] (282.62,113.73) --
	(479.48,113.73);

\path[draw=drawColor,line width= 0.2pt,line join=round] (282.62,155.15) --
	(479.48,155.15);

\path[draw=drawColor,line width= 0.2pt,line join=round] (282.62,196.57) --
	(479.48,196.57);

\path[draw=drawColor,line width= 0.2pt,line join=round] (291.56, 32.43) --
	(291.56,217.40);

\path[draw=drawColor,line width= 0.2pt,line join=round] (340.93, 32.43) --
	(340.93,217.40);

\path[draw=drawColor,line width= 0.2pt,line join=round] (365.62, 32.43) --
	(365.62,217.40);

\path[draw=drawColor,line width= 0.2pt,line join=round] (390.31, 32.43) --
	(390.31,217.40);

\path[draw=drawColor,line width= 0.2pt,line join=round] (439.68, 32.43) --
	(439.68,217.40);

\path[draw=drawColor,line width= 0.2pt,line join=round] (470.53, 32.43) --
	(470.53,217.40);
\definecolor{drawColor}{RGB}{0,0,0}

\path[draw=drawColor,line width= 1.7pt,line join=round] (291.56, 47.77) --
	(293.35, 48.78) --
	(295.14, 49.78) --
	(296.93, 50.78) --
	(298.72, 51.76) --
	(300.51, 52.74) --
	(302.30, 53.71) --
	(304.09, 54.67) --
	(305.88, 55.62) --
	(307.67, 56.57) --
	(309.46, 57.51) --
	(311.25, 58.43) --
	(313.04, 59.35) --
	(314.83, 60.27) --
	(316.62, 61.17) --
	(318.41, 62.07) --
	(320.20, 62.96) --
	(321.99, 63.84) --
	(323.78, 64.71) --
	(325.57, 65.58) --
	(327.36, 66.44) --
	(329.15, 67.29) --
	(330.94, 68.13) --
	(332.73, 68.96) --
	(334.52, 69.79) --
	(336.31, 70.61) --
	(338.10, 71.42) --
	(339.89, 72.23) --
	(341.67, 73.02) --
	(343.46, 73.81) --
	(345.25, 74.60) --
	(347.04, 75.37) --
	(348.83, 76.14) --
	(350.62, 76.90) --
	(352.41, 77.65) --
	(354.20, 78.40) --
	(355.99, 79.14) --
	(357.78, 79.87) --
	(359.57, 80.59) --
	(361.36, 81.31) --
	(363.15, 82.02) --
	(364.94, 82.72) --
	(366.73, 83.42) --
	(368.52, 84.11) --
	(370.31, 84.79) --
	(372.10, 85.46) --
	(373.89, 86.13) --
	(375.68, 86.79) --
	(377.47, 87.44) --
	(379.26, 88.09) --
	(381.05, 88.73) --
	(382.84, 89.37) --
	(384.63, 89.99) --
	(386.42, 90.61) --
	(388.21, 91.23) --
	(390.00, 91.83) --
	(391.79, 92.43) --
	(393.58, 93.03) --
	(395.37, 93.61) --
	(397.16, 94.19) --
	(398.95, 94.77) --
	(400.73, 95.34) --
	(402.52, 95.90) --
	(404.31, 96.45) --
	(406.10, 97.00) --
	(407.89, 97.54) --
	(409.68, 98.07) --
	(411.47, 98.60) --
	(413.26, 99.13) --
	(415.05, 99.64) --
	(416.84,100.15) --
	(418.63,100.66) --
	(420.42,101.15) --
	(422.21,101.64) --
	(424.00,102.13) --
	(425.79,102.61) --
	(427.58,103.08) --
	(429.37,103.55) --
	(431.16,104.01) --
	(432.95,104.46) --
	(434.74,104.91) --
	(436.53,105.35) --
	(438.32,105.79) --
	(440.11,106.22) --
	(441.90,106.64) --
	(443.69,107.06) --
	(445.48,107.48) --
	(447.27,107.88) --
	(449.06,108.28) --
	(450.85,108.68) --
	(452.64,109.07) --
	(454.43,109.45) --
	(456.22,109.83) --
	(458.00,110.20) --
	(459.79,110.57) --
	(461.58,110.93) --
	(463.37,111.29) --
	(465.16,111.64) --
	(466.95,111.98) --
	(468.74,112.32) --
	(470.53,112.65);

\path[draw=drawColor,line width= 1.7pt,dash pattern=on 2pt off 2pt ,line join=round] (291.56, 61.02) --
	(293.35, 62.03) --
	(295.14, 63.03) --
	(296.93, 64.02) --
	(298.72, 65.01) --
	(300.51, 65.99) --
	(302.30, 66.96) --
	(304.09, 67.92) --
	(305.88, 68.87) --
	(307.67, 69.81) --
	(309.46, 70.75) --
	(311.25, 71.68) --
	(313.04, 72.60) --
	(314.83, 73.51) --
	(316.62, 74.42) --
	(318.41, 75.31) --
	(320.20, 76.20) --
	(321.99, 77.08) --
	(323.78, 77.96) --
	(325.57, 78.82) --
	(327.36, 79.68) --
	(329.15, 80.53) --
	(330.94, 81.37) --
	(332.73, 82.21) --
	(334.52, 83.04) --
	(336.31, 83.86) --
	(338.10, 84.67) --
	(339.89, 85.47) --
	(341.67, 86.27) --
	(343.46, 87.06) --
	(345.25, 87.84) --
	(347.04, 88.62) --
	(348.83, 89.38) --
	(350.62, 90.14) --
	(352.41, 90.90) --
	(354.20, 91.64) --
	(355.99, 92.38) --
	(357.78, 93.11) --
	(359.57, 93.84) --
	(361.36, 94.55) --
	(363.15, 95.26) --
	(364.94, 95.97) --
	(366.73, 96.66) --
	(368.52, 97.35) --
	(370.31, 98.03) --
	(372.10, 98.71) --
	(373.89, 99.38) --
	(375.68,100.04) --
	(377.47,100.69) --
	(379.26,101.34) --
	(381.05,101.98) --
	(382.84,102.61) --
	(384.63,103.24) --
	(386.42,103.86) --
	(388.21,104.47) --
	(390.00,105.08) --
	(391.79,105.68) --
	(393.58,106.27) --
	(395.37,106.86) --
	(397.16,107.44) --
	(398.95,108.01) --
	(400.73,108.58) --
	(402.52,109.14) --
	(404.31,109.70) --
	(406.10,110.24) --
	(407.89,110.79) --
	(409.68,111.32) --
	(411.47,111.85) --
	(413.26,112.37) --
	(415.05,112.89) --
	(416.84,113.40) --
	(418.63,113.90) --
	(420.42,114.40) --
	(422.21,114.89) --
	(424.00,115.37) --
	(425.79,115.85) --
	(427.58,116.33) --
	(429.37,116.79) --
	(431.16,117.25) --
	(432.95,117.71) --
	(434.74,118.16) --
	(436.53,118.60) --
	(438.32,119.04) --
	(440.11,119.47) --
	(441.90,119.89) --
	(443.69,120.31) --
	(445.48,120.72) --
	(447.27,121.13) --
	(449.06,121.53) --
	(450.85,121.92) --
	(452.64,122.31) --
	(454.43,122.70) --
	(456.22,123.08) --
	(458.00,123.45) --
	(459.79,123.82) --
	(461.58,124.18) --
	(463.37,124.53) --
	(465.16,124.88) --
	(466.95,125.23) --
	(468.74,125.56) --
	(470.53,125.90);

\path[draw=drawColor,line width= 1.7pt,dash pattern=on 4pt off 2pt ,line join=round] (291.56, 73.98) --
	(293.35, 74.99) --
	(295.14, 75.99) --
	(296.93, 76.98) --
	(298.72, 77.97) --
	(300.51, 78.94) --
	(302.30, 79.91) --
	(304.09, 80.88) --
	(305.88, 81.83) --
	(307.67, 82.77) --
	(309.46, 83.71) --
	(311.25, 84.64) --
	(313.04, 85.56) --
	(314.83, 86.47) --
	(316.62, 87.38) --
	(318.41, 88.27) --
	(320.20, 89.16) --
	(321.99, 90.04) --
	(323.78, 90.92) --
	(325.57, 91.78) --
	(327.36, 92.64) --
	(329.15, 93.49) --
	(330.94, 94.33) --
	(332.73, 95.17) --
	(334.52, 95.99) --
	(336.31, 96.81) --
	(338.10, 97.63) --
	(339.89, 98.43) --
	(341.67, 99.23) --
	(343.46,100.02) --
	(345.25,100.80) --
	(347.04,101.57) --
	(348.83,102.34) --
	(350.62,103.10) --
	(352.41,103.86) --
	(354.20,104.60) --
	(355.99,105.34) --
	(357.78,106.07) --
	(359.57,106.79) --
	(361.36,107.51) --
	(363.15,108.22) --
	(364.94,108.92) --
	(366.73,109.62) --
	(368.52,110.31) --
	(370.31,110.99) --
	(372.10,111.67) --
	(373.89,112.33) --
	(375.68,112.99) --
	(377.47,113.65) --
	(379.26,114.30) --
	(381.05,114.94) --
	(382.84,115.57) --
	(384.63,116.20) --
	(386.42,116.82) --
	(388.21,117.43) --
	(390.00,118.04) --
	(391.79,118.64) --
	(393.58,119.23) --
	(395.37,119.82) --
	(397.16,120.40) --
	(398.95,120.97) --
	(400.73,121.54) --
	(402.52,122.10) --
	(404.31,122.65) --
	(406.10,123.20) --
	(407.89,123.74) --
	(409.68,124.28) --
	(411.47,124.81) --
	(413.26,125.33) --
	(415.05,125.85) --
	(416.84,126.36) --
	(418.63,126.86) --
	(420.42,127.36) --
	(422.21,127.85) --
	(424.00,128.33) --
	(425.79,128.81) --
	(427.58,129.28) --
	(429.37,129.75) --
	(431.16,130.21) --
	(432.95,130.67) --
	(434.74,131.11) --
	(436.53,131.56) --
	(438.32,131.99) --
	(440.11,132.42) --
	(441.90,132.85) --
	(443.69,133.27) --
	(445.48,133.68) --
	(447.27,134.09) --
	(449.06,134.49) --
	(450.85,134.88) --
	(452.64,135.27) --
	(454.43,135.66) --
	(456.22,136.03) --
	(458.00,136.41) --
	(459.79,136.77) --
	(461.58,137.13) --
	(463.37,137.49) --
	(465.16,137.84) --
	(466.95,138.18) --
	(468.74,138.52) --
	(470.53,138.86);

\path[draw=drawColor,line width= 0.9pt,line join=round,line cap=round] (282.62, 32.43) rectangle (479.48,217.40);
\end{scope}
\begin{scope}
\path[clip] (  0.00,  0.00) rectangle (483.70,227.34);
\definecolor{drawColor}{RGB}{0,0,0}

\node[text=drawColor,anchor=base east,inner sep=0pt, outer sep=0pt, scale=  1.00] at (277.22, 67.36) {\(-5\)};

\node[text=drawColor,anchor=base east,inner sep=0pt, outer sep=0pt, scale=  1.00] at (277.22,108.78) {\(0\)};

\node[text=drawColor,anchor=base east,inner sep=0pt, outer sep=0pt, scale=  1.00] at (277.22,150.19) {\(5\)};

\node[text=drawColor,anchor=base east,inner sep=0pt, outer sep=0pt, scale=  1.00] at (277.22,191.61) {\(10\)};
\end{scope}
\begin{scope}
\path[clip] (  0.00,  0.00) rectangle (483.70,227.34);
\definecolor{drawColor}{RGB}{0,0,0}

\path[draw=drawColor,line width= 0.6pt,line join=round] (279.62, 72.32) --
	(282.62, 72.32);

\path[draw=drawColor,line width= 0.6pt,line join=round] (279.62,113.73) --
	(282.62,113.73);

\path[draw=drawColor,line width= 0.6pt,line join=round] (279.62,155.15) --
	(282.62,155.15);

\path[draw=drawColor,line width= 0.6pt,line join=round] (279.62,196.57) --
	(282.62,196.57);
\end{scope}
\begin{scope}
\path[clip] (  0.00,  0.00) rectangle (483.70,227.34);
\definecolor{drawColor}{RGB}{0,0,0}

\path[draw=drawColor,line width= 0.6pt,line join=round] (291.56, 29.43) --
	(291.56, 32.43);

\path[draw=drawColor,line width= 0.6pt,line join=round] (340.93, 29.43) --
	(340.93, 32.43);

\path[draw=drawColor,line width= 0.6pt,line join=round] (365.62, 29.43) --
	(365.62, 32.43);

\path[draw=drawColor,line width= 0.6pt,line join=round] (390.31, 29.43) --
	(390.31, 32.43);

\path[draw=drawColor,line width= 0.6pt,line join=round] (439.68, 29.43) --
	(439.68, 32.43);

\path[draw=drawColor,line width= 0.6pt,line join=round] (470.53, 29.43) --
	(470.53, 32.43);
\end{scope}
\begin{scope}
\path[clip] (  0.00,  0.00) rectangle (483.70,227.34);
\definecolor{drawColor}{RGB}{0,0,0}

\node[text=drawColor,anchor=base,inner sep=0pt, outer sep=0pt, scale=  1.00] at (291.56, 17.12) {\(-60\)};

\node[text=drawColor,anchor=base,inner sep=0pt, outer sep=0pt, scale=  1.00] at (340.93, 17.12) {\(-20\)};

\node[text=drawColor,anchor=base,inner sep=0pt, outer sep=0pt, scale=  1.00] at (365.62, 17.12) {\(0\)};

\node[text=drawColor,anchor=base,inner sep=0pt, outer sep=0pt, scale=  1.00] at (390.31, 17.12) {\(20\)};

\node[text=drawColor,anchor=base,inner sep=0pt, outer sep=0pt, scale=  1.00] at (439.68, 17.12) {\(60\)};

\node[text=drawColor,anchor=base,inner sep=0pt, outer sep=0pt, scale=  1.00] at (470.53, 17.12) {\(85\)};
\end{scope}
\begin{scope}
\path[clip] (  0.00,  0.00) rectangle (483.70,227.34);
\definecolor{drawColor}{RGB}{0,0,0}

\node[text=drawColor,anchor=base,inner sep=0pt, outer sep=0pt, scale=  1.00] at (381.05,  2.40) {Температура, {\textdegree}C};
\end{scope}
\begin{scope}
\path[clip] (  0.00,  0.00) rectangle (483.70,227.34);
\definecolor{drawColor}{RGB}{0,0,0}

\node[text=drawColor,rotate= 90.00,anchor=base,inner sep=0pt, outer sep=0pt, scale=  1.00] at (254.17,124.92) {\(\sigma_{si}\), МПа};
\end{scope}
\begin{scope}
\path[clip] (  0.00,  0.00) rectangle (483.70,227.34);
\definecolor{drawColor}{RGB}{0,0,0}
\definecolor{fillColor}{RGB}{255,255,255}

\path[draw=drawColor,line width= 0.6pt,line join=round,line cap=round,fill=fillColor] (371.16,149.72) rectangle (475.54,213.70);
\end{scope}
\begin{scope}
\path[clip] (  0.00,  0.00) rectangle (483.70,227.34);
\definecolor{drawColor}{RGB}{0,0,0}

\node[text=drawColor,anchor=base west,inner sep=0pt, outer sep=0pt, scale=  1.00] at (375.42,199.52) {Соединено при:\hspace*{1.3em}};
\end{scope}
\begin{scope}
\path[clip] (  0.00,  0.00) rectangle (483.70,227.34);
\definecolor{drawColor}{RGB}{255,255,255}
\definecolor{fillColor}{RGB}{255,255,255}

\path[draw=drawColor,line width= 0.6pt,line join=round,line cap=round,fill=fillColor] (375.42,180.97) rectangle (402.40,194.46);
\end{scope}
\begin{scope}
\path[clip] (  0.00,  0.00) rectangle (483.70,227.34);
\definecolor{drawColor}{RGB}{0,0,0}

\path[draw=drawColor,line width= 1.7pt,line join=round] (378.12,187.71) -- (399.71,187.71);
\end{scope}
\begin{scope}
\path[clip] (  0.00,  0.00) rectangle (483.70,227.34);
\definecolor{drawColor}{RGB}{0,0,0}

\path[draw=drawColor,line width= 1.7pt,line join=round] (378.12,187.71) -- (399.71,187.71);
\end{scope}
\begin{scope}
\path[clip] (  0.00,  0.00) rectangle (483.70,227.34);
\definecolor{drawColor}{RGB}{0,0,0}

\path[draw=drawColor,line width= 1.7pt,line join=round] (378.12,187.71) -- (399.71,187.71);
\end{scope}
\begin{scope}
\path[clip] (  0.00,  0.00) rectangle (483.70,227.34);
\definecolor{drawColor}{RGB}{255,255,255}
\definecolor{fillColor}{RGB}{255,255,255}

\path[draw=drawColor,line width= 0.6pt,line join=round,line cap=round,fill=fillColor] (375.42,167.48) rectangle (402.40,180.97);
\end{scope}
\begin{scope}
\path[clip] (  0.00,  0.00) rectangle (483.70,227.34);
\definecolor{drawColor}{RGB}{0,0,0}

\path[draw=drawColor,line width= 1.7pt,dash pattern=on 2pt off 2pt ,line join=round] (378.12,174.22) -- (399.71,174.22);
\end{scope}
\begin{scope}
\path[clip] (  0.00,  0.00) rectangle (483.70,227.34);
\definecolor{drawColor}{RGB}{0,0,0}

\path[draw=drawColor,line width= 1.7pt,dash pattern=on 2pt off 2pt ,line join=round] (378.12,174.22) -- (399.71,174.22);
\end{scope}
\begin{scope}
\path[clip] (  0.00,  0.00) rectangle (483.70,227.34);
\definecolor{drawColor}{RGB}{0,0,0}

\path[draw=drawColor,line width= 1.7pt,dash pattern=on 2pt off 2pt ,line join=round] (378.12,174.22) -- (399.71,174.22);
\end{scope}
\begin{scope}
\path[clip] (  0.00,  0.00) rectangle (483.70,227.34);
\definecolor{drawColor}{RGB}{255,255,255}
\definecolor{fillColor}{RGB}{255,255,255}

\path[draw=drawColor,line width= 0.6pt,line join=round,line cap=round,fill=fillColor] (375.42,153.99) rectangle (402.40,167.48);
\end{scope}
\begin{scope}
\path[clip] (  0.00,  0.00) rectangle (483.70,227.34);
\definecolor{drawColor}{RGB}{0,0,0}

\path[draw=drawColor,line width= 1.7pt,dash pattern=on 4pt off 2pt ,line join=round] (378.12,160.73) -- (399.71,160.73);
\end{scope}
\begin{scope}
\path[clip] (  0.00,  0.00) rectangle (483.70,227.34);
\definecolor{drawColor}{RGB}{0,0,0}

\path[draw=drawColor,line width= 1.7pt,dash pattern=on 4pt off 2pt ,line join=round] (378.12,160.73) -- (399.71,160.73);
\end{scope}
\begin{scope}
\path[clip] (  0.00,  0.00) rectangle (483.70,227.34);
\definecolor{drawColor}{RGB}{0,0,0}

\path[draw=drawColor,line width= 1.7pt,dash pattern=on 4pt off 2pt ,line join=round] (378.12,160.73) -- (399.71,160.73);
\end{scope}
\begin{scope}
\path[clip] (  0.00,  0.00) rectangle (483.70,227.34);
\definecolor{drawColor}{RGB}{0,0,0}

\node[text=drawColor,anchor=base east,inner sep=0pt, outer sep=0pt, scale=  1.00] at (445.46,182.75) {300 {\textdegree}C};
\end{scope}
\begin{scope}
\path[clip] (  0.00,  0.00) rectangle (483.70,227.34);
\definecolor{drawColor}{RGB}{0,0,0}

\node[text=drawColor,anchor=base east,inner sep=0pt, outer sep=0pt, scale=  1.00] at (445.46,169.26) {375 {\textdegree}C};
\end{scope}
\begin{scope}
\path[clip] (  0.00,  0.00) rectangle (483.70,227.34);
\definecolor{drawColor}{RGB}{0,0,0}

\node[text=drawColor,anchor=base east,inner sep=0pt, outer sep=0pt, scale=  1.00] at (445.46,155.77) {450 {\textdegree}C};
\end{scope}
\begin{scope}
\path[clip] (  0.00,  0.00) rectangle (483.70,227.34);
\definecolor{drawColor}{RGB}{0,0,0}

\node[text=drawColor,anchor=base west,inner sep=0pt, outer sep=0pt, scale=  1.00] at (285.98,202.22) {\bfseries ЛК5};
\end{scope}
\end{tikzpicture}

%% file: Dissertation/images_tikz/disser_glass_stress_multilayer_ot_h_tw.tikz
% Created by tikzDevice version 0.10.1 on 2016-10-01 21:04:21
% !TEX encoding = UTF-8 Unicode
\begin{tikzpicture}[x=0.985pt,y=0.985pt]
\definecolor{fillColor}{RGB}{255,255,255}
\path[use as bounding box,fill=fillColor] (0,0) rectangle (241.85,217.66);
\begin{scope}
\path[clip] (  0.00,  0.00) rectangle (241.85,217.66);
\definecolor{drawColor}{RGB}{255,255,255}

\path[draw=drawColor,line width= 0.6pt,line join=round,line cap=round,fill=fillColor] ( -0.00,  0.00) rectangle (241.85,217.66);
\end{scope}
\begin{scope}
\path[clip] ( 45.96, 34.42) rectangle (237.63,217.66);
\definecolor{fillColor}{RGB}{255,255,255}

\path[fill=fillColor] ( 45.96, 34.42) rectangle (237.63,217.66);
\definecolor{drawColor}{gray}{0.98}

\path[draw=drawColor,line width= 0.6pt,line join=round] ( 45.96, 64.36) --
	(237.63, 64.36);

\path[draw=drawColor,line width= 0.6pt,line join=round] ( 45.96, 96.62) --
	(237.63, 96.62);

\path[draw=drawColor,line width= 0.6pt,line join=round] ( 45.96,128.88) --
	(237.63,128.88);

\path[draw=drawColor,line width= 0.6pt,line join=round] ( 45.96,161.14) --
	(237.63,161.14);

\path[draw=drawColor,line width= 0.6pt,line join=round] ( 45.96,193.40) --
	(237.63,193.40);

\path[draw=drawColor,line width= 0.6pt,line join=round] ( 72.10, 34.42) --
	( 72.10,217.66);

\path[draw=drawColor,line width= 0.6pt,line join=round] (106.95, 34.42) --
	(106.95,217.66);

\path[draw=drawColor,line width= 0.6pt,line join=round] (141.80, 34.42) --
	(141.80,217.66);

\path[draw=drawColor,line width= 0.6pt,line join=round] (176.65, 34.42) --
	(176.65,217.66);

\path[draw=drawColor,line width= 0.6pt,line join=round] (211.50, 34.42) --
	(211.50,217.66);
\definecolor{drawColor}{gray}{0.80}

\path[draw=drawColor,line width= 0.3pt,line join=round] ( 45.96, 48.23) --
	(237.63, 48.23);

\path[draw=drawColor,line width= 0.3pt,line join=round] ( 45.96, 80.49) --
	(237.63, 80.49);

\path[draw=drawColor,line width= 0.3pt,line join=round] ( 45.96,112.75) --
	(237.63,112.75);

\path[draw=drawColor,line width= 0.3pt,line join=round] ( 45.96,145.01) --
	(237.63,145.01);

\path[draw=drawColor,line width= 0.3pt,line join=round] ( 45.96,177.27) --
	(237.63,177.27);

\path[draw=drawColor,line width= 0.3pt,line join=round] ( 45.96,209.53) --
	(237.63,209.53);

\path[draw=drawColor,line width= 0.3pt,line join=round] ( 54.67, 34.42) --
	( 54.67,217.66);

\path[draw=drawColor,line width= 0.3pt,line join=round] ( 89.52, 34.42) --
	( 89.52,217.66);

\path[draw=drawColor,line width= 0.3pt,line join=round] (124.37, 34.42) --
	(124.37,217.66);

\path[draw=drawColor,line width= 0.3pt,line join=round] (159.22, 34.42) --
	(159.22,217.66);

\path[draw=drawColor,line width= 0.3pt,line join=round] (194.07, 34.42) --
	(194.07,217.66);

\path[draw=drawColor,line width= 0.3pt,line join=round] (228.92, 34.42) --
	(228.92,217.66);
\definecolor{drawColor}{RGB}{0,0,0}

\path[draw=drawColor,line width= 1.4pt,line join=round] ( 54.67,145.01) --
	( 55.25,145.54) --
	( 55.83,146.06) --
	( 56.42,146.59) --
	( 57.00,147.10) --
	( 57.58,147.61) --
	( 58.16,148.11) --
	( 58.74,148.60) --
	( 59.32,149.08) --
	( 59.90,149.54) --
	( 60.48,149.99) --
	( 61.06,150.43) --
	( 61.64,150.85) --
	( 62.22,151.26) --
	( 62.80,151.66) --
	( 63.39,152.03) --
	( 63.97,152.39) --
	( 64.55,152.74) --
	( 65.13,153.07) --
	( 65.71,153.38) --
	( 66.29,153.67) --
	( 66.87,153.95) --
	( 67.45,154.22) --
	( 68.03,154.47) --
	( 68.61,154.70) --
	( 69.19,154.92) --
	( 69.77,155.12) --
	( 70.36,155.31) --
	( 70.94,155.49) --
	( 71.52,155.65) --
	( 72.10,155.80) --
	( 72.68,155.94) --
	( 73.26,156.06) --
	( 73.84,156.18) --
	( 74.42,156.28) --
	( 75.00,156.37) --
	( 75.58,156.46) --
	( 76.16,156.53) --
	( 76.74,156.59) --
	( 77.33,156.64) --
	( 77.91,156.69) --
	( 78.49,156.73) --
	( 79.07,156.76) --
	( 79.65,156.78) --
	( 80.23,156.79) --
	( 80.81,156.80) --
	( 81.39,156.80) --
	( 81.97,156.79) --
	( 82.55,156.78) --
	( 83.13,156.77) --
	( 83.71,156.74) --
	( 84.30,156.72) --
	( 84.88,156.69) --
	( 85.46,156.65) --
	( 86.04,156.61) --
	( 86.62,156.56) --
	( 87.20,156.52) --
	( 87.78,156.46) --
	( 88.36,156.41) --
	( 88.94,156.35) --
	( 89.52,156.29) --
	( 90.10,156.22) --
	( 90.68,156.16) --
	( 91.27,156.09) --
	( 91.85,156.01) --
	( 92.43,155.94) --
	( 93.01,155.86) --
	( 93.59,155.79) --
	( 94.17,155.70) --
	( 94.75,155.62) --
	( 95.33,155.54) --
	( 95.91,155.45) --
	( 96.49,155.37) --
	( 97.07,155.28) --
	( 97.65,155.19) --
	( 98.24,155.10) --
	( 98.82,155.01) --
	( 99.40,154.91) --
	( 99.98,154.82) --
	(100.56,154.73) --
	(101.14,154.63) --
	(101.72,154.54) --
	(102.30,154.44) --
	(102.88,154.34) --
	(103.46,154.25) --
	(104.04,154.15) --
	(104.62,154.05) --
	(105.20,153.95) --
	(105.79,153.85) --
	(106.37,153.75) --
	(106.95,153.65) --
	(107.53,153.55) --
	(108.11,153.45) --
	(108.69,153.35) --
	(109.27,153.25) --
	(109.85,153.15) --
	(110.43,153.05) --
	(111.01,152.95) --
	(111.59,152.84) --
	(112.17,152.74) --
	(112.76,152.64) --
	(113.34,152.54) --
	(113.92,152.44) --
	(114.50,152.34) --
	(115.08,152.24) --
	(115.66,152.14) --
	(116.24,152.04) --
	(116.82,151.94) --
	(117.40,151.84) --
	(117.98,151.74) --
	(118.56,151.64) --
	(119.14,151.54) --
	(119.73,151.44) --
	(120.31,151.34) --
	(120.89,151.24) --
	(121.47,151.14) --
	(122.05,151.04) --
	(122.63,150.95) --
	(123.21,150.85) --
	(123.79,150.75) --
	(124.37,150.65) --
	(124.95,150.56) --
	(125.53,150.46) --
	(126.11,150.36) --
	(126.70,150.27) --
	(127.28,150.17) --
	(127.86,150.07) --
	(128.44,149.98) --
	(129.02,149.88) --
	(129.60,149.79) --
	(130.18,149.70) --
	(130.76,149.60) --
	(131.34,149.51) --
	(131.92,149.41) --
	(132.50,149.32) --
	(133.08,149.23) --
	(133.67,149.14) --
	(134.25,149.04) --
	(134.83,148.95) --
	(135.41,148.86) --
	(135.99,148.77) --
	(136.57,148.68) --
	(137.15,148.59) --
	(137.73,148.50) --
	(138.31,148.41) --
	(138.89,148.32) --
	(139.47,148.23) --
	(140.05,148.14) --
	(140.64,148.06) --
	(141.22,147.97) --
	(141.80,147.88) --
	(142.38,147.79) --
	(142.96,147.71) --
	(143.54,147.62) --
	(144.12,147.53) --
	(144.70,147.45) --
	(145.28,147.36) --
	(145.86,147.28) --
	(146.44,147.19) --
	(147.02,147.11) --
	(147.61,147.03) --
	(148.19,146.94) --
	(148.77,146.86) --
	(149.35,146.78) --
	(149.93,146.69) --
	(150.51,146.61) --
	(151.09,146.53) --
	(151.67,146.45) --
	(152.25,146.37) --
	(152.83,146.29) --
	(153.41,146.21) --
	(153.99,146.13) --
	(154.57,146.05) --
	(155.16,145.97) --
	(155.74,145.89) --
	(156.32,145.81) --
	(156.90,145.73) --
	(157.48,145.65) --
	(158.06,145.57) --
	(158.64,145.50) --
	(159.22,145.42) --
	(159.80,145.34) --
	(160.38,145.26) --
	(160.96,145.19) --
	(161.54,145.11) --
	(162.13,145.04) --
	(162.71,144.96) --
	(163.29,144.89) --
	(163.87,144.81) --
	(164.45,144.74) --
	(165.03,144.66) --
	(165.61,144.59) --
	(166.19,144.51) --
	(166.77,144.44) --
	(167.35,144.37) --
	(167.93,144.30) --
	(168.51,144.22) --
	(169.10,144.15) --
	(169.68,144.08) --
	(170.26,144.01) --
	(170.84,143.94) --
	(171.42,143.87) --
	(172.00,143.79) --
	(172.58,143.72) --
	(173.16,143.65) --
	(173.74,143.58) --
	(174.32,143.51) --
	(174.90,143.45) --
	(175.48,143.38) --
	(176.07,143.31) --
	(176.65,143.24) --
	(177.23,143.17) --
	(177.81,143.10) --
	(178.39,143.04) --
	(178.97,142.97) --
	(179.55,142.90) --
	(180.13,142.83) --
	(180.71,142.77) --
	(181.29,142.70) --
	(181.87,142.63) --
	(182.45,142.57) --
	(183.04,142.50) --
	(183.62,142.44) --
	(184.20,142.37) --
	(184.78,142.31) --
	(185.36,142.24) --
	(185.94,142.18) --
	(186.52,142.11) --
	(187.10,142.05) --
	(187.68,141.99) --
	(188.26,141.92) --
	(188.84,141.86) --
	(189.42,141.80) --
	(190.01,141.73) --
	(190.59,141.67) --
	(191.17,141.61) --
	(191.75,141.55) --
	(192.33,141.49) --
	(192.91,141.42) --
	(193.49,141.36) --
	(194.07,141.30) --
	(194.65,141.24) --
	(195.23,141.18) --
	(195.81,141.12) --
	(196.39,141.06) --
	(196.98,141.00) --
	(197.56,140.94) --
	(198.14,140.88) --
	(198.72,140.82) --
	(199.30,140.76) --
	(199.88,140.70) --
	(200.46,140.64) --
	(201.04,140.59) --
	(201.62,140.53) --
	(202.20,140.47) --
	(202.78,140.41) --
	(203.36,140.35) --
	(203.94,140.30) --
	(204.53,140.24) --
	(205.11,140.18) --
	(205.69,140.12) --
	(206.27,140.07) --
	(206.85,140.01) --
	(207.43,139.95) --
	(208.01,139.90) --
	(208.59,139.84) --
	(209.17,139.79) --
	(209.75,139.73) --
	(210.33,139.68) --
	(210.91,139.62) --
	(211.50,139.57) --
	(212.08,139.51) --
	(212.66,139.46) --
	(213.24,139.40) --
	(213.82,139.35) --
	(214.40,139.29) --
	(214.98,139.24) --
	(215.56,139.19) --
	(216.14,139.13) --
	(216.72,139.08) --
	(217.30,139.03) --
	(217.88,138.97) --
	(218.47,138.92) --
	(219.05,138.87) --
	(219.63,138.82) --
	(220.21,138.76) --
	(220.79,138.71) --
	(221.37,138.66) --
	(221.95,138.61) --
	(222.53,138.56) --
	(223.11,138.51) --
	(223.69,138.45) --
	(224.27,138.40) --
	(224.85,138.35) --
	(225.44,138.30) --
	(226.02,138.25) --
	(226.60,138.20) --
	(227.18,138.15) --
	(227.76,138.10) --
	(228.34,138.05) --
	(228.92,138.00);

\path[draw=drawColor,line width= 1.4pt,dash pattern=on 7pt off 3pt ,line join=round] ( 54.67,145.01) --
	( 55.25,147.96) --
	( 55.83,150.88) --
	( 56.42,153.79) --
	( 57.00,156.65) --
	( 57.58,159.47) --
	( 58.16,162.23) --
	( 58.74,164.94) --
	( 59.32,167.57) --
	( 59.90,170.14) --
	( 60.48,172.63) --
	( 61.06,175.03) --
	( 61.64,177.36) --
	( 62.22,179.60) --
	( 62.80,181.74) --
	( 63.39,183.80) --
	( 63.97,185.77) --
	( 64.55,187.65) --
	( 65.13,189.43) --
	( 65.71,191.13) --
	( 66.29,192.73) --
	( 66.87,194.25) --
	( 67.45,195.67) --
	( 68.03,197.02) --
	( 68.61,198.27) --
	( 69.19,199.45) --
	( 69.77,200.54) --
	( 70.36,201.56) --
	( 70.94,202.50) --
	( 71.52,203.37) --
	( 72.10,204.17) --
	( 72.68,204.90) --
	( 73.26,205.56) --
	( 73.84,206.17) --
	( 74.42,206.71) --
	( 75.00,207.19) --
	( 75.58,207.61) --
	( 76.16,207.98) --
	( 76.74,208.31) --
	( 77.33,208.58) --
	( 77.91,208.80) --
	( 78.49,208.98) --
	( 79.07,209.12) --
	( 79.65,209.21) --
	( 80.23,209.27) --
	( 80.81,209.29) --
	( 81.39,209.28) --
	( 81.97,209.23) --
	( 82.55,209.15) --
	( 83.13,209.04) --
	( 83.71,208.91) --
	( 84.30,208.74) --
	( 84.88,208.55) --
	( 85.46,208.34) --
	( 86.04,208.10) --
	( 86.62,207.84) --
	( 87.20,207.56) --
	( 87.78,207.26) --
	( 88.36,206.95) --
	( 88.94,206.61) --
	( 89.52,206.26) --
	( 90.10,205.89) --
	( 90.68,205.51) --
	( 91.27,205.12) --
	( 91.85,204.71) --
	( 92.43,204.29) --
	( 93.01,203.86) --
	( 93.59,203.42) --
	( 94.17,202.97) --
	( 94.75,202.51) --
	( 95.33,202.05) --
	( 95.91,201.57) --
	( 96.49,201.09) --
	( 97.07,200.59) --
	( 97.65,200.10) --
	( 98.24,199.59) --
	( 98.82,199.08) --
	( 99.40,198.57) --
	( 99.98,198.05) --
	(100.56,197.53) --
	(101.14,197.00) --
	(101.72,196.47) --
	(102.30,195.93) --
	(102.88,195.39) --
	(103.46,194.85) --
	(104.04,194.31) --
	(104.62,193.76) --
	(105.20,193.21) --
	(105.79,192.66) --
	(106.37,192.11) --
	(106.95,191.56) --
	(107.53,191.01) --
	(108.11,190.45) --
	(108.69,189.89) --
	(109.27,189.34) --
	(109.85,188.78) --
	(110.43,188.22) --
	(111.01,187.67) --
	(111.59,187.11) --
	(112.17,186.55) --
	(112.76,185.99) --
	(113.34,185.44) --
	(113.92,184.88) --
	(114.50,184.32) --
	(115.08,183.77) --
	(115.66,183.21) --
	(116.24,182.66) --
	(116.82,182.10) --
	(117.40,181.55) --
	(117.98,181.00) --
	(118.56,180.45) --
	(119.14,179.90) --
	(119.73,179.35) --
	(120.31,178.81) --
	(120.89,178.26) --
	(121.47,177.72) --
	(122.05,177.18) --
	(122.63,176.64) --
	(123.21,176.10) --
	(123.79,175.56) --
	(124.37,175.02) --
	(124.95,174.49) --
	(125.53,173.95) --
	(126.11,173.42) --
	(126.70,172.89) --
	(127.28,172.37) --
	(127.86,171.84) --
	(128.44,171.32) --
	(129.02,170.79) --
	(129.60,170.27) --
	(130.18,169.75) --
	(130.76,169.24) --
	(131.34,168.72) --
	(131.92,168.21) --
	(132.50,167.70) --
	(133.08,167.19) --
	(133.67,166.68) --
	(134.25,166.17) --
	(134.83,165.67) --
	(135.41,165.17) --
	(135.99,164.67) --
	(136.57,164.17) --
	(137.15,163.67) --
	(137.73,163.18) --
	(138.31,162.69) --
	(138.89,162.20) --
	(139.47,161.71) --
	(140.05,161.22) --
	(140.64,160.74) --
	(141.22,160.26) --
	(141.80,159.78) --
	(142.38,159.30) --
	(142.96,158.82) --
	(143.54,158.35) --
	(144.12,157.87) --
	(144.70,157.40) --
	(145.28,156.93) --
	(145.86,156.47) --
	(146.44,156.00) --
	(147.02,155.54) --
	(147.61,155.08) --
	(148.19,154.62) --
	(148.77,154.16) --
	(149.35,153.71) --
	(149.93,153.25) --
	(150.51,152.80) --
	(151.09,152.35) --
	(151.67,151.90) --
	(152.25,151.46) --
	(152.83,151.01) --
	(153.41,150.57) --
	(153.99,150.13) --
	(154.57,149.69) --
	(155.16,149.25) --
	(155.74,148.82) --
	(156.32,148.39) --
	(156.90,147.95) --
	(157.48,147.52) --
	(158.06,147.10) --
	(158.64,146.67) --
	(159.22,146.25) --
	(159.80,145.82) --
	(160.38,145.40) --
	(160.96,144.98) --
	(161.54,144.57) --
	(162.13,144.15) --
	(162.71,143.74) --
	(163.29,143.32) --
	(163.87,142.91) --
	(164.45,142.50) --
	(165.03,142.10) --
	(165.61,141.69) --
	(166.19,141.29) --
	(166.77,140.88) --
	(167.35,140.48) --
	(167.93,140.08) --
	(168.51,139.69) --
	(169.10,139.29) --
	(169.68,138.90) --
	(170.26,138.50) --
	(170.84,138.11) --
	(171.42,137.72) --
	(172.00,137.33) --
	(172.58,136.95) --
	(173.16,136.56) --
	(173.74,136.18) --
	(174.32,135.80) --
	(174.90,135.42) --
	(175.48,135.04) --
	(176.07,134.66) --
	(176.65,134.29) --
	(177.23,133.91) --
	(177.81,133.54) --
	(178.39,133.17) --
	(178.97,132.80) --
	(179.55,132.43) --
	(180.13,132.06) --
	(180.71,131.70) --
	(181.29,131.33) --
	(181.87,130.97) --
	(182.45,130.61) --
	(183.04,130.25) --
	(183.62,129.89) --
	(184.20,129.53) --
	(184.78,129.17) --
	(185.36,128.82) --
	(185.94,128.47) --
	(186.52,128.11) --
	(187.10,127.76) --
	(187.68,127.41) --
	(188.26,127.07) --
	(188.84,126.72) --
	(189.42,126.38) --
	(190.01,126.03) --
	(190.59,125.69) --
	(191.17,125.35) --
	(191.75,125.01) --
	(192.33,124.67) --
	(192.91,124.33) --
	(193.49,123.99) --
	(194.07,123.66) --
	(194.65,123.32) --
	(195.23,122.99) --
	(195.81,122.66) --
	(196.39,122.33) --
	(196.98,122.00) --
	(197.56,121.67) --
	(198.14,121.35) --
	(198.72,121.02) --
	(199.30,120.70) --
	(199.88,120.37) --
	(200.46,120.05) --
	(201.04,119.73) --
	(201.62,119.41) --
	(202.20,119.09) --
	(202.78,118.78) --
	(203.36,118.46) --
	(203.94,118.15) --
	(204.53,117.83) --
	(205.11,117.52) --
	(205.69,117.21) --
	(206.27,116.90) --
	(206.85,116.59) --
	(207.43,116.28) --
	(208.01,115.97) --
	(208.59,115.67) --
	(209.17,115.36) --
	(209.75,115.06) --
	(210.33,114.75) --
	(210.91,114.45) --
	(211.50,114.15) --
	(212.08,113.85) --
	(212.66,113.55) --
	(213.24,113.25) --
	(213.82,112.96) --
	(214.40,112.66) --
	(214.98,112.37) --
	(215.56,112.07) --
	(216.14,111.78) --
	(216.72,111.49) --
	(217.30,111.20) --
	(217.88,110.91) --
	(218.47,110.62) --
	(219.05,110.33) --
	(219.63,110.04) --
	(220.21,109.76) --
	(220.79,109.47) --
	(221.37,109.19) --
	(221.95,108.91) --
	(222.53,108.62) --
	(223.11,108.34) --
	(223.69,108.06) --
	(224.27,107.78) --
	(224.85,107.50) --
	(225.44,107.23) --
	(226.02,106.95) --
	(226.60,106.67) --
	(227.18,106.40) --
	(227.76,106.13) --
	(228.34,105.85) --
	(228.92,105.58);
\definecolor{drawColor}{RGB}{149,149,149}

\path[draw=drawColor,line width= 1.4pt,line join=round] ( 54.67,145.01) --
	( 55.25,144.02) --
	( 55.83,143.04) --
	( 56.42,142.07) --
	( 57.00,141.10) --
	( 57.58,140.15) --
	( 58.16,139.22) --
	( 58.74,138.31) --
	( 59.32,137.41) --
	( 59.90,136.55) --
	( 60.48,135.70) --
	( 61.06,134.88) --
	( 61.64,134.09) --
	( 62.22,133.33) --
	( 62.80,132.60) --
	( 63.39,131.89) --
	( 63.97,131.22) --
	( 64.55,130.57) --
	( 65.13,129.96) --
	( 65.71,129.38) --
	( 66.29,128.83) --
	( 66.87,128.30) --
	( 67.45,127.81) --
	( 68.03,127.35) --
	( 68.61,126.91) --
	( 69.19,126.50) --
	( 69.77,126.12) --
	( 70.36,125.77) --
	( 70.94,125.44) --
	( 71.52,125.13) --
	( 72.10,124.85) --
	( 72.68,124.59) --
	( 73.26,124.36) --
	( 73.84,124.15) --
	( 74.42,123.95) --
	( 75.00,123.78) --
	( 75.58,123.63) --
	( 76.16,123.49) --
	( 76.74,123.38) --
	( 77.33,123.28) --
	( 77.91,123.19) --
	( 78.49,123.12) --
	( 79.07,123.07) --
	( 79.65,123.03) --
	( 80.23,123.00) --
	( 80.81,122.99) --
	( 81.39,122.99) --
	( 81.97,123.00) --
	( 82.55,123.02) --
	( 83.13,123.05) --
	( 83.71,123.09) --
	( 84.30,123.14) --
	( 84.88,123.20) --
	( 85.46,123.27) --
	( 86.04,123.34) --
	( 86.62,123.43) --
	( 87.20,123.52) --
	( 87.78,123.61) --
	( 88.36,123.72) --
	( 88.94,123.83) --
	( 89.52,123.94) --
	( 90.10,124.06) --
	( 90.68,124.19) --
	( 91.27,124.32) --
	( 91.85,124.45) --
	( 92.43,124.59) --
	( 93.01,124.73) --
	( 93.59,124.88) --
	( 94.17,125.03) --
	( 94.75,125.19) --
	( 95.33,125.34) --
	( 95.91,125.50) --
	( 96.49,125.66) --
	( 97.07,125.83) --
	( 97.65,125.99) --
	( 98.24,126.16) --
	( 98.82,126.33) --
	( 99.40,126.51) --
	( 99.98,126.68) --
	(100.56,126.86) --
	(101.14,127.04) --
	(101.72,127.21) --
	(102.30,127.39) --
	(102.88,127.58) --
	(103.46,127.76) --
	(104.04,127.94) --
	(104.62,128.13) --
	(105.20,128.31) --
	(105.79,128.50) --
	(106.37,128.68) --
	(106.95,128.87) --
	(107.53,129.06) --
	(108.11,129.24) --
	(108.69,129.43) --
	(109.27,129.62) --
	(109.85,129.81) --
	(110.43,130.00) --
	(111.01,130.18) --
	(111.59,130.37) --
	(112.17,130.56) --
	(112.76,130.75) --
	(113.34,130.94) --
	(113.92,131.13) --
	(114.50,131.32) --
	(115.08,131.50) --
	(115.66,131.69) --
	(116.24,131.88) --
	(116.82,132.07) --
	(117.40,132.25) --
	(117.98,132.44) --
	(118.56,132.63) --
	(119.14,132.81) --
	(119.73,133.00) --
	(120.31,133.18) --
	(120.89,133.37) --
	(121.47,133.55) --
	(122.05,133.74) --
	(122.63,133.92) --
	(123.21,134.10) --
	(123.79,134.29) --
	(124.37,134.47) --
	(124.95,134.65) --
	(125.53,134.83) --
	(126.11,135.01) --
	(126.70,135.19) --
	(127.28,135.37) --
	(127.86,135.55) --
	(128.44,135.73) --
	(129.02,135.90) --
	(129.60,136.08) --
	(130.18,136.26) --
	(130.76,136.43) --
	(131.34,136.61) --
	(131.92,136.78) --
	(132.50,136.95) --
	(133.08,137.13) --
	(133.67,137.30) --
	(134.25,137.47) --
	(134.83,137.64) --
	(135.41,137.81) --
	(135.99,137.98) --
	(136.57,138.15) --
	(137.15,138.32) --
	(137.73,138.49) --
	(138.31,138.66) --
	(138.89,138.82) --
	(139.47,138.99) --
	(140.05,139.15) --
	(140.64,139.32) --
	(141.22,139.48) --
	(141.80,139.65) --
	(142.38,139.81) --
	(142.96,139.97) --
	(143.54,140.13) --
	(144.12,140.29) --
	(144.70,140.45) --
	(145.28,140.61) --
	(145.86,140.77) --
	(146.44,140.93) --
	(147.02,141.08) --
	(147.61,141.24) --
	(148.19,141.40) --
	(148.77,141.55) --
	(149.35,141.71) --
	(149.93,141.86) --
	(150.51,142.02) --
	(151.09,142.17) --
	(151.67,142.32) --
	(152.25,142.47) --
	(152.83,142.62) --
	(153.41,142.77) --
	(153.99,142.92) --
	(154.57,143.07) --
	(155.16,143.22) --
	(155.74,143.37) --
	(156.32,143.52) --
	(156.90,143.66) --
	(157.48,143.81) --
	(158.06,143.96) --
	(158.64,144.10) --
	(159.22,144.24) --
	(159.80,144.39) --
	(160.38,144.53) --
	(160.96,144.67) --
	(161.54,144.82) --
	(162.13,144.96) --
	(162.71,145.10) --
	(163.29,145.24) --
	(163.87,145.38) --
	(164.45,145.52) --
	(165.03,145.66) --
	(165.61,145.79) --
	(166.19,145.93) --
	(166.77,146.07) --
	(167.35,146.21) --
	(167.93,146.34) --
	(168.51,146.48) --
	(169.10,146.61) --
	(169.68,146.75) --
	(170.26,146.88) --
	(170.84,147.01) --
	(171.42,147.14) --
	(172.00,147.28) --
	(172.58,147.41) --
	(173.16,147.54) --
	(173.74,147.67) --
	(174.32,147.80) --
	(174.90,147.93) --
	(175.48,148.06) --
	(176.07,148.19) --
	(176.65,148.31) --
	(177.23,148.44) --
	(177.81,148.57) --
	(178.39,148.70) --
	(178.97,148.82) --
	(179.55,148.95) --
	(180.13,149.07) --
	(180.71,149.20) --
	(181.29,149.32) --
	(181.87,149.44) --
	(182.45,149.57) --
	(183.04,149.69) --
	(183.62,149.81) --
	(184.20,149.93) --
	(184.78,150.06) --
	(185.36,150.18) --
	(185.94,150.30) --
	(186.52,150.42) --
	(187.10,150.54) --
	(187.68,150.65) --
	(188.26,150.77) --
	(188.84,150.89) --
	(189.42,151.01) --
	(190.01,151.13) --
	(190.59,151.24) --
	(191.17,151.36) --
	(191.75,151.47) --
	(192.33,151.59) --
	(192.91,151.71) --
	(193.49,151.82) --
	(194.07,151.93) --
	(194.65,152.05) --
	(195.23,152.16) --
	(195.81,152.27) --
	(196.39,152.39) --
	(196.98,152.50) --
	(197.56,152.61) --
	(198.14,152.72) --
	(198.72,152.83) --
	(199.30,152.94) --
	(199.88,153.05) --
	(200.46,153.16) --
	(201.04,153.27) --
	(201.62,153.38) --
	(202.20,153.49) --
	(202.78,153.60) --
	(203.36,153.71) --
	(203.94,153.81) --
	(204.53,153.92) --
	(205.11,154.03) --
	(205.69,154.13) --
	(206.27,154.24) --
	(206.85,154.34) --
	(207.43,154.45) --
	(208.01,154.55) --
	(208.59,154.66) --
	(209.17,154.76) --
	(209.75,154.87) --
	(210.33,154.97) --
	(210.91,155.07) --
	(211.50,155.18) --
	(212.08,155.28) --
	(212.66,155.38) --
	(213.24,155.48) --
	(213.82,155.58) --
	(214.40,155.68) --
	(214.98,155.78) --
	(215.56,155.88) --
	(216.14,155.98) --
	(216.72,156.08) --
	(217.30,156.18) --
	(217.88,156.28) --
	(218.47,156.38) --
	(219.05,156.48) --
	(219.63,156.58) --
	(220.21,156.67) --
	(220.79,156.77) --
	(221.37,156.87) --
	(221.95,156.96) --
	(222.53,157.06) --
	(223.11,157.16) --
	(223.69,157.25) --
	(224.27,157.35) --
	(224.85,157.44) --
	(225.44,157.54) --
	(226.02,157.63) --
	(226.60,157.73) --
	(227.18,157.82) --
	(227.76,157.91) --
	(228.34,158.01) --
	(228.92,158.10);

\path[draw=drawColor,line width= 1.4pt,dash pattern=on 7pt off 3pt ,line join=round] ( 54.67,145.01) --
	( 55.25,145.98) --
	( 55.83,146.95) --
	( 56.42,147.91) --
	( 57.00,148.86) --
	( 57.58,149.79) --
	( 58.16,150.70) --
	( 58.74,151.59) --
	( 59.32,152.47) --
	( 59.90,153.31) --
	( 60.48,154.14) --
	( 61.06,154.93) --
	( 61.64,155.70) --
	( 62.22,156.44) --
	( 62.80,157.15) --
	( 63.39,157.83) --
	( 63.97,158.48) --
	( 64.55,159.10) --
	( 65.13,159.69) --
	( 65.71,160.25) --
	( 66.29,160.78) --
	( 66.87,161.28) --
	( 67.45,161.75) --
	( 68.03,162.19) --
	( 68.61,162.61) --
	( 69.19,163.00) --
	( 69.77,163.36) --
	( 70.36,163.70) --
	( 70.94,164.01) --
	( 71.52,164.30) --
	( 72.10,164.56) --
	( 72.68,164.80) --
	( 73.26,165.02) --
	( 73.84,165.22) --
	( 74.42,165.40) --
	( 75.00,165.56) --
	( 75.58,165.70) --
	( 76.16,165.82) --
	( 76.74,165.92) --
	( 77.33,166.01) --
	( 77.91,166.09) --
	( 78.49,166.15) --
	( 79.07,166.19) --
	( 79.65,166.23) --
	( 80.23,166.24) --
	( 80.81,166.25) --
	( 81.39,166.25) --
	( 81.97,166.23) --
	( 82.55,166.20) --
	( 83.13,166.17) --
	( 83.71,166.12) --
	( 84.30,166.07) --
	( 84.88,166.01) --
	( 85.46,165.94) --
	( 86.04,165.86) --
	( 86.62,165.77) --
	( 87.20,165.68) --
	( 87.78,165.58) --
	( 88.36,165.48) --
	( 88.94,165.37) --
	( 89.52,165.25) --
	( 90.10,165.13) --
	( 90.68,165.00) --
	( 91.27,164.87) --
	( 91.85,164.74) --
	( 92.43,164.60) --
	( 93.01,164.46) --
	( 93.59,164.31) --
	( 94.17,164.16) --
	( 94.75,164.01) --
	( 95.33,163.86) --
	( 95.91,163.70) --
	( 96.49,163.54) --
	( 97.07,163.38) --
	( 97.65,163.21) --
	( 98.24,163.05) --
	( 98.82,162.88) --
	( 99.40,162.71) --
	( 99.98,162.54) --
	(100.56,162.36) --
	(101.14,162.19) --
	(101.72,162.01) --
	(102.30,161.84) --
	(102.88,161.66) --
	(103.46,161.48) --
	(104.04,161.30) --
	(104.62,161.12) --
	(105.20,160.94) --
	(105.79,160.76) --
	(106.37,160.57) --
	(106.95,160.39) --
	(107.53,160.21) --
	(108.11,160.02) --
	(108.69,159.84) --
	(109.27,159.66) --
	(109.85,159.47) --
	(110.43,159.29) --
	(111.01,159.10) --
	(111.59,158.92) --
	(112.17,158.74) --
	(112.76,158.55) --
	(113.34,158.37) --
	(113.92,158.18) --
	(114.50,158.00) --
	(115.08,157.82) --
	(115.66,157.63) --
	(116.24,157.45) --
	(116.82,157.27) --
	(117.40,157.08) --
	(117.98,156.90) --
	(118.56,156.72) --
	(119.14,156.54) --
	(119.73,156.36) --
	(120.31,156.18) --
	(120.89,156.00) --
	(121.47,155.82) --
	(122.05,155.64) --
	(122.63,155.46) --
	(123.21,155.28) --
	(123.79,155.10) --
	(124.37,154.93) --
	(124.95,154.75) --
	(125.53,154.57) --
	(126.11,154.40) --
	(126.70,154.22) --
	(127.28,154.05) --
	(127.86,153.88) --
	(128.44,153.70) --
	(129.02,153.53) --
	(129.60,153.36) --
	(130.18,153.19) --
	(130.76,153.01) --
	(131.34,152.84) --
	(131.92,152.67) --
	(132.50,152.51) --
	(133.08,152.34) --
	(133.67,152.17) --
	(134.25,152.00) --
	(134.83,151.84) --
	(135.41,151.67) --
	(135.99,151.51) --
	(136.57,151.34) --
	(137.15,151.18) --
	(137.73,151.01) --
	(138.31,150.85) --
	(138.89,150.69) --
	(139.47,150.53) --
	(140.05,150.37) --
	(140.64,150.21) --
	(141.22,150.05) --
	(141.80,149.89) --
	(142.38,149.73) --
	(142.96,149.57) --
	(143.54,149.42) --
	(144.12,149.26) --
	(144.70,149.10) --
	(145.28,148.95) --
	(145.86,148.80) --
	(146.44,148.64) --
	(147.02,148.49) --
	(147.61,148.34) --
	(148.19,148.18) --
	(148.77,148.03) --
	(149.35,147.88) --
	(149.93,147.73) --
	(150.51,147.58) --
	(151.09,147.44) --
	(151.67,147.29) --
	(152.25,147.14) --
	(152.83,146.99) --
	(153.41,146.85) --
	(153.99,146.70) --
	(154.57,146.56) --
	(155.16,146.41) --
	(155.74,146.27) --
	(156.32,146.13) --
	(156.90,145.98) --
	(157.48,145.84) --
	(158.06,145.70) --
	(158.64,145.56) --
	(159.22,145.42) --
	(159.80,145.28) --
	(160.38,145.14) --
	(160.96,145.00) --
	(161.54,144.86) --
	(162.13,144.73) --
	(162.71,144.59) --
	(163.29,144.45) --
	(163.87,144.32) --
	(164.45,144.18) --
	(165.03,144.05) --
	(165.61,143.91) --
	(166.19,143.78) --
	(166.77,143.65) --
	(167.35,143.51) --
	(167.93,143.38) --
	(168.51,143.25) --
	(169.10,143.12) --
	(169.68,142.99) --
	(170.26,142.86) --
	(170.84,142.73) --
	(171.42,142.60) --
	(172.00,142.47) --
	(172.58,142.35) --
	(173.16,142.22) --
	(173.74,142.09) --
	(174.32,141.97) --
	(174.90,141.84) --
	(175.48,141.71) --
	(176.07,141.59) --
	(176.65,141.47) --
	(177.23,141.34) --
	(177.81,141.22) --
	(178.39,141.10) --
	(178.97,140.97) --
	(179.55,140.85) --
	(180.13,140.73) --
	(180.71,140.61) --
	(181.29,140.49) --
	(181.87,140.37) --
	(182.45,140.25) --
	(183.04,140.13) --
	(183.62,140.01) --
	(184.20,139.89) --
	(184.78,139.78) --
	(185.36,139.66) --
	(185.94,139.54) --
	(186.52,139.43) --
	(187.10,139.31) --
	(187.68,139.20) --
	(188.26,139.08) --
	(188.84,138.97) --
	(189.42,138.85) --
	(190.01,138.74) --
	(190.59,138.62) --
	(191.17,138.51) --
	(191.75,138.40) --
	(192.33,138.29) --
	(192.91,138.18) --
	(193.49,138.06) --
	(194.07,137.95) --
	(194.65,137.84) --
	(195.23,137.73) --
	(195.81,137.62) --
	(196.39,137.52) --
	(196.98,137.41) --
	(197.56,137.30) --
	(198.14,137.19) --
	(198.72,137.08) --
	(199.30,136.98) --
	(199.88,136.87) --
	(200.46,136.76) --
	(201.04,136.66) --
	(201.62,136.55) --
	(202.20,136.45) --
	(202.78,136.34) --
	(203.36,136.24) --
	(203.94,136.13) --
	(204.53,136.03) --
	(205.11,135.93) --
	(205.69,135.82) --
	(206.27,135.72) --
	(206.85,135.62) --
	(207.43,135.52) --
	(208.01,135.41) --
	(208.59,135.31) --
	(209.17,135.21) --
	(209.75,135.11) --
	(210.33,135.01) --
	(210.91,134.91) --
	(211.50,134.81) --
	(212.08,134.71) --
	(212.66,134.61) --
	(213.24,134.52) --
	(213.82,134.42) --
	(214.40,134.32) --
	(214.98,134.22) --
	(215.56,134.13) --
	(216.14,134.03) --
	(216.72,133.93) --
	(217.30,133.84) --
	(217.88,133.74) --
	(218.47,133.64) --
	(219.05,133.55) --
	(219.63,133.45) --
	(220.21,133.36) --
	(220.79,133.27) --
	(221.37,133.17) --
	(221.95,133.08) --
	(222.53,132.99) --
	(223.11,132.89) --
	(223.69,132.80) --
	(224.27,132.71) --
	(224.85,132.62) --
	(225.44,132.52) --
	(226.02,132.43) --
	(226.60,132.34) --
	(227.18,132.25) --
	(227.76,132.16) --
	(228.34,132.07) --
	(228.92,131.98);
\definecolor{drawColor}{gray}{0.80}

\path[draw=drawColor,line width= 1.4pt,line join=round] ( 54.67,145.01) --
	( 55.25,143.52) --
	( 55.83,142.04) --
	( 56.42,140.57) --
	( 57.00,139.12) --
	( 57.58,137.69) --
	( 58.16,136.29) --
	( 58.74,134.91) --
	( 59.32,133.57) --
	( 59.90,132.26) --
	( 60.48,130.98) --
	( 61.06,129.75) --
	( 61.64,128.56) --
	( 62.22,127.41) --
	( 62.80,126.31) --
	( 63.39,125.24) --
	( 63.97,124.23) --
	( 64.55,123.26) --
	( 65.13,122.34) --
	( 65.71,121.46) --
	( 66.29,120.63) --
	( 66.87,119.84) --
	( 67.45,119.09) --
	( 68.03,118.40) --
	( 68.61,117.74) --
	( 69.19,117.12) --
	( 69.77,116.55) --
	( 70.36,116.01) --
	( 70.94,115.52) --
	( 71.52,115.06) --
	( 72.10,114.64) --
	( 72.68,114.25) --
	( 73.26,113.90) --
	( 73.84,113.57) --
	( 74.42,113.28) --
	( 75.00,113.02) --
	( 75.58,112.79) --
	( 76.16,112.59) --
	( 76.74,112.41) --
	( 77.33,112.26) --
	( 77.91,112.14) --
	( 78.49,112.03) --
	( 79.07,111.95) --
	( 79.65,111.89) --
	( 80.23,111.85) --
	( 80.81,111.83) --
	( 81.39,111.83) --
	( 81.97,111.84) --
	( 82.55,111.87) --
	( 83.13,111.92) --
	( 83.71,111.98) --
	( 84.30,112.06) --
	( 84.88,112.15) --
	( 85.46,112.25) --
	( 86.04,112.36) --
	( 86.62,112.49) --
	( 87.20,112.63) --
	( 87.78,112.77) --
	( 88.36,112.93) --
	( 88.94,113.09) --
	( 89.52,113.27) --
	( 90.10,113.45) --
	( 90.68,113.64) --
	( 91.27,113.83) --
	( 91.85,114.04) --
	( 92.43,114.25) --
	( 93.01,114.46) --
	( 93.59,114.68) --
	( 94.17,114.91) --
	( 94.75,115.14) --
	( 95.33,115.38) --
	( 95.91,115.62) --
	( 96.49,115.86) --
	( 97.07,116.11) --
	( 97.65,116.36) --
	( 98.24,116.61) --
	( 98.82,116.87) --
	( 99.40,117.13) --
	( 99.98,117.39) --
	(100.56,117.66) --
	(101.14,117.93) --
	(101.72,118.20) --
	(102.30,118.47) --
	(102.88,118.74) --
	(103.46,119.02) --
	(104.04,119.29) --
	(104.62,119.57) --
	(105.20,119.85) --
	(105.79,120.13) --
	(106.37,120.41) --
	(106.95,120.69) --
	(107.53,120.97) --
	(108.11,121.25) --
	(108.69,121.54) --
	(109.27,121.82) --
	(109.85,122.10) --
	(110.43,122.39) --
	(111.01,122.67) --
	(111.59,122.96) --
	(112.17,123.24) --
	(112.76,123.53) --
	(113.34,123.81) --
	(113.92,124.09) --
	(114.50,124.38) --
	(115.08,124.66) --
	(115.66,124.94) --
	(116.24,125.23) --
	(116.82,125.51) --
	(117.40,125.79) --
	(117.98,126.07) --
	(118.56,126.35) --
	(119.14,126.63) --
	(119.73,126.91) --
	(120.31,127.19) --
	(120.89,127.47) --
	(121.47,127.75) --
	(122.05,128.03) --
	(122.63,128.30) --
	(123.21,128.58) --
	(123.79,128.85) --
	(124.37,129.13) --
	(124.95,129.40) --
	(125.53,129.67) --
	(126.11,129.94) --
	(126.70,130.22) --
	(127.28,130.49) --
	(127.86,130.75) --
	(128.44,131.02) --
	(129.02,131.29) --
	(129.60,131.56) --
	(130.18,131.82) --
	(130.76,132.09) --
	(131.34,132.35) --
	(131.92,132.61) --
	(132.50,132.87) --
	(133.08,133.13) --
	(133.67,133.39) --
	(134.25,133.65) --
	(134.83,133.91) --
	(135.41,134.17) --
	(135.99,134.42) --
	(136.57,134.68) --
	(137.15,134.93) --
	(137.73,135.18) --
	(138.31,135.44) --
	(138.89,135.69) --
	(139.47,135.94) --
	(140.05,136.19) --
	(140.64,136.43) --
	(141.22,136.68) --
	(141.80,136.93) --
	(142.38,137.17) --
	(142.96,137.42) --
	(143.54,137.66) --
	(144.12,137.90) --
	(144.70,138.14) --
	(145.28,138.38) --
	(145.86,138.62) --
	(146.44,138.86) --
	(147.02,139.10) --
	(147.61,139.33) --
	(148.19,139.57) --
	(148.77,139.80) --
	(149.35,140.04) --
	(149.93,140.27) --
	(150.51,140.50) --
	(151.09,140.73) --
	(151.67,140.96) --
	(152.25,141.19) --
	(152.83,141.41) --
	(153.41,141.64) --
	(153.99,141.87) --
	(154.57,142.09) --
	(155.16,142.32) --
	(155.74,142.54) --
	(156.32,142.76) --
	(156.90,142.98) --
	(157.48,143.20) --
	(158.06,143.42) --
	(158.64,143.64) --
	(159.22,143.86) --
	(159.80,144.07) --
	(160.38,144.29) --
	(160.96,144.50) --
	(161.54,144.72) --
	(162.13,144.93) --
	(162.71,145.14) --
	(163.29,145.36) --
	(163.87,145.57) --
	(164.45,145.78) --
	(165.03,145.98) --
	(165.61,146.19) --
	(166.19,146.40) --
	(166.77,146.61) --
	(167.35,146.81) --
	(167.93,147.02) --
	(168.51,147.22) --
	(169.10,147.42) --
	(169.68,147.63) --
	(170.26,147.83) --
	(170.84,148.03) --
	(171.42,148.23) --
	(172.00,148.43) --
	(172.58,148.62) --
	(173.16,148.82) --
	(173.74,149.02) --
	(174.32,149.21) --
	(174.90,149.41) --
	(175.48,149.60) --
	(176.07,149.80) --
	(176.65,149.99) --
	(177.23,150.18) --
	(177.81,150.37) --
	(178.39,150.56) --
	(178.97,150.75) --
	(179.55,150.94) --
	(180.13,151.13) --
	(180.71,151.32) --
	(181.29,151.51) --
	(181.87,151.69) --
	(182.45,151.88) --
	(183.04,152.06) --
	(183.62,152.25) --
	(184.20,152.43) --
	(184.78,152.61) --
	(185.36,152.79) --
	(185.94,152.98) --
	(186.52,153.16) --
	(187.10,153.34) --
	(187.68,153.52) --
	(188.26,153.69) --
	(188.84,153.87) --
	(189.42,154.05) --
	(190.01,154.23) --
	(190.59,154.40) --
	(191.17,154.58) --
	(191.75,154.75) --
	(192.33,154.93) --
	(192.91,155.10) --
	(193.49,155.27) --
	(194.07,155.44) --
	(194.65,155.61) --
	(195.23,155.79) --
	(195.81,155.96) --
	(196.39,156.13) --
	(196.98,156.29) --
	(197.56,156.46) --
	(198.14,156.63) --
	(198.72,156.80) --
	(199.30,156.96) --
	(199.88,157.13) --
	(200.46,157.30) --
	(201.04,157.46) --
	(201.62,157.62) --
	(202.20,157.79) --
	(202.78,157.95) --
	(203.36,158.11) --
	(203.94,158.27) --
	(204.53,158.44) --
	(205.11,158.60) --
	(205.69,158.76) --
	(206.27,158.92) --
	(206.85,159.08) --
	(207.43,159.23) --
	(208.01,159.39) --
	(208.59,159.55) --
	(209.17,159.71) --
	(209.75,159.86) --
	(210.33,160.02) --
	(210.91,160.17) --
	(211.50,160.33) --
	(212.08,160.48) --
	(212.66,160.63) --
	(213.24,160.79) --
	(213.82,160.94) --
	(214.40,161.09) --
	(214.98,161.24) --
	(215.56,161.39) --
	(216.14,161.55) --
	(216.72,161.70) --
	(217.30,161.84) --
	(217.88,161.99) --
	(218.47,162.14) --
	(219.05,162.29) --
	(219.63,162.44) --
	(220.21,162.58) --
	(220.79,162.73) --
	(221.37,162.88) --
	(221.95,163.02) --
	(222.53,163.17) --
	(223.11,163.31) --
	(223.69,163.46) --
	(224.27,163.60) --
	(224.85,163.74) --
	(225.44,163.89) --
	(226.02,164.03) --
	(226.60,164.17) --
	(227.18,164.31) --
	(227.76,164.45) --
	(228.34,164.59) --
	(228.92,164.73);

\path[draw=drawColor,line width= 1.4pt,dash pattern=on 7pt off 3pt ,line join=round] ( 54.67,145.01) --
	( 55.25,145.06) --
	( 55.83,145.11) --
	( 56.42,145.15) --
	( 57.00,145.20) --
	( 57.58,145.25) --
	( 58.16,145.29) --
	( 58.74,145.34) --
	( 59.32,145.38) --
	( 59.90,145.42) --
	( 60.48,145.46) --
	( 61.06,145.50) --
	( 61.64,145.54) --
	( 62.22,145.58) --
	( 62.80,145.61) --
	( 63.39,145.65) --
	( 63.97,145.68) --
	( 64.55,145.71) --
	( 65.13,145.74) --
	( 65.71,145.77) --
	( 66.29,145.79) --
	( 66.87,145.82) --
	( 67.45,145.84) --
	( 68.03,145.86) --
	( 68.61,145.88) --
	( 69.19,145.90) --
	( 69.77,145.92) --
	( 70.36,145.94) --
	( 70.94,145.95) --
	( 71.52,145.97) --
	( 72.10,145.98) --
	( 72.68,145.99) --
	( 73.26,146.00) --
	( 73.84,146.01) --
	( 74.42,146.02) --
	( 75.00,146.03) --
	( 75.58,146.04) --
	( 76.16,146.04) --
	( 76.74,146.05) --
	( 77.33,146.05) --
	( 77.91,146.06) --
	( 78.49,146.06) --
	( 79.07,146.06) --
	( 79.65,146.06) --
	( 80.23,146.06) --
	( 80.81,146.06) --
	( 81.39,146.06) --
	( 81.97,146.06) --
	( 82.55,146.06) --
	( 83.13,146.06) --
	( 83.71,146.06) --
	( 84.30,146.06) --
	( 84.88,146.05) --
	( 85.46,146.05) --
	( 86.04,146.04) --
	( 86.62,146.04) --
	( 87.20,146.04) --
	( 87.78,146.03) --
	( 88.36,146.03) --
	( 88.94,146.02) --
	( 89.52,146.01) --
	( 90.10,146.01) --
	( 90.68,146.00) --
	( 91.27,146.00) --
	( 91.85,145.99) --
	( 92.43,145.98) --
	( 93.01,145.98) --
	( 93.59,145.97) --
	( 94.17,145.96) --
	( 94.75,145.95) --
	( 95.33,145.95) --
	( 95.91,145.94) --
	( 96.49,145.93) --
	( 97.07,145.92) --
	( 97.65,145.91) --
	( 98.24,145.91) --
	( 98.82,145.90) --
	( 99.40,145.89) --
	( 99.98,145.88) --
	(100.56,145.87) --
	(101.14,145.86) --
	(101.72,145.85) --
	(102.30,145.85) --
	(102.88,145.84) --
	(103.46,145.83) --
	(104.04,145.82) --
	(104.62,145.81) --
	(105.20,145.80) --
	(105.79,145.79) --
	(106.37,145.78) --
	(106.95,145.77) --
	(107.53,145.76) --
	(108.11,145.76) --
	(108.69,145.75) --
	(109.27,145.74) --
	(109.85,145.73) --
	(110.43,145.72) --
	(111.01,145.71) --
	(111.59,145.70) --
	(112.17,145.69) --
	(112.76,145.68) --
	(113.34,145.67) --
	(113.92,145.66) --
	(114.50,145.65) --
	(115.08,145.65) --
	(115.66,145.64) --
	(116.24,145.63) --
	(116.82,145.62) --
	(117.40,145.61) --
	(117.98,145.60) --
	(118.56,145.59) --
	(119.14,145.58) --
	(119.73,145.57) --
	(120.31,145.56) --
	(120.89,145.56) --
	(121.47,145.55) --
	(122.05,145.54) --
	(122.63,145.53) --
	(123.21,145.52) --
	(123.79,145.51) --
	(124.37,145.50) --
	(124.95,145.49) --
	(125.53,145.48) --
	(126.11,145.48) --
	(126.70,145.47) --
	(127.28,145.46) --
	(127.86,145.45) --
	(128.44,145.44) --
	(129.02,145.43) --
	(129.60,145.42) --
	(130.18,145.42) --
	(130.76,145.41) --
	(131.34,145.40) --
	(131.92,145.39) --
	(132.50,145.38) --
	(133.08,145.37) --
	(133.67,145.36) --
	(134.25,145.36) --
	(134.83,145.35) --
	(135.41,145.34) --
	(135.99,145.33) --
	(136.57,145.32) --
	(137.15,145.32) --
	(137.73,145.31) --
	(138.31,145.30) --
	(138.89,145.29) --
	(139.47,145.28) --
	(140.05,145.28) --
	(140.64,145.27) --
	(141.22,145.26) --
	(141.80,145.25) --
	(142.38,145.24) --
	(142.96,145.24) --
	(143.54,145.23) --
	(144.12,145.22) --
	(144.70,145.21) --
	(145.28,145.20) --
	(145.86,145.20) --
	(146.44,145.19) --
	(147.02,145.18) --
	(147.61,145.17) --
	(148.19,145.17) --
	(148.77,145.16) --
	(149.35,145.15) --
	(149.93,145.14) --
	(150.51,145.14) --
	(151.09,145.13) --
	(151.67,145.12) --
	(152.25,145.11) --
	(152.83,145.11) --
	(153.41,145.10) --
	(153.99,145.09) --
	(154.57,145.09) --
	(155.16,145.08) --
	(155.74,145.07) --
	(156.32,145.06) --
	(156.90,145.06) --
	(157.48,145.05) --
	(158.06,145.04) --
	(158.64,145.04) --
	(159.22,145.03) --
	(159.80,145.02) --
	(160.38,145.02) --
	(160.96,145.01) --
	(161.54,145.00) --
	(162.13,145.00) --
	(162.71,144.99) --
	(163.29,144.98) --
	(163.87,144.97) --
	(164.45,144.97) --
	(165.03,144.96) --
	(165.61,144.95) --
	(166.19,144.95) --
	(166.77,144.94) --
	(167.35,144.93) --
	(167.93,144.93) --
	(168.51,144.92) --
	(169.10,144.92) --
	(169.68,144.91) --
	(170.26,144.90) --
	(170.84,144.90) --
	(171.42,144.89) --
	(172.00,144.88) --
	(172.58,144.88) --
	(173.16,144.87) --
	(173.74,144.86) --
	(174.32,144.86) --
	(174.90,144.85) --
	(175.48,144.85) --
	(176.07,144.84) --
	(176.65,144.83) --
	(177.23,144.83) --
	(177.81,144.82) --
	(178.39,144.81) --
	(178.97,144.81) --
	(179.55,144.80) --
	(180.13,144.80) --
	(180.71,144.79) --
	(181.29,144.78) --
	(181.87,144.78) --
	(182.45,144.77) --
	(183.04,144.77) --
	(183.62,144.76) --
	(184.20,144.76) --
	(184.78,144.75) --
	(185.36,144.74) --
	(185.94,144.74) --
	(186.52,144.73) --
	(187.10,144.73) --
	(187.68,144.72) --
	(188.26,144.71) --
	(188.84,144.71) --
	(189.42,144.70) --
	(190.01,144.70) --
	(190.59,144.69) --
	(191.17,144.69) --
	(191.75,144.68) --
	(192.33,144.68) --
	(192.91,144.67) --
	(193.49,144.66) --
	(194.07,144.66) --
	(194.65,144.65) --
	(195.23,144.65) --
	(195.81,144.64) --
	(196.39,144.64) --
	(196.98,144.63) --
	(197.56,144.63) --
	(198.14,144.62) --
	(198.72,144.62) --
	(199.30,144.61) --
	(199.88,144.60) --
	(200.46,144.60) --
	(201.04,144.59) --
	(201.62,144.59) --
	(202.20,144.58) --
	(202.78,144.58) --
	(203.36,144.57) --
	(203.94,144.57) --
	(204.53,144.56) --
	(205.11,144.56) --
	(205.69,144.55) --
	(206.27,144.55) --
	(206.85,144.54) --
	(207.43,144.54) --
	(208.01,144.53) --
	(208.59,144.53) --
	(209.17,144.52) --
	(209.75,144.52) --
	(210.33,144.51) --
	(210.91,144.51) --
	(211.50,144.50) --
	(212.08,144.50) --
	(212.66,144.49) --
	(213.24,144.49) --
	(213.82,144.48) --
	(214.40,144.48) --
	(214.98,144.47) --
	(215.56,144.47) --
	(216.14,144.46) --
	(216.72,144.46) --
	(217.30,144.45) --
	(217.88,144.45) --
	(218.47,144.44) --
	(219.05,144.44) --
	(219.63,144.44) --
	(220.21,144.43) --
	(220.79,144.43) --
	(221.37,144.42) --
	(221.95,144.42) --
	(222.53,144.41) --
	(223.11,144.41) --
	(223.69,144.40) --
	(224.27,144.40) --
	(224.85,144.39) --
	(225.44,144.39) --
	(226.02,144.38) --
	(226.60,144.38) --
	(227.18,144.38) --
	(227.76,144.37) --
	(228.34,144.37) --
	(228.92,144.36);
\definecolor{drawColor}{RGB}{0,0,0}

\path[draw=drawColor,line width= 0.9pt,line join=round,line cap=round] ( 45.96, 34.42) rectangle (237.63,217.66);
\end{scope}
\begin{scope}
\path[clip] (  0.00,  0.00) rectangle (241.85,217.66);
\definecolor{drawColor}{RGB}{0,0,0}

\node[text=drawColor,anchor=base east,inner sep=0pt, outer sep=0pt, scale=  0.86] at ( 40.56, 43.97) {\(-15\)};

\node[text=drawColor,anchor=base east,inner sep=0pt, outer sep=0pt, scale=  0.86] at ( 40.56, 76.23) {\(-10\)};

\node[text=drawColor,anchor=base east,inner sep=0pt, outer sep=0pt, scale=  0.86] at ( 40.56,108.49) {\(-5\)};

\node[text=drawColor,anchor=base east,inner sep=0pt, outer sep=0pt, scale=  0.86] at ( 40.56,140.74) {\(0\)};

\node[text=drawColor,anchor=base east,inner sep=0pt, outer sep=0pt, scale=  0.86] at ( 40.56,173.00) {\(5\)};

\node[text=drawColor,anchor=base east,inner sep=0pt, outer sep=0pt, scale=  0.86] at ( 40.56,205.26) {\(10\)};
\end{scope}
\begin{scope}
\path[clip] (  0.00,  0.00) rectangle (241.85,217.66);
\definecolor{drawColor}{RGB}{0,0,0}

\path[draw=drawColor,line width= 0.6pt,line join=round] ( 42.96, 48.23) --
	( 45.96, 48.23);

\path[draw=drawColor,line width= 0.6pt,line join=round] ( 42.96, 80.49) --
	( 45.96, 80.49);

\path[draw=drawColor,line width= 0.6pt,line join=round] ( 42.96,112.75) --
	( 45.96,112.75);

\path[draw=drawColor,line width= 0.6pt,line join=round] ( 42.96,145.01) --
	( 45.96,145.01);

\path[draw=drawColor,line width= 0.6pt,line join=round] ( 42.96,177.27) --
	( 45.96,177.27);

\path[draw=drawColor,line width= 0.6pt,line join=round] ( 42.96,209.53) --
	( 45.96,209.53);
\end{scope}
\begin{scope}
\path[clip] (  0.00,  0.00) rectangle (241.85,217.66);
\definecolor{drawColor}{RGB}{0,0,0}

\path[draw=drawColor,line width= 0.6pt,line join=round] ( 54.67, 31.42) --
	( 54.67, 34.42);

\path[draw=drawColor,line width= 0.6pt,line join=round] ( 89.52, 31.42) --
	( 89.52, 34.42);

\path[draw=drawColor,line width= 0.6pt,line join=round] (124.37, 31.42) --
	(124.37, 34.42);

\path[draw=drawColor,line width= 0.6pt,line join=round] (159.22, 31.42) --
	(159.22, 34.42);

\path[draw=drawColor,line width= 0.6pt,line join=round] (194.07, 31.42) --
	(194.07, 34.42);

\path[draw=drawColor,line width= 0.6pt,line join=round] (228.92, 31.42) --
	(228.92, 34.42);
\end{scope}
\begin{scope}
\path[clip] (  0.00,  0.00) rectangle (241.85,217.66);
\definecolor{drawColor}{RGB}{0,0,0}

\node[text=drawColor,anchor=base,inner sep=0pt, outer sep=0pt, scale=  0.86] at ( 54.67, 20.49) {0};

\node[text=drawColor,anchor=base,inner sep=0pt, outer sep=0pt, scale=  0.86] at ( 89.52, 20.49) {1};

\node[text=drawColor,anchor=base,inner sep=0pt, outer sep=0pt, scale=  0.86] at (124.37, 20.49) {2};

\node[text=drawColor,anchor=base,inner sep=0pt, outer sep=0pt, scale=  0.86] at (159.22, 20.49) {3};

\node[text=drawColor,anchor=base,inner sep=0pt, outer sep=0pt, scale=  0.86] at (194.07, 20.49) {4};

\node[text=drawColor,anchor=base,inner sep=0pt, outer sep=0pt, scale=  0.86] at (228.92, 20.49) {5};
\end{scope}
\begin{scope}
\path[clip] (  0.00,  0.00) rectangle (241.85,217.66);
\definecolor{drawColor}{RGB}{0,0,0}

\node[text=drawColor,anchor=base,inner sep=0pt, outer sep=0pt, scale=  1.00] at (141.80,  5.77) {\(h_g\)/\(h_{si}\)};
\end{scope}
\begin{scope}
\path[clip] (  0.00,  0.00) rectangle (241.85,217.66);
\definecolor{drawColor}{RGB}{0,0,0}

\node[text=drawColor,rotate= 90.00,anchor=base,inner sep=0pt, outer sep=0pt, scale=  1.00] at ( 14.00,126.04) {\(\sigma_x^T\), МПа};
\end{scope}
\begin{scope}
\path[clip] (  0.00,  0.00) rectangle (241.85,217.66);
\definecolor{drawColor}{RGB}{0,0,0}
\definecolor{fillColor}{RGB}{255,255,255}

\path[draw=drawColor,line width= 0.6pt,line join=round,line cap=round,fill=fillColor] ( 46.99, 41.31) rectangle (123.97, 95.37);
\end{scope}
\begin{scope}
\path[clip] (  0.00,  0.00) rectangle (241.85,217.66);
\definecolor{drawColor}{RGB}{255,255,255}
\definecolor{fillColor}{RGB}{255,255,255}

\path[draw=drawColor,line width= 0.6pt,line join=round,line cap=round,fill=fillColor] ( 51.26, 72.56) rectangle ( 78.24, 86.05);
\end{scope}
\begin{scope}
\path[clip] (  0.00,  0.00) rectangle (241.85,217.66);
\definecolor{drawColor}{RGB}{0,0,0}

\path[draw=drawColor,line width= 1.4pt,line join=round] ( 53.96, 79.30) -- ( 75.54, 79.30);
\end{scope}
\begin{scope}
\path[clip] (  0.00,  0.00) rectangle (241.85,217.66);
\definecolor{drawColor}{RGB}{255,255,255}
\definecolor{fillColor}{RGB}{255,255,255}

\path[draw=drawColor,line width= 0.6pt,line join=round,line cap=round,fill=fillColor] ( 51.26, 59.07) rectangle ( 78.24, 72.56);
\end{scope}
\begin{scope}
\path[clip] (  0.00,  0.00) rectangle (241.85,217.66);
\definecolor{drawColor}{RGB}{149,149,149}

\path[draw=drawColor,line width= 1.4pt,line join=round] ( 53.96, 65.81) -- ( 75.54, 65.81);
\end{scope}
\begin{scope}
\path[clip] (  0.00,  0.00) rectangle (241.85,217.66);
\definecolor{drawColor}{RGB}{255,255,255}
\definecolor{fillColor}{RGB}{255,255,255}

\path[draw=drawColor,line width= 0.6pt,line join=round,line cap=round,fill=fillColor] ( 51.26, 45.58) rectangle ( 78.24, 59.07);
\end{scope}
\begin{scope}
\path[clip] (  0.00,  0.00) rectangle (241.85,217.66);
\definecolor{drawColor}{gray}{0.80}

\path[draw=drawColor,line width= 1.4pt,line join=round] ( 53.96, 52.32) -- ( 75.54, 52.32);
\end{scope}
\begin{scope}
\path[clip] (  0.00,  0.00) rectangle (241.85,217.66);
\definecolor{drawColor}{RGB}{0,0,0}

\node[text=drawColor,anchor=base east,inner sep=0pt, outer sep=0pt, scale=  0.86] at (119.70, 75.04) {\(-60\) {\textdegree}C};
\end{scope}
\begin{scope}
\path[clip] (  0.00,  0.00) rectangle (241.85,217.66);
\definecolor{drawColor}{RGB}{0,0,0}

\node[text=drawColor,anchor=base east,inner sep=0pt, outer sep=0pt, scale=  0.86] at (119.70, 61.55) {\(20\) {\textdegree}C};
\end{scope}
\begin{scope}
\path[clip] (  0.00,  0.00) rectangle (241.85,217.66);
\definecolor{drawColor}{RGB}{0,0,0}

\node[text=drawColor,anchor=base east,inner sep=0pt, outer sep=0pt, scale=  0.86] at (119.70, 48.06) {\(85\) {\textdegree}C};
\end{scope}
\begin{scope}
\path[clip] (  0.00,  0.00) rectangle (241.85,217.66);
\definecolor{drawColor}{RGB}{0,0,0}
\definecolor{fillColor}{RGB}{255,255,255}

\path[draw=drawColor,line width= 0.6pt,line join=round,line cap=round,fill=fillColor] (127.34, 48.05) rectangle (228.74, 88.63);
\end{scope}
\begin{scope}
\path[clip] (  0.00,  0.00) rectangle (241.85,217.66);
\definecolor{drawColor}{RGB}{255,255,255}
\definecolor{fillColor}{RGB}{255,255,255}

\path[draw=drawColor,line width= 0.6pt,line join=round,line cap=round,fill=fillColor] (131.61, 65.81) rectangle (158.59, 79.30);
\end{scope}
\begin{scope}
\path[clip] (  0.00,  0.00) rectangle (241.85,217.66);
\definecolor{drawColor}{RGB}{0,0,0}

\path[draw=drawColor,line width= 1.4pt,line join=round] (134.31, 72.56) -- (155.89, 72.56);
\end{scope}
\begin{scope}
\path[clip] (  0.00,  0.00) rectangle (241.85,217.66);
\definecolor{drawColor}{RGB}{255,255,255}
\definecolor{fillColor}{RGB}{255,255,255}

\path[draw=drawColor,line width= 0.6pt,line join=round,line cap=round,fill=fillColor] (131.61, 52.32) rectangle (158.59, 65.81);
\end{scope}
\begin{scope}
\path[clip] (  0.00,  0.00) rectangle (241.85,217.66);
\definecolor{drawColor}{RGB}{0,0,0}

\path[draw=drawColor,line width= 1.4pt,dash pattern=on 7pt off 3pt ,line join=round] (134.31, 59.07) -- (155.89, 59.07);
\end{scope}
\begin{scope}
\path[clip] (  0.00,  0.00) rectangle (241.85,217.66);
\definecolor{drawColor}{RGB}{0,0,0}

\node[text=drawColor,anchor=base east,inner sep=0pt, outer sep=0pt, scale=  0.86] at (224.47, 68.29) {Borofloat 33};
\end{scope}
\begin{scope}
\path[clip] (  0.00,  0.00) rectangle (241.85,217.66);
\definecolor{drawColor}{RGB}{0,0,0}

\node[text=drawColor,anchor=base east,inner sep=0pt, outer sep=0pt, scale=  0.86] at (224.47, 54.80) {ЛК5};
\end{scope}
\end{tikzpicture}

%% file: Dissertation/images_tikz/disser_glass_stress_multilayer_ot_h_tb.tikz
% Created by tikzDevice version 0.10.1 on 2016-10-01 21:04:38
% !TEX encoding = UTF-8 Unicode
\begin{tikzpicture}[x=0.985pt,y=0.985pt]
\definecolor{fillColor}{RGB}{255,255,255}
\path[use as bounding box,fill=fillColor] (0,0) rectangle (241.85,217.66);
\begin{scope}
\path[clip] (  0.00,  0.00) rectangle (241.85,217.66);
\definecolor{drawColor}{RGB}{255,255,255}

\path[draw=drawColor,line width= 0.6pt,line join=round,line cap=round,fill=fillColor] ( -0.00,  0.00) rectangle (241.85,217.66);
\end{scope}
\begin{scope}
\path[clip] ( 45.96, 34.42) rectangle (237.63,217.66);
\definecolor{fillColor}{RGB}{255,255,255}

\path[fill=fillColor] ( 45.96, 34.42) rectangle (237.63,217.66);
\definecolor{drawColor}{gray}{0.98}

\path[draw=drawColor,line width= 0.6pt,line join=round] ( 45.96, 64.36) --
	(237.63, 64.36);

\path[draw=drawColor,line width= 0.6pt,line join=round] ( 45.96, 96.62) --
	(237.63, 96.62);

\path[draw=drawColor,line width= 0.6pt,line join=round] ( 45.96,128.88) --
	(237.63,128.88);

\path[draw=drawColor,line width= 0.6pt,line join=round] ( 45.96,161.14) --
	(237.63,161.14);

\path[draw=drawColor,line width= 0.6pt,line join=round] ( 45.96,193.40) --
	(237.63,193.40);

\path[draw=drawColor,line width= 0.6pt,line join=round] ( 72.10, 34.42) --
	( 72.10,217.66);

\path[draw=drawColor,line width= 0.6pt,line join=round] (106.95, 34.42) --
	(106.95,217.66);

\path[draw=drawColor,line width= 0.6pt,line join=round] (141.80, 34.42) --
	(141.80,217.66);

\path[draw=drawColor,line width= 0.6pt,line join=round] (176.65, 34.42) --
	(176.65,217.66);

\path[draw=drawColor,line width= 0.6pt,line join=round] (211.50, 34.42) --
	(211.50,217.66);
\definecolor{drawColor}{gray}{0.80}

\path[draw=drawColor,line width= 0.3pt,line join=round] ( 45.96, 48.23) --
	(237.63, 48.23);

\path[draw=drawColor,line width= 0.3pt,line join=round] ( 45.96, 80.49) --
	(237.63, 80.49);

\path[draw=drawColor,line width= 0.3pt,line join=round] ( 45.96,112.75) --
	(237.63,112.75);

\path[draw=drawColor,line width= 0.3pt,line join=round] ( 45.96,145.01) --
	(237.63,145.01);

\path[draw=drawColor,line width= 0.3pt,line join=round] ( 45.96,177.27) --
	(237.63,177.27);

\path[draw=drawColor,line width= 0.3pt,line join=round] ( 45.96,209.53) --
	(237.63,209.53);

\path[draw=drawColor,line width= 0.3pt,line join=round] ( 54.67, 34.42) --
	( 54.67,217.66);

\path[draw=drawColor,line width= 0.3pt,line join=round] ( 89.52, 34.42) --
	( 89.52,217.66);

\path[draw=drawColor,line width= 0.3pt,line join=round] (124.37, 34.42) --
	(124.37,217.66);

\path[draw=drawColor,line width= 0.3pt,line join=round] (159.22, 34.42) --
	(159.22,217.66);

\path[draw=drawColor,line width= 0.3pt,line join=round] (194.07, 34.42) --
	(194.07,217.66);

\path[draw=drawColor,line width= 0.3pt,line join=round] (228.92, 34.42) --
	(228.92,217.66);
\definecolor{drawColor}{RGB}{0,0,0}

\path[draw=drawColor,line width= 1.4pt,line join=round] ( 54.67,145.01) --
	( 55.25,143.21) --
	( 55.83,141.41) --
	( 56.42,139.63) --
	( 57.00,137.87) --
	( 57.58,136.14) --
	( 58.16,134.44) --
	( 58.74,132.77) --
	( 59.32,131.14) --
	( 59.90,129.55) --
	( 60.48,128.01) --
	( 61.06,126.52) --
	( 61.64,125.07) --
	( 62.22,123.68) --
	( 62.80,122.34) --
	( 63.39,121.05) --
	( 63.97,119.82) --
	( 64.55,118.65) --
	( 65.13,117.53) --
	( 65.71,116.47) --
	( 66.29,115.46) --
	( 66.87,114.50) --
	( 67.45,113.60) --
	( 68.03,112.75) --
	( 68.61,111.96) --
	( 69.19,111.21) --
	( 69.77,110.51) --
	( 70.36,109.87) --
	( 70.94,109.27) --
	( 71.52,108.71) --
	( 72.10,108.20) --
	( 72.68,107.73) --
	( 73.26,107.30) --
	( 73.84,106.91) --
	( 74.42,106.56) --
	( 75.00,106.24) --
	( 75.58,105.96) --
	( 76.16,105.72) --
	( 76.74,105.50) --
	( 77.33,105.32) --
	( 77.91,105.17) --
	( 78.49,105.04) --
	( 79.07,104.94) --
	( 79.65,104.87) --
	( 80.23,104.82) --
	( 80.81,104.80) --
	( 81.39,104.79) --
	( 81.97,104.81) --
	( 82.55,104.85) --
	( 83.13,104.91) --
	( 83.71,104.98) --
	( 84.30,105.07) --
	( 84.88,105.18) --
	( 85.46,105.31) --
	( 86.04,105.44) --
	( 86.62,105.60) --
	( 87.20,105.76) --
	( 87.78,105.94) --
	( 88.36,106.13) --
	( 88.94,106.33) --
	( 89.52,106.54) --
	( 90.10,106.76) --
	( 90.68,106.99) --
	( 91.27,107.22) --
	( 91.85,107.47) --
	( 92.43,107.72) --
	( 93.01,107.99) --
	( 93.59,108.25) --
	( 94.17,108.53) --
	( 94.75,108.81) --
	( 95.33,109.09) --
	( 95.91,109.38) --
	( 96.49,109.68) --
	( 97.07,109.98) --
	( 97.65,110.28) --
	( 98.24,110.59) --
	( 98.82,110.91) --
	( 99.40,111.22) --
	( 99.98,111.54) --
	(100.56,111.86) --
	(101.14,112.19) --
	(101.72,112.51) --
	(102.30,112.84) --
	(102.88,113.17) --
	(103.46,113.51) --
	(104.04,113.84) --
	(104.62,114.18) --
	(105.20,114.51) --
	(105.79,114.85) --
	(106.37,115.19) --
	(106.95,115.53) --
	(107.53,115.88) --
	(108.11,116.22) --
	(108.69,116.56) --
	(109.27,116.91) --
	(109.85,117.25) --
	(110.43,117.59) --
	(111.01,117.94) --
	(111.59,118.28) --
	(112.17,118.63) --
	(112.76,118.97) --
	(113.34,119.32) --
	(113.92,119.66) --
	(114.50,120.00) --
	(115.08,120.35) --
	(115.66,120.69) --
	(116.24,121.03) --
	(116.82,121.38) --
	(117.40,121.72) --
	(117.98,122.06) --
	(118.56,122.40) --
	(119.14,122.74) --
	(119.73,123.08) --
	(120.31,123.42) --
	(120.89,123.75) --
	(121.47,124.09) --
	(122.05,124.43) --
	(122.63,124.76) --
	(123.21,125.09) --
	(123.79,125.43) --
	(124.37,125.76) --
	(124.95,126.09) --
	(125.53,126.42) --
	(126.11,126.75) --
	(126.70,127.08) --
	(127.28,127.41) --
	(127.86,127.73) --
	(128.44,128.06) --
	(129.02,128.38) --
	(129.60,128.70) --
	(130.18,129.03) --
	(130.76,129.35) --
	(131.34,129.67) --
	(131.92,129.98) --
	(132.50,130.30) --
	(133.08,130.62) --
	(133.67,130.93) --
	(134.25,131.24) --
	(134.83,131.56) --
	(135.41,131.87) --
	(135.99,132.18) --
	(136.57,132.49) --
	(137.15,132.79) --
	(137.73,133.10) --
	(138.31,133.41) --
	(138.89,133.71) --
	(139.47,134.01) --
	(140.05,134.32) --
	(140.64,134.62) --
	(141.22,134.92) --
	(141.80,135.21) --
	(142.38,135.51) --
	(142.96,135.81) --
	(143.54,136.10) --
	(144.12,136.39) --
	(144.70,136.69) --
	(145.28,136.98) --
	(145.86,137.27) --
	(146.44,137.56) --
	(147.02,137.84) --
	(147.61,138.13) --
	(148.19,138.41) --
	(148.77,138.70) --
	(149.35,138.98) --
	(149.93,139.26) --
	(150.51,139.54) --
	(151.09,139.82) --
	(151.67,140.10) --
	(152.25,140.38) --
	(152.83,140.65) --
	(153.41,140.93) --
	(153.99,141.20) --
	(154.57,141.47) --
	(155.16,141.75) --
	(155.74,142.02) --
	(156.32,142.28) --
	(156.90,142.55) --
	(157.48,142.82) --
	(158.06,143.09) --
	(158.64,143.35) --
	(159.22,143.61) --
	(159.80,143.88) --
	(160.38,144.14) --
	(160.96,144.40) --
	(161.54,144.66) --
	(162.13,144.92) --
	(162.71,145.17) --
	(163.29,145.43) --
	(163.87,145.68) --
	(164.45,145.94) --
	(165.03,146.19) --
	(165.61,146.44) --
	(166.19,146.69) --
	(166.77,146.94) --
	(167.35,147.19) --
	(167.93,147.44) --
	(168.51,147.69) --
	(169.10,147.93) --
	(169.68,148.18) --
	(170.26,148.42) --
	(170.84,148.67) --
	(171.42,148.91) --
	(172.00,149.15) --
	(172.58,149.39) --
	(173.16,149.63) --
	(173.74,149.87) --
	(174.32,150.11) --
	(174.90,150.34) --
	(175.48,150.58) --
	(176.07,150.81) --
	(176.65,151.05) --
	(177.23,151.28) --
	(177.81,151.51) --
	(178.39,151.74) --
	(178.97,151.97) --
	(179.55,152.20) --
	(180.13,152.43) --
	(180.71,152.66) --
	(181.29,152.88) --
	(181.87,153.11) --
	(182.45,153.33) --
	(183.04,153.56) --
	(183.62,153.78) --
	(184.20,154.00) --
	(184.78,154.22) --
	(185.36,154.44) --
	(185.94,154.66) --
	(186.52,154.88) --
	(187.10,155.10) --
	(187.68,155.32) --
	(188.26,155.54) --
	(188.84,155.75) --
	(189.42,155.97) --
	(190.01,156.18) --
	(190.59,156.39) --
	(191.17,156.61) --
	(191.75,156.82) --
	(192.33,157.03) --
	(192.91,157.24) --
	(193.49,157.45) --
	(194.07,157.66) --
	(194.65,157.86) --
	(195.23,158.07) --
	(195.81,158.28) --
	(196.39,158.48) --
	(196.98,158.69) --
	(197.56,158.89) --
	(198.14,159.09) --
	(198.72,159.30) --
	(199.30,159.50) --
	(199.88,159.70) --
	(200.46,159.90) --
	(201.04,160.10) --
	(201.62,160.30) --
	(202.20,160.50) --
	(202.78,160.69) --
	(203.36,160.89) --
	(203.94,161.09) --
	(204.53,161.28) --
	(205.11,161.48) --
	(205.69,161.67) --
	(206.27,161.86) --
	(206.85,162.06) --
	(207.43,162.25) --
	(208.01,162.44) --
	(208.59,162.63) --
	(209.17,162.82) --
	(209.75,163.01) --
	(210.33,163.20) --
	(210.91,163.39) --
	(211.50,163.57) --
	(212.08,163.76) --
	(212.66,163.95) --
	(213.24,164.13) --
	(213.82,164.32) --
	(214.40,164.50) --
	(214.98,164.69) --
	(215.56,164.87) --
	(216.14,165.05) --
	(216.72,165.23) --
	(217.30,165.41) --
	(217.88,165.59) --
	(218.47,165.77) --
	(219.05,165.95) --
	(219.63,166.13) --
	(220.21,166.31) --
	(220.79,166.49) --
	(221.37,166.67) --
	(221.95,166.84) --
	(222.53,167.02) --
	(223.11,167.19) --
	(223.69,167.37) --
	(224.27,167.54) --
	(224.85,167.71) --
	(225.44,167.89) --
	(226.02,168.06) --
	(226.60,168.23) --
	(227.18,168.40) --
	(227.76,168.57) --
	(228.34,168.74) --
	(228.92,168.91);

\path[draw=drawColor,line width= 1.4pt,dash pattern=on 7pt off 3pt ,line join=round] ( 54.67,145.01) --
	( 55.25,145.60) --
	( 55.83,146.19) --
	( 56.42,146.77) --
	( 57.00,147.34) --
	( 57.58,147.90) --
	( 58.16,148.46) --
	( 58.74,149.00) --
	( 59.32,149.53) --
	( 59.90,150.04) --
	( 60.48,150.54) --
	( 61.06,151.02) --
	( 61.64,151.48) --
	( 62.22,151.93) --
	( 62.80,152.36) --
	( 63.39,152.77) --
	( 63.97,153.17) --
	( 64.55,153.54) --
	( 65.13,153.90) --
	( 65.71,154.24) --
	( 66.29,154.56) --
	( 66.87,154.86) --
	( 67.45,155.15) --
	( 68.03,155.42) --
	( 68.61,155.67) --
	( 69.19,155.91) --
	( 69.77,156.13) --
	( 70.36,156.33) --
	( 70.94,156.52) --
	( 71.52,156.69) --
	( 72.10,156.85) --
	( 72.68,157.00) --
	( 73.26,157.13) --
	( 73.84,157.25) --
	( 74.42,157.36) --
	( 75.00,157.46) --
	( 75.58,157.54) --
	( 76.16,157.62) --
	( 76.74,157.68) --
	( 77.33,157.73) --
	( 77.91,157.78) --
	( 78.49,157.81) --
	( 79.07,157.84) --
	( 79.65,157.86) --
	( 80.23,157.87) --
	( 80.81,157.88) --
	( 81.39,157.87) --
	( 81.97,157.86) --
	( 82.55,157.85) --
	( 83.13,157.83) --
	( 83.71,157.80) --
	( 84.30,157.77) --
	( 84.88,157.73) --
	( 85.46,157.69) --
	( 86.04,157.64) --
	( 86.62,157.59) --
	( 87.20,157.53) --
	( 87.78,157.47) --
	( 88.36,157.41) --
	( 88.94,157.34) --
	( 89.52,157.27) --
	( 90.10,157.20) --
	( 90.68,157.12) --
	( 91.27,157.04) --
	( 91.85,156.96) --
	( 92.43,156.88) --
	( 93.01,156.79) --
	( 93.59,156.70) --
	( 94.17,156.61) --
	( 94.75,156.52) --
	( 95.33,156.43) --
	( 95.91,156.33) --
	( 96.49,156.23) --
	( 97.07,156.14) --
	( 97.65,156.04) --
	( 98.24,155.94) --
	( 98.82,155.83) --
	( 99.40,155.73) --
	( 99.98,155.63) --
	(100.56,155.52) --
	(101.14,155.42) --
	(101.72,155.31) --
	(102.30,155.20) --
	(102.88,155.09) --
	(103.46,154.99) --
	(104.04,154.88) --
	(104.62,154.77) --
	(105.20,154.66) --
	(105.79,154.55) --
	(106.37,154.44) --
	(106.95,154.33) --
	(107.53,154.22) --
	(108.11,154.11) --
	(108.69,153.99) --
	(109.27,153.88) --
	(109.85,153.77) --
	(110.43,153.66) --
	(111.01,153.55) --
	(111.59,153.44) --
	(112.17,153.32) --
	(112.76,153.21) --
	(113.34,153.10) --
	(113.92,152.99) --
	(114.50,152.88) --
	(115.08,152.77) --
	(115.66,152.66) --
	(116.24,152.55) --
	(116.82,152.43) --
	(117.40,152.32) --
	(117.98,152.21) --
	(118.56,152.10) --
	(119.14,151.99) --
	(119.73,151.88) --
	(120.31,151.77) --
	(120.89,151.67) --
	(121.47,151.56) --
	(122.05,151.45) --
	(122.63,151.34) --
	(123.21,151.23) --
	(123.79,151.12) --
	(124.37,151.02) --
	(124.95,150.91) --
	(125.53,150.80) --
	(126.11,150.70) --
	(126.70,150.59) --
	(127.28,150.49) --
	(127.86,150.38) --
	(128.44,150.27) --
	(129.02,150.17) --
	(129.60,150.07) --
	(130.18,149.96) --
	(130.76,149.86) --
	(131.34,149.76) --
	(131.92,149.65) --
	(132.50,149.55) --
	(133.08,149.45) --
	(133.67,149.35) --
	(134.25,149.25) --
	(134.83,149.14) --
	(135.41,149.04) --
	(135.99,148.94) --
	(136.57,148.84) --
	(137.15,148.75) --
	(137.73,148.65) --
	(138.31,148.55) --
	(138.89,148.45) --
	(139.47,148.35) --
	(140.05,148.25) --
	(140.64,148.16) --
	(141.22,148.06) --
	(141.80,147.97) --
	(142.38,147.87) --
	(142.96,147.77) --
	(143.54,147.68) --
	(144.12,147.58) --
	(144.70,147.49) --
	(145.28,147.40) --
	(145.86,147.30) --
	(146.44,147.21) --
	(147.02,147.12) --
	(147.61,147.02) --
	(148.19,146.93) --
	(148.77,146.84) --
	(149.35,146.75) --
	(149.93,146.66) --
	(150.51,146.57) --
	(151.09,146.48) --
	(151.67,146.39) --
	(152.25,146.30) --
	(152.83,146.21) --
	(153.41,146.12) --
	(153.99,146.03) --
	(154.57,145.95) --
	(155.16,145.86) --
	(155.74,145.77) --
	(156.32,145.69) --
	(156.90,145.60) --
	(157.48,145.51) --
	(158.06,145.43) --
	(158.64,145.34) --
	(159.22,145.26) --
	(159.80,145.17) --
	(160.38,145.09) --
	(160.96,145.00) --
	(161.54,144.92) --
	(162.13,144.84) --
	(162.71,144.75) --
	(163.29,144.67) --
	(163.87,144.59) --
	(164.45,144.51) --
	(165.03,144.43) --
	(165.61,144.34) --
	(166.19,144.26) --
	(166.77,144.18) --
	(167.35,144.10) --
	(167.93,144.02) --
	(168.51,143.94) --
	(169.10,143.86) --
	(169.68,143.79) --
	(170.26,143.71) --
	(170.84,143.63) --
	(171.42,143.55) --
	(172.00,143.47) --
	(172.58,143.40) --
	(173.16,143.32) --
	(173.74,143.24) --
	(174.32,143.17) --
	(174.90,143.09) --
	(175.48,143.01) --
	(176.07,142.94) --
	(176.65,142.86) --
	(177.23,142.79) --
	(177.81,142.71) --
	(178.39,142.64) --
	(178.97,142.56) --
	(179.55,142.49) --
	(180.13,142.42) --
	(180.71,142.34) --
	(181.29,142.27) --
	(181.87,142.20) --
	(182.45,142.13) --
	(183.04,142.05) --
	(183.62,141.98) --
	(184.20,141.91) --
	(184.78,141.84) --
	(185.36,141.77) --
	(185.94,141.70) --
	(186.52,141.63) --
	(187.10,141.56) --
	(187.68,141.49) --
	(188.26,141.42) --
	(188.84,141.35) --
	(189.42,141.28) --
	(190.01,141.21) --
	(190.59,141.14) --
	(191.17,141.07) --
	(191.75,141.01) --
	(192.33,140.94) --
	(192.91,140.87) --
	(193.49,140.80) --
	(194.07,140.74) --
	(194.65,140.67) --
	(195.23,140.60) --
	(195.81,140.54) --
	(196.39,140.47) --
	(196.98,140.40) --
	(197.56,140.34) --
	(198.14,140.27) --
	(198.72,140.21) --
	(199.30,140.14) --
	(199.88,140.08) --
	(200.46,140.01) --
	(201.04,139.95) --
	(201.62,139.89) --
	(202.20,139.82) --
	(202.78,139.76) --
	(203.36,139.69) --
	(203.94,139.63) --
	(204.53,139.57) --
	(205.11,139.51) --
	(205.69,139.44) --
	(206.27,139.38) --
	(206.85,139.32) --
	(207.43,139.26) --
	(208.01,139.20) --
	(208.59,139.14) --
	(209.17,139.07) --
	(209.75,139.01) --
	(210.33,138.95) --
	(210.91,138.89) --
	(211.50,138.83) --
	(212.08,138.77) --
	(212.66,138.71) --
	(213.24,138.65) --
	(213.82,138.59) --
	(214.40,138.53) --
	(214.98,138.47) --
	(215.56,138.42) --
	(216.14,138.36) --
	(216.72,138.30) --
	(217.30,138.24) --
	(217.88,138.18) --
	(218.47,138.13) --
	(219.05,138.07) --
	(219.63,138.01) --
	(220.21,137.95) --
	(220.79,137.90) --
	(221.37,137.84) --
	(221.95,137.78) --
	(222.53,137.73) --
	(223.11,137.67) --
	(223.69,137.61) --
	(224.27,137.56) --
	(224.85,137.50) --
	(225.44,137.45) --
	(226.02,137.39) --
	(226.60,137.34) --
	(227.18,137.28) --
	(227.76,137.23) --
	(228.34,137.17) --
	(228.92,137.12);
\definecolor{drawColor}{RGB}{149,149,149}

\path[draw=drawColor,line width= 1.4pt,line join=round] ( 54.67,145.01) --
	( 55.25,142.29) --
	( 55.83,139.58) --
	( 56.42,136.89) --
	( 57.00,134.23) --
	( 57.58,131.61) --
	( 58.16,129.03) --
	( 58.74,126.51) --
	( 59.32,124.05) --
	( 59.90,121.65) --
	( 60.48,119.32) --
	( 61.06,117.06) --
	( 61.64,114.88) --
	( 62.22,112.78) --
	( 62.80,110.75) --
	( 63.39,108.81) --
	( 63.97,106.95) --
	( 64.55,105.18) --
	( 65.13,103.49) --
	( 65.71,101.88) --
	( 66.29,100.35) --
	( 66.87, 98.91) --
	( 67.45, 97.55) --
	( 68.03, 96.27) --
	( 68.61, 95.06) --
	( 69.19, 93.94) --
	( 69.77, 92.88) --
	( 70.36, 91.90) --
	( 70.94, 91.00) --
	( 71.52, 90.16) --
	( 72.10, 89.38) --
	( 72.68, 88.67) --
	( 73.26, 88.02) --
	( 73.84, 87.44) --
	( 74.42, 86.91) --
	( 75.00, 86.43) --
	( 75.58, 86.01) --
	( 76.16, 85.64) --
	( 76.74, 85.31) --
	( 77.33, 85.04) --
	( 77.91, 84.80) --
	( 78.49, 84.62) --
	( 79.07, 84.47) --
	( 79.65, 84.36) --
	( 80.23, 84.28) --
	( 80.81, 84.24) --
	( 81.39, 84.24) --
	( 81.97, 84.27) --
	( 82.55, 84.32) --
	( 83.13, 84.41) --
	( 83.71, 84.52) --
	( 84.30, 84.66) --
	( 84.88, 84.83) --
	( 85.46, 85.01) --
	( 86.04, 85.22) --
	( 86.62, 85.45) --
	( 87.20, 85.70) --
	( 87.78, 85.97) --
	( 88.36, 86.25) --
	( 88.94, 86.56) --
	( 89.52, 86.87) --
	( 90.10, 87.21) --
	( 90.68, 87.55) --
	( 91.27, 87.91) --
	( 91.85, 88.28) --
	( 92.43, 88.67) --
	( 93.01, 89.06) --
	( 93.59, 89.47) --
	( 94.17, 89.88) --
	( 94.75, 90.30) --
	( 95.33, 90.74) --
	( 95.91, 91.18) --
	( 96.49, 91.62) --
	( 97.07, 92.08) --
	( 97.65, 92.54) --
	( 98.24, 93.00) --
	( 98.82, 93.47) --
	( 99.40, 93.95) --
	( 99.98, 94.43) --
	(100.56, 94.92) --
	(101.14, 95.41) --
	(101.72, 95.90) --
	(102.30, 96.40) --
	(102.88, 96.90) --
	(103.46, 97.40) --
	(104.04, 97.91) --
	(104.62, 98.42) --
	(105.20, 98.93) --
	(105.79, 99.44) --
	(106.37, 99.95) --
	(106.95,100.47) --
	(107.53,100.99) --
	(108.11,101.50) --
	(108.69,102.02) --
	(109.27,102.54) --
	(109.85,103.06) --
	(110.43,103.58) --
	(111.01,104.10) --
	(111.59,104.62) --
	(112.17,105.14) --
	(112.76,105.66) --
	(113.34,106.18) --
	(113.92,106.70) --
	(114.50,107.22) --
	(115.08,107.74) --
	(115.66,108.26) --
	(116.24,108.78) --
	(116.82,109.30) --
	(117.40,109.81) --
	(117.98,110.33) --
	(118.56,110.84) --
	(119.14,111.36) --
	(119.73,111.87) --
	(120.31,112.38) --
	(120.89,112.89) --
	(121.47,113.40) --
	(122.05,113.91) --
	(122.63,114.41) --
	(123.21,114.92) --
	(123.79,115.42) --
	(124.37,115.92) --
	(124.95,116.42) --
	(125.53,116.92) --
	(126.11,117.42) --
	(126.70,117.91) --
	(127.28,118.41) --
	(127.86,118.90) --
	(128.44,119.39) --
	(129.02,119.88) --
	(129.60,120.37) --
	(130.18,120.86) --
	(130.76,121.34) --
	(131.34,121.82) --
	(131.92,122.30) --
	(132.50,122.78) --
	(133.08,123.26) --
	(133.67,123.74) --
	(134.25,124.21) --
	(134.83,124.68) --
	(135.41,125.15) --
	(135.99,125.62) --
	(136.57,126.09) --
	(137.15,126.55) --
	(137.73,127.02) --
	(138.31,127.48) --
	(138.89,127.94) --
	(139.47,128.39) --
	(140.05,128.85) --
	(140.64,129.30) --
	(141.22,129.76) --
	(141.80,130.21) --
	(142.38,130.66) --
	(142.96,131.10) --
	(143.54,131.55) --
	(144.12,131.99) --
	(144.70,132.43) --
	(145.28,132.87) --
	(145.86,133.31) --
	(146.44,133.75) --
	(147.02,134.18) --
	(147.61,134.61) --
	(148.19,135.04) --
	(148.77,135.47) --
	(149.35,135.90) --
	(149.93,136.32) --
	(150.51,136.75) --
	(151.09,137.17) --
	(151.67,137.59) --
	(152.25,138.01) --
	(152.83,138.43) --
	(153.41,138.84) --
	(153.99,139.25) --
	(154.57,139.67) --
	(155.16,140.08) --
	(155.74,140.49) --
	(156.32,140.89) --
	(156.90,141.30) --
	(157.48,141.70) --
	(158.06,142.10) --
	(158.64,142.50) --
	(159.22,142.90) --
	(159.80,143.30) --
	(160.38,143.69) --
	(160.96,144.09) --
	(161.54,144.48) --
	(162.13,144.87) --
	(162.71,145.26) --
	(163.29,145.64) --
	(163.87,146.03) --
	(164.45,146.41) --
	(165.03,146.80) --
	(165.61,147.18) --
	(166.19,147.56) --
	(166.77,147.93) --
	(167.35,148.31) --
	(167.93,148.69) --
	(168.51,149.06) --
	(169.10,149.43) --
	(169.68,149.80) --
	(170.26,150.17) --
	(170.84,150.54) --
	(171.42,150.90) --
	(172.00,151.27) --
	(172.58,151.63) --
	(173.16,151.99) --
	(173.74,152.35) --
	(174.32,152.71) --
	(174.90,153.07) --
	(175.48,153.42) --
	(176.07,153.78) --
	(176.65,154.13) --
	(177.23,154.48) --
	(177.81,154.83) --
	(178.39,155.18) --
	(178.97,155.53) --
	(179.55,155.88) --
	(180.13,156.22) --
	(180.71,156.57) --
	(181.29,156.91) --
	(181.87,157.25) --
	(182.45,157.59) --
	(183.04,157.93) --
	(183.62,158.26) --
	(184.20,158.60) --
	(184.78,158.93) --
	(185.36,159.27) --
	(185.94,159.60) --
	(186.52,159.93) --
	(187.10,160.26) --
	(187.68,160.59) --
	(188.26,160.92) --
	(188.84,161.24) --
	(189.42,161.57) --
	(190.01,161.89) --
	(190.59,162.21) --
	(191.17,162.53) --
	(191.75,162.85) --
	(192.33,163.17) --
	(192.91,163.49) --
	(193.49,163.80) --
	(194.07,164.12) --
	(194.65,164.43) --
	(195.23,164.75) --
	(195.81,165.06) --
	(196.39,165.37) --
	(196.98,165.68) --
	(197.56,165.99) --
	(198.14,166.29) --
	(198.72,166.60) --
	(199.30,166.90) --
	(199.88,167.21) --
	(200.46,167.51) --
	(201.04,167.81) --
	(201.62,168.11) --
	(202.20,168.41) --
	(202.78,168.71) --
	(203.36,169.01) --
	(203.94,169.30) --
	(204.53,169.60) --
	(205.11,169.89) --
	(205.69,170.19) --
	(206.27,170.48) --
	(206.85,170.77) --
	(207.43,171.06) --
	(208.01,171.35) --
	(208.59,171.64) --
	(209.17,171.93) --
	(209.75,172.21) --
	(210.33,172.50) --
	(210.91,172.78) --
	(211.50,173.06) --
	(212.08,173.35) --
	(212.66,173.63) --
	(213.24,173.91) --
	(213.82,174.19) --
	(214.40,174.47) --
	(214.98,174.74) --
	(215.56,175.02) --
	(216.14,175.29) --
	(216.72,175.57) --
	(217.30,175.84) --
	(217.88,176.12) --
	(218.47,176.39) --
	(219.05,176.66) --
	(219.63,176.93) --
	(220.21,177.20) --
	(220.79,177.47) --
	(221.37,177.73) --
	(221.95,178.00) --
	(222.53,178.27) --
	(223.11,178.53) --
	(223.69,178.79) --
	(224.27,179.06) --
	(224.85,179.32) --
	(225.44,179.58) --
	(226.02,179.84) --
	(226.60,180.10) --
	(227.18,180.36) --
	(227.76,180.62) --
	(228.34,180.88) --
	(228.92,181.13);

\path[draw=drawColor,line width= 1.4pt,dash pattern=on 7pt off 3pt ,line join=round] ( 54.67,145.01) --
	( 55.25,145.18) --
	( 55.83,145.36) --
	( 56.42,145.53) --
	( 57.00,145.70) --
	( 57.58,145.87) --
	( 58.16,146.03) --
	( 58.74,146.19) --
	( 59.32,146.35) --
	( 59.90,146.50) --
	( 60.48,146.65) --
	( 61.06,146.79) --
	( 61.64,146.93) --
	( 62.22,147.06) --
	( 62.80,147.19) --
	( 63.39,147.31) --
	( 63.97,147.43) --
	( 64.55,147.54) --
	( 65.13,147.65) --
	( 65.71,147.75) --
	( 66.29,147.84) --
	( 66.87,147.93) --
	( 67.45,148.02) --
	( 68.03,148.10) --
	( 68.61,148.17) --
	( 69.19,148.24) --
	( 69.77,148.31) --
	( 70.36,148.37) --
	( 70.94,148.42) --
	( 71.52,148.48) --
	( 72.10,148.52) --
	( 72.68,148.57) --
	( 73.26,148.61) --
	( 73.84,148.64) --
	( 74.42,148.67) --
	( 75.00,148.70) --
	( 75.58,148.73) --
	( 76.16,148.75) --
	( 76.74,148.77) --
	( 77.33,148.79) --
	( 77.91,148.80) --
	( 78.49,148.81) --
	( 79.07,148.82) --
	( 79.65,148.82) --
	( 80.23,148.83) --
	( 80.81,148.83) --
	( 81.39,148.83) --
	( 81.97,148.82) --
	( 82.55,148.82) --
	( 83.13,148.81) --
	( 83.71,148.81) --
	( 84.30,148.80) --
	( 84.88,148.78) --
	( 85.46,148.77) --
	( 86.04,148.76) --
	( 86.62,148.74) --
	( 87.20,148.73) --
	( 87.78,148.71) --
	( 88.36,148.69) --
	( 88.94,148.67) --
	( 89.52,148.65) --
	( 90.10,148.63) --
	( 90.68,148.60) --
	( 91.27,148.58) --
	( 91.85,148.56) --
	( 92.43,148.53) --
	( 93.01,148.51) --
	( 93.59,148.48) --
	( 94.17,148.45) --
	( 94.75,148.43) --
	( 95.33,148.40) --
	( 95.91,148.37) --
	( 96.49,148.34) --
	( 97.07,148.31) --
	( 97.65,148.28) --
	( 98.24,148.25) --
	( 98.82,148.22) --
	( 99.40,148.19) --
	( 99.98,148.16) --
	(100.56,148.13) --
	(101.14,148.10) --
	(101.72,148.07) --
	(102.30,148.03) --
	(102.88,148.00) --
	(103.46,147.97) --
	(104.04,147.94) --
	(104.62,147.91) --
	(105.20,147.87) --
	(105.79,147.84) --
	(106.37,147.81) --
	(106.95,147.77) --
	(107.53,147.74) --
	(108.11,147.71) --
	(108.69,147.68) --
	(109.27,147.64) --
	(109.85,147.61) --
	(110.43,147.58) --
	(111.01,147.54) --
	(111.59,147.51) --
	(112.17,147.48) --
	(112.76,147.44) --
	(113.34,147.41) --
	(113.92,147.38) --
	(114.50,147.34) --
	(115.08,147.31) --
	(115.66,147.28) --
	(116.24,147.25) --
	(116.82,147.21) --
	(117.40,147.18) --
	(117.98,147.15) --
	(118.56,147.11) --
	(119.14,147.08) --
	(119.73,147.05) --
	(120.31,147.02) --
	(120.89,146.98) --
	(121.47,146.95) --
	(122.05,146.92) --
	(122.63,146.89) --
	(123.21,146.86) --
	(123.79,146.82) --
	(124.37,146.79) --
	(124.95,146.76) --
	(125.53,146.73) --
	(126.11,146.70) --
	(126.70,146.67) --
	(127.28,146.63) --
	(127.86,146.60) --
	(128.44,146.57) --
	(129.02,146.54) --
	(129.60,146.51) --
	(130.18,146.48) --
	(130.76,146.45) --
	(131.34,146.42) --
	(131.92,146.39) --
	(132.50,146.36) --
	(133.08,146.33) --
	(133.67,146.30) --
	(134.25,146.27) --
	(134.83,146.24) --
	(135.41,146.21) --
	(135.99,146.18) --
	(136.57,146.15) --
	(137.15,146.12) --
	(137.73,146.09) --
	(138.31,146.06) --
	(138.89,146.03) --
	(139.47,146.00) --
	(140.05,145.97) --
	(140.64,145.94) --
	(141.22,145.91) --
	(141.80,145.89) --
	(142.38,145.86) --
	(142.96,145.83) --
	(143.54,145.80) --
	(144.12,145.77) --
	(144.70,145.75) --
	(145.28,145.72) --
	(145.86,145.69) --
	(146.44,145.66) --
	(147.02,145.63) --
	(147.61,145.61) --
	(148.19,145.58) --
	(148.77,145.55) --
	(149.35,145.53) --
	(149.93,145.50) --
	(150.51,145.47) --
	(151.09,145.45) --
	(151.67,145.42) --
	(152.25,145.39) --
	(152.83,145.37) --
	(153.41,145.34) --
	(153.99,145.31) --
	(154.57,145.29) --
	(155.16,145.26) --
	(155.74,145.24) --
	(156.32,145.21) --
	(156.90,145.18) --
	(157.48,145.16) --
	(158.06,145.13) --
	(158.64,145.11) --
	(159.22,145.08) --
	(159.80,145.06) --
	(160.38,145.03) --
	(160.96,145.01) --
	(161.54,144.98) --
	(162.13,144.96) --
	(162.71,144.93) --
	(163.29,144.91) --
	(163.87,144.88) --
	(164.45,144.86) --
	(165.03,144.84) --
	(165.61,144.81) --
	(166.19,144.79) --
	(166.77,144.76) --
	(167.35,144.74) --
	(167.93,144.72) --
	(168.51,144.69) --
	(169.10,144.67) --
	(169.68,144.65) --
	(170.26,144.62) --
	(170.84,144.60) --
	(171.42,144.58) --
	(172.00,144.55) --
	(172.58,144.53) --
	(173.16,144.51) --
	(173.74,144.48) --
	(174.32,144.46) --
	(174.90,144.44) --
	(175.48,144.42) --
	(176.07,144.39) --
	(176.65,144.37) --
	(177.23,144.35) --
	(177.81,144.33) --
	(178.39,144.31) --
	(178.97,144.28) --
	(179.55,144.26) --
	(180.13,144.24) --
	(180.71,144.22) --
	(181.29,144.20) --
	(181.87,144.17) --
	(182.45,144.15) --
	(183.04,144.13) --
	(183.62,144.11) --
	(184.20,144.09) --
	(184.78,144.07) --
	(185.36,144.05) --
	(185.94,144.03) --
	(186.52,144.01) --
	(187.10,143.98) --
	(187.68,143.96) --
	(188.26,143.94) --
	(188.84,143.92) --
	(189.42,143.90) --
	(190.01,143.88) --
	(190.59,143.86) --
	(191.17,143.84) --
	(191.75,143.82) --
	(192.33,143.80) --
	(192.91,143.78) --
	(193.49,143.76) --
	(194.07,143.74) --
	(194.65,143.72) --
	(195.23,143.70) --
	(195.81,143.68) --
	(196.39,143.66) --
	(196.98,143.64) --
	(197.56,143.62) --
	(198.14,143.60) --
	(198.72,143.58) --
	(199.30,143.56) --
	(199.88,143.55) --
	(200.46,143.53) --
	(201.04,143.51) --
	(201.62,143.49) --
	(202.20,143.47) --
	(202.78,143.45) --
	(203.36,143.43) --
	(203.94,143.41) --
	(204.53,143.39) --
	(205.11,143.38) --
	(205.69,143.36) --
	(206.27,143.34) --
	(206.85,143.32) --
	(207.43,143.30) --
	(208.01,143.28) --
	(208.59,143.27) --
	(209.17,143.25) --
	(209.75,143.23) --
	(210.33,143.21) --
	(210.91,143.19) --
	(211.50,143.18) --
	(212.08,143.16) --
	(212.66,143.14) --
	(213.24,143.12) --
	(213.82,143.10) --
	(214.40,143.09) --
	(214.98,143.07) --
	(215.56,143.05) --
	(216.14,143.04) --
	(216.72,143.02) --
	(217.30,143.00) --
	(217.88,142.98) --
	(218.47,142.97) --
	(219.05,142.95) --
	(219.63,142.93) --
	(220.21,142.91) --
	(220.79,142.90) --
	(221.37,142.88) --
	(221.95,142.86) --
	(222.53,142.85) --
	(223.11,142.83) --
	(223.69,142.81) --
	(224.27,142.80) --
	(224.85,142.78) --
	(225.44,142.76) --
	(226.02,142.75) --
	(226.60,142.73) --
	(227.18,142.72) --
	(227.76,142.70) --
	(228.34,142.68) --
	(228.92,142.67);
\definecolor{drawColor}{gray}{0.80}

\path[draw=drawColor,line width= 1.4pt,line join=round] ( 54.67,145.01) --
	( 55.25,141.30) --
	( 55.83,137.60) --
	( 56.42,133.94) --
	( 57.00,130.31) --
	( 57.58,126.74) --
	( 58.16,123.23) --
	( 58.74,119.79) --
	( 59.32,116.44) --
	( 59.90,113.17) --
	( 60.48,109.99) --
	( 61.06,106.91) --
	( 61.64,103.93) --
	( 62.22,101.07) --
	( 62.80, 98.31) --
	( 63.39, 95.66) --
	( 63.97, 93.12) --
	( 64.55, 90.70) --
	( 65.13, 88.40) --
	( 65.71, 86.21) --
	( 66.29, 84.13) --
	( 66.87, 82.16) --
	( 67.45, 80.30) --
	( 68.03, 78.56) --
	( 68.61, 76.92) --
	( 69.19, 75.38) --
	( 69.77, 73.94) --
	( 70.36, 72.61) --
	( 70.94, 71.37) --
	( 71.52, 70.23) --
	( 72.10, 69.17) --
	( 72.68, 68.20) --
	( 73.26, 67.32) --
	( 73.84, 66.52) --
	( 74.42, 65.80) --
	( 75.00, 65.15) --
	( 75.58, 64.57) --
	( 76.16, 64.06) --
	( 76.74, 63.62) --
	( 77.33, 63.25) --
	( 77.91, 62.93) --
	( 78.49, 62.67) --
	( 79.07, 62.47) --
	( 79.65, 62.32) --
	( 80.23, 62.22) --
	( 80.81, 62.17) --
	( 81.39, 62.16) --
	( 81.97, 62.20) --
	( 82.55, 62.27) --
	( 83.13, 62.39) --
	( 83.71, 62.55) --
	( 84.30, 62.74) --
	( 84.88, 62.96) --
	( 85.46, 63.21) --
	( 86.04, 63.50) --
	( 86.62, 63.81) --
	( 87.20, 64.15) --
	( 87.78, 64.52) --
	( 88.36, 64.91) --
	( 88.94, 65.32) --
	( 89.52, 65.75) --
	( 90.10, 66.20) --
	( 90.68, 66.68) --
	( 91.27, 67.17) --
	( 91.85, 67.67) --
	( 92.43, 68.20) --
	( 93.01, 68.73) --
	( 93.59, 69.29) --
	( 94.17, 69.85) --
	( 94.75, 70.43) --
	( 95.33, 71.02) --
	( 95.91, 71.62) --
	( 96.49, 72.22) --
	( 97.07, 72.84) --
	( 97.65, 73.47) --
	( 98.24, 74.11) --
	( 98.82, 74.75) --
	( 99.40, 75.40) --
	( 99.98, 76.06) --
	(100.56, 76.72) --
	(101.14, 77.39) --
	(101.72, 78.06) --
	(102.30, 78.74) --
	(102.88, 79.42) --
	(103.46, 80.11) --
	(104.04, 80.80) --
	(104.62, 81.49) --
	(105.20, 82.19) --
	(105.79, 82.88) --
	(106.37, 83.58) --
	(106.95, 84.29) --
	(107.53, 84.99) --
	(108.11, 85.70) --
	(108.69, 86.40) --
	(109.27, 87.11) --
	(109.85, 87.82) --
	(110.43, 88.53) --
	(111.01, 89.24) --
	(111.59, 89.95) --
	(112.17, 90.66) --
	(112.76, 91.37) --
	(113.34, 92.08) --
	(113.92, 92.79) --
	(114.50, 93.49) --
	(115.08, 94.20) --
	(115.66, 94.91) --
	(116.24, 95.61) --
	(116.82, 96.32) --
	(117.40, 97.02) --
	(117.98, 97.73) --
	(118.56, 98.43) --
	(119.14, 99.13) --
	(119.73, 99.83) --
	(120.31,100.52) --
	(120.89,101.22) --
	(121.47,101.91) --
	(122.05,102.60) --
	(122.63,103.29) --
	(123.21,103.98) --
	(123.79,104.67) --
	(124.37,105.35) --
	(124.95,106.04) --
	(125.53,106.72) --
	(126.11,107.39) --
	(126.70,108.07) --
	(127.28,108.74) --
	(127.86,109.42) --
	(128.44,110.09) --
	(129.02,110.75) --
	(129.60,111.42) --
	(130.18,112.08) --
	(130.76,112.74) --
	(131.34,113.40) --
	(131.92,114.05) --
	(132.50,114.71) --
	(133.08,115.36) --
	(133.67,116.01) --
	(134.25,116.65) --
	(134.83,117.30) --
	(135.41,117.94) --
	(135.99,118.58) --
	(136.57,119.21) --
	(137.15,119.85) --
	(137.73,120.48) --
	(138.31,121.11) --
	(138.89,121.73) --
	(139.47,122.36) --
	(140.05,122.98) --
	(140.64,123.60) --
	(141.22,124.21) --
	(141.80,124.83) --
	(142.38,125.44) --
	(142.96,126.05) --
	(143.54,126.66) --
	(144.12,127.26) --
	(144.70,127.86) --
	(145.28,128.46) --
	(145.86,129.06) --
	(146.44,129.65) --
	(147.02,130.25) --
	(147.61,130.83) --
	(148.19,131.42) --
	(148.77,132.01) --
	(149.35,132.59) --
	(149.93,133.17) --
	(150.51,133.75) --
	(151.09,134.32) --
	(151.67,134.90) --
	(152.25,135.47) --
	(152.83,136.03) --
	(153.41,136.60) --
	(153.99,137.16) --
	(154.57,137.73) --
	(155.16,138.28) --
	(155.74,138.84) --
	(156.32,139.40) --
	(156.90,139.95) --
	(157.48,140.50) --
	(158.06,141.05) --
	(158.64,141.59) --
	(159.22,142.13) --
	(159.80,142.67) --
	(160.38,143.21) --
	(160.96,143.75) --
	(161.54,144.28) --
	(162.13,144.82) --
	(162.71,145.35) --
	(163.29,145.87) --
	(163.87,146.40) --
	(164.45,146.92) --
	(165.03,147.45) --
	(165.61,147.96) --
	(166.19,148.48) --
	(166.77,149.00) --
	(167.35,149.51) --
	(167.93,150.02) --
	(168.51,150.53) --
	(169.10,151.04) --
	(169.68,151.54) --
	(170.26,152.04) --
	(170.84,152.55) --
	(171.42,153.04) --
	(172.00,153.54) --
	(172.58,154.04) --
	(173.16,154.53) --
	(173.74,155.02) --
	(174.32,155.51) --
	(174.90,156.00) --
	(175.48,156.48) --
	(176.07,156.97) --
	(176.65,157.45) --
	(177.23,157.93) --
	(177.81,158.40) --
	(178.39,158.88) --
	(178.97,159.35) --
	(179.55,159.83) --
	(180.13,160.30) --
	(180.71,160.76) --
	(181.29,161.23) --
	(181.87,161.70) --
	(182.45,162.16) --
	(183.04,162.62) --
	(183.62,163.08) --
	(184.20,163.54) --
	(184.78,163.99) --
	(185.36,164.45) --
	(185.94,164.90) --
	(186.52,165.35) --
	(187.10,165.80) --
	(187.68,166.25) --
	(188.26,166.69) --
	(188.84,167.14) --
	(189.42,167.58) --
	(190.01,168.02) --
	(190.59,168.46) --
	(191.17,168.90) --
	(191.75,169.33) --
	(192.33,169.77) --
	(192.91,170.20) --
	(193.49,170.63) --
	(194.07,171.06) --
	(194.65,171.49) --
	(195.23,171.92) --
	(195.81,172.34) --
	(196.39,172.77) --
	(196.98,173.19) --
	(197.56,173.61) --
	(198.14,174.03) --
	(198.72,174.44) --
	(199.30,174.86) --
	(199.88,175.27) --
	(200.46,175.69) --
	(201.04,176.10) --
	(201.62,176.51) --
	(202.20,176.92) --
	(202.78,177.32) --
	(203.36,177.73) --
	(203.94,178.13) --
	(204.53,178.53) --
	(205.11,178.94) --
	(205.69,179.34) --
	(206.27,179.73) --
	(206.85,180.13) --
	(207.43,180.53) --
	(208.01,180.92) --
	(208.59,181.31) --
	(209.17,181.70) --
	(209.75,182.09) --
	(210.33,182.48) --
	(210.91,182.87) --
	(211.50,183.26) --
	(212.08,183.64) --
	(212.66,184.03) --
	(213.24,184.41) --
	(213.82,184.79) --
	(214.40,185.17) --
	(214.98,185.55) --
	(215.56,185.92) --
	(216.14,186.30) --
	(216.72,186.67) --
	(217.30,187.05) --
	(217.88,187.42) --
	(218.47,187.79) --
	(219.05,188.16) --
	(219.63,188.53) --
	(220.21,188.89) --
	(220.79,189.26) --
	(221.37,189.62) --
	(221.95,189.99) --
	(222.53,190.35) --
	(223.11,190.71) --
	(223.69,191.07) --
	(224.27,191.43) --
	(224.85,191.79) --
	(225.44,192.14) --
	(226.02,192.50) --
	(226.60,192.85) --
	(227.18,193.20) --
	(227.76,193.56) --
	(228.34,193.91) --
	(228.92,194.26);

\path[draw=drawColor,line width= 1.4pt,dash pattern=on 7pt off 3pt ,line join=round] ( 54.67,145.01) --
	( 55.25,144.81) --
	( 55.83,144.62) --
	( 56.42,144.42) --
	( 57.00,144.23) --
	( 57.58,144.04) --
	( 58.16,143.86) --
	( 58.74,143.68) --
	( 59.32,143.50) --
	( 59.90,143.33) --
	( 60.48,143.16) --
	( 61.06,143.00) --
	( 61.64,142.85) --
	( 62.22,142.70) --
	( 62.80,142.55) --
	( 63.39,142.42) --
	( 63.97,142.29) --
	( 64.55,142.16) --
	( 65.13,142.04) --
	( 65.71,141.93) --
	( 66.29,141.82) --
	( 66.87,141.72) --
	( 67.45,141.62) --
	( 68.03,141.53) --
	( 68.61,141.45) --
	( 69.19,141.37) --
	( 69.77,141.30) --
	( 70.36,141.23) --
	( 70.94,141.17) --
	( 71.52,141.11) --
	( 72.10,141.06) --
	( 72.68,141.01) --
	( 73.26,140.96) --
	( 73.84,140.92) --
	( 74.42,140.89) --
	( 75.00,140.86) --
	( 75.58,140.83) --
	( 76.16,140.80) --
	( 76.74,140.78) --
	( 77.33,140.76) --
	( 77.91,140.75) --
	( 78.49,140.74) --
	( 79.07,140.73) --
	( 79.65,140.72) --
	( 80.23,140.72) --
	( 80.81,140.71) --
	( 81.39,140.72) --
	( 81.97,140.72) --
	( 82.55,140.72) --
	( 83.13,140.73) --
	( 83.71,140.74) --
	( 84.30,140.75) --
	( 84.88,140.76) --
	( 85.46,140.78) --
	( 86.04,140.79) --
	( 86.62,140.81) --
	( 87.20,140.83) --
	( 87.78,140.85) --
	( 88.36,140.87) --
	( 88.94,140.89) --
	( 89.52,140.92) --
	( 90.10,140.94) --
	( 90.68,140.97) --
	( 91.27,140.99) --
	( 91.85,141.02) --
	( 92.43,141.05) --
	( 93.01,141.08) --
	( 93.59,141.11) --
	( 94.17,141.14) --
	( 94.75,141.17) --
	( 95.33,141.20) --
	( 95.91,141.23) --
	( 96.49,141.26) --
	( 97.07,141.30) --
	( 97.65,141.33) --
	( 98.24,141.36) --
	( 98.82,141.40) --
	( 99.40,141.43) --
	( 99.98,141.47) --
	(100.56,141.50) --
	(101.14,141.54) --
	(101.72,141.57) --
	(102.30,141.61) --
	(102.88,141.64) --
	(103.46,141.68) --
	(104.04,141.72) --
	(104.62,141.75) --
	(105.20,141.79) --
	(105.79,141.83) --
	(106.37,141.86) --
	(106.95,141.90) --
	(107.53,141.94) --
	(108.11,141.97) --
	(108.69,142.01) --
	(109.27,142.05) --
	(109.85,142.08) --
	(110.43,142.12) --
	(111.01,142.16) --
	(111.59,142.20) --
	(112.17,142.23) --
	(112.76,142.27) --
	(113.34,142.31) --
	(113.92,142.35) --
	(114.50,142.38) --
	(115.08,142.42) --
	(115.66,142.46) --
	(116.24,142.49) --
	(116.82,142.53) --
	(117.40,142.57) --
	(117.98,142.60) --
	(118.56,142.64) --
	(119.14,142.68) --
	(119.73,142.71) --
	(120.31,142.75) --
	(120.89,142.79) --
	(121.47,142.82) --
	(122.05,142.86) --
	(122.63,142.90) --
	(123.21,142.93) --
	(123.79,142.97) --
	(124.37,143.00) --
	(124.95,143.04) --
	(125.53,143.08) --
	(126.11,143.11) --
	(126.70,143.15) --
	(127.28,143.18) --
	(127.86,143.22) --
	(128.44,143.25) --
	(129.02,143.29) --
	(129.60,143.32) --
	(130.18,143.36) --
	(130.76,143.39) --
	(131.34,143.43) --
	(131.92,143.46) --
	(132.50,143.49) --
	(133.08,143.53) --
	(133.67,143.56) --
	(134.25,143.60) --
	(134.83,143.63) --
	(135.41,143.66) --
	(135.99,143.70) --
	(136.57,143.73) --
	(137.15,143.76) --
	(137.73,143.80) --
	(138.31,143.83) --
	(138.89,143.86) --
	(139.47,143.89) --
	(140.05,143.93) --
	(140.64,143.96) --
	(141.22,143.99) --
	(141.80,144.02) --
	(142.38,144.05) --
	(142.96,144.09) --
	(143.54,144.12) --
	(144.12,144.15) --
	(144.70,144.18) --
	(145.28,144.21) --
	(145.86,144.24) --
	(146.44,144.27) --
	(147.02,144.31) --
	(147.61,144.34) --
	(148.19,144.37) --
	(148.77,144.40) --
	(149.35,144.43) --
	(149.93,144.46) --
	(150.51,144.49) --
	(151.09,144.52) --
	(151.67,144.55) --
	(152.25,144.58) --
	(152.83,144.61) --
	(153.41,144.64) --
	(153.99,144.67) --
	(154.57,144.70) --
	(155.16,144.73) --
	(155.74,144.75) --
	(156.32,144.78) --
	(156.90,144.81) --
	(157.48,144.84) --
	(158.06,144.87) --
	(158.64,144.90) --
	(159.22,144.93) --
	(159.80,144.95) --
	(160.38,144.98) --
	(160.96,145.01) --
	(161.54,145.04) --
	(162.13,145.07) --
	(162.71,145.09) --
	(163.29,145.12) --
	(163.87,145.15) --
	(164.45,145.18) --
	(165.03,145.20) --
	(165.61,145.23) --
	(166.19,145.26) --
	(166.77,145.28) --
	(167.35,145.31) --
	(167.93,145.34) --
	(168.51,145.36) --
	(169.10,145.39) --
	(169.68,145.42) --
	(170.26,145.44) --
	(170.84,145.47) --
	(171.42,145.50) --
	(172.00,145.52) --
	(172.58,145.55) --
	(173.16,145.57) --
	(173.74,145.60) --
	(174.32,145.62) --
	(174.90,145.65) --
	(175.48,145.68) --
	(176.07,145.70) --
	(176.65,145.73) --
	(177.23,145.75) --
	(177.81,145.78) --
	(178.39,145.80) --
	(178.97,145.83) --
	(179.55,145.85) --
	(180.13,145.87) --
	(180.71,145.90) --
	(181.29,145.92) --
	(181.87,145.95) --
	(182.45,145.97) --
	(183.04,146.00) --
	(183.62,146.02) --
	(184.20,146.04) --
	(184.78,146.07) --
	(185.36,146.09) --
	(185.94,146.11) --
	(186.52,146.14) --
	(187.10,146.16) --
	(187.68,146.18) --
	(188.26,146.21) --
	(188.84,146.23) --
	(189.42,146.25) --
	(190.01,146.28) --
	(190.59,146.30) --
	(191.17,146.32) --
	(191.75,146.35) --
	(192.33,146.37) --
	(192.91,146.39) --
	(193.49,146.41) --
	(194.07,146.44) --
	(194.65,146.46) --
	(195.23,146.48) --
	(195.81,146.50) --
	(196.39,146.52) --
	(196.98,146.55) --
	(197.56,146.57) --
	(198.14,146.59) --
	(198.72,146.61) --
	(199.30,146.63) --
	(199.88,146.65) --
	(200.46,146.68) --
	(201.04,146.70) --
	(201.62,146.72) --
	(202.20,146.74) --
	(202.78,146.76) --
	(203.36,146.78) --
	(203.94,146.80) --
	(204.53,146.82) --
	(205.11,146.85) --
	(205.69,146.87) --
	(206.27,146.89) --
	(206.85,146.91) --
	(207.43,146.93) --
	(208.01,146.95) --
	(208.59,146.97) --
	(209.17,146.99) --
	(209.75,147.01) --
	(210.33,147.03) --
	(210.91,147.05) --
	(211.50,147.07) --
	(212.08,147.09) --
	(212.66,147.11) --
	(213.24,147.13) --
	(213.82,147.15) --
	(214.40,147.17) --
	(214.98,147.19) --
	(215.56,147.21) --
	(216.14,147.23) --
	(216.72,147.25) --
	(217.30,147.27) --
	(217.88,147.29) --
	(218.47,147.31) --
	(219.05,147.33) --
	(219.63,147.35) --
	(220.21,147.36) --
	(220.79,147.38) --
	(221.37,147.40) --
	(221.95,147.42) --
	(222.53,147.44) --
	(223.11,147.46) --
	(223.69,147.48) --
	(224.27,147.50) --
	(224.85,147.51) --
	(225.44,147.53) --
	(226.02,147.55) --
	(226.60,147.57) --
	(227.18,147.59) --
	(227.76,147.61) --
	(228.34,147.63) --
	(228.92,147.64);
\definecolor{drawColor}{RGB}{0,0,0}

\path[draw=drawColor,line width= 0.9pt,line join=round,line cap=round] ( 45.96, 34.42) rectangle (237.63,217.66);
\end{scope}
\begin{scope}
\path[clip] (  0.00,  0.00) rectangle (241.85,217.66);
\definecolor{drawColor}{RGB}{0,0,0}

\node[text=drawColor,anchor=base east,inner sep=0pt, outer sep=0pt, scale=  0.86] at ( 40.56, 43.97) {\(-15\)};

\node[text=drawColor,anchor=base east,inner sep=0pt, outer sep=0pt, scale=  0.86] at ( 40.56, 76.23) {\(-10\)};

\node[text=drawColor,anchor=base east,inner sep=0pt, outer sep=0pt, scale=  0.86] at ( 40.56,108.49) {\(-5\)};

\node[text=drawColor,anchor=base east,inner sep=0pt, outer sep=0pt, scale=  0.86] at ( 40.56,140.74) {\(0\)};

\node[text=drawColor,anchor=base east,inner sep=0pt, outer sep=0pt, scale=  0.86] at ( 40.56,173.00) {\(5\)};

\node[text=drawColor,anchor=base east,inner sep=0pt, outer sep=0pt, scale=  0.86] at ( 40.56,205.26) {\(10\)};
\end{scope}
\begin{scope}
\path[clip] (  0.00,  0.00) rectangle (241.85,217.66);
\definecolor{drawColor}{RGB}{0,0,0}

\path[draw=drawColor,line width= 0.6pt,line join=round] ( 42.96, 48.23) --
	( 45.96, 48.23);

\path[draw=drawColor,line width= 0.6pt,line join=round] ( 42.96, 80.49) --
	( 45.96, 80.49);

\path[draw=drawColor,line width= 0.6pt,line join=round] ( 42.96,112.75) --
	( 45.96,112.75);

\path[draw=drawColor,line width= 0.6pt,line join=round] ( 42.96,145.01) --
	( 45.96,145.01);

\path[draw=drawColor,line width= 0.6pt,line join=round] ( 42.96,177.27) --
	( 45.96,177.27);

\path[draw=drawColor,line width= 0.6pt,line join=round] ( 42.96,209.53) --
	( 45.96,209.53);
\end{scope}
\begin{scope}
\path[clip] (  0.00,  0.00) rectangle (241.85,217.66);
\definecolor{drawColor}{RGB}{0,0,0}

\path[draw=drawColor,line width= 0.6pt,line join=round] ( 54.67, 31.42) --
	( 54.67, 34.42);

\path[draw=drawColor,line width= 0.6pt,line join=round] ( 89.52, 31.42) --
	( 89.52, 34.42);

\path[draw=drawColor,line width= 0.6pt,line join=round] (124.37, 31.42) --
	(124.37, 34.42);

\path[draw=drawColor,line width= 0.6pt,line join=round] (159.22, 31.42) --
	(159.22, 34.42);

\path[draw=drawColor,line width= 0.6pt,line join=round] (194.07, 31.42) --
	(194.07, 34.42);

\path[draw=drawColor,line width= 0.6pt,line join=round] (228.92, 31.42) --
	(228.92, 34.42);
\end{scope}
\begin{scope}
\path[clip] (  0.00,  0.00) rectangle (241.85,217.66);
\definecolor{drawColor}{RGB}{0,0,0}

\node[text=drawColor,anchor=base,inner sep=0pt, outer sep=0pt, scale=  0.86] at ( 54.67, 20.49) {0};

\node[text=drawColor,anchor=base,inner sep=0pt, outer sep=0pt, scale=  0.86] at ( 89.52, 20.49) {1};

\node[text=drawColor,anchor=base,inner sep=0pt, outer sep=0pt, scale=  0.86] at (124.37, 20.49) {2};

\node[text=drawColor,anchor=base,inner sep=0pt, outer sep=0pt, scale=  0.86] at (159.22, 20.49) {3};

\node[text=drawColor,anchor=base,inner sep=0pt, outer sep=0pt, scale=  0.86] at (194.07, 20.49) {4};

\node[text=drawColor,anchor=base,inner sep=0pt, outer sep=0pt, scale=  0.86] at (228.92, 20.49) {5};
\end{scope}
\begin{scope}
\path[clip] (  0.00,  0.00) rectangle (241.85,217.66);
\definecolor{drawColor}{RGB}{0,0,0}

\node[text=drawColor,anchor=base,inner sep=0pt, outer sep=0pt, scale=  1.00] at (141.80,  5.77) {\(h_g\)/\(h_{si}\)};
\end{scope}
\begin{scope}
\path[clip] (  0.00,  0.00) rectangle (241.85,217.66);
\definecolor{drawColor}{RGB}{0,0,0}

\node[text=drawColor,rotate= 90.00,anchor=base,inner sep=0pt, outer sep=0pt, scale=  1.00] at ( 14.00,126.04) {\(\sigma_x^T\), МПа};
\end{scope}
\begin{scope}
\path[clip] (  0.00,  0.00) rectangle (241.85,217.66);
\definecolor{drawColor}{RGB}{0,0,0}
\definecolor{fillColor}{RGB}{255,255,255}

\path[draw=drawColor,line width= 0.6pt,line join=round,line cap=round,fill=fillColor] (155.33, 85.26) rectangle (228.74,139.32);
\end{scope}
\begin{scope}
\path[clip] (  0.00,  0.00) rectangle (241.85,217.66);
\definecolor{drawColor}{RGB}{255,255,255}
\definecolor{fillColor}{RGB}{255,255,255}

\path[draw=drawColor,line width= 0.6pt,line join=round,line cap=round,fill=fillColor] (159.60,116.51) rectangle (186.58,130.00);
\end{scope}
\begin{scope}
\path[clip] (  0.00,  0.00) rectangle (241.85,217.66);
\definecolor{drawColor}{RGB}{0,0,0}

\path[draw=drawColor,line width= 1.4pt,line join=round] (162.30,123.25) -- (183.88,123.25);
\end{scope}
\begin{scope}
\path[clip] (  0.00,  0.00) rectangle (241.85,217.66);
\definecolor{drawColor}{RGB}{255,255,255}
\definecolor{fillColor}{RGB}{255,255,255}

\path[draw=drawColor,line width= 0.6pt,line join=round,line cap=round,fill=fillColor] (159.60,103.02) rectangle (186.58,116.51);
\end{scope}
\begin{scope}
\path[clip] (  0.00,  0.00) rectangle (241.85,217.66);
\definecolor{drawColor}{RGB}{149,149,149}

\path[draw=drawColor,line width= 1.4pt,line join=round] (162.30,109.76) -- (183.88,109.76);
\end{scope}
\begin{scope}
\path[clip] (  0.00,  0.00) rectangle (241.85,217.66);
\definecolor{drawColor}{RGB}{255,255,255}
\definecolor{fillColor}{RGB}{255,255,255}

\path[draw=drawColor,line width= 0.6pt,line join=round,line cap=round,fill=fillColor] (159.60, 89.53) rectangle (186.58,103.02);
\end{scope}
\begin{scope}
\path[clip] (  0.00,  0.00) rectangle (241.85,217.66);
\definecolor{drawColor}{gray}{0.80}

\path[draw=drawColor,line width= 1.4pt,line join=round] (162.30, 96.27) -- (183.88, 96.27);
\end{scope}
\begin{scope}
\path[clip] (  0.00,  0.00) rectangle (241.85,217.66);
\definecolor{drawColor}{RGB}{0,0,0}

\node[text=drawColor,anchor=base east,inner sep=0pt, outer sep=0pt, scale=  0.86] at (224.47,118.99) {\(350\) {\textdegree}C};
\end{scope}
\begin{scope}
\path[clip] (  0.00,  0.00) rectangle (241.85,217.66);
\definecolor{drawColor}{RGB}{0,0,0}

\node[text=drawColor,anchor=base east,inner sep=0pt, outer sep=0pt, scale=  0.86] at (224.47,105.50) {\(400\) {\textdegree}C};
\end{scope}
\begin{scope}
\path[clip] (  0.00,  0.00) rectangle (241.85,217.66);
\definecolor{drawColor}{RGB}{0,0,0}

\node[text=drawColor,anchor=base east,inner sep=0pt, outer sep=0pt, scale=  0.86] at (224.47, 92.01) {\(450\) {\textdegree}C};
\end{scope}
\begin{scope}
\path[clip] (  0.00,  0.00) rectangle (241.85,217.66);
\definecolor{drawColor}{RGB}{0,0,0}
\definecolor{fillColor}{RGB}{255,255,255}

\path[draw=drawColor,line width= 0.6pt,line join=round,line cap=round,fill=fillColor] (127.34, 41.31) rectangle (228.74, 81.88);
\end{scope}
\begin{scope}
\path[clip] (  0.00,  0.00) rectangle (241.85,217.66);
\definecolor{drawColor}{RGB}{255,255,255}
\definecolor{fillColor}{RGB}{255,255,255}

\path[draw=drawColor,line width= 0.6pt,line join=round,line cap=round,fill=fillColor] (131.61, 59.07) rectangle (158.59, 72.56);
\end{scope}
\begin{scope}
\path[clip] (  0.00,  0.00) rectangle (241.85,217.66);
\definecolor{drawColor}{RGB}{0,0,0}

\path[draw=drawColor,line width= 1.4pt,line join=round] (134.31, 65.81) -- (155.89, 65.81);
\end{scope}
\begin{scope}
\path[clip] (  0.00,  0.00) rectangle (241.85,217.66);
\definecolor{drawColor}{RGB}{255,255,255}
\definecolor{fillColor}{RGB}{255,255,255}

\path[draw=drawColor,line width= 0.6pt,line join=round,line cap=round,fill=fillColor] (131.61, 45.58) rectangle (158.59, 59.07);
\end{scope}
\begin{scope}
\path[clip] (  0.00,  0.00) rectangle (241.85,217.66);
\definecolor{drawColor}{RGB}{0,0,0}

\path[draw=drawColor,line width= 1.4pt,dash pattern=on 7pt off 3pt ,line join=round] (134.31, 52.32) -- (155.89, 52.32);
\end{scope}
\begin{scope}
\path[clip] (  0.00,  0.00) rectangle (241.85,217.66);
\definecolor{drawColor}{RGB}{0,0,0}

\node[text=drawColor,anchor=base east,inner sep=0pt, outer sep=0pt, scale=  0.86] at (224.47, 61.55) {Borofloat 33};
\end{scope}
\begin{scope}
\path[clip] (  0.00,  0.00) rectangle (241.85,217.66);
\definecolor{drawColor}{RGB}{0,0,0}

\node[text=drawColor,anchor=base east,inner sep=0pt, outer sep=0pt, scale=  0.86] at (224.47, 48.06) {ЛК5};
\end{scope}
\end{tikzpicture}

%% file: Dissertation/images_tikz/disser_stressX_distrib_600.tikz
% Created by tikzDevice version 0.10.1 on 2016-09-27 23:29:32
% !TEX encoding = UTF-8 Unicode
\begin{tikzpicture}[x=1pt,y=1pt]
\definecolor{fillColor}{RGB}{255,255,255}
\path[use as bounding box,fill=fillColor] (0,0) rectangle (284.53,170.72);
\begin{scope}
\path[clip] (  0.00,  0.00) rectangle (284.53,170.72);
\definecolor{drawColor}{RGB}{255,255,255}

\path[draw=drawColor,line width= 0.6pt,line join=round,line cap=round,fill=fillColor] (  0.00,  0.00) rectangle (284.53,170.72);
\end{scope}
\begin{scope}
\path[clip] ( 40.77, 32.43) rectangle (284.53,170.72);
\definecolor{fillColor}{RGB}{255,255,255}

\path[fill=fillColor] ( 40.77, 32.43) rectangle (284.53,170.72);
\definecolor{drawColor}{gray}{0.98}

\path[draw=drawColor,line width= 0.6pt,line join=round] ( 40.77, 54.61) --
	(284.53, 54.61);

\path[draw=drawColor,line width= 0.6pt,line join=round] ( 40.77, 95.11) --
	(284.53, 95.11);

\path[draw=drawColor,line width= 0.6pt,line join=round] ( 40.77,135.61) --
	(284.53,135.61);

\path[draw=drawColor,line width= 0.6pt,line join=round] ( 53.94, 32.43) --
	( 53.94,170.72);

\path[draw=drawColor,line width= 0.6pt,line join=round] (116.65, 32.43) --
	(116.65,170.72);

\path[draw=drawColor,line width= 0.6pt,line join=round] (179.37, 32.43) --
	(179.37,170.72);

\path[draw=drawColor,line width= 0.6pt,line join=round] (242.09, 32.43) --
	(242.09,170.72);
\definecolor{drawColor}{gray}{0.90}

\path[draw=drawColor,line width= 0.2pt,line join=round] ( 40.77, 34.36) --
	(284.53, 34.36);

\path[draw=drawColor,line width= 0.2pt,line join=round] ( 40.77, 74.86) --
	(284.53, 74.86);

\path[draw=drawColor,line width= 0.2pt,line join=round] ( 40.77,115.36) --
	(284.53,115.36);

\path[draw=drawColor,line width= 0.2pt,line join=round] ( 40.77,155.87) --
	(284.53,155.87);

\path[draw=drawColor,line width= 0.2pt,line join=round] ( 85.30, 32.43) --
	( 85.30,170.72);

\path[draw=drawColor,line width= 0.2pt,line join=round] (148.01, 32.43) --
	(148.01,170.72);

\path[draw=drawColor,line width= 0.2pt,line join=round] (210.73, 32.43) --
	(210.73,170.72);

\path[draw=drawColor,line width= 0.2pt,line join=round] (273.45, 32.43) --
	(273.45,170.72);
\definecolor{drawColor}{RGB}{0,0,0}

\path[draw=drawColor,line width= 1.3pt,line join=round] ( 51.85,157.39) --
	( 99.93, 98.07) --
	(148.01, 38.72) --
	(148.01,164.43) --
	(160.56,157.29) --
	(173.10,150.15) --
	(185.64,143.00) --
	(198.19,135.86) --
	(210.73,128.71) --
	(223.27,121.56) --
	(235.82,114.42) --
	(248.36,107.27) --
	(260.90,100.12) --
	(273.45, 92.97);

\path[draw=drawColor,line width= 0.9pt,line join=round,line cap=round] ( 40.77, 32.43) rectangle (284.53,170.72);
\end{scope}
\begin{scope}
\path[clip] (  0.00,  0.00) rectangle (284.53,170.72);
\definecolor{drawColor}{RGB}{0,0,0}

\node[text=drawColor,anchor=base east,inner sep=0pt, outer sep=0pt, scale=  1.00] at ( 35.37, 29.40) {\(-4\)};

\node[text=drawColor,anchor=base east,inner sep=0pt, outer sep=0pt, scale=  1.00] at ( 35.37, 69.90) {\(-2\)};

\node[text=drawColor,anchor=base east,inner sep=0pt, outer sep=0pt, scale=  1.00] at ( 35.37,110.40) {\(0\)};

\node[text=drawColor,anchor=base east,inner sep=0pt, outer sep=0pt, scale=  1.00] at ( 35.37,150.91) {\(2\)};
\end{scope}
\begin{scope}
\path[clip] (  0.00,  0.00) rectangle (284.53,170.72);
\definecolor{drawColor}{RGB}{0,0,0}

\path[draw=drawColor,line width= 0.6pt,line join=round] ( 37.77, 34.36) --
	( 40.77, 34.36);

\path[draw=drawColor,line width= 0.6pt,line join=round] ( 37.77, 74.86) --
	( 40.77, 74.86);

\path[draw=drawColor,line width= 0.6pt,line join=round] ( 37.77,115.36) --
	( 40.77,115.36);

\path[draw=drawColor,line width= 0.6pt,line join=round] ( 37.77,155.87) --
	( 40.77,155.87);
\end{scope}
\begin{scope}
\path[clip] (  0.00,  0.00) rectangle (284.53,170.72);
\definecolor{drawColor}{RGB}{0,0,0}

\path[draw=drawColor,line width= 0.6pt,line join=round] ( 85.30, 29.43) --
	( 85.30, 32.43);

\path[draw=drawColor,line width= 0.6pt,line join=round] (148.01, 29.43) --
	(148.01, 32.43);

\path[draw=drawColor,line width= 0.6pt,line join=round] (210.73, 29.43) --
	(210.73, 32.43);

\path[draw=drawColor,line width= 0.6pt,line join=round] (273.45, 29.43) --
	(273.45, 32.43);
\end{scope}
\begin{scope}
\path[clip] (  0.00,  0.00) rectangle (284.53,170.72);
\definecolor{drawColor}{RGB}{0,0,0}

\node[text=drawColor,anchor=base,inner sep=0pt, outer sep=0pt, scale=  1.00] at ( 85.30, 17.12) {\(-300\)};

\node[text=drawColor,anchor=base,inner sep=0pt, outer sep=0pt, scale=  1.00] at (148.01, 17.12) {\(0\)};

\node[text=drawColor,anchor=base,inner sep=0pt, outer sep=0pt, scale=  1.00] at (210.73, 17.12) {\(300\)};

\node[text=drawColor,anchor=base,inner sep=0pt, outer sep=0pt, scale=  1.00] at (273.45, 17.12) {\(600\)};
\end{scope}
\begin{scope}
\path[clip] (  0.00,  0.00) rectangle (284.53,170.72);
\definecolor{drawColor}{RGB}{0,0,0}

\node[text=drawColor,anchor=base,inner sep=0pt, outer sep=0pt, scale=  1.00] at (162.65,  2.40) {\(z\), мкм};
\end{scope}
\begin{scope}
\path[clip] (  0.00,  0.00) rectangle (284.53,170.72);
\definecolor{drawColor}{RGB}{0,0,0}

\node[text=drawColor,rotate= 90.00,anchor=base,inner sep=0pt, outer sep=0pt, scale=  1.00] at ( 12.32,101.57) {\(\sigma_x^T\), МПа};
\end{scope}
\end{tikzpicture}

%% file: Dissertation/images_tikz/disser_tis_g.tikz
\begin{tikzpicture}[x=1pt,y=1pt]
\definecolor{fillColor}{RGB}{255,255,255}
\path[use as bounding box,fill=fillColor] (0,0) rectangle (426.79,256.07);
\begin{scope}
\path[clip] (  0.00,  0.00) rectangle (426.79,256.07);
\definecolor{drawColor}{RGB}{255,255,255}

\path[draw=drawColor,line width= 0.6pt,line join=round,line cap=round,fill=fillColor] (  0.00,  0.00) rectangle (426.79,256.07);
\end{scope}
\begin{scope}
\path[clip] ( 36.22, 32.43) rectangle (426.79,242.11);
\definecolor{fillColor}{RGB}{255,255,255}

\path[fill=fillColor] ( 36.22, 32.43) rectangle (426.79,242.11);
\definecolor{drawColor}{gray}{0.98}

\path[draw=drawColor,line width= 0.6pt,line join=round] ( 36.22, 64.93) --
	(426.79, 64.93);

\path[draw=drawColor,line width= 0.6pt,line join=round] ( 36.22,132.24) --
	(426.79,132.24);

\path[draw=drawColor,line width= 0.6pt,line join=round] ( 36.22,199.55) --
	(426.79,199.55);

\path[draw=drawColor,line width= 0.6pt,line join=round] ( 73.00, 32.43) --
	( 73.00,242.11);

\path[draw=drawColor,line width= 0.6pt,line join=round] (104.70, 32.43) --
	(104.70,242.11);

\path[draw=drawColor,line width= 0.6pt,line join=round] (136.40, 32.43) --
	(136.40,242.11);

\path[draw=drawColor,line width= 0.6pt,line join=round] (187.12, 32.43) --
	(187.12,242.11);

\path[draw=drawColor,line width= 0.6pt,line join=round] (250.53, 32.43) --
	(250.53,242.11);

\path[draw=drawColor,line width= 0.6pt,line join=round] (313.93, 32.43) --
	(313.93,242.11);

\path[draw=drawColor,line width= 0.6pt,line join=round] (377.34, 32.43) --
	(377.34,242.11);
\definecolor{drawColor}{gray}{0.90}

\path[draw=drawColor,line width= 0.2pt,line join=round] ( 36.22, 98.59) --
	(426.79, 98.59);

\path[draw=drawColor,line width= 0.2pt,line join=round] ( 36.22,165.90) --
	(426.79,165.90);

\path[draw=drawColor,line width= 0.2pt,line join=round] ( 36.22,233.21) --
	(426.79,233.21);

\path[draw=drawColor,line width= 0.2pt,line join=round] ( 53.98, 32.43) --
	( 53.98,242.11);

\path[draw=drawColor,line width= 0.2pt,line join=round] ( 92.02, 32.43) --
	( 92.02,242.11);

\path[draw=drawColor,line width= 0.2pt,line join=round] (117.38, 32.43) --
	(117.38,242.11);

\path[draw=drawColor,line width= 0.2pt,line join=round] (155.42, 32.43) --
	(155.42,242.11);

\path[draw=drawColor,line width= 0.2pt,line join=round] (218.83, 32.43) --
	(218.83,242.11);

\path[draw=drawColor,line width= 0.2pt,line join=round] (282.23, 32.43) --
	(282.23,242.11);

\path[draw=drawColor,line width= 0.2pt,line join=round] (345.63, 32.43) --
	(345.63,242.11);

\path[draw=drawColor,line width= 0.2pt,line join=round] (409.04, 32.43) --
	(409.04,242.11);
\definecolor{drawColor}{RGB}{0,0,0}

\path[draw=drawColor,line width= 1.4pt,dash pattern=on 4pt off 4pt ,line join=round] ( 53.98, 41.96) --
	( 86.95, 64.32) --
	( 92.02, 70.56) --
	(104.70, 87.28) --
	(117.38,100.68) --
	(123.72,109.64) --
	(168.10,155.99) --
	(180.78,165.96) --
	(193.47,174.76) --
	(206.15,182.50) --
	(218.83,189.32) --
	(282.23,212.97) --
	(409.04,232.58);

\path[draw=drawColor,line width= 1.4pt,line join=round] ( 53.98, 79.84) --
	( 86.95, 87.25) --
	( 92.02, 89.32) --
	(104.70, 94.86) --
	(117.38, 99.29) --
	(123.72,102.27) --
	(168.10,117.61) --
	(180.78,120.92) --
	(193.47,123.83) --
	(206.15,126.40) --
	(218.83,128.66) --
	(282.23,136.48) --
	(409.04,142.99);

\path[draw=drawColor,line width= 1.4pt,dash pattern=on 1pt off 3pt on 4pt off 3pt ,line join=round] ( 53.98, 97.60) --
	( 86.95, 97.99) --
	( 92.02, 98.10) --
	(104.70, 98.39) --
	(117.38, 98.62) --
	(123.72, 98.78) --
	(168.10, 99.59) --
	(180.78, 99.76) --
	(193.47, 99.91) --
	(206.15,100.05) --
	(218.83,100.17) --
	(282.23,100.58) --
	(409.04,100.92);

\path[draw=drawColor,line width= 0.9pt,line join=round,line cap=round] ( 36.22, 32.43) rectangle (426.79,242.11);
\end{scope}
\begin{scope}
\path[clip] (  0.00,  0.00) rectangle (426.79,256.07);
\definecolor{drawColor}{RGB}{0,0,0}

\node[text=drawColor,anchor=base east,inner sep=0pt, outer sep=0pt, scale=  1.00] at ( 30.82, 93.63) {0};

\node[text=drawColor,anchor=base east,inner sep=0pt, outer sep=0pt, scale=  1.00] at ( 30.82,160.94) {5};

\node[text=drawColor,anchor=base east,inner sep=0pt, outer sep=0pt, scale=  1.00] at ( 30.82,228.25) {10};
\end{scope}
\begin{scope}
\path[clip] (  0.00,  0.00) rectangle (426.79,256.07);
\definecolor{drawColor}{RGB}{0,0,0}

\path[draw=drawColor,line width= 0.6pt,line join=round] ( 33.22, 98.59) --
	( 36.22, 98.59);

\path[draw=drawColor,line width= 0.6pt,line join=round] ( 33.22,165.90) --
	( 36.22,165.90);

\path[draw=drawColor,line width= 0.6pt,line join=round] ( 33.22,233.21) --
	( 36.22,233.21);
\end{scope}
\begin{scope}
\path[clip] (  0.00,  0.00) rectangle (426.79,256.07);
\definecolor{drawColor}{RGB}{0,0,0}

\path[draw=drawColor,line width= 0.6pt,line join=round] ( 53.98, 29.43) --
	( 53.98, 32.43);

\path[draw=drawColor,line width= 0.6pt,line join=round] ( 92.02, 29.43) --
	( 92.02, 32.43);

\path[draw=drawColor,line width= 0.6pt,line join=round] (117.38, 29.43) --
	(117.38, 32.43);

\path[draw=drawColor,line width= 0.6pt,line join=round] (155.42, 29.43) --
	(155.42, 32.43);

\path[draw=drawColor,line width= 0.6pt,line join=round] (218.83, 29.43) --
	(218.83, 32.43);

\path[draw=drawColor,line width= 0.6pt,line join=round] (282.23, 29.43) --
	(282.23, 32.43);

\path[draw=drawColor,line width= 0.6pt,line join=round] (345.63, 29.43) --
	(345.63, 32.43);

\path[draw=drawColor,line width= 0.6pt,line join=round] (409.04, 29.43) --
	(409.04, 32.43);
\end{scope}
\begin{scope}
\path[clip] (  0.00,  0.00) rectangle (426.79,256.07);
\definecolor{drawColor}{RGB}{0,0,0}

\node[text=drawColor,anchor=base,inner sep=0pt, outer sep=0pt, scale=  1.00] at ( 53.98, 17.12) {200};

\node[text=drawColor,anchor=base,inner sep=0pt, outer sep=0pt, scale=  1.00] at ( 92.02, 17.12) {500};

\node[text=drawColor,anchor=base,inner sep=0pt, outer sep=0pt, scale=  1.00] at (117.38, 17.12) {700};

\node[text=drawColor,anchor=base,inner sep=0pt, outer sep=0pt, scale=  1.00] at (155.42, 17.12) {1000};

\node[text=drawColor,anchor=base,inner sep=0pt, outer sep=0pt, scale=  1.00] at (218.83, 17.12) {1500};

\node[text=drawColor,anchor=base,inner sep=0pt, outer sep=0pt, scale=  1.00] at (282.23, 17.12) {2000};

\node[text=drawColor,anchor=base,inner sep=0pt, outer sep=0pt, scale=  1.00] at (345.63, 17.12) {2500};

\node[text=drawColor,anchor=base,inner sep=0pt, outer sep=0pt, scale=  1.00] at (409.04, 17.12) {3000};
\end{scope}
\begin{scope}
\path[clip] (  0.00,  0.00) rectangle (426.79,256.07);
\definecolor{drawColor}{RGB}{0,0,0}

\node[text=drawColor,anchor=base,inner sep=0pt, outer sep=0pt, scale=  1.00] at (231.51,  2.40) {Толщина стекла, мкм};
\end{scope}
\begin{scope}
\path[clip] (  0.00,  0.00) rectangle (426.79,256.07);
\definecolor{drawColor}{RGB}{0,0,0}

\node[text=drawColor,rotate= 90.00,anchor=base,inner sep=0pt, outer sep=0pt, scale=  1.00] at ( 12.32,137.27) {Напряжение, МПа};
\end{scope}
\begin{scope}
\path[clip] (  0.00,  0.00) rectangle (426.79,256.07);
\definecolor{drawColor}{RGB}{0,0,0}
\definecolor{fillColor}{RGB}{255,255,255}

\path[draw=drawColor,line width= 0.6pt,line join=round,line cap=round,fill=fillColor] ( 44.04,173.93) rectangle (164.60,237.92);
\end{scope}
\begin{scope}
\path[clip] (  0.00,  0.00) rectangle (426.79,256.07);
\definecolor{drawColor}{RGB}{0,0,0}

\node[text=drawColor,anchor=base west,inner sep=0pt, outer sep=0pt, scale=  1.00] at ( 48.30,223.73) {Температура, {\textdegree}C:\hspace*{0.8em}};
\end{scope}
\begin{scope}
\path[clip] (  0.00,  0.00) rectangle (426.79,256.07);
\definecolor{drawColor}{RGB}{255,255,255}
\definecolor{fillColor}{RGB}{255,255,255}

\path[draw=drawColor,line width= 0.6pt,line join=round,line cap=round,fill=fillColor] ( 48.30,205.18) rectangle ( 82.03,218.67);
\end{scope}
\begin{scope}
\path[clip] (  0.00,  0.00) rectangle (426.79,256.07);
\definecolor{drawColor}{RGB}{0,0,0}

\path[draw=drawColor,line width= 1.4pt,dash pattern=on 4pt off 4pt ,line join=round] ( 51.68,211.93) -- ( 78.66,211.93);
\end{scope}
\begin{scope}
\path[clip] (  0.00,  0.00) rectangle (426.79,256.07);
\definecolor{drawColor}{RGB}{255,255,255}
\definecolor{fillColor}{RGB}{255,255,255}

\path[draw=drawColor,line width= 0.6pt,line join=round,line cap=round,fill=fillColor] ( 48.30,191.69) rectangle ( 82.03,205.18);
\end{scope}
\begin{scope}
\path[clip] (  0.00,  0.00) rectangle (426.79,256.07);
\definecolor{drawColor}{RGB}{0,0,0}

\path[draw=drawColor,line width= 1.4pt,line join=round] ( 51.68,198.44) -- ( 78.66,198.44);
\end{scope}
\begin{scope}
\path[clip] (  0.00,  0.00) rectangle (426.79,256.07);
\definecolor{drawColor}{RGB}{255,255,255}
\definecolor{fillColor}{RGB}{255,255,255}

\path[draw=drawColor,line width= 0.6pt,line join=round,line cap=round,fill=fillColor] ( 48.30,178.20) rectangle ( 82.03,191.69);
\end{scope}
\begin{scope}
\path[clip] (  0.00,  0.00) rectangle (426.79,256.07);
\definecolor{drawColor}{RGB}{0,0,0}

\path[draw=drawColor,line width= 1.4pt,dash pattern=on 1pt off 3pt on 4pt off 3pt ,line join=round] ( 51.68,184.95) -- ( 78.66,184.95);
\end{scope}
\begin{scope}
\path[clip] (  0.00,  0.00) rectangle (426.79,256.07);
\definecolor{drawColor}{RGB}{0,0,0}

\node[text=drawColor,anchor=base east,inner sep=0pt, outer sep=0pt, scale=  1.00] at (109.47,206.97) {\(-\)60};
\end{scope}
\begin{scope}
\path[clip] (  0.00,  0.00) rectangle (426.79,256.07);
\definecolor{drawColor}{RGB}{0,0,0}

\node[text=drawColor,anchor=base east,inner sep=0pt, outer sep=0pt, scale=  1.00] at (109.47,193.48) {20};
\end{scope}
\begin{scope}
\path[clip] (  0.00,  0.00) rectangle (426.79,256.07);
\definecolor{drawColor}{RGB}{0,0,0}

\node[text=drawColor,anchor=base east,inner sep=0pt, outer sep=0pt, scale=  1.00] at (109.47,179.99) {85};
\end{scope}
\begin{scope}
\path[clip] (  0.00,  0.00) rectangle (426.79,256.07);
\definecolor{drawColor}{RGB}{0,0,0}

\node[text=drawColor,anchor=base,inner sep=0pt, outer sep=0pt, scale=  1.00] at (231.51,245.53) {Третий инвариант тензора девиатора напряжений};
\end{scope}
\end{tikzpicture}

%% file: Dissertation/part4.tex
\chapter{Проверка выдвинутых положений и рекомендации разработчикам}

В данной главе описаны эксперименты,
проведённые для подтверждения утверждений из~предыдущих глав. Описаны
исследования и практический опыт по снижению остаточных напряжений
посредством термообработки. Предложена методика минимизации остаточных
напряжений, представленная в виде шагов для разработчика прибора.
Также описаны сложности, на которые стоит обратить внимание при
разработке современной электронной техники с~применением технологии
электростатического соединения.

\section{Определение остаточных напряжений в стекле}
\subsection{Описание метода поляризационно-оптического измерения разности хода лучей}\label{chap:measure_glass_stress}
<<Метод основан на явлении двулучепреломления, которое наблюдается в~напряжённом стекле при прохождении через него луча линейно-поляризованного света, и заключается в разложении луча на два: обыкновенный и~необыкновенный, распространяющиеся с различными скоростями и~вследствие этого имеющие при выходе из напряжённого
стекла разность хода>>~\cite{gost_metod_opred_dvulucheprelom}.

<<Метод включает качественное, полуколичественное и количественное определение
напряжения, исходя из разности хода поляризованного света, проходящего через
образец>>~\cite{gost_metod_opred_dvulucheprelom}.
При качественном и полуколичественном определении <<испытание проводят
способом сравнения, который состоит в~оценке на~полярископе общего
распределения напряжений в~изделии и~в~оценке значения разности
хода>>~\cite{gost_metod_opred_dvulucheprelom} сравнением наблюдаемого цвета одним из~двух способов:
\begin{itemize}
    \item <<с данными
    таблицы интерференционных цветов>>~\cite{gost_metod_opred_dvulucheprelom};
    \item <<с соответствующим цветом
    правильно ориентированного ступенчатого клина, или же одинакового
    изделия с количественно оценёнными разностями хода на обозначенных
    местах>>~\cite{gost_metod_opred_dvulucheprelom}.
\end{itemize}

При количественном определении напряжений <<испытание проводят способом
компенсации, который состоит в количественном определении значения разности хода
с помощью поляриметра, снабжённого компенсатором. В~качестве компенсатора
применяют четвертьволновую фазовую пластинку с~поворотным анализатором (компенсатор
Сенармона)>>~\cite{gost_metod_opred_dvulucheprelom}.
Четвертьволновая фазовая пластинка (фазовая пластинка $\lambda/$4) "---
<<устройство, создающее разность фаз между ортогональными линейно-поляризованными
составляющими оптического излучения определенной длины волны, равную
\((2n+1)\frac{\pi}{2}\), что соответствует разности хода между этими
составляющими, равной \((2n+1)\frac{\lambda}{4}\),
где \( n \) "--- целое число>>~\cite{gost_metod_opred_dvulucheprelom},
\( \lambda \) "--- длина волны света, нм.

<<При измерении разности хода с помощью поляриметра лимб анализатора устанавливают
на нулевую отметку, вводят пластину $\lambda/$4 и зелёный светофильтр.
Анализатор поворачивают на несколько градусов в обе стороны и~для компенсации
выбирают то направление поворота анализатора, при котором тёмные полосы
приближаются к месту измерения. Медленно вращают анализатор до~тех пор, пока
наблюдаемые в поле зрения тёмные полосы не сольются в~одну утолщённую. Затем
анализатор слегка поворачивают в~обратную сторону, в~результате чего образуется
небольшой просвет между полосами, которые затем снова доводят до
соприкосновения. Угол поворота анализатора определяют по~лимбу анализатора. Для
расчёта разности хода применяют среднее значение результатов трёх
отсчётов>>~\cite{gost_metod_opred_dvulucheprelom}.

<<Разность хода ($\Delta$) в нанометрах вычисляют по формуле:
\begin{equation*}
    \Delta =\frac{\lambda \phi }{180}=3\phi,
\end{equation*}
где $\lambda$ "--- длина волны света, равная 540~нм при условии применения зелёного светофильтра;
$\phi$ "--- угол поворота анализатора>>~\cite{gost_metod_opred_dvulucheprelom}.

\begin{samepage}
<<Удельную разность хода ($\Delta'$) в
10\textsuperscript{$-$6}
вычисляют по формуле:
\begin{equation*}
    \Delta'=\frac{\Delta}{l},
\end{equation*}
где $\Delta $ "--- разность хода, нм;
$ l $ "--- длина пути луча в напряжённом стекле, нм>>~\cite{gost_metod_opred_dvulucheprelom}.

\subsection{Результаты и обсуждение}
Были проведены эксперименты по соединению пластин кремния и стекла.
Использовались пластины кремния марки КЭС (кремний электронного типа проводимости, легирующий элемент "--- сурьма)
диаметром 60 мм ориентации \{100\} с~удельным сопротивлением 0,01 Ом$\cdot$см и прямоугольные пластины стекла ЛК5
размерами 30\(\,\times\,\)50\(\,\times\,\)4,5~мм. Соединения были проведены при
температуре от~330 до~350~{\textdegree}C. Измерения величины двулучепреломления в
стекле проводились в~соответствии с~\cite{gost_metod_opred_dvulucheprelom} на
полярископе\nb-поляриметре ПКС\nb-250.
\end{samepage}

В соединениях стекло-кремний наблюдалась смена знака напряжений в~стекле от
растягивающих к сжимающим по мере удаления от границы соединения, что хорошо
соотносится с расчётными результатами, приведёнными
на~Рисунке~\ref{fig:tm_stress_si_compos_2fig_bf33_lk5}
на странице~\pageref{fig:tm_stress_si_compos_2fig_bf33_lk5}.
После соединения стекло"--~кремний"--~стекло напряжения
в стекле по~мере удаления от
плоскости соединения не~изменялись, что можно считать качественной проверкой
результатов моделирования, приведённых на Рисунке~\ref{fig:sigma_z_gsg_lk5}
на странице~\pageref{fig:sigma_z_gsg_lk5}.

\section{Определение остаточных напряжений в кремнии}
\subsection{Описание метода спектроскопии комбинационного рассеяния света (эффект Раман)}
Методы рамановской спектроскопии (РС) (в отечественной литературе
используется также "--- спектроскопия комбинационного рассеяния (КР) света)
применяются для исследования спектров электронных возбуждений и~оптических
фононов в разных веществах.
Метод РС основан на эффекте неупругого рассеяния света на возбуждениях
системы при воздействии лазерного излучения.
При рождении возбуждения частота рассеянного света оказывается сдвинутой в
красную область относительно частоты лазера на~величину энергии рождённого
возбуждения (стоксовое КР света), а при поглощении соответствующего
возбуждения системы "--- сдвинутой в синюю область на~величину энергии
поглощённого возбуждения (антистоксовое КР~света).
В~частности, по~отношению интенсивностей стоксовой и антистоксовой
компонент КР света на~модах оптических фононов можно определять
бесконтактным образом локальную температуру образца с высоким
пространственным разрешением.

<<Монохроматический свет, падающий на образец, может быть отражён, поглощён или рассеян. Процесс рассеяния света может быть упругим (без изменения частоты света) и неупругим (с изменением его частоты).
Упругое рассеяние света называется рэлеевским. Оно является преобладающим: в~среднем лишь один фотон из 10 миллионов рассеивается неупруго. При рэлеевском рассеянии частота рассеянного света в точности равна частоте света падающего.
Неупругое рассеяние света называется комбинационным, или рамановским. При комбинационном рассеянии света частота рассеянного света может как уменьшаться (Стоксово рассеяние "--- S), так и увеличиваться (анти-Стоксово рассеяние "--- AS)>>~\cite{belogorohov2014_dis}.

<<В рамановском спектре за ,,ноль`` принимают частоту рэлеевского рассеяния (то есть частоту источника излучения "--- лазера L), а частоту линии (Raman shift) в спектре вычисляют вычитанием частоты Стоксовой (анти-Стоксовой) линии из частоты рэлеевского излучения>>~\cite{belogorohov2014_dis}.

<<Спектры КР очень чувствительны к природе химических связей как в~органических молекулах и полимерных материалах, так и в неорганических кристаллических решётках и кластерах. По этой причине каждое определённое вещество, каждый материал, обладает своим собственным, индивидуальным КР~спектром, который является для него аналогом ,,отпечатка пальцев``>>~\cite{belogorohov2014_dis}.

Остаточные напряжения часто являются доминирующим фактором, приводящим к
спектральному сдвигу рамановского пика в кремнии.
В случае одноосного напряжённого состояния сжимающие напряжения (compressive
stress) приводят к увеличению рамановской частоты (положительному сдвигу частоты
относительно частоты ненапряжённого состояния), а растягивающие напряжения
(tensile stress) "--- к уменьшению частоты~\cite{DeWolf1996}. В случае более
сложных напряжённых состояний "--- влияние комбинированное по~шести
направлениям. При трактовке результатов измерения важно иметь чёткую
гипотезу о характере измеряемого напряжённого состояния, иначе погрешности
оценок могут превышать
сотни~МПа~\cite{Srikar2003critreview}.

<<Преимущества метода КР:
\begin{itemize}
    \item неразрушающий характер метода;
    \item бесконтактный метод;
    \item не требует специальной подготовки образцов;
    \item позволяет проводить анализ твёрдых материалов и жидкостей,
    в~определённых случаях также газов;
    \item достаточно быстрый анализ (от секунд до минут);
    \item возможность картирования образцов с высоким латеральным разрешением до 1,5 мкм (как правило, в этом случае КР спектрометр совмещается с оптическим микроскопом)>>~\cite{mironov2011_issledovanie_metodom_raman}.
\end{itemize}

\subsection{Оценка механических напряжений методом спектроскопии комбинационного рассеяния}
Основы пересчёта смещений пиков КР в значения воздействующих деформаций и
напряжений были подробно описаны в~\cite{Ganesan1970}, частные случаи для разных
материалов были рассчитаны или измерены в других работах.
Здесь остановимся лишь на нескольких моментах для случая пластин кремния
ориентации \{100\}, в которых измерения проводились со стороны поверхности,
перпендикулярной кристаллографическому направлению (001) (типовой случай для
пластин кремния, применяемых в микроэлектронике).
Изменения \(\mathbf{\Delta K}\) в тензоре силовых констант, вызванные
деформациями в случае кремния описываются формулой, записанной для главных осей
кремния~\cite{Narayanan1997}:
\begin{equation}
    \begin{pmatrix}
    \Delta K_{11}\\
    \Delta K_{22}\\
    \Delta K_{33}\\
    2\Delta K_{23}\\
    2\Delta K_{13}\\
    2\Delta K_{12}
    \end{pmatrix}
    =
    \begin{pmatrix}
    p & q & q & 0 & 0 & 0 \\
    q & p & q & 0 & 0 & 0 \\
    q & q & p & 0 & 0 & 0 \\
    0 & 0 & 0 & r & 0 & 0 \\
    0 & 0 & 0 & 0 & r & 0 \\
    0 & 0 & 0 & 0 & 0 & r
    \end{pmatrix}
    \begin{pmatrix}
    \epsilon_{11}\\
    \epsilon_{22}\\
    \epsilon_{33}\\
    \epsilon_{23}\\
    \epsilon_{13}\\
    \epsilon_{12}
    \end{pmatrix}\!,
\end{equation}
где \(p, q, r\) "--- потенциалы деформации фононов. Наиболее актуальные
результаты измерений этих величин для кремния опубликованы в
работе~\cite{Anastassakis1990}:%
\[ \frac{p}{\omega_0^2} = -1,85 \pm 0,06, \]%
\[ \frac{q}{\omega_0^2} = -2,31 \pm 0,06, \]%
\[ \frac{r}{\omega_0^2} = -0,71 \pm 0,02, \]%
где \(\omega_0\) "--- рамановский пик ненапряжённого кремния.

Связь смещений и деформаций определена решениями характеристического уравнения,
равенства детерминанта матрицы нулю~\cite{DeWolf1996}:
\begin{equation}\label{eq:ramanseculequation}
    \setlength{\arraycolsep}{-6pt}% default is 6pt
    \begin{vmatrix}
    p \epsilon_{11} + q (\epsilon_{22} + \epsilon_{33}) - \lambda & 2r \epsilon_{12} & 2r \epsilon_{13} \\
    2r \epsilon_{12} & p \epsilon_{22} + q (\epsilon_{33} + \epsilon_{11}) - \lambda & 2r \epsilon_{23} \\
    2r \epsilon_{13} & 2r \epsilon_{23} & p \epsilon_{33} + q (\epsilon_{11} + \epsilon_{22}) - \lambda
    \end{vmatrix}
    =
    0.
\end{equation}
Вывод уравнения~\eqref{eq:ramanseculequation} подробно рассмотрен в работе~\cite{DeWolf2015relation}.

Характеристические числа \(\lambda_i\) матрицы  \(\mathbf{\Delta K}\) связаны со
смещениями рамановских пиков следующими уравнениями:
\begin{equation}
\lambda_i = \omega_i^2 - \omega_0^2,\quad i = 1, 2, 3
\end{equation}
\begin{equation}\label{eq:ramandeltaomega}
\Delta\omega_i = \omega_i - \omega_0 \cong \frac{\lambda_i}{2\omega_0}.
\end{equation}

Деформации связаны с напряжениями через матрицу, обратную матрице жёсткости
\(\mathbf{S} = \mathbf{C}^{-1}\), для кристалла кремния она имеет всего три
независимых значения:
\begin{equation}\label{eq:def_stress_relation}
    \begin{pmatrix}
    \epsilon_{11}\\
    \epsilon_{22}\\
    \epsilon_{33}\\
    2\epsilon_{23}\\
    2\epsilon_{13}\\
    2\epsilon_{12}
    \end{pmatrix}
    =
    \begin{pmatrix}
    S_{11} & S_{12} & S_{12} & 0 & 0 & 0 \\
    S_{12} & S_{11} & S_{12} & 0 & 0 & 0 \\
    S_{12} & S_{12} & S_{11} & 0 & 0 & 0 \\
    0 & 0 & 0 & S_{44} & 0 & 0 \\
    0 & 0 & 0 & 0 & S_{44} & 0 \\
    0 & 0 & 0 & 0 & 0 & S_{44}
    \end{pmatrix}
    \begin{pmatrix}
    \sigma_{11}\\
    \sigma_{22}\\
    \sigma_{33}\\
    \sigma_{23}\\
    \sigma_{13}\\
    \sigma_{12}
    \end{pmatrix}\!.
\end{equation}

Решая совместно уравнения \eqref{eq:ramandeltaomega},
\eqref{eq:ramanseculequation} и  \eqref{eq:def_stress_relation} можно получить
связь смещений трёх мод рамановских пиков в предположении о воздействии
одноосного напряжения в направлении (100) (одноосного напряжённого
состояния)~\cite{DeWolf1999}:
\begin{equation}
    \begin{aligned}
    \Delta\omega_1 &= \frac{1}{2\omega_0} (p S_{11} + 2q S_{12})\sigma,\\
    \Delta\omega_2 &= \frac{1}{2\omega_0} (p S_{12} + q(S_{11} + S_{12})) \sigma,\\
    \Delta\omega_3 &= \frac{1}{2\omega_0} (p S_{12} + q(S_{11} + S_{12})) \sigma.
    \end{aligned}
\end{equation}
При проведении измерения в направлении (001) наблюдается только третья мода и
потому связь одноосного напряжения и измеренного смещения выражается формулой:
\begin{equation}\label{eq:ramansigma_ot_smesh_uniaxial}
\sigma = \frac{2\omega_0}{p S_{12} + q(S_{11} + S_{12})} \Delta\omega_3.
\end{equation}

Аналогичным образом выводится зависимость для двухосного напряжённого состояния
в осях (100) и (010) (в векторе напряжений ненулевые только компоненты
\(\sigma_{11}\) и \(\sigma_{22}\))~\cite{DeWolf1996integcirc}:
\begin{equation}\label{eq:ramansigma_ot_smesh_biaxial}
\sigma_{11} + \sigma_{22} = \frac{2\omega_0}{p S_{12} + q(S_{11} + S_{12})} \Delta\omega_3.
\end{equation}

В связи с тем, что есть разные источники, измерившие и оценившие величины,
входящие в коэффициент перед смещением в
формулах~\eqref{eq:ramansigma_ot_smesh_uniaxial}
и~\eqref{eq:ramansigma_ot_smesh_biaxial}, сам коэффициент может принимать
значение от~минус~430 до~минус~500~МПа/см\textsuperscript{$-$1}~\cite{Naka2015}.
В этой диссертационной работе упругие
свойства пластины кремния ориентации \{100\} взяты
из~\cites[42]{Bao_part_Mech_Beam_Diaphragm_Structures}, потенциалы
деформации фононов из~\cite{Anastassakis1990} и рамановский пик
ненапряжённого кремния $\omega_0=$ 520,3~см\textsuperscript{$-$1}~\cite{Loechelt1999},
что даёт значение данного
коэффициента минус \(439,7\)~МПа/см\textsuperscript{$-$1}.

Оценка одноосного напряжённого состояния в направлении (110) показана в
работе~\cite{DeWolf1996}. Оценка напряжённых состояний в пластинах кремния
ориентации \{111\} показана в работе~\cite{Narayanan1997}.

\subsection{Результаты и обсуждение}
Измеряли пластины диаметром 100~мм кремния марки КЭФ (кремний электронного типа
проводимости, легирующий элемент "--- фосфор) односторонней (ОП) и двусторонней
полировки (ДП) с~удельным сопротивлением 4,5~Ом$\cdot$см в свободном состоянии и
после соединения с пластинами стекла марки Borofloat~33.
Проводили измерения на дисперсионном Раман микроскопе Nicolet DXR Spectrometer.
Для возбуждения использовался лазер с~длиной волны 633~нм. Апертура
устанавливалась в 25 мкм с формой отверстия. Использовался объектив 10х/0.25 BD
и решётка разрешением 1200 линий/мм.

Определение пиков полученных спектров проводилось с помощью поставленной с оборудованием программы обработки результатов измерений OMNIC. Полученные результаты сведены в Таблицу~\ref{tab:raman_results_vniia}.

\begin{table} [ht]
    \centering%
    \caption[Raman peaks of silicon bonded to glass measured by
    dispersive Raman microscope]{Положение рамановского пика кремния,
    соединённого со стеклом (дисперсионный Раман микроскоп)}%
    \label{tab:raman_results_vniia}% label всегда желательно идти после caption
    \renewcommand{\arraystretch}{1.3}%% Увеличение расстояния между рядами, для улучшения восприятия.
    \def\tabularxcolumn#1{m{#1}}
    \begin{SingleSpace}
    \begin{tabularx}{\textwidth}{@{}
    >{\raggedright}X
    >{\centering}m{0.16\textwidth}
    >{\centering}m{0.16\textwidth}
    >{\centering}m{0.165\textwidth}
    >{\centering\arraybackslash}m{0.18\textwidth}%
    @{}}
        \toprule     %%% верхняя линейка
        Марка кремния &
        Толщина Si, мкм &
        Толщина Borofloat~33, мкм &
        Температура соединения,~{\textdegree}C &
        Измеренный пик, см\textsuperscript{$-$1}\\
        \midrule
        100 КЭФ4,5 ОП & 470 & -- & -- &
        521,07\\
        100 КЭФ4,5 ДП & 460 & 600 & 300 &
        521,17\\
        100 КЭФ4,5 ОП & 470 & 400 & 400 &
        521,24\\
        \bottomrule %%% нижняя линейка& &
\end{tabularx}%
\end{SingleSpace}
\end{table}

В связи с тем, что в данной комбинации лазера и решётки не удастся получить разрешение выше, чем 1,6 см\textsuperscript{$-$1}, приведённые результаты могут служить лишь частичным качественным подтверждением увеличения
сжимающих напряжений
на поверхности кремния с увеличением температуры соединения в~представленных образцах.

Также были проведены измерения на ИК-Фурье спектрометре Thermo Scientific Nicolet iS50 FT-IR с насадкой iS50 Raman.
Для возбуждения использовался лазер с длиной волны 1064~нм (ближний инфракрасный спектр).
Мощность лазера была задана 0,1~Вт.

Определение пиков, полученных спектров проводилось с помощью поставленной с оборудованием программы обработки результатов измерений OMNIC.
Полученные результаты сведены в Таблицу~\ref{tab:raman_results_vniia_ft}.

Полученные методом ИК-Фурье спектроскопии результаты также можно
считать качественным подтверждением снижения растягивающих напряжений
на поверхности кремния с увеличением температуры соединения
в~представленных образцах.

\begin{table} [!htb]
    \centering%
    \caption[Raman peaks of silicon bonded to glass measured by FT-IR
    Raman spectrometer]{Положение рамановского пика кремния,
    соединённого со стеклом (ИК-Фурье Раман спектрометр)}%
    \label{tab:raman_results_vniia_ft}% label всегда желательно идти после caption
    \renewcommand{\arraystretch}{1.3}%% Увеличение расстояния между рядами, для улучшения восприятия.
    \def\tabularxcolumn#1{m{#1}}
    \begin{SingleSpace}
    \begin{tabularx}{\textwidth}{@{}
    >{\raggedright}X
    >{\centering}m{0.16\textwidth}
    >{\centering}m{0.16\textwidth}
    >{\centering}m{0.165\textwidth}
    >{\centering\arraybackslash}m{0.18\textwidth}%
    @{}}
        \toprule     %%% верхняя линейка
        Марка кремния &
        Толщина Si, мкм &
        Толщина Borofloat~33, мкм &
        Температура соединения,~{\textdegree}C &
        Измеренный пик, см\textsuperscript{$-$1}\\
        \midrule
        100 КЭФ4,5 ОП & 470 & --  & --  &
        520,79\\
        100 КЭФ4,5 ДП & 460 & 600 & 300 &
        520,58\\
        100 КЭФ4,5 ОП & 470 & 400 & 400 &
        520,75\\
        \bottomrule %%% нижняя линейка& &
    \end{tabularx}%
    \end{SingleSpace}
\end{table}

\section{Проведение процесса с~ограничением по~току}

\begin{figure}[htbp]
    \centering
    \begingroup%
      \makeatletter%
      \providecommand\color[2][]{%
        \errmessage{(Inkscape) Color is used for the text in Inkscape, but the package 'color.sty' is not loaded}%
        \renewcommand\color[2][]{}%
      }%
      \providecommand\transparent[1]{%
        \errmessage{(Inkscape) Transparency is used (non-zero) for the text in Inkscape, but the package 'transparent.sty' is not loaded}%
        \renewcommand\transparent[1]{}%
      }%
      \providecommand\rotatebox[2]{#2}%
      \ifx\svgwidth\undefined%
        \setlength{\unitlength}{0.65\textwidth}%
        \ifx\svgscale\undefined%
          \relax%
        \else%
          \setlength{\unitlength}{\unitlength * \real{\svgscale}}%
        \fi%
      \else%
        \setlength{\unitlength}{\svgwidth}%
      \fi%
      \global\let\svgwidth\undefined%
      \global\let\svgscale\undefined%
      \makeatother%
      \begin{picture}(1,0.76651135)%
        \put(0,0){\includegraphics[width=\unitlength]{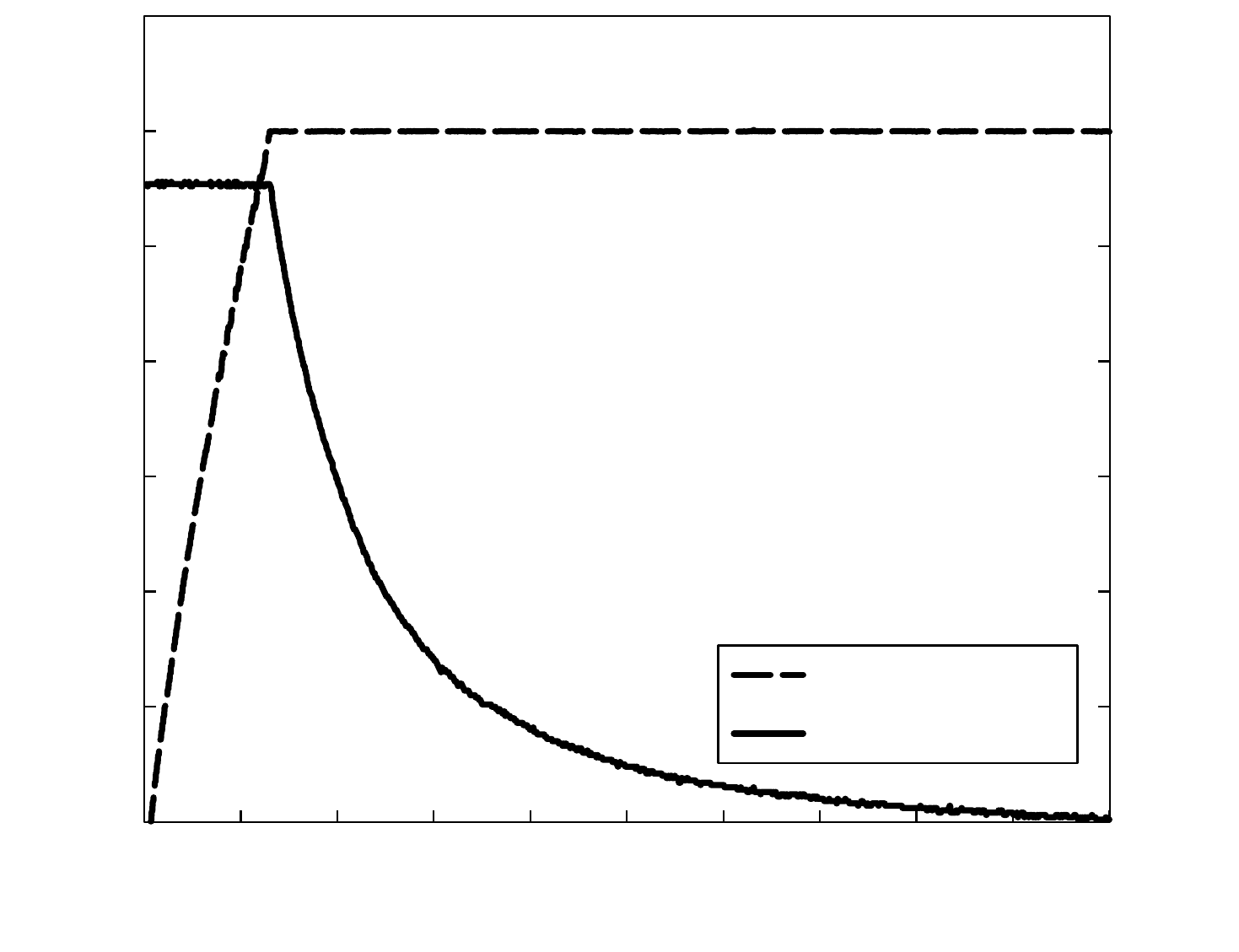}}%
        \put(0.91208943,0.09259998){\color[named]{black}\makebox(0,0)[lb]{\smash{0}}}%
        \put(0.91208943,0.18566494){\color[named]{black}\makebox(0,0)[lb]{\smash{0,5}}}%
        \put(0.91208943,0.27876279){\color[named]{black}\makebox(0,0)[lb]{\smash{1}}}%
        \put(0.91208943,0.37182771){\color[named]{black}\makebox(0,0)[lb]{\smash{1,5}}}%
        \put(0.91208943,0.46489266){\color[named]{black}\makebox(0,0)[lb]{\smash{2}}}%
        \put(0.91208943,0.55799055){\color[named]{black}\makebox(0,0)[lb]{\smash{2,5}}}%
        \put(0.91208943,0.65105543){\color[named]{black}\makebox(0,0)[lb]{\smash{3}}}%
        \put(0.91208943,0.74412039){\color[named]{black}\makebox(0,0)[lb]{\smash{3,5}}}%
        \put(0.10,0.09259998){\color[named]{black}\makebox(0,0)[rb]{\smash{400}}}%
        \put(0.10,0.18566494){\color[named]{black}\makebox(0,0)[rb]{\smash{500}}}%
        \put(0.10,0.27876279){\color[named]{black}\makebox(0,0)[rb]{\smash{600}}}%
        \put(0.10,0.37182771){\color[named]{black}\makebox(0,0)[rb]{\smash{700}}}%
        \put(0.10,0.46489266){\color[named]{black}\makebox(0,0)[rb]{\smash{800}}}%
        \put(0.10,0.55799055){\color[named]{black}\makebox(0,0)[rb]{\smash{900}}}%
        \put(0.10,0.65105543){\color[named]{black}\makebox(0,0)[rb]{\smash{1000}}}%
        \put(0.10,0.74412039){\color[named]{black}\makebox(0,0)[rb]{\smash{1100}}}%
        \put(0.11657128,0.055963){\color[named]{black}\makebox(0,0)[b]{\smash{0}}}%
        \put(0.19530171,0.055963){\color[named]{black}\makebox(0,0)[b]{\smash{3}}}%
        \put(0.27254373,0.055963){\color[named]{black}\makebox(0,0)[b]{\smash{6}}}%
        \put(0.35069174,0.055963){\color[named]{black}\makebox(0,0)[b]{\smash{9}}}%
        \put(0.42739995,0.055963){\color[named]{black}\makebox(0,0)[b]{\smash{12}}}%
        \put(0.50582292,0.055963){\color[named]{black}\makebox(0,0)[b]{\smash{15}}}%
        \put(0.58370406,0.055963){\color[named]{black}\makebox(0,0)[b]{\smash{18}}}%
        \put(0.66437591,0.055963){\color[named]{black}\makebox(0,0)[b]{\smash{21}}}%
        \put(0.74094652,0.055963){\color[named]{black}\makebox(0,0)[b]{\smash{24}}}%
        \put(0.81912689,0.055963){\color[named]{black}\makebox(0,0)[b]{\smash{27}}}%
        \put(0.89668439,0.055963){\color[named]{black}\makebox(0,0)[b]{\smash{30}}}%
        \put(1.01,0.4){\color[named]{black}\rotatebox{90}{\makebox(0,0)[b]{\smash{Ток, мА}}}}%
        \put(0.0,0.4){\color[named]{black}\rotatebox{90}{\makebox(0,0)[b]{\smash{Напряжение, В}}}}%
        \put(0.50676936,0.0070862){\color[named]{black}\makebox(0,0)[b]{\smash{Время, мин}}}%
        \put(0.65544289,0.21014445){\color[rgb]{0,0,0}\makebox(0,0)[lb]{\small\smash{Напряжение}}}%
        \put(0.65544289,0.16261668){\color[rgb]{0,0,0}\makebox(0,0)[lb]{\small\smash{Ток}}}%
      \end{picture}%
    \endgroup%
    \caption[Current and voltage during the current limited anodic bonding
    process]{Ток и напряжение в~процессе электростатического соединения
    в~режиме ограничения тока}
    \label{fig:current_limited_graph}
\end{figure}

В подразделе~\ref{chap:local_heating} были рассмотрены литературные данные по причинам и последствиям проведения процесса анодной посадки с ограничением по~току. На Рисунке~\ref{fig:current_limited_graph} приведены графики изменения тока и напряжения в процессе соединения пластин кремния и стекла. Диаметр пластин 100~мм. Марка стекла "--- Borofloat~33, толщина 700~мкм. Марка кремния "--- КДБ (кремний дырочного типа проводимости, легирующий элемент "--- бор), толщина 460~мкм.
Температура проведения процесса 300~{\textdegree}C. Напряжение 1000~В.
Величина тока ограничена 2,8~мА.
Показатели тока и напряжения снимались через аналоговый выход
\mbox{0--10~В} источника высокого напряжения FUG HCE350\nb-2000.

\section{Снижение остаточных напряжений}
Помимо выбора оптимальной температуры соединения существуют работы,
показывающие, что за счёт термообработки, следующей непосредственно за~процессом
соединения, возможно снизить возникающие остаточные напряжения. В
подразделе~\ref{chap_temp_inconsistence} описаны статьи, посвящённые работам
по~управлению величиной прогиба соединённых пластин за счёт термообработки при
температурах от~450 до~560~{\textdegree}C. Согласно опубликованным материалам,
это возможно благодаря вязкоупругим свойствам стекла.

Автор данной диссертационной работы проводил  исследования влияния
термообработки на остаточные напряжения. Использовались кристаллы кремния
размером 4\(\,\times\,\)4~мм и толщиной 0,4~мм с вытравленной полостью. Они были
предварительно электростатически соединены с пьедесталами из~стекла ЛК5 толщиной
5~мм при температурах от 300 до 400~{\textdegree}C. Ширина обода контакта
кремния со стеклом составляла 0,7~мм. В некоторых стеклянных пьедесталах были
сформированы отверстия диаметром 2~мм.

Режимы последующей термообработки приведены на Рисунке~\ref{fig:annealing_results}.
Первый вариант термообработки "--- нагрев до температуры
(415\(\pm\)30)~{\textdegree}C, выдержка 60~минут и~затем неуправляемое охлаждение.
Второй вариант термообработки "--- нагрев до температуры
(440\(\pm\)30)~{\textdegree}C, выдержка 15 минут и~затем управляемое охлаждение со
скоростью не превышающей 3~{\textdegree}C/мин. Охлаждение проводилось
до~температуры 200~{\textdegree}C, на которой проходила выдержка 30~минут. Затем
образцы охлаждались в неконтролируемом режиме.
Контроль и управление нагревом осуществлялось ПИД\nb-регулятором Термодат~13К2
на основании данных термопар типа ТХК или терморезисторов Honeywell
\mbox{HEL-707-T-0-12-00}.

Остаточные напряжения в стекле до и после термообработки измерялись
в~соответствии с процедурами, описанными в
подразделе~\ref{chap:measure_glass_stress} на~полярископе\nb-поляриметре
ПКС\nb-250. Съём показаний с полярископа осуществляла
М.\,Г.~Лукоперова.
Результаты обработки измерений приведены на Рисунке~\ref{fig:annealing_results}.
Среднеарифметическое значение остаточных напряжений в~стекле было снижено более
чем в два раза в каждом из вариантов термообработки.

\begin{figure}[!htb]
    \centering%
    \noindent%
    \ifdefmacro{\tikzsetnextfilename}{\tikzsetnextfilename{annealing_results_p1}}{}%
    \input{Dissertation/images_tikz/disser_tt1+tt2.tikz}%
    \par%
    \noindent%
    \ifdefmacro{\tikzsetnextfilename}{\tikzsetnextfilename{annealing_results_p2}}{}%
    \input{Dissertation/images_tikz/disser_tt_result.tikz}%
    \par%
    \caption[Residual stress decrease by thermal treatment]{Снижение остаточных
        напряжений термообработкой}
    \label{fig:annealing_results}
\end{figure}

Были проведены исследования возможности снижения остаточных напряжений в
процессе остывания сборок непосредственно по окончании процесса, за~счёт
контролируемой скорости охлаждения. По сравнению с отдельной термообработкой
происходит выигрыш во времени на охлаждение после процесса и~дальнейший нагрев с
выдержкой, но используется более щадящий режим охлаждения.

Исследовалось влияние контролируемого охлаждения со скоростью не~выше
2~{\textdegree}C/мин сразу после окончания подачи высокого напряжения
до~температуры 200~{\textdegree}C, а затем неуправляемого охлаждения. Ограничением
такого способа является снижение его потенциального воздействия со~снижением
температуры проведения процесса.

На Рисунке~\ref{fig:cooling_control_temp} приведена кривая исследованного
процесса управляемого охлаждения. Образцы, аналогичные исследовавшимся при
отдельной термообработке, выдерживались не менее 15 минут при температуре
(440$\pm$30)~{\textdegree}C, затем проводилось электростатическое соединение,
занимавшее от~10~до~15~минут. По~окончании подачи высокого напряжения начиналось
контролируемое охлаждение со~скоростью не выше 2~{\textdegree}C/мин до
200~{\textdegree}C, после чего охлаждением не~управляли.

\begin{figure}[!htb]
    \centering%
    \noindent%
    \ifdefmacro{\tikzsetnextfilename}{\tikzsetnextfilename{cooling_control_temp_p1}}{}%
    \input{Dissertation/images_tikz/disser_treat_g.tikz}%
    \par%
    \noindent%
    \ifdefmacro{\tikzsetnextfilename}{\tikzsetnextfilename{cooling_control_temp_p2}}{}%
    \input{Dissertation/images_tikz/disser_tt_graph2.tikz}%
    \par%
    \caption[Residual stress decrease as a result of the controllable speed
    of cooling]{Влияние скорости охлаждения}
    \label{fig:cooling_control_temp}
\end{figure}

На сходных с описанными выше образцами проводились процессы с~неуправляемым
охлаждением (36 образцов) и с управляемым охлаждением (36~образцов) при
одинаковой температуре сращивания.
Среднеарифметическая величина снижения остаточных напряжений в стекле составила
63~\% от~исходного значения.

Результаты экспериментов продемонстрировали возможность снижения внутренних
механических напряжений при помощи термообработки и~возможность совмещения
процессов соединения и последующего отжига. Для стекла ЛК5 возможно снижать
напряжения за счёт охлаждения со скоростью \mbox{2~{\textdegree}C/мин}.
Предположительно, этот вывод можно распространить и~на~другие боросиликатные
стёкла, совместимые с анодной посадкой: Schott Borofloat~33 и~Corning~7740.

Проведённые эксперименты не продемонстрировали возможности изменения изгиба
пластин при обозначенном режиме охлаждения. Тем не~менее при~возникновении
такой необходимости рекомендуется рассмотреть возможности многочасовой
термообработки при режимах, описанных
в~\cite{Kim2015warpage}~и~\cite{Harz1996Curvature_changing}.

\section{Шаги минимизации остаточных напряжений}\label{metod_minim_ost_napr}

На основании описанных в главе 3 примеров применения моделей оценки остаточных напряжений можно предложить следующие шаги минимизации напряжений при разработке конструкции приборов электронной техники:
\begin{enumerate}
    \item Определить рабочий диапазон температур прибора и верхнюю границу температуры соединения.
    \item Получить термомеханические характеристики кремния и стекла, доступных к~применению.
    \item\label{shag_vybora_stekla} Выбрать марку стекла, сравнив эти характеристики.
    \item По модели двух тонких слоёв примерно определить температуру проведения процесса.
    \item При необходимости, учесть несимметричность распределения напряжений в рабочем диапазоне температур, соответствующим образом подобрав температуру соединения.
    \item Выбрать толщину стекла
    (с учётом поправки на площадь соединения),
    если конструкция позволяет.
    \item По возможности, провести более точное моделирование методом конечных элементов.
    \item При необходимости повторить шаги, начиная с~\ref{shag_vybora_stekla}.
\end{enumerate}

Суть шагов сводится к расчётному учёту важных для прибора характеристик с постепенно возрастающей вычислительной сложностью. Следует помнить, что от партии к партии свойства стекла могут отличаться. У стёкол отечественного производства такие отличия могут быть значительными~\cite{timoshenkov2009_issled_stekol}.

\ifnumequal{\value{usealtfont}}{2}{%
    \clearpage
}{}
\section{Технологические рекомендации по~снижению остаточных напряжений}

На основании литературных данных и опыта, описанного в этой главе, можно дать следующие технологические рекомендации по снижению остаточных напряжений, возникающих в процессе электростатического соединения:
\begin{enumerate}
    \item Убедиться в адекватности измеряемой датчиками температуры,
    при необходимости воспользоваться поправками.
    \item Выдерживать детали при температуре соединения не менее 15
    минут для выравнивания температурного поля.
    \item Использовать ограничение тока в процессе проведения
    электростатического соединения для снижения риска локального
    перегрева стыка кремний"--~стекло.
    \item Сразу после окончания подачи высокого напряжения проводить
    контролируемое охлаждение сборки со скоростью не выше
    2~{\textdegree}C/мин начиная с температуры соединения или с более
    высокой температуры (предварительно нагрев до неё, и выдержав на
    ней не менее 30~мин).
\end{enumerate}

\section{Выводы по главе 4}
\begin{enumerate}
    \item Для стёкол Borofloat 33 и ЛК5 возможно снижать остаточные
    напряжения за счёт охлаждения со скоростью 2~{\textdegree}C/мин.
    Вероятно, этот вывод можно распространить и на стёкла других
    марок.
    \item Для минимизации остаточных напряжений рекомендуется
    применять разработанную методику выбора параметров процесса
    электростатического соединения и подбора конструктивных решений
    (подраздел~\ref{metod_minim_ost_napr}).
    \item Для снижения риска локального перегрева стыка
    кремний\nb-стекло рекомендуется использовать ограничение плотности
    тока 0,4 А/м{\textsuperscript{2}} в~процессе проведения
    электростатического соединения пластин.
\end{enumerate}

%% file: Dissertation/images_tikz/disser_tt1+tt2.tikz
\begin{tikzpicture}[x=1pt,y=1pt]
\definecolor{fillColor}{RGB}{255,255,255}
\path[use as bounding box,fill=fillColor] (0,0) rectangle (426.79,192.06);
\begin{scope}
\path[clip] (  0.00,  0.00) rectangle (213.40,192.06);
\definecolor{drawColor}{RGB}{255,255,255}

\path[draw=drawColor,line width= 0.6pt,line join=round,line cap=round,fill=fillColor] ( -0.00,  0.00) rectangle (213.40,192.06);
\end{scope}
\begin{scope}
\path[clip] ( 39.23, 29.60) rectangle (206.65,192.06);
\definecolor{fillColor}{RGB}{255,255,255}

\path[fill=fillColor] ( 39.23, 29.60) rectangle (206.65,192.06);
\definecolor{drawColor}{gray}{0.98}

\path[draw=drawColor,line width= 0.6pt,line join=round] ( 39.23, 48.78) --
	(206.65, 48.78);

\path[draw=drawColor,line width= 0.6pt,line join=round] ( 39.23, 80.97) --
	(206.65, 80.97);

\path[draw=drawColor,line width= 0.6pt,line join=round] ( 39.23,116.73) --
	(206.65,116.73);

\path[draw=drawColor,line width= 0.6pt,line join=round] ( 39.23,152.49) --
	(206.65,152.49);

\path[draw=drawColor,line width= 0.6pt,line join=round] ( 39.23,177.52) --
	(206.65,177.52);

\path[draw=drawColor,line width= 0.6pt,line join=round] ( 72.78, 29.60) --
	( 72.78,192.06);

\path[draw=drawColor,line width= 0.6pt,line join=round] (124.67, 29.60) --
	(124.67,192.06);

\path[draw=drawColor,line width= 0.6pt,line join=round] (176.56, 29.60) --
	(176.56,192.06);
\definecolor{drawColor}{gray}{0.80}

\path[draw=drawColor,line width= 0.3pt,line join=round] ( 39.23, 34.48) --
	(206.65, 34.48);

\path[draw=drawColor,line width= 0.3pt,line join=round] ( 39.23, 63.09) --
	(206.65, 63.09);

\path[draw=drawColor,line width= 0.3pt,line join=round] ( 39.23, 98.85) --
	(206.65, 98.85);

\path[draw=drawColor,line width= 0.3pt,line join=round] ( 39.23,134.61) --
	(206.65,134.61);

\path[draw=drawColor,line width= 0.3pt,line join=round] ( 39.23,170.37) --
	(206.65,170.37);

\path[draw=drawColor,line width= 0.3pt,line join=round] ( 39.23,184.67) --
	(206.65,184.67);

\path[draw=drawColor,line width= 0.3pt,line join=round] ( 46.84, 29.60) --
	( 46.84,192.06);

\path[draw=drawColor,line width= 0.3pt,line join=round] ( 98.72, 29.60) --
	( 98.72,192.06);

\path[draw=drawColor,line width= 0.3pt,line join=round] (150.61, 29.60) --
	(150.61,192.06);

\path[draw=drawColor,line width= 0.3pt,line join=round] (202.50, 29.60) --
	(202.50,192.06);
\definecolor{drawColor}{RGB}{0,0,0}

\path[draw=drawColor,line width= 1.3pt,line join=round] ( 46.84, 36.98) --
	( 47.34, 41.48) --
	( 47.85, 45.97) --
	( 48.35, 50.46) --
	( 48.85, 54.95) --
	( 49.36, 59.44) --
	( 49.86, 63.94) --
	( 50.37, 68.43) --
	( 50.87, 72.92) --
	( 51.37, 77.41) --
	( 51.88, 81.91) --
	( 52.38, 86.40) --
	( 52.89, 90.89) --
	( 53.39, 95.38) --
	( 53.89, 99.87) --
	( 54.40,104.37) --
	( 54.90,108.86) --
	( 55.40,113.35) --
	( 55.91,117.84) --
	( 56.41,122.34) --
	( 56.92,126.83) --
	( 57.42,131.32) --
	( 57.92,135.81) --
	( 58.43,140.30) --
	( 58.93,144.80) --
	( 59.44,149.29) --
	( 59.94,153.78) --
	( 60.44,158.27) --
	( 60.95,162.77) --
	( 61.45,167.26) --
	( 61.96,171.75) --
	( 62.46,175.73) --
	( 62.96,175.73) --
	( 63.47,175.73) --
	( 63.97,175.73) --
	( 64.48,175.73) --
	( 64.98,175.73) --
	( 65.48,175.73) --
	( 65.99,175.73) --
	( 66.49,175.73) --
	( 67.00,175.73) --
	( 67.50,175.73) --
	( 68.00,175.73) --
	( 68.51,175.73) --
	( 69.01,175.73) --
	( 69.52,175.73) --
	( 70.02,175.73) --
	( 70.52,175.73) --
	( 71.03,175.73) --
	( 71.53,175.73) --
	( 72.04,175.73) --
	( 72.54,175.73) --
	( 73.04,175.73) --
	( 73.55,175.73) --
	( 74.05,175.73) --
	( 74.56,175.73) --
	( 75.06,175.73) --
	( 75.56,175.73) --
	( 76.07,175.73) --
	( 76.57,175.73) --
	( 77.08,175.73) --
	( 77.58,175.73) --
	( 78.08,175.73) --
	( 78.59,175.73) --
	( 79.09,175.73) --
	( 79.60,175.73) --
	( 80.10,175.73) --
	( 80.60,175.73) --
	( 81.11,175.73) --
	( 81.61,175.73) --
	( 82.12,175.73) --
	( 82.62,175.73) --
	( 83.12,175.73) --
	( 83.63,175.73) --
	( 84.13,175.73) --
	( 84.64,175.73) --
	( 85.14,175.73) --
	( 85.64,175.73) --
	( 86.15,175.73) --
	( 86.65,175.73) --
	( 87.16,175.73) --
	( 87.66,175.73) --
	( 88.16,175.73) --
	( 88.67,175.73) --
	( 89.17,175.73) --
	( 89.68,175.73) --
	( 90.18,175.73) --
	( 90.68,175.73) --
	( 91.19,175.73) --
	( 91.69,175.73) --
	( 92.20,175.73) --
	( 92.70,175.73) --
	( 93.20,175.73) --
	( 93.71,175.73) --
	( 94.21,175.73) --
	( 94.72,175.73) --
	( 95.22,175.73) --
	( 95.72,175.73) --
	( 96.23,175.73) --
	( 96.73,175.73) --
	( 97.24,175.73) --
	( 97.74,175.73) --
	( 98.24,175.73) --
	( 98.75,175.73) --
	( 99.25,175.73) --
	( 99.76,175.73) --
	(100.26,175.73) --
	(100.76,175.73) --
	(101.27,175.73) --
	(101.77,175.73) --
	(102.28,175.73) --
	(102.78,175.73) --
	(103.28,175.73) --
	(103.79,175.73) --
	(104.29,175.73) --
	(104.80,175.73) --
	(105.30,175.73) --
	(105.80,175.73) --
	(106.31,175.73) --
	(106.81,175.73) --
	(107.32,175.73) --
	(107.82,175.73) --
	(108.32,175.73) --
	(108.83,175.73) --
	(109.33,175.73) --
	(109.84,175.73) --
	(110.34,175.73) --
	(110.84,175.73) --
	(111.35,175.73) --
	(111.85,175.73) --
	(112.36,175.73) --
	(112.86,175.73) --
	(113.36,175.73) --
	(113.87,175.73) --
	(114.37,175.08) --
	(114.88,171.02) --
	(115.38,167.05) --
	(115.88,163.18) --
	(116.39,159.40) --
	(116.89,155.70) --
	(117.39,152.10) --
	(117.90,148.58) --
	(118.40,145.15) --
	(118.91,141.81) --
	(119.41,138.54) --
	(119.91,135.36) --
	(120.42,132.26) --
	(120.92,129.24) --
	(121.43,126.29) --
	(121.93,123.43) --
	(122.43,120.63) --
	(122.94,117.91) --
	(123.44,115.26) --
	(123.95,112.68) --
	(124.45,110.18) --
	(124.95,107.74) --
	(125.46,105.36) --
	(125.96,103.06) --
	(126.47,100.82) --
	(126.97, 98.64) --
	(127.47, 96.52) --
	(127.98, 94.47) --
	(128.48, 92.47) --
	(128.99, 90.53) --
	(129.49, 88.66) --
	(129.99, 86.83) --
	(130.50, 85.06) --
	(131.00, 83.35) --
	(131.51, 81.69) --
	(132.01, 80.08) --
	(132.51, 78.52) --
	(133.02, 77.02) --
	(133.52, 75.56) --
	(134.03, 74.14) --
	(134.53, 72.78) --
	(135.03, 71.46) --
	(135.54, 70.18) --
	(136.04, 68.95) --
	(136.55, 67.76) --
	(137.05, 66.61) --
	(137.55, 65.51) --
	(138.06, 64.44) --
	(138.56, 63.41) --
	(139.07, 62.42) --
	(139.57, 61.46) --
	(140.07, 60.54) --
	(140.58, 59.66) --
	(141.08, 58.81) --
	(141.59, 57.99) --
	(142.09, 57.20) --
	(142.59, 56.45) --
	(143.10, 55.73) --
	(143.60, 55.03) --
	(144.11, 54.37) --
	(144.61, 53.73) --
	(145.11, 53.12) --
	(145.62, 52.54) --
	(146.12, 51.98) --
	(146.63, 51.45) --
	(147.13, 50.94) --
	(147.63, 50.46) --
	(148.14, 50.00) --
	(148.64, 49.56) --
	(149.15, 49.14) --
	(149.65, 48.74) --
	(150.15, 48.36) --
	(150.66, 48.00) --
	(151.16, 47.66) --
	(151.67, 47.34) --
	(152.17, 47.04) --
	(152.67, 46.75) --
	(153.18, 46.47) --
	(153.68, 46.22) --
	(154.19, 45.97) --
	(154.69, 45.75) --
	(155.19, 45.53) --
	(155.70, 45.33) --
	(156.20, 45.14) --
	(156.71, 44.97) --
	(157.21, 44.80) --
	(157.71, 44.65) --
	(158.22, 44.50) --
	(158.72, 44.37) --
	(159.23, 44.24) --
	(159.73, 44.13) --
	(160.23, 44.02) --
	(160.74, 43.92) --
	(161.24, 43.83) --
	(161.75, 43.75) --
	(162.25, 43.67) --
	(162.75, 43.60) --
	(163.26, 43.54) --
	(163.76, 43.48) --
	(164.27, 43.43) --
	(164.77, 43.38) --
	(165.27, 43.34) --
	(165.78, 43.30);

\path[draw=drawColor,line width= 0.9pt,line join=round,line cap=round] ( 39.23, 29.60) rectangle (206.65,192.06);
\end{scope}
\begin{scope}
\path[clip] (  0.00,  0.00) rectangle (426.79,192.06);
\definecolor{drawColor}{RGB}{0,0,0}

\node[text=drawColor,anchor=base east,inner sep=0pt, outer sep=0pt, scale=  0.86] at ( 33.83, 30.23) {\(20\)};

\node[text=drawColor,anchor=base east,inner sep=0pt, outer sep=0pt, scale=  0.86] at ( 33.83, 58.84) {\(100\)};

\node[text=drawColor,anchor=base east,inner sep=0pt, outer sep=0pt, scale=  0.86] at ( 33.83, 94.60) {\(200\)};

\node[text=drawColor,anchor=base east,inner sep=0pt, outer sep=0pt, scale=  0.86] at ( 33.83,130.36) {\(300\)};

\node[text=drawColor,anchor=base east,inner sep=0pt, outer sep=0pt, scale=  0.86] at ( 33.83,166.12) {\(400\)};

\node[text=drawColor,anchor=base east,inner sep=0pt, outer sep=0pt, scale=  0.86] at ( 33.83,180.42) {\(440\)};
\end{scope}
\begin{scope}
\path[clip] (  0.00,  0.00) rectangle (426.79,192.06);
\definecolor{drawColor}{RGB}{0,0,0}

\path[draw=drawColor,line width= 0.6pt,line join=round] ( 36.23, 34.48) --
	( 39.23, 34.48);

\path[draw=drawColor,line width= 0.6pt,line join=round] ( 36.23, 63.09) --
	( 39.23, 63.09);

\path[draw=drawColor,line width= 0.6pt,line join=round] ( 36.23, 98.85) --
	( 39.23, 98.85);

\path[draw=drawColor,line width= 0.6pt,line join=round] ( 36.23,134.61) --
	( 39.23,134.61);

\path[draw=drawColor,line width= 0.6pt,line join=round] ( 36.23,170.37) --
	( 39.23,170.37);

\path[draw=drawColor,line width= 0.6pt,line join=round] ( 36.23,184.67) --
	( 39.23,184.67);
\end{scope}
\begin{scope}
\path[clip] (  0.00,  0.00) rectangle (426.79,192.06);
\definecolor{drawColor}{RGB}{0,0,0}

\path[draw=drawColor,line width= 0.6pt,line join=round] ( 46.84, 26.60) --
	( 46.84, 29.60);

\path[draw=drawColor,line width= 0.6pt,line join=round] ( 98.72, 26.60) --
	( 98.72, 29.60);

\path[draw=drawColor,line width= 0.6pt,line join=round] (150.61, 26.60) --
	(150.61, 29.60);

\path[draw=drawColor,line width= 0.6pt,line join=round] (202.50, 26.60) --
	(202.50, 29.60);
\end{scope}
\begin{scope}
\path[clip] (  0.00,  0.00) rectangle (426.79,192.06);
\definecolor{drawColor}{RGB}{0,0,0}

\node[text=drawColor,anchor=base,inner sep=0pt, outer sep=0pt, scale=  0.86] at ( 46.84, 15.70) {\(0\)};

\node[text=drawColor,anchor=base,inner sep=0pt, outer sep=0pt, scale=  0.86] at ( 98.72, 15.70) {\(60\)};

\node[text=drawColor,anchor=base,inner sep=0pt, outer sep=0pt, scale=  0.86] at (150.61, 15.70) {\(120\)};

\node[text=drawColor,anchor=base,inner sep=0pt, outer sep=0pt, scale=  0.86] at (202.50, 15.70) {\(180\)};
\end{scope}
\begin{scope}
\path[clip] (  0.00,  0.00) rectangle (426.79,192.06);
\definecolor{drawColor}{RGB}{0,0,0}

\node[text=drawColor,anchor=base,inner sep=0pt, outer sep=0pt, scale=  0.86] at (122.94,  2.40) {Время, мин};
\end{scope}
\begin{scope}
\path[clip] (  0.00,  0.00) rectangle (426.79,192.06);
\definecolor{drawColor}{RGB}{0,0,0}

\node[text=drawColor,rotate= 90.00,anchor=base,inner sep=0pt, outer sep=0pt, scale=  0.86] at ( 10.90,110.83) {Температура, {\textdegree}C\hspace*{1.7em}};
\end{scope}
\begin{scope}
\path[clip] (213.40,  0.00) rectangle (426.79,192.06);
\definecolor{drawColor}{RGB}{255,255,255}
\definecolor{fillColor}{RGB}{255,255,255}

\path[draw=drawColor,line width= 0.6pt,line join=round,line cap=round,fill=fillColor] (213.40,  0.00) rectangle (426.79,192.06);
\end{scope}
\begin{scope}
\path[clip] (252.62, 29.60) rectangle (420.05,192.06);
\definecolor{fillColor}{RGB}{255,255,255}

\path[fill=fillColor] (252.62, 29.60) rectangle (420.05,192.06);
\definecolor{drawColor}{gray}{0.98}

\path[draw=drawColor,line width= 0.6pt,line join=round] (252.62, 48.78) --
	(420.05, 48.78);

\path[draw=drawColor,line width= 0.6pt,line join=round] (252.62, 80.97) --
	(420.05, 80.97);

\path[draw=drawColor,line width= 0.6pt,line join=round] (252.62,116.73) --
	(420.05,116.73);

\path[draw=drawColor,line width= 0.6pt,line join=round] (252.62,152.49) --
	(420.05,152.49);

\path[draw=drawColor,line width= 0.6pt,line join=round] (252.62,177.52) --
	(420.05,177.52);

\path[draw=drawColor,line width= 0.6pt,line join=round] (286.18, 29.60) --
	(286.18,192.06);

\path[draw=drawColor,line width= 0.6pt,line join=round] (338.06, 29.60) --
	(338.06,192.06);

\path[draw=drawColor,line width= 0.6pt,line join=round] (389.95, 29.60) --
	(389.95,192.06);
\definecolor{drawColor}{gray}{0.80}

\path[draw=drawColor,line width= 0.3pt,line join=round] (252.62, 34.48) --
	(420.05, 34.48);

\path[draw=drawColor,line width= 0.3pt,line join=round] (252.62, 63.09) --
	(420.05, 63.09);

\path[draw=drawColor,line width= 0.3pt,line join=round] (252.62, 98.85) --
	(420.05, 98.85);

\path[draw=drawColor,line width= 0.3pt,line join=round] (252.62,134.61) --
	(420.05,134.61);

\path[draw=drawColor,line width= 0.3pt,line join=round] (252.62,170.37) --
	(420.05,170.37);

\path[draw=drawColor,line width= 0.3pt,line join=round] (252.62,184.67) --
	(420.05,184.67);

\path[draw=drawColor,line width= 0.3pt,line join=round] (260.23, 29.60) --
	(260.23,192.06);

\path[draw=drawColor,line width= 0.3pt,line join=round] (312.12, 29.60) --
	(312.12,192.06);

\path[draw=drawColor,line width= 0.3pt,line join=round] (364.01, 29.60) --
	(364.01,192.06);

\path[draw=drawColor,line width= 0.3pt,line join=round] (415.90, 29.60) --
	(415.90,192.06);
\definecolor{drawColor}{RGB}{0,0,0}

\path[draw=drawColor,line width= 1.3pt,line join=round] (260.23, 36.98) --
	(260.74, 42.05) --
	(261.24, 47.11) --
	(261.74, 52.17) --
	(262.25, 57.24) --
	(262.75, 62.30) --
	(263.26, 67.36) --
	(263.76, 72.42) --
	(264.26, 77.49) --
	(264.77, 82.55) --
	(265.27, 87.61) --
	(265.78, 92.68) --
	(266.28, 97.74) --
	(266.78,102.80) --
	(267.29,107.86) --
	(267.79,112.93) --
	(268.30,117.99) --
	(268.80,123.05) --
	(269.30,128.12) --
	(269.81,133.18) --
	(270.31,138.24) --
	(270.82,143.31) --
	(271.32,148.37) --
	(271.82,153.43) --
	(272.33,158.49) --
	(272.83,163.56) --
	(273.34,168.62) --
	(273.84,173.68) --
	(274.34,178.75) --
	(274.85,183.81) --
	(275.35,184.67) --
	(275.86,184.67) --
	(276.36,184.67) --
	(276.86,184.67) --
	(277.37,184.67) --
	(277.87,184.67) --
	(278.38,184.67) --
	(278.88,184.67) --
	(279.38,184.67) --
	(279.89,184.67) --
	(280.39,184.67) --
	(280.90,184.67) --
	(281.40,184.67) --
	(281.90,184.67) --
	(282.41,184.67) --
	(282.91,184.67) --
	(283.42,184.67) --
	(283.92,184.67) --
	(284.42,184.67) --
	(284.93,184.67) --
	(285.43,184.67) --
	(285.94,184.67) --
	(286.44,184.67) --
	(286.94,184.67) --
	(287.45,184.67) --
	(287.95,184.67) --
	(288.46,184.67) --
	(288.96,184.67) --
	(289.46,184.67) --
	(289.97,184.67) --
	(290.47,184.67) --
	(290.98,184.08) --
	(291.48,183.46) --
	(291.98,182.83) --
	(292.49,182.21) --
	(292.99,181.58) --
	(293.50,180.96) --
	(294.00,180.33) --
	(294.50,179.71) --
	(295.01,179.08) --
	(295.51,178.46) --
	(296.02,177.83) --
	(296.52,177.20) --
	(297.02,176.58) --
	(297.53,175.95) --
	(298.03,175.33) --
	(298.54,174.70) --
	(299.04,174.08) --
	(299.54,173.45) --
	(300.05,172.83) --
	(300.55,172.20) --
	(301.06,171.58) --
	(301.56,170.95) --
	(302.06,170.33) --
	(302.57,169.70) --
	(303.07,169.08) --
	(303.58,168.45) --
	(304.08,167.83) --
	(304.58,167.20) --
	(305.09,166.58) --
	(305.59,165.95) --
	(306.10,165.33) --
	(306.60,164.70) --
	(307.10,164.08) --
	(307.61,163.45) --
	(308.11,162.83) --
	(308.62,162.20) --
	(309.12,161.57) --
	(309.62,160.95) --
	(310.13,160.32) --
	(310.63,159.70) --
	(311.14,159.07) --
	(311.64,158.45) --
	(312.14,157.82) --
	(312.65,157.20) --
	(313.15,156.57) --
	(313.66,155.95) --
	(314.16,155.32) --
	(314.66,154.70) --
	(315.17,154.07) --
	(315.67,153.45) --
	(316.18,152.82) --
	(316.68,152.20) --
	(317.18,151.57) --
	(317.69,150.95) --
	(318.19,150.32) --
	(318.70,149.70) --
	(319.20,149.07) --
	(319.70,148.45) --
	(320.21,147.82) --
	(320.71,147.19) --
	(321.21,146.57) --
	(321.72,145.94) --
	(322.22,145.32) --
	(322.73,144.69) --
	(323.23,144.07) --
	(323.73,143.44) --
	(324.24,142.82) --
	(324.74,142.19) --
	(325.25,141.57) --
	(325.75,140.94) --
	(326.25,140.32) --
	(326.76,139.69) --
	(327.26,139.07) --
	(327.77,138.44) --
	(328.27,137.82) --
	(328.77,137.19) --
	(329.28,136.57) --
	(329.78,135.94) --
	(330.29,135.32) --
	(330.79,134.69) --
	(331.29,134.07) --
	(331.80,133.44) --
	(332.30,132.82) --
	(332.81,132.19) --
	(333.31,131.56) --
	(333.81,130.94) --
	(334.32,130.31) --
	(334.82,129.69) --
	(335.33,129.06) --
	(335.83,128.44) --
	(336.33,127.81) --
	(336.84,127.19) --
	(337.34,126.56) --
	(337.85,125.94) --
	(338.35,125.31) --
	(338.85,124.69) --
	(339.36,124.06) --
	(339.86,123.44) --
	(340.37,122.81) --
	(340.87,122.19) --
	(341.37,121.56) --
	(341.88,120.94) --
	(342.38,120.31) --
	(342.89,119.69) --
	(343.39,119.06) --
	(343.89,118.44) --
	(344.40,117.81) --
	(344.90,117.19) --
	(345.41,116.56) --
	(345.91,115.93) --
	(346.41,115.31) --
	(346.92,114.68) --
	(347.42,114.06) --
	(347.93,113.43) --
	(348.43,112.81) --
	(348.93,112.18) --
	(349.44,111.56) --
	(349.94,110.93) --
	(350.45,110.31) --
	(350.95,109.68) --
	(351.45,109.06) --
	(351.96,108.43) --
	(352.46,107.81) --
	(352.97,107.18) --
	(353.47,106.56) --
	(353.97,105.93) --
	(354.48,105.31) --
	(354.98,104.68) --
	(355.49,104.06) --
	(355.99,103.43) --
	(356.49,102.81) --
	(357.00,102.18) --
	(357.50,101.55) --
	(358.01,100.93) --
	(358.51,100.30) --
	(359.01, 99.68) --
	(359.52, 99.05) --
	(360.02, 98.85) --
	(360.53, 98.85) --
	(361.03, 98.85) --
	(361.53, 98.85) --
	(362.04, 98.85) --
	(362.54, 98.85) --
	(363.05, 98.85) --
	(363.55, 98.85) --
	(364.05, 98.85) --
	(364.56, 98.85) --
	(365.06, 98.85) --
	(365.57, 98.85) --
	(366.07, 98.85) --
	(366.57, 98.85) --
	(367.08, 98.85) --
	(367.58, 98.85) --
	(368.09, 98.85) --
	(368.59, 98.85) --
	(369.09, 98.85) --
	(369.60, 98.85) --
	(370.10, 98.85) --
	(370.61, 98.85) --
	(371.11, 98.85) --
	(371.61, 98.85) --
	(372.12, 98.85) --
	(372.62, 98.85) --
	(373.13, 98.85) --
	(373.63, 98.85) --
	(374.13, 98.85) --
	(374.64, 98.85) --
	(375.14, 98.85) --
	(375.65, 98.85) --
	(376.15, 98.85) --
	(376.65, 98.85) --
	(377.16, 98.85) --
	(377.66, 98.85) --
	(378.17, 98.85) --
	(378.67, 98.85) --
	(379.17, 98.85) --
	(379.68, 98.85) --
	(380.18, 98.85) --
	(380.69, 98.85) --
	(381.19, 98.85) --
	(381.69, 98.85) --
	(382.20, 98.72) --
	(382.70, 96.51) --
	(383.20, 94.36) --
	(383.71, 92.27) --
	(384.21, 90.24) --
	(384.72, 88.28) --
	(385.22, 86.37) --
	(385.72, 84.52) --
	(386.23, 82.73) --
	(386.73, 81.00) --
	(387.24, 79.32) --
	(387.74, 77.69) --
	(388.24, 76.12) --
	(388.75, 74.59) --
	(389.25, 73.12) --
	(389.76, 71.70) --
	(390.26, 70.32) --
	(390.76, 68.99) --
	(391.27, 67.70) --
	(391.77, 66.46) --
	(392.28, 65.27) --
	(392.78, 64.11) --
	(393.28, 63.00) --
	(393.79, 61.93) --
	(394.29, 60.90) --
	(394.80, 59.90) --
	(395.30, 58.95) --
	(395.80, 58.03) --
	(396.31, 57.14) --
	(396.81, 56.29) --
	(397.32, 55.48) --
	(397.82, 54.70) --
	(398.32, 53.94) --
	(398.83, 53.23) --
	(399.33, 52.54) --
	(399.84, 51.88) --
	(400.34, 51.24) --
	(400.84, 50.64) --
	(401.35, 50.06) --
	(401.85, 49.51) --
	(402.36, 48.99) --
	(402.86, 48.49) --
	(403.36, 48.01) --
	(403.87, 47.55) --
	(404.37, 47.12) --
	(404.88, 46.71) --
	(405.38, 46.32) --
	(405.88, 45.95) --
	(406.39, 45.60) --
	(406.89, 45.27) --
	(407.40, 44.95) --
	(407.90, 44.66) --
	(408.40, 44.38) --
	(408.91, 44.11) --
	(409.41, 43.86) --
	(409.92, 43.63) --
	(410.42, 43.41) --
	(410.92, 43.21) --
	(411.43, 43.01) --
	(411.93, 42.83) --
	(412.44, 42.66);

\path[draw=drawColor,line width= 0.9pt,line join=round,line cap=round] (252.62, 29.60) rectangle (420.05,192.06);
\end{scope}
\begin{scope}
\path[clip] (  0.00,  0.00) rectangle (426.79,192.06);
\definecolor{drawColor}{RGB}{0,0,0}

\node[text=drawColor,anchor=base east,inner sep=0pt, outer sep=0pt, scale=  0.86] at (247.22, 30.23) {\(20\)};

\node[text=drawColor,anchor=base east,inner sep=0pt, outer sep=0pt, scale=  0.86] at (247.22, 58.84) {\(100\)};

\node[text=drawColor,anchor=base east,inner sep=0pt, outer sep=0pt, scale=  0.86] at (247.22, 94.60) {\(200\)};

\node[text=drawColor,anchor=base east,inner sep=0pt, outer sep=0pt, scale=  0.86] at (247.22,130.36) {\(300\)};

\node[text=drawColor,anchor=base east,inner sep=0pt, outer sep=0pt, scale=  0.86] at (247.22,166.12) {\(400\)};

\node[text=drawColor,anchor=base east,inner sep=0pt, outer sep=0pt, scale=  0.86] at (247.22,180.42) {\(440\)};
\end{scope}
\begin{scope}
\path[clip] (  0.00,  0.00) rectangle (426.79,192.06);
\definecolor{drawColor}{RGB}{0,0,0}

\path[draw=drawColor,line width= 0.6pt,line join=round] (249.62, 34.48) --
	(252.62, 34.48);

\path[draw=drawColor,line width= 0.6pt,line join=round] (249.62, 63.09) --
	(252.62, 63.09);

\path[draw=drawColor,line width= 0.6pt,line join=round] (249.62, 98.85) --
	(252.62, 98.85);

\path[draw=drawColor,line width= 0.6pt,line join=round] (249.62,134.61) --
	(252.62,134.61);

\path[draw=drawColor,line width= 0.6pt,line join=round] (249.62,170.37) --
	(252.62,170.37);

\path[draw=drawColor,line width= 0.6pt,line join=round] (249.62,184.67) --
	(252.62,184.67);
\end{scope}
\begin{scope}
\path[clip] (  0.00,  0.00) rectangle (426.79,192.06);
\definecolor{drawColor}{RGB}{0,0,0}

\path[draw=drawColor,line width= 0.6pt,line join=round] (260.23, 26.60) --
	(260.23, 29.60);

\path[draw=drawColor,line width= 0.6pt,line join=round] (312.12, 26.60) --
	(312.12, 29.60);

\path[draw=drawColor,line width= 0.6pt,line join=round] (364.01, 26.60) --
	(364.01, 29.60);

\path[draw=drawColor,line width= 0.6pt,line join=round] (415.90, 26.60) --
	(415.90, 29.60);
\end{scope}
\begin{scope}
\path[clip] (  0.00,  0.00) rectangle (426.79,192.06);
\definecolor{drawColor}{RGB}{0,0,0}

\node[text=drawColor,anchor=base,inner sep=0pt, outer sep=0pt, scale=  0.86] at (260.23, 15.70) {\(0\)};

\node[text=drawColor,anchor=base,inner sep=0pt, outer sep=0pt, scale=  0.86] at (312.12, 15.70) {\(60\)};

\node[text=drawColor,anchor=base,inner sep=0pt, outer sep=0pt, scale=  0.86] at (364.01, 15.70) {\(120\)};

\node[text=drawColor,anchor=base,inner sep=0pt, outer sep=0pt, scale=  0.86] at (415.90, 15.70) {\(180\)};
\end{scope}
\begin{scope}
\path[clip] (  0.00,  0.00) rectangle (426.79,192.06);
\definecolor{drawColor}{RGB}{0,0,0}

\node[text=drawColor,anchor=base,inner sep=0pt, outer sep=0pt, scale=  0.86] at (336.33,  2.40) {Время, мин};
\end{scope}
\begin{scope}
\path[clip] (  0.00,  0.00) rectangle (426.79,192.06);
\definecolor{drawColor}{RGB}{0,0,0}

\node[text=drawColor,rotate= 90.00,anchor=base,inner sep=0pt, outer sep=0pt, scale=  0.86] at (224.30,110.83) {Температура, {\textdegree}C\hspace*{1.7em}};
\end{scope}
\end{tikzpicture}

%% file: Dissertation/images_tikz/disser_tt_result.tikz
\begin{tikzpicture}[x=1pt,y=1pt]
\definecolor{fillColor}{RGB}{255,255,255}
\path[use as bounding box,fill=fillColor] (0,0) rectangle (426.79,213.40);
\begin{scope}
\path[clip] (  0.00,  0.00) rectangle (426.79,213.40);
\definecolor{drawColor}{RGB}{255,255,255}

\path[draw=drawColor,line width= 0.6pt,line join=round,line cap=round,fill=fillColor] (  0.00,  0.00) rectangle (426.79,213.40);
\end{scope}
\begin{scope}
\path[clip] ( 39.98, 31.05) rectangle (226.17,195.49);
\definecolor{fillColor}{RGB}{255,255,255}

\path[fill=fillColor] ( 39.98, 31.05) rectangle (226.17,195.49);
\definecolor{drawColor}{gray}{0.98}

\path[draw=drawColor,line width= 0.6pt,line join=round] ( 39.98, 56.57) --
	(226.17, 56.57);

\path[draw=drawColor,line width= 0.6pt,line join=round] ( 39.98, 92.66) --
	(226.17, 92.66);

\path[draw=drawColor,line width= 0.6pt,line join=round] ( 39.98,128.75) --
	(226.17,128.75);

\path[draw=drawColor,line width= 0.6pt,line join=round] ( 39.98,164.84) --
	(226.17,164.84);
\definecolor{drawColor}{gray}{0.80}

\path[draw=drawColor,line width= 0.3pt,line join=round] ( 39.98, 38.52) --
	(226.17, 38.52);

\path[draw=drawColor,line width= 0.3pt,line join=round] ( 39.98, 74.61) --
	(226.17, 74.61);

\path[draw=drawColor,line width= 0.3pt,line join=round] ( 39.98,110.70) --
	(226.17,110.70);

\path[draw=drawColor,line width= 0.3pt,line join=round] ( 39.98,146.80) --
	(226.17,146.80);

\path[draw=drawColor,line width= 0.3pt,line join=round] ( 39.98,182.89) --
	(226.17,182.89);

\path[draw=drawColor,line width= 0.3pt,line join=round] ( 90.76, 31.05) --
	( 90.76,195.49);

\path[draw=drawColor,line width= 0.3pt,line join=round] (175.39, 31.05) --
	(175.39,195.49);
\definecolor{fillColor}{RGB}{128,128,128}

\path[fill=fillColor] ( 52.67, 38.52) rectangle (128.84,134.68);
\definecolor{fillColor}{gray}{0.80}

\path[fill=fillColor] (137.30, 38.52) rectangle (213.47, 83.94);
\definecolor{drawColor}{RGB}{0,0,0}

\node[text=drawColor,anchor=base,inner sep=0pt, outer sep=0pt, scale=  1.00] at ( 90.76, 81.62) {100 \%};

\node[text=drawColor,anchor=base,inner sep=0pt, outer sep=0pt, scale=  1.00] at (175.39, 56.25) {47 \%};

\path[draw=drawColor,line width= 0.9pt,line join=round,line cap=round] ( 39.98, 31.05) rectangle (226.17,195.49);
\end{scope}
\begin{scope}
\path[clip] (232.17, 31.05) rectangle (418.36,195.49);
\definecolor{fillColor}{RGB}{255,255,255}

\path[fill=fillColor] (232.17, 31.05) rectangle (418.36,195.49);
\definecolor{drawColor}{gray}{0.98}

\path[draw=drawColor,line width= 0.6pt,line join=round] (232.17, 56.57) --
	(418.36, 56.57);

\path[draw=drawColor,line width= 0.6pt,line join=round] (232.17, 92.66) --
	(418.36, 92.66);

\path[draw=drawColor,line width= 0.6pt,line join=round] (232.17,128.75) --
	(418.36,128.75);

\path[draw=drawColor,line width= 0.6pt,line join=round] (232.17,164.84) --
	(418.36,164.84);
\definecolor{drawColor}{gray}{0.80}

\path[draw=drawColor,line width= 0.3pt,line join=round] (232.17, 38.52) --
	(418.36, 38.52);

\path[draw=drawColor,line width= 0.3pt,line join=round] (232.17, 74.61) --
	(418.36, 74.61);

\path[draw=drawColor,line width= 0.3pt,line join=round] (232.17,110.70) --
	(418.36,110.70);

\path[draw=drawColor,line width= 0.3pt,line join=round] (232.17,146.80) --
	(418.36,146.80);

\path[draw=drawColor,line width= 0.3pt,line join=round] (232.17,182.89) --
	(418.36,182.89);

\path[draw=drawColor,line width= 0.3pt,line join=round] (282.95, 31.05) --
	(282.95,195.49);

\path[draw=drawColor,line width= 0.3pt,line join=round] (367.58, 31.05) --
	(367.58,195.49);
\definecolor{fillColor}{RGB}{128,128,128}

\path[fill=fillColor] (244.86, 38.52) rectangle (321.03,188.02);
\definecolor{fillColor}{gray}{0.80}

\path[fill=fillColor] (329.50, 38.52) rectangle (405.66,104.99);
\definecolor{drawColor}{RGB}{0,0,0}

\node[text=drawColor,anchor=base,inner sep=0pt, outer sep=0pt, scale=  1.00] at (282.95,108.29) {100 \%};

\node[text=drawColor,anchor=base,inner sep=0pt, outer sep=0pt, scale=  1.00] at (367.58, 66.78) {44 \%};

\path[draw=drawColor,line width= 0.9pt,line join=round,line cap=round] (232.17, 31.05) rectangle (418.36,195.49);
\end{scope}
\begin{scope}
\path[clip] ( 39.98,195.49) rectangle (226.17,210.02);
\definecolor{drawColor}{gray}{0.40}
\definecolor{fillColor}{gray}{0.90}

\path[draw=drawColor,line width= 0.9pt,line join=round,line cap=round,fill=fillColor] ( 39.98,195.49) rectangle (226.17,210.02);
\definecolor{drawColor}{gray}{0.10}

\node[text=drawColor,anchor=base,inner sep=0pt, outer sep=0pt, scale=  0.86] at (133.07,198.49) {Выдержка 415 {\textdegree}C 60 мин};
\end{scope}
\begin{scope}
\path[clip] (232.17,195.49) rectangle (418.36,210.02);
\definecolor{drawColor}{gray}{0.40}
\definecolor{fillColor}{gray}{0.90}

\path[draw=drawColor,line width= 0.9pt,line join=round,line cap=round,fill=fillColor] (232.17,195.49) rectangle (418.36,210.02);
\definecolor{drawColor}{gray}{0.10}

\node[text=drawColor,anchor=base,inner sep=0pt, outer sep=0pt, scale=  0.86] at (325.26,198.49) {Выдержка 440 {\textdegree}C 15 мин};
\end{scope}
\begin{scope}
\path[clip] (  0.00,  0.00) rectangle (426.79,213.40);
\definecolor{drawColor}{RGB}{0,0,0}

\node[text=drawColor,anchor=base east,inner sep=0pt, outer sep=0pt, scale=  0.86] at ( 34.58, 34.26) {0};

\node[text=drawColor,anchor=base east,inner sep=0pt, outer sep=0pt, scale=  0.86] at ( 34.58, 70.35) {0,2};

\node[text=drawColor,anchor=base east,inner sep=0pt, outer sep=0pt, scale=  0.86] at ( 34.58,106.44) {0,4};

\node[text=drawColor,anchor=base east,inner sep=0pt, outer sep=0pt, scale=  0.86] at ( 34.58,142.53) {0,6};

\node[text=drawColor,anchor=base east,inner sep=0pt, outer sep=0pt, scale=  0.86] at ( 34.58,178.62) {0,8};
\end{scope}
\begin{scope}
\path[clip] (  0.00,  0.00) rectangle (426.79,213.40);
\definecolor{drawColor}{RGB}{0,0,0}

\path[draw=drawColor,line width= 0.6pt,line join=round] ( 36.98, 38.52) --
	( 39.98, 38.52);

\path[draw=drawColor,line width= 0.6pt,line join=round] ( 36.98, 74.61) --
	( 39.98, 74.61);

\path[draw=drawColor,line width= 0.6pt,line join=round] ( 36.98,110.70) --
	( 39.98,110.70);

\path[draw=drawColor,line width= 0.6pt,line join=round] ( 36.98,146.80) --
	( 39.98,146.80);

\path[draw=drawColor,line width= 0.6pt,line join=round] ( 36.98,182.89) --
	( 39.98,182.89);
\end{scope}
\begin{scope}
\path[clip] (  0.00,  0.00) rectangle (426.79,213.40);
\definecolor{drawColor}{RGB}{0,0,0}

\path[draw=drawColor,line width= 0.6pt,line join=round] ( 90.76, 28.05) --
	( 90.76, 31.05);

\path[draw=drawColor,line width= 0.6pt,line join=round] (175.39, 28.05) --
	(175.39, 31.05);
\end{scope}
\begin{scope}
\path[clip] (  0.00,  0.00) rectangle (426.79,213.40);
\definecolor{drawColor}{RGB}{0,0,0}

\node[text=drawColor,anchor=base,inner sep=0pt, outer sep=0pt, scale=  0.86] at ( 90.76, 17.12) {Нет};

\node[text=drawColor,anchor=base,inner sep=0pt, outer sep=0pt, scale=  0.86] at (175.39, 17.12) {Да};
\end{scope}
\begin{scope}
\path[clip] (  0.00,  0.00) rectangle (426.79,213.40);
\definecolor{drawColor}{RGB}{0,0,0}

\path[draw=drawColor,line width= 0.6pt,line join=round] (282.95, 28.05) --
	(282.95, 31.05);

\path[draw=drawColor,line width= 0.6pt,line join=round] (367.58, 28.05) --
	(367.58, 31.05);
\end{scope}
\begin{scope}
\path[clip] (  0.00,  0.00) rectangle (426.79,213.40);
\definecolor{drawColor}{RGB}{0,0,0}

\node[text=drawColor,anchor=base,inner sep=0pt, outer sep=0pt, scale=  0.86] at (282.95, 17.12) {Нет};

\node[text=drawColor,anchor=base,inner sep=0pt, outer sep=0pt, scale=  0.86] at (367.58, 17.12) {Да};
\end{scope}
\begin{scope}
\path[clip] (  0.00,  0.00) rectangle (426.79,213.40);
\definecolor{drawColor}{RGB}{0,0,0}

\node[text=drawColor,anchor=base,inner sep=0pt, outer sep=0pt, scale=  1.00] at (229.17,  2.40) {Термообработка};
\end{scope}
\begin{scope}
\path[clip] (  0.00,  0.00) rectangle (426.79,213.40);
\definecolor{drawColor}{RGB}{0,0,0}

\node[text=drawColor,rotate= 90.00,anchor=base,inner sep=0pt, outer sep=0pt, scale=  1.00] at ( 12.32,113.27) {Напряжение, МПа\hspace*{1.5em}};
\end{scope}
\end{tikzpicture}

%% file: Dissertation/images_tikz/disser_treat_g.tikz
\begin{tikzpicture}[x=1pt,y=1pt]
\definecolor{fillColor}{RGB}{255,255,255}
\path[use as bounding box,fill=fillColor] (0,0) rectangle (398.34,199.17);
\begin{scope}
\path[clip] (  0.00,  0.00) rectangle (398.34,199.17);
\definecolor{drawColor}{RGB}{255,255,255}

\path[draw=drawColor,line width= 0.6pt,line join=round,line cap=round,fill=fillColor] (  0.00,  0.00) rectangle (398.34,199.17);
\end{scope}
\begin{scope}
\path[clip] ( 39.23, 29.60) rectangle (394.12,199.17);
\definecolor{fillColor}{RGB}{255,255,255}

\path[fill=fillColor] ( 39.23, 29.60) rectangle (394.12,199.17);
\definecolor{drawColor}{gray}{0.98}

\path[draw=drawColor,line width= 0.6pt,line join=round] ( 39.23, 51.99) --
	(394.12, 51.99);

\path[draw=drawColor,line width= 0.6pt,line join=round] ( 39.23, 85.02) --
	(394.12, 85.02);

\path[draw=drawColor,line width= 0.6pt,line join=round] ( 39.23,121.72) --
	(394.12,121.72);

\path[draw=drawColor,line width= 0.6pt,line join=round] ( 39.23,158.43) --
	(394.12,158.43);

\path[draw=drawColor,line width= 0.6pt,line join=round] ( 39.23,184.12) --
	(394.12,184.12);

\path[draw=drawColor,line width= 0.6pt,line join=round] ( 75.52, 29.60) --
	( 75.52,199.17);

\path[draw=drawColor,line width= 0.6pt,line join=round] (115.85, 29.60) --
	(115.85,199.17);

\path[draw=drawColor,line width= 0.6pt,line join=round] (156.18, 29.60) --
	(156.18,199.17);

\path[draw=drawColor,line width= 0.6pt,line join=round] (196.51, 29.60) --
	(196.51,199.17);

\path[draw=drawColor,line width= 0.6pt,line join=round] (236.84, 29.60) --
	(236.84,199.17);

\path[draw=drawColor,line width= 0.6pt,line join=round] (277.17, 29.60) --
	(277.17,199.17);

\path[draw=drawColor,line width= 0.6pt,line join=round] (317.50, 29.60) --
	(317.50,199.17);

\path[draw=drawColor,line width= 0.6pt,line join=round] (357.83, 29.60) --
	(357.83,199.17);
\definecolor{drawColor}{gray}{0.80}

\path[draw=drawColor,line width= 0.3pt,line join=round] ( 39.23, 37.31) --
	(394.12, 37.31);

\path[draw=drawColor,line width= 0.3pt,line join=round] ( 39.23, 66.67) --
	(394.12, 66.67);

\path[draw=drawColor,line width= 0.3pt,line join=round] ( 39.23,103.37) --
	(394.12,103.37);

\path[draw=drawColor,line width= 0.3pt,line join=round] ( 39.23,140.08) --
	(394.12,140.08);

\path[draw=drawColor,line width= 0.3pt,line join=round] ( 39.23,176.78) --
	(394.12,176.78);

\path[draw=drawColor,line width= 0.3pt,line join=round] ( 39.23,191.46) --
	(394.12,191.46);

\path[draw=drawColor,line width= 0.3pt,line join=round] ( 55.36, 29.60) --
	( 55.36,199.17);

\path[draw=drawColor,line width= 0.3pt,line join=round] ( 95.69, 29.60) --
	( 95.69,199.17);

\path[draw=drawColor,line width= 0.3pt,line join=round] (136.02, 29.60) --
	(136.02,199.17);

\path[draw=drawColor,line width= 0.3pt,line join=round] (176.35, 29.60) --
	(176.35,199.17);

\path[draw=drawColor,line width= 0.3pt,line join=round] (216.67, 29.60) --
	(216.67,199.17);

\path[draw=drawColor,line width= 0.3pt,line join=round] (257.00, 29.60) --
	(257.00,199.17);

\path[draw=drawColor,line width= 0.3pt,line join=round] (297.33, 29.60) --
	(297.33,199.17);

\path[draw=drawColor,line width= 0.3pt,line join=round] (337.66, 29.60) --
	(337.66,199.17);

\path[draw=drawColor,line width= 0.3pt,line join=round] (377.99, 29.60) --
	(377.99,199.17);
\definecolor{drawColor}{RGB}{0,0,0}

\path[draw=drawColor,line width= 1.7pt,dash pattern=on 7pt off 3pt ,line join=round] ( 55.36, 37.31) --
	( 56.96, 49.58) --
	( 58.57, 61.85) --
	( 60.17, 74.12) --
	( 61.78, 86.39) --
	( 63.38, 98.66) --
	( 64.99,110.93) --
	( 66.59,123.20) --
	( 68.20,135.48) --
	( 69.80,147.75) --
	( 71.41,160.02) --
	( 73.02,172.29) --
	( 74.62,184.56) --
	( 76.23,191.46) --
	( 77.83,191.46) --
	( 79.44,191.46) --
	( 81.04,191.46) --
	( 82.65,191.46) --
	( 84.25,191.46) --
	( 85.86,191.46) --
	( 87.46,191.46) --
	( 89.07,191.46) --
	( 90.67,191.46) --
	( 92.28,191.46) --
	( 93.88,191.46) --
	( 95.49,191.46) --
	( 97.09,191.46) --
	( 98.70,191.46) --
	(100.30,191.46) --
	(101.91,191.46) --
	(103.51,191.46) --
	(105.12,191.46) --
	(106.72,191.46) --
	(108.33,191.46) --
	(109.93,191.46) --
	(111.54,191.46) --
	(113.14,191.46) --
	(114.75,191.46) --
	(116.35,191.46) --
	(117.96,191.46) --
	(119.56,191.46) --
	(121.17,191.46) --
	(122.77,191.46) --
	(124.38,191.46) --
	(125.98,191.46) --
	(127.59,191.46) --
	(129.19,191.46) --
	(130.80,191.46) --
	(132.41,191.46) --
	(134.01,191.46) --
	(135.62,191.46) --
	(137.22,191.46) --
	(138.83,191.46) --
	(140.43,191.46) --
	(142.04,191.46) --
	(143.64,191.46) --
	(145.25,191.46) --
	(146.85,191.46) --
	(148.46,191.46) --
	(150.06,191.46) --
	(151.67,191.46) --
	(153.27,191.46) --
	(154.88,191.46) --
	(156.48,190.64) --
	(158.09,186.29) --
	(159.69,182.04) --
	(161.30,177.89) --
	(162.90,173.82) --
	(164.51,169.85) --
	(166.11,165.96) --
	(167.72,162.16) --
	(169.32,158.45) --
	(170.93,154.82) --
	(172.53,151.27) --
	(174.14,147.81) --
	(175.74,144.43) --
	(177.35,141.12) --
	(178.95,137.90) --
	(180.56,134.75) --
	(182.16,131.68) --
	(183.77,128.68) --
	(185.37,125.76) --
	(186.98,122.91) --
	(188.59,120.13) --
	(190.19,117.42) --
	(191.80,114.78) --
	(193.40,112.20) --
	(195.01,109.69) --
	(196.61,107.25) --
	(198.22,104.87) --
	(199.82,102.55) --
	(201.43,100.30) --
	(203.03, 98.11) --
	(204.64, 95.97) --
	(206.24, 93.90) --
	(207.85, 91.88) --
	(209.45, 89.92) --
	(211.06, 88.01) --
	(212.66, 86.16) --
	(214.27, 84.36) --
	(215.87, 82.61) --
	(217.48, 80.91) --
	(219.08, 79.27) --
	(220.69, 77.67) --
	(222.29, 76.12) --
	(223.90, 74.62) --
	(225.50, 73.17) --
	(227.11, 71.76) --
	(228.71, 70.39) --
	(230.32, 69.07) --
	(231.92, 67.79) --
	(233.53, 66.55) --
	(235.13, 65.35) --
	(236.74, 64.19) --
	(238.34, 63.07) --
	(239.95, 61.99) --
	(241.55, 60.95) --
	(243.16, 59.94) --
	(244.76, 58.97) --
	(246.37, 58.03) --
	(247.98, 57.12) --
	(249.58, 56.25) --
	(251.19, 55.41) --
	(252.79, 54.60) --
	(254.40, 53.83) --
	(256.00, 53.08) --
	(257.61, 52.36) --
	(259.21, 51.67) --
	(260.82, 51.00) --
	(262.42, 50.37) --
	(264.03, 49.76) --
	(265.63, 49.17) --
	(267.24, 48.61) --
	(268.84, 48.07) --
	(270.45, 47.56) --
	(272.05, 47.07) --
	(273.66, 46.60) --
	(275.26, 46.15) --
	(276.87, 45.72) --
	(278.47, 45.31) --
	(280.08, 44.93) --
	(281.68, 44.56) --
	(283.29, 44.20) --
	(284.89, 43.87) --
	(286.50, 43.55) --
	(288.10, 43.25) --
	(289.71, 42.96) --
	(291.31, 42.69) --
	(292.92, 42.44) --
	(294.52, 42.20) --
	(296.13, 41.97) --
	(297.73, 41.75) --
	(299.34, 41.55) --
	(300.94, 41.36) --
	(302.55, 41.18) --
	(304.15, 41.01) --
	(305.76, 40.86) --
	(307.37, 40.71) --
	(308.97, 40.57) --
	(310.58, 40.44) --
	(312.18, 40.32) --
	(313.79, 40.21) --
	(315.39, 40.11) --
	(317.00, 40.01) --
	(318.60, 39.93) --
	(320.21, 39.84) --
	(321.81, 39.77) --
	(323.42, 39.70) --
	(325.02, 39.64) --
	(326.63, 39.58) --
	(328.23, 39.53) --
	(329.84, 39.48) --
	(331.44, 39.44) --
	(333.05, 39.40) --
	(334.65, 39.36) --
	(336.26, 39.33) --
	(337.86, 39.31) --
	(339.47, 39.28) --
	(341.07, 39.26) --
	(342.68, 39.24) --
	(344.28, 39.22) --
	(345.89, 39.21) --
	(347.49, 39.20) --
	(349.10, 39.19) --
	(350.70, 39.18) --
	(352.31, 39.17) --
	(353.91, 39.16) --
	(355.52, 39.16) --
	(357.12, 39.15) --
	(358.73, 39.15) --
	(360.33, 39.15) --
	(361.94, 39.15) --
	(363.54, 39.14) --
	(365.15, 39.14) --
	(366.76, 39.14) --
	(368.36, 39.14) --
	(369.97, 39.14) --
	(371.57, 39.14) --
	(373.18, 39.14) --
	(374.78, 39.14) --
	(376.39, 39.14) --
	(377.99, 39.14);

\path[draw=drawColor,line width= 1.7pt,line join=round] ( 55.36, 37.31) --
	( 56.96, 49.58) --
	( 58.57, 61.85) --
	( 60.17, 74.12) --
	( 61.78, 86.39) --
	( 63.38, 98.66) --
	( 64.99,110.93) --
	( 66.59,123.20) --
	( 68.20,135.48) --
	( 69.80,147.75) --
	( 71.41,160.02) --
	( 73.02,172.29) --
	( 74.62,184.56) --
	( 76.23,191.46) --
	( 77.83,191.46) --
	( 79.44,191.46) --
	( 81.04,191.46) --
	( 82.65,191.46) --
	( 84.25,191.46) --
	( 85.86,191.46) --
	( 87.46,191.46) --
	( 89.07,191.46) --
	( 90.67,191.46) --
	( 92.28,191.46) --
	( 93.88,191.46) --
	( 95.49,191.46) --
	( 97.09,191.46) --
	( 98.70,191.46) --
	(100.30,191.46) --
	(101.91,191.46) --
	(103.51,191.46) --
	(105.12,191.46) --
	(106.72,191.46) --
	(108.33,191.46) --
	(109.93,191.46) --
	(111.54,191.46) --
	(113.14,191.46) --
	(114.75,191.46) --
	(116.35,191.46) --
	(117.96,191.46) --
	(119.56,191.46) --
	(121.17,191.46) --
	(122.77,191.46) --
	(124.38,191.46) --
	(125.98,191.46) --
	(127.59,191.46) --
	(129.19,191.46) --
	(130.80,191.46) --
	(132.41,191.46) --
	(134.01,191.46) --
	(135.62,191.46) --
	(137.22,191.46) --
	(138.83,191.46) --
	(140.43,191.46) --
	(142.04,191.46) --
	(143.64,191.46) --
	(145.25,191.46) --
	(146.85,191.46) --
	(148.46,191.46) --
	(150.06,191.46) --
	(151.67,191.46) --
	(153.27,191.46) --
	(154.88,191.46) --
	(156.48,191.30) --
	(158.09,190.42) --
	(159.69,189.54) --
	(161.30,188.67) --
	(162.90,187.79) --
	(164.51,186.91) --
	(166.11,186.04) --
	(167.72,185.16) --
	(169.32,184.29) --
	(170.93,183.41) --
	(172.53,182.53) --
	(174.14,181.66) --
	(175.74,180.78) --
	(177.35,179.90) --
	(178.95,179.03) --
	(180.56,178.15) --
	(182.16,177.27) --
	(183.77,176.40) --
	(185.37,175.52) --
	(186.98,174.64) --
	(188.59,173.77) --
	(190.19,172.89) --
	(191.80,172.01) --
	(193.40,171.14) --
	(195.01,170.26) --
	(196.61,169.38) --
	(198.22,168.51) --
	(199.82,167.63) --
	(201.43,166.76) --
	(203.03,165.88) --
	(204.64,165.00) --
	(206.24,164.13) --
	(207.85,163.25) --
	(209.45,162.37) --
	(211.06,161.50) --
	(212.66,160.62) --
	(214.27,159.74) --
	(215.87,158.87) --
	(217.48,157.99) --
	(219.08,157.11) --
	(220.69,156.24) --
	(222.29,155.36) --
	(223.90,154.48) --
	(225.50,153.61) --
	(227.11,152.73) --
	(228.71,151.85) --
	(230.32,150.98) --
	(231.92,150.10) --
	(233.53,149.23) --
	(235.13,148.35) --
	(236.74,147.47) --
	(238.34,146.60) --
	(239.95,145.72) --
	(241.55,144.84) --
	(243.16,143.97) --
	(244.76,143.09) --
	(246.37,142.21) --
	(247.98,141.34) --
	(249.58,140.46) --
	(251.19,139.58) --
	(252.79,138.71) --
	(254.40,137.83) --
	(256.00,136.95) --
	(257.61,136.08) --
	(259.21,135.20) --
	(260.82,134.32) --
	(262.42,133.45) --
	(264.03,132.57) --
	(265.63,131.70) --
	(267.24,130.82) --
	(268.84,129.94) --
	(270.45,129.07) --
	(272.05,128.19) --
	(273.66,127.31) --
	(275.26,126.44) --
	(276.87,125.56) --
	(278.47,124.68) --
	(280.08,123.81) --
	(281.68,122.93) --
	(283.29,122.05) --
	(284.89,121.18) --
	(286.50,120.30) --
	(288.10,119.42) --
	(289.71,118.55) --
	(291.31,117.67) --
	(292.92,116.79) --
	(294.52,115.92) --
	(296.13,115.04) --
	(297.73,114.17) --
	(299.34,113.29) --
	(300.94,112.41) --
	(302.55,111.54) --
	(304.15,110.66) --
	(305.76,109.78) --
	(307.37,108.91) --
	(308.97,108.03) --
	(310.58,107.15) --
	(312.18,106.28) --
	(313.79,105.40) --
	(315.39,104.52) --
	(317.00,103.65) --
	(318.60,100.61) --
	(320.21, 96.74) --
	(321.81, 93.07) --
	(323.42, 89.57) --
	(325.02, 86.24) --
	(326.63, 83.08) --
	(328.23, 80.09) --
	(329.84, 77.24) --
	(331.44, 74.55) --
	(333.05, 72.01) --
	(334.65, 69.60) --
	(336.26, 67.33) --
	(337.86, 65.19) --
	(339.47, 63.17) --
	(341.07, 61.28) --
	(342.68, 59.49) --
	(344.28, 57.82) --
	(345.89, 56.25) --
	(347.49, 54.78) --
	(349.10, 53.41) --
	(350.70, 52.14) --
	(352.31, 50.94) --
	(353.91, 49.84) --
	(355.52, 48.81) --
	(357.12, 47.86) --
	(358.73, 46.98) --
	(360.33, 46.17) --
	(361.94, 45.42) --
	(363.54, 44.74) --
	(365.15, 44.11) --
	(366.76, 43.54) --
	(368.36, 43.01) --
	(369.97, 42.54) --
	(371.57, 42.11) --
	(373.18, 41.72) --
	(374.78, 41.38) --
	(376.39, 41.06) --
	(377.99, 40.79);

\path[draw=drawColor,line width= 0.9pt,line join=round,line cap=round] ( 39.23, 29.60) rectangle (394.12,199.17);
\end{scope}
\begin{scope}
\path[clip] (  0.00,  0.00) rectangle (398.34,199.17);
\definecolor{drawColor}{RGB}{0,0,0}

\node[text=drawColor,anchor=base east,inner sep=0pt, outer sep=0pt, scale=  0.86] at ( 33.83, 33.06) {\(20\)};

\node[text=drawColor,anchor=base east,inner sep=0pt, outer sep=0pt, scale=  0.86] at ( 33.83, 62.42) {\(100\)};

\node[text=drawColor,anchor=base east,inner sep=0pt, outer sep=0pt, scale=  0.86] at ( 33.83, 99.12) {\(200\)};

\node[text=drawColor,anchor=base east,inner sep=0pt, outer sep=0pt, scale=  0.86] at ( 33.83,135.83) {\(300\)};

\node[text=drawColor,anchor=base east,inner sep=0pt, outer sep=0pt, scale=  0.86] at ( 33.83,172.53) {\(400\)};

\node[text=drawColor,anchor=base east,inner sep=0pt, outer sep=0pt, scale=  0.86] at ( 33.83,187.21) {\(440\)};
\end{scope}
\begin{scope}
\path[clip] (  0.00,  0.00) rectangle (398.34,199.17);
\definecolor{drawColor}{RGB}{0,0,0}

\path[draw=drawColor,line width= 0.6pt,line join=round] ( 36.23, 37.31) --
	( 39.23, 37.31);

\path[draw=drawColor,line width= 0.6pt,line join=round] ( 36.23, 66.67) --
	( 39.23, 66.67);

\path[draw=drawColor,line width= 0.6pt,line join=round] ( 36.23,103.37) --
	( 39.23,103.37);

\path[draw=drawColor,line width= 0.6pt,line join=round] ( 36.23,140.08) --
	( 39.23,140.08);

\path[draw=drawColor,line width= 0.6pt,line join=round] ( 36.23,176.78) --
	( 39.23,176.78);

\path[draw=drawColor,line width= 0.6pt,line join=round] ( 36.23,191.46) --
	( 39.23,191.46);
\end{scope}
\begin{scope}
\path[clip] (  0.00,  0.00) rectangle (398.34,199.17);
\definecolor{drawColor}{RGB}{0,0,0}

\path[draw=drawColor,line width= 0.6pt,line join=round] ( 55.36, 26.60) --
	( 55.36, 29.60);

\path[draw=drawColor,line width= 0.6pt,line join=round] ( 95.69, 26.60) --
	( 95.69, 29.60);

\path[draw=drawColor,line width= 0.6pt,line join=round] (136.02, 26.60) --
	(136.02, 29.60);

\path[draw=drawColor,line width= 0.6pt,line join=round] (176.35, 26.60) --
	(176.35, 29.60);

\path[draw=drawColor,line width= 0.6pt,line join=round] (216.67, 26.60) --
	(216.67, 29.60);

\path[draw=drawColor,line width= 0.6pt,line join=round] (257.00, 26.60) --
	(257.00, 29.60);

\path[draw=drawColor,line width= 0.6pt,line join=round] (297.33, 26.60) --
	(297.33, 29.60);

\path[draw=drawColor,line width= 0.6pt,line join=round] (337.66, 26.60) --
	(337.66, 29.60);

\path[draw=drawColor,line width= 0.6pt,line join=round] (377.99, 26.60) --
	(377.99, 29.60);
\end{scope}
\begin{scope}
\path[clip] (  0.00,  0.00) rectangle (398.34,199.17);
\definecolor{drawColor}{RGB}{0,0,0}

\node[text=drawColor,anchor=base,inner sep=0pt, outer sep=0pt, scale=  0.86] at ( 55.36, 15.70) {\(0\)};

\node[text=drawColor,anchor=base,inner sep=0pt, outer sep=0pt, scale=  0.86] at ( 95.69, 15.70) {\(30\)};

\node[text=drawColor,anchor=base,inner sep=0pt, outer sep=0pt, scale=  0.86] at (136.02, 15.70) {\(60\)};

\node[text=drawColor,anchor=base,inner sep=0pt, outer sep=0pt, scale=  0.86] at (176.35, 15.70) {\(90\)};

\node[text=drawColor,anchor=base,inner sep=0pt, outer sep=0pt, scale=  0.86] at (216.67, 15.70) {\(120\)};

\node[text=drawColor,anchor=base,inner sep=0pt, outer sep=0pt, scale=  0.86] at (257.00, 15.70) {\(150\)};

\node[text=drawColor,anchor=base,inner sep=0pt, outer sep=0pt, scale=  0.86] at (297.33, 15.70) {\(180\)};

\node[text=drawColor,anchor=base,inner sep=0pt, outer sep=0pt, scale=  0.86] at (337.66, 15.70) {\(210\)};

\node[text=drawColor,anchor=base,inner sep=0pt, outer sep=0pt, scale=  0.86] at (377.99, 15.70) {\(240\)};
\end{scope}
\begin{scope}
\path[clip] (  0.00,  0.00) rectangle (398.34,199.17);
\definecolor{drawColor}{RGB}{0,0,0}

\node[text=drawColor,anchor=base,inner sep=0pt, outer sep=0pt, scale=  0.86] at (216.67,  2.40) {Время, мин};
\end{scope}
\begin{scope}
\path[clip] (  0.00,  0.00) rectangle (398.34,199.17);
\definecolor{drawColor}{RGB}{0,0,0}

\node[text=drawColor,rotate= 90.00,anchor=base,inner sep=0pt, outer sep=0pt, scale=  0.86] at ( 10.90,114.38) {Температура, {\textdegree}C};
\end{scope}
\begin{scope}
\path[clip] (  0.00,  0.00) rectangle (398.34,199.17);
\definecolor{drawColor}{RGB}{0,0,0}
\definecolor{fillColor}{RGB}{255,255,255}

\path[draw=drawColor,line width= 0.6pt,line join=round,line cap=round,fill=fillColor] ( 67.62, 36.38) rectangle (208, 85.46);
\end{scope}
\begin{scope}
\path[clip] (  0.00,  0.00) rectangle (398.34,199.17);
\definecolor{drawColor}{RGB}{0,0,0}

\node[text=drawColor,anchor=base west,inner sep=0pt, outer sep=0pt, scale=  0.86] at ( 71.89, 72.69) {Охлаждение:};
\end{scope}
\begin{scope}
\path[clip] (  0.00,  0.00) rectangle (398.34,199.17);
\definecolor{drawColor}{RGB}{255,255,255}
\definecolor{fillColor}{RGB}{255,255,255}

\path[draw=drawColor,line width= 0.6pt,line join=round,line cap=round,fill=fillColor] ( 71.89, 54.14) rectangle ( 98.87, 67.63);
\end{scope}
\begin{scope}
\path[clip] (  0.00,  0.00) rectangle (398.34,199.17);
\definecolor{drawColor}{RGB}{0,0,0}

\path[draw=drawColor,line width= 1.7pt,line join=round] ( 74.58, 60.89) -- ( 96.17, 60.89);
\end{scope}
\begin{scope}
\path[clip] (  0.00,  0.00) rectangle (398.34,199.17);
\definecolor{drawColor}{RGB}{0,0,0}

\path[draw=drawColor,line width= 1.7pt,line join=round] ( 74.58, 60.89) -- ( 96.17, 60.89);
\end{scope}
\begin{scope}
\path[clip] (  0.00,  0.00) rectangle (398.34,199.17);
\definecolor{drawColor}{RGB}{255,255,255}
\definecolor{fillColor}{RGB}{255,255,255}

\path[draw=drawColor,line width= 0.6pt,line join=round,line cap=round,fill=fillColor] ( 71.89, 40.65) rectangle ( 98.87, 54.14);
\end{scope}
\begin{scope}
\path[clip] (  0.00,  0.00) rectangle (398.34,199.17);
\definecolor{drawColor}{RGB}{0,0,0}

\path[draw=drawColor,line width= 1.7pt,dash pattern=on 7pt off 3pt ,line join=round] ( 74.58, 47.40) -- ( 96.17, 47.40);
\end{scope}
\begin{scope}
\path[clip] (  0.00,  0.00) rectangle (398.34,199.17);
\definecolor{drawColor}{RGB}{0,0,0}

\path[draw=drawColor,line width= 1.7pt,dash pattern=on 7pt off 3pt ,line join=round] ( 74.58, 47.40) -- ( 96.17, 47.40);
\end{scope}
\begin{scope}
\path[clip] (  0.00,  0.00) rectangle (398.34,199.17);
\definecolor{drawColor}{RGB}{0,0,0}

\node[text=drawColor,anchor=base west,inner sep=0pt, outer sep=0pt, scale=  0.86] at (101.40, 56.64) {контролируемое};
\end{scope}
\begin{scope}
\path[clip] (  0.00,  0.00) rectangle (398.34,199.17);
\definecolor{drawColor}{RGB}{0,0,0}

\node[text=drawColor,anchor=base west,inner sep=0pt, outer sep=0pt, scale=  0.86] at (101.40, 43.15) {неконтролируемое\hspace*{1.7em}};
\end{scope}
\end{tikzpicture}

%% file: Dissertation/images_tikz/disser_tt_graph2.tikz
\begin{tikzpicture}[x=1pt,y=1pt]
\definecolor{fillColor}{RGB}{255,255,255}
\path[use as bounding box,fill=fillColor] (0,0) rectangle (398.34,159.34);
\begin{scope}
\path[clip] (  0.00,  0.00) rectangle (398.34,159.34);
\definecolor{drawColor}{RGB}{255,255,255}

\path[draw=drawColor,line width= 0.6pt,line join=round,line cap=round,fill=fillColor] ( -0.00,  0.00) rectangle (398.34,159.34);
\end{scope}
\begin{scope}
\path[clip] ( 38.51, 29.60) rectangle (394.12,155.96);
\definecolor{fillColor}{RGB}{255,255,255}

\path[fill=fillColor] ( 38.51, 29.60) rectangle (394.12,155.96);
\definecolor{drawColor}{gray}{0.98}

\path[draw=drawColor,line width= 0.6pt,line join=round] ( 38.51, 53.58) --
	(394.12, 53.58);

\path[draw=drawColor,line width= 0.6pt,line join=round] ( 38.51, 90.04) --
	(394.12, 90.04);

\path[draw=drawColor,line width= 0.6pt,line join=round] ( 38.51,126.51) --
	(394.12,126.51);
\definecolor{drawColor}{gray}{0.80}

\path[draw=drawColor,line width= 0.3pt,line join=round] ( 38.51, 35.34) --
	(394.12, 35.34);

\path[draw=drawColor,line width= 0.3pt,line join=round] ( 38.51, 71.81) --
	(394.12, 71.81);

\path[draw=drawColor,line width= 0.3pt,line join=round] ( 38.51,108.28) --
	(394.12,108.28);

\path[draw=drawColor,line width= 0.3pt,line join=round] ( 38.51,144.74) --
	(394.12,144.74);

\path[draw=drawColor,line width= 0.3pt,line join=round] (135.49, 29.60) --
	(135.49,155.96);

\path[draw=drawColor,line width= 0.3pt,line join=round] (297.14, 29.60) --
	(297.14,155.96);
\definecolor{fillColor}{RGB}{128,128,128}

\path[fill=fillColor] ( 62.76, 35.34) rectangle (208.23,150.22);
\definecolor{fillColor}{gray}{0.80}

\path[fill=fillColor] (224.40, 35.34) rectangle (369.88,107.76);
\definecolor{drawColor}{RGB}{0,0,0}

\node[text=drawColor,anchor=base,inner sep=0pt, outer sep=0pt, scale=  1.00] at (135.49, 87.80) {100 \%};

\node[text=drawColor,anchor=base,inner sep=0pt, outer sep=0pt, scale=  1.00] at (297.14, 66.57) {63 \%};

\path[draw=drawColor,line width= 0.9pt,line join=round,line cap=round] ( 38.51, 29.60) rectangle (394.12,155.96);
\end{scope}
\begin{scope}
\path[clip] (  0.00,  0.00) rectangle (398.34,159.34);
\definecolor{drawColor}{RGB}{0,0,0}

\node[text=drawColor,anchor=base east,inner sep=0pt, outer sep=0pt, scale=  0.86] at ( 33.11, 31.09) {0};

\node[text=drawColor,anchor=base east,inner sep=0pt, outer sep=0pt, scale=  0.86] at ( 33.11, 67.56) {0,2};

\node[text=drawColor,anchor=base east,inner sep=0pt, outer sep=0pt, scale=  0.86] at ( 33.11,104.03) {0,4};

\node[text=drawColor,anchor=base east,inner sep=0pt, outer sep=0pt, scale=  0.86] at ( 33.11,140.49) {0,6};
\end{scope}
\begin{scope}
\path[clip] (  0.00,  0.00) rectangle (398.34,159.34);
\definecolor{drawColor}{RGB}{0,0,0}

\path[draw=drawColor,line width= 0.6pt,line join=round] ( 35.51, 35.34) --
	( 38.51, 35.34);

\path[draw=drawColor,line width= 0.6pt,line join=round] ( 35.51, 71.81) --
	( 38.51, 71.81);

\path[draw=drawColor,line width= 0.6pt,line join=round] ( 35.51,108.28) --
	( 38.51,108.28);

\path[draw=drawColor,line width= 0.6pt,line join=round] ( 35.51,144.74) --
	( 38.51,144.74);
\end{scope}
\begin{scope}
\path[clip] (  0.00,  0.00) rectangle (398.34,159.34);
\definecolor{drawColor}{RGB}{0,0,0}

\path[draw=drawColor,line width= 0.6pt,line join=round] (135.49, 26.60) --
	(135.49, 29.60);

\path[draw=drawColor,line width= 0.6pt,line join=round] (297.14, 26.60) --
	(297.14, 29.60);
\end{scope}
\begin{scope}
\path[clip] (  0.00,  0.00) rectangle (398.34,159.34);
\definecolor{drawColor}{RGB}{0,0,0}

\node[text=drawColor,anchor=base,inner sep=0pt, outer sep=0pt, scale=  0.86] at (135.49, 15.70) {Неконтролируемое};

\node[text=drawColor,anchor=base,inner sep=0pt, outer sep=0pt, scale=  0.86] at (297.14, 15.70) {Контролируемое};
\end{scope}
\begin{scope}
\path[clip] (  0.00,  0.00) rectangle (398.34,159.34);
\definecolor{drawColor}{RGB}{0,0,0}

\node[text=drawColor,anchor=base,inner sep=0pt, outer sep=0pt, scale=  0.86] at (216.32,  2.40) {Охлаждение};
\end{scope}
\begin{scope}
\path[clip] (  0.00,  0.00) rectangle (398.34,159.34);
\definecolor{drawColor}{RGB}{0,0,0}

\node[text=drawColor,rotate= 90.00,anchor=base,inner sep=0pt, outer sep=0pt, scale=  0.86] at ( 10.90, 92.78) {Напряжение, МПа\hspace*{1.5em}};
\end{scope}
\end{tikzpicture}

%% file: Dissertation/conclusion.tex
\chapter*{Общие выводы и заключение}						% Заголовок
\addcontentsline{toc}{chapter}{Общие выводы и заключение}	% Добавляем его в оглавление

%% Согласно ГОСТ Р 7.0.11-2011:
%% 5.3.3 В заключении диссертации излагают итоги выполненного исследования, рекомендации, перспективы дальнейшей разработки темы.
%% 9.2.3 В заключении автореферата диссертации излагают итоги данного исследования, рекомендации и перспективы дальнейшей разработки темы.
%% Поэтому имеет смысл сделать эту часть общей и загрузить из одного файла в автореферат и в диссертацию:

\input{common/concl}

%% file: common/concl.tex
\input{common/vyvods}
\beforevyvods

\aftervyvods

\section*{Общие выводы по работе}

\begin{enumerate}[labelindent=!, leftmargin=\parindent]
\vyvodsall
\end{enumerate}

%% file: Dissertation/acronyms.tex
\chapter*{Список сокращений}             % Заголовок
\addcontentsline{toc}{chapter}{Список сокращений}  % Добавляем его в оглавление
\noindent%
\begin{tabu} to \textwidth {l X}

\textbf{КМОП} & комплементарная структура металл-оксид-полупроводник.\\

\textbf{КР} & комбинационное рассеяние.\\

\textbf{МЭМС} & микроэлектромеханическая система.\\

\textbf{ПИД-регулятор} & пропорционально-интегрально-дифференцирующий регулятор.\\
\textbf{РС} & Рамановская спектроскопия.\\
\textbf{РФЭС} & рентгеновская фотоэлектронная спектроскопия.\\

\textbf{ТКЛР} & температурный коэффициент линейного расширения.\\
\textbf{MSE} & среднеквадратичная ошибка (mean square error).\\

\textbf{RSS} & сумма квадратов регрессионных остатков (residual sum of~squares).\\

\textbf{SER} & стандартная ошибка регрессии (standard error of the regression).\\

\end{tabu}

%% file: Dissertation/references.tex
\clearpage                                  % В том числе гарантирует, что список литературы в оглавлении будет с правильным номером страницы
\urlstyle{rm}                               % ссылки URL обычным шрифтом
\ifdefmacro{\microtypesetup}{\microtypesetup{protrusion=false}}{} % не рекомендуется применять пакет микротипографики к автоматически генерируемому списку литературы
\insertbibliofull%                           % Подключаем Bib-базы
\ifdefmacro{\microtypesetup}{\microtypesetup{protrusion=true}}{}
\urlstyle{tt}                               % возвращаем установки шрифта ссылок URL